\documentclass[11pt]{article}
\usepackage{makeidx}
\usepackage{amsfonts}
\usepackage{amssymb}
\usepackage{graphicx}
\usepackage{amsmath}
\usepackage{geometry}
\usepackage{appendix}

\let\dprod\prod
\DeclareMathSizes{10.95}{10}{7}{6}
\input{tcilatex}
\begin{document}

\title{Statistical Field Theory and Neural Structures Dynamics II: Signals
Propagation, Interferences, Bound States.}
\author{Pierre Gosselin\thanks{%
Pierre Gosselin: Institut Fourier, UMR 5582 CNRS-UGA, Universit\'{e}
Grenoble Alpes, BP 74, 38402 Saint Martin d'H\`{e}res, France\ E-Mail:
Pierre.Gosselin@univ-grenoble-alpes.fr} \and A\"{\i}leen Lotz\thanks{%
A\"{\i}leen Lotz: Cerca Trova, BP 114, 38001 Grenoble Cedex 1, France.\
E-mail: a.lotz@cercatrova.eu}}
\maketitle

\begin{abstract}
We continue our study of a field formalism for large sets of interacting
neurons, together with their connectivity functions. Expanding upon the
foundation laid in (\cite{GLr}), we formulate an effective formalism for the
connectivity field in the presence of external sources. We proceed to deduce
the propagation of external signals within the system. This enables us to
investigate the activation and association of groups of bound cells.
\end{abstract}

\section{Introduction}

In this series of papers, we develop a field-theoretic approach to study the
dynamics of connectivities in a system of interacting spiking neurons. To
achieve this, in (\cite{GLr}), we established a two-field model that
describes both the dynamics of neural activity and the connectivity between
points in the network. This field theory is the outcome of a two-step
process and is based on a method originally developed in (\cite{Kl}) and
subsequently adapted for complex interacting systems in \cite{GL1}\cite{GL2}%
\cite{GL3}, and \cite{GL4}. In the first step, we extend the standard
formalism of dynamic equations for a large assembly of interacting neurons,
as outlined in (\cite{V11}), to include a dynamic system accounting for the
evolving nature of neural connectivity. We employ the formalism for
connectivity functions presented in (\cite{IFR}), which is rewritten in a
format suitable for translation into field theory. In the second step, we
transform this two sets of dynamic equations into a second-quantized
Euclidean field theory, as detailed in (see \cite{GL1}\cite{GL2}\cite{GL3}
for the method). The action functional of this field theory depends on two
fields.The first field, analogous to the one introduced in (\cite{GL}),
characterizes the assembly of neurons, while the second field delineates the
dynamics of connectivity between cells. Both fields are subject to self-
interactions, depicting interactions across the network, and also interact
mutually with one another, encapsulating the interdependencies between
neural activities and connectivities. This field-based description
encompasses both collective and individual aspects of the system. The system
with these two fields is delineated by a field action functional that
records the interactions at the microscopic level. This action functional
comprehensively encapsulates the dynamics of the entire system.

This field-theoretic framework enables us to derive the system's effective
action, as well as the corresponding background field, namely, the minimum
of the effective action. This background field characterizes the collective
state of the system. The field framework allows us to compute firing rates,
i.e., neural activity, at each point in the system in a given background
state. Additionally, we can derive the propagation of perturbations in
neural activity from one point to another. In a prior work (\cite{GL}), we
demonstrated the existence of persistent nonlinear traveling waves along the
network by considering the field action for neurons alone. In (\cite{GLr}),
where the field for connectivity functions is included in the system, our
description enables the derivation of both background fields for neural
interactions and connectivities, which minimize the action functional. These
background fields represent the collective configurations of the system and
dictate the potential static equilibria for neural activities and
connectivities. These equilibria serve as the structural foundation of the
system, governing fluctuations and the propagation of signals within it.
They depend on internal parameters of the system and external stimuli. We
showed the existence of several possible background states and their
corresponding connectivities, the thread being mainly organized into groups
of interconnected points.

Assuming that the timescale of connectivities is slower than that of
individual cells, we have demonstrated how repeated activations at certain
points can propagate along the thread, gradually altering the connectivity
functions. Foroscillatory perturbations, the oscillatory response may
exhibit interference phenomena. At points of constructive interference, both
the background state for connectivities and average connectivities undergo
modifications. These long-term modifications manifest as emerging states
characterized by enhanced connectivities between specific points. These
states are reflective of external activations and can be regarded as records
of these activations. They are slowly fading over time but can be
reactivated by external perturbations. Furthermore, the association of such
emerging states arises if their activation occurs at similar times. The
resultant state is a combination of two states, describable as a
modification of the initial background state at several points. Activating
one of the two states may reactivate their combination. Therefore,
regardless of the cause of their activation, these enhanced connectivity
states exhibit the characteristics of interacting partial neuronal
assemblies.

Nonetheless, these results were derived solely by working with the
connectivities field. We made use of the findings from (\cite{GL}) and did
not establish our results based on interactions between the neuronal field
and the connectivity field. The objective of th present work is to
incorporate the interactions between these two fields and ultimately derive
an effective action for the connectivities. Any modification in terms of
cell activity resulting from external signals or cell interactions will then
be then inherently encompassed within the effective action for the
connectivity field.

This effective formalism enables us to contemplate the dynamics of the
connectivity system as alterations of the connectivity field induced by
external perturbations. The outcomes from (\cite{GLr}) are thus recovered as
transitions between initial and final states of this field. The outcomes
from (\cite{GLr}), such as the emergence of combined structures and the
reactivation of one structure by another, thus occur within a coherent field
description of the connectivity system.

This paper is organised as follows: In Part I, we provide an overview of the
model and results from (\cite{GLr}), Sections 2 and 3 revisit the individual
dynamics of interacting neurons and the field-theoretic formulation of the
model, respectively. In Section 4, we review the characteristics of the
background states and qualitatively discuss the influence of external
perturbations on these states.

To develop an effective field theory for the connectivity field, Part II
integrates the degrees of freedom of the neuronal field in the presence of
external sources out. Section 5 details the modifications to the neuronal
field's path integral induced by the presence of sources. Section 6 computes
the saddle-path neural activity in the presence of sources and derives the
effects of interferences on this activity. Section 7 deduces the impact of
interferences on the emergence of bound states.

In Part III, building on the previous sections, we derive an effective field
theory for the connectivity field. Section 8 outlines the effective theory
and the associated Green functions for the bound states. These Green
functions are then utilized to investigate modifications in the connectivity
background state. In Section 9, we present various applications, including
activations, associations, and reactivations of structures as externally
induced transitions of the background states. We uncover the effects of
interferences as consequences of these transitions. Section 10 is for the
conclusion.

\part*{I. Model, static background fields and external perturbations}

We present the dynamical model for cells activity and connectivities between
a large set of cells. We recall the field translation of this model. We
present the result of prtI: the static background fields along with the
associated equilibrium connectivities and activities, and the perturbations
associated wth external sources. Details are given in prtI.

\section{A dynamical system of interacting cells.}

Following \cite{GL1}\cite{GL2}\cite{GL3}, we describe a system of a large
number of neurons ($N>>1$). We define their individual equations and the
dynamics for the connectivity functions.

\subsection{Individual dynamics}

We follow the description of \cite{V11} for coupled quadratic
integrate-and-fire (QIF) neurons, but use the additional hypothesis that
each neuron is characterized by its position in some spatial range.

Each neuron's potential $X_{i}\left( t\right) $ satisfies the differential
equation:%
\begin{equation}
\dot{X}_{i}\left( t\right) =\gamma X_{i}^{2}\left( t\right) +J_{i}\left(
t\right)  \label{ptn}
\end{equation}%
for $X_{i}\left( t\right) <X_{p}$, where $X_{p}$ denotes the potential level
of a spike. When $X=X_{p}$, the potential is reset to its resting value $%
X_{i}\left( t\right) =X_{r}<X_{p}$. For the sake of simplicity, following (%
\cite{V11}) we have chosen the squared form $\gamma X_{i}^{2}\left( t\right) 
$ in (\ref{ptn}). However any form $f\left( X_{i}\left( t\right) \right) $\
could be used. The current of signals reaching cell $i$ at time $t$ is
written $J_{i}\left( t\right) $.

Our purpose is to find the system dynamics in terms of the spikes'
frequencies. First, we consider the time for the $n$-th spike of cell $i$, $%
\theta _{n}^{\left( i\right) }$. This is written as a function of $n$, $%
\theta ^{\left( i\right) }\left( n\right) $. Then, a continuous
approximation $n\rightarrow t$ allows to write the spike time variable as $%
\theta ^{\left( i\right) }\left( t\right) $. We thus have replaced:

\begin{equation*}
\theta _{n}^{\left( i\right) }\rightarrow \theta ^{\left( i\right) }\left(
n\right) \rightarrow \theta ^{\left( i\right) }\left( t\right)
\end{equation*}%
The continuous approximation could be removed, but is convenient and
simplifies the notations and computations. We assume now that the timespans
between two spikes are relatively small. The time between two spikes for
cell $i$ is obtained by writing (\ref{ptn}) as:%
\begin{equation*}
\frac{dX_{i}\left( t\right) }{dt}=\gamma X_{i}^{2}\left( t\right)
+J_{i}\left( t\right)
\end{equation*}%
and by inverting this relation to write:%
\begin{equation*}
dt=\frac{dX_{i}}{\gamma X_{i}^{2}+J^{\left( i\right) }\left( \theta ^{\left(
i\right) }\left( n-1\right) \right) }
\end{equation*}%
Integrating the potential between two spikes thus yields:%
\begin{equation*}
\theta ^{\left( i\right) }\left( n\right) -\theta ^{\left( i\right) }\left(
n-1\right) \simeq \int_{X_{r}}^{X_{p}}\frac{dX}{\gamma X^{2}+J^{\left(
i\right) }\left( \theta ^{\left( i\right) }\left( n-1\right) \right) }
\end{equation*}%
Replacing $J^{\left( i\right) }\left( \theta ^{\left( i\right) }\left(
n-1\right) \right) $ by its average value during the small time period $%
\theta ^{\left( i\right) }\left( n\right) -\theta ^{\left( i\right) }\left(
n-1\right) $, we can consider $J^{\left( i\right) }\left( \theta ^{\left(
i\right) }\left( n-1\right) \right) $ as constant in first approximation,
and we find:%
\begin{equation}
\theta ^{\left( i\right) }\left( n\right) -\theta ^{\left( i\right) }\left(
n-1\right) \equiv G\left( \theta ^{\left( i\right) }\left( n-1\right)
\right) =\frac{\arctan \left( \frac{\left( \frac{1}{X_{r}}-\frac{1}{X_{p}}%
\right) \sqrt{\frac{J^{\left( i\right) }\left( \theta ^{\left( i\right)
}\left( n-1\right) \right) }{\gamma }}}{1+\frac{J^{\left( n\right) }\left(
\theta ^{\left( n-1\right) }\right) }{\gamma X_{r}X_{p}}}\right) }{\sqrt{%
\gamma J^{\left( i\right) }\left( \theta ^{\left( i\right) }\left(
n-1\right) \right) }}  \label{spt}
\end{equation}%
The activity or firing rate at $t$, $\omega _{i}\left( t\right) $, is
defined by the inverse time span (\ref{spt}) between two spikes:%
\begin{equation}
\omega _{i}\left( t\right) =\frac{1}{G\left( \theta ^{\left( i\right)
}\left( n-1\right) \right) }\equiv F\left( \theta ^{\left( i\right) }\left(
n-1\right) \right)  \label{MGF}
\end{equation}%
Since we consider small time intervals between two spikes, we can write:%
\begin{equation}
\theta ^{\left( i\right) }\left( n\right) -\theta ^{\left( i\right) }\left(
n-1\right) \simeq \frac{d}{dt}\theta ^{\left( i\right) }\left( t\right)
-\omega _{i}^{-1}\left( t\right) =\varepsilon _{i}\left( t\right)
\label{dnm}
\end{equation}%
where the white noise perturbation $\varepsilon _{i}\left( t\right) $ for
each period was added to account for any internal uncertainty in the time
span $\theta ^{\left( i\right) }\left( n\right) -\theta ^{\left( i\right)
}\left( n-1\right) $. This white noise is independent from the instantaneous
inverse activity $\omega _{i}^{-1}\left( t\right) $. We assume these $%
\varepsilon _{i}\left( t\right) $ to have variance $\sigma ^{2}$, so that
equation (\ref{dnm}) writes: 
\begin{equation}
\frac{d}{dt}\theta ^{\left( i\right) }\left( t\right) -G\left( \theta
^{\left( i\right) }\left( t\right) ,J^{\left( i\right) }\left( \theta
^{\left( i\right) }\left( t\right) \right) \right) =\varepsilon _{i}\left(
t\right)  \label{dnq}
\end{equation}%
The $\omega _{i}\left( t\right) $ are computed by considering the overall
current which, using the discrete time notation, is given by:%
\begin{eqnarray}
\hat{J}^{\left( i\right) }\left( \left( n-1\right) \right) &=&J^{\left(
i\right) }\left( \left( n-1\right) \right)  \label{crt} \\
&&+\frac{\kappa }{N}\sum_{j,m}\frac{\omega _{j}\left( m\right) }{\omega
_{i}\left( n-1\right) }\delta \left( \theta ^{\left( i\right) }\left(
n-1\right) -\theta ^{\left( j\right) }\left( m\right) -\frac{\left\vert
Z_{i}-Z_{j}\right\vert }{c}\right) T_{ij}\left( \left( n-1,Z_{i}\right)
,\left( m,Z_{j}\right) \right)  \notag
\end{eqnarray}%
The quantity $J^{\left( i\right) }\left( \left( n-1\right) \right) $ denotes
an external current. The term inside the sum is the average current sent to $%
i$ by neuron $j$ during the short time span $\theta ^{\left( i\right)
}\left( n\right) -\theta ^{\left( i\right) }\left( n-1\right) $. The
function $T_{ij}\left( \left( n-1,Z_{i}\right) ,\left( m,Z_{j}\right)
\right) $ is the connectivity (or transfer) function between cells $j$ and $%
i $. It measures the level of connectivity between $i$ and $j$.

In this paper, the connectivity function is a dynamical object whose dynamic
equations are described in the next paragraph. We will work in the
continuous approximation, so that formula (\ref{crt}) is replaced by:%
\begin{equation}
\hat{J}^{\left( i\right) }\left( t\right) =J^{\left( i\right) }\left(
t\right) +\frac{\kappa }{N}\int \sum_{j}\frac{\omega _{j}\left( s\right) }{%
\omega _{i}\left( t\right) }\delta \left( \theta ^{\left( i\right) }\left(
t\right) -\theta ^{\left( j\right) }\left( s\right) -\frac{\left\vert
Z_{i}-Z_{j}\right\vert }{c}\right) T_{ij}\left( \left( t,Z_{i}\right)
,\left( s,Z_{j}\right) \right) ds  \label{crT}
\end{equation}

Formula (\ref{crT}) shows that the dynamic equation (\ref{dnm}) has to be
coupled with the neurons activities equation:%
\begin{equation}
\omega _{i}\left( t\right) =G\left( \theta ^{\left( i\right) }\left(
t\right) ,\hat{J}\left( \theta ^{\left( i\right) }\left( t\right) \right)
\right) +\upsilon _{i}\left( t\right)  \label{cstrt}
\end{equation}%
and $J^{\left( i\right) }\left( t\right) $ is defined by (\ref{crT}). A
white noise $\upsilon _{i}\left( t\right) $ accounts for the possible
deviations from this relation, due to some internal or external causes for
each cell. We assume that the variances of $\upsilon _{i}\left( t\right) $
are constant, and equal to $\eta ^{2}$, such that $\eta ^{2}<<\sigma ^{2}$.

\subsection{Connectivity functions dynamics}

We describe the dynamics for the connectivity functions $T_{ij}\left( \left(
n-1,Z_{i}\right) ,\left( m,Z_{j}\right) \right) $ between cells. To do so we
adapt the description of (\cite{IFR}) to our context. In this work, the
connectivity functions depend on some intermediate variables and do not
present any space index. The connectivity between neurons $i$ and $j$
satisifies a differential equation:%
\begin{equation}
\frac{dT_{ij}}{dt}=-\frac{T_{ij}\left( t\right) }{\tau }+\lambda \hat{T}%
_{ij}\left( t\right) \sum_{l}\delta \left( t-\Delta t_{ij}-t_{j}^{l}\right)
\label{spn}
\end{equation}%
where $\hat{T}_{ij}\left( t\right) $ represents the variation in
connectivity, due to the synaptic interactions between the two neurons. The
delay $\Delta t_{ij}$ is the time of arrival at neuton $i$ for a spike of
neuron $j$. The time $t_{j}^{l}$ accounts for time of neuron $j$ spikes. The
sum:%
\begin{equation*}
\sum_{l}\delta \left( t-\Delta t_{ij}-t_{j}^{l}\right)
\end{equation*}%
counts the number of spikes emitted by neuron $j$ and arriving at time $t$
at neuron $i$.

The variation in connectivity satisfies itself an equation:%
\begin{equation}
\frac{d\hat{T}_{ij}}{dt}=\rho \left( 1-\hat{T}_{ij}\left( t\right) \right)
C_{ij}\left( t\right) \sum_{k}\delta \left( t-t_{i}^{k}\right) -\hat{T}%
_{ij}\left( t\right) D_{i}\left( t\right) \sum_{l}\delta \left( t-\Delta
t_{ij}-t_{j}^{l}\right)  \label{SPt}
\end{equation}%
where $C_{ij}\left( t\right) $ and $D_{i}\left( t\right) $\ measure the
cumulated postsynaptic and presynaptic activity. The sum:%
\begin{equation*}
\sum_{k}\delta \left( t-t_{i}^{k}\right)
\end{equation*}%
counts the number of spikes emitted at time $t$. Quantities $C_{ij}\left(
t\right) $ and $D_{i}\left( t\right) $ follow the dynamics:%
\begin{equation}
\frac{dC_{ij}}{dt}=-\frac{C_{ij}\left( t\right) }{\tau _{C}}+\alpha
_{C}\left( 1-C_{ij}\left( t\right) \right) \sum_{l}\delta \left( t-\Delta
t_{ij}-t_{j}^{l}\right)  \label{spr}
\end{equation}%
\begin{equation}
\frac{dD_{i}}{dt}=-\frac{D_{i}\left( t\right) }{\tau _{D}}+\alpha _{C}\left(
1-D_{i}\left( t\right) \right) \sum_{k}\delta \left( t-t_{i}^{k}\right)
\label{spf}
\end{equation}

To translate these equations in our set up, we have to consider connectivity
functions of the form:%
\begin{equation*}
T_{ij}\left( \left( n_{i},Z_{i}\right) ,\left( n_{j},Z_{j}\right) \right)
\end{equation*}%
that include the positions of neurons $i$ and $j$ and the parameter $n_{i}$
and $n_{j}$ which are our counting variables of neurons spikes. However,
equations (\ref{spn}), (\ref{SPt}), (\ref{spr}), (\ref{spf}) include a time
variable.

In our formalism, the time variable $\theta ^{\left( i\right) }\left(
n_{i}\right) $ is the time at which neuron $i$ produces its $n_{i}$-th
spike. We should write classical equations depending on these variables.

Moreover, the number of spikes $\sum_{l}\delta \left( t-\Delta
t_{ij}-t_{j}^{l}\right) $ emitted by cell $j$ at time $t_{j}^{l}$ and the
number of spikes $\sum_{k}\delta \left( t-t_{i}^{k}\right) $ emitted by cell 
$i$ at time $t$ are proportional to $\delta \left( \theta ^{\left( j\right)
}\left( n_{j}\right) -\left( t-\Delta t_{ij}\right) \right) \omega
_{j}\left( n_{j}\right) $ and $\delta \left( \theta ^{\left( i\right)
}\left( n_{i}\right) -t\right) \omega _{j}\left( n_{i}\right) $
respectively. Given the introduction of a spatial indices, we have the
relation:%
\begin{equation*}
\Delta t_{ij}=\frac{\left\vert Z_{i}-Z_{j}\right\vert }{c}
\end{equation*}%
and the first $\delta $ function writes:%
\begin{equation*}
\delta \left( \theta ^{\left( j\right) }\left( n_{j}\right) -\left( t-\Delta
t_{ij}\right) \right) =\delta \left( \theta ^{\left( j\right) }\left(
n_{j}\right) -\left( \theta ^{\left( i\right) }\left( n_{i}\right) -\frac{%
\left\vert Z_{i}-Z_{j}\right\vert }{c}\right) \right) \delta \left( \theta
^{\left( i\right) }\left( n_{i}\right) -t\right)
\end{equation*}

As a consequence, we will write first the connectivity functions from $i$ to 
$j$ as:%
\begin{equation*}
T\left( \left( Z_{i},\theta ^{\left( i\right) }\left( n_{i}\right) ,\omega
_{i}\left( n_{i}\right) \right) ,\left( Z_{j},\theta ^{\left( j\right)
}\left( n_{j}\right) ,\omega _{j}\left( n_{j}\right) \right) \right)
\end{equation*}%
This function, together with the variation in connectivity:%
\begin{equation*}
\hat{T}\left( \left( Z_{i},\theta ^{\left( i\right) }\left( n_{i}\right)
,\omega _{i}\left( n_{i}\right) \right) ,\left( Z_{j},\theta ^{\left(
j\right) }\left( n_{j}\right) ,\omega _{j}\left( n_{j}\right) \right) \right)
\end{equation*}%
along with the auxiliary variables:%
\begin{equation*}
C\left( \left( Z_{i},\theta ^{\left( i\right) }\left( n_{i}\right) ,\omega
_{i}\left( n_{i}\right) \right) ,\left( Z_{j},\theta ^{\left( j\right)
}\left( n_{j}\right) ,\omega _{j}\left( n_{j}\right) \right) \right)
\end{equation*}%
and:%
\begin{equation*}
D\left( \left( Z_{i},\theta ^{\left( i\right) }\left( n_{i}\right) ,\omega
_{i}\left( n_{i}\right) \right) \right)
\end{equation*}%
satisfy the following translations of equations (\ref{spn}), (\ref{SPt}), (%
\ref{spr}), (\ref{spf}): 
\begin{eqnarray}
&&\nabla _{\theta ^{\left( i\right) }\left( n_{i}\right) }T\left( \left(
Z_{i},\theta ^{\left( i\right) }\left( n_{i}\right) ,\omega _{i}\left(
n_{i}\right) \right) ,\left( Z_{j},\theta ^{\left( j\right) }\left(
n_{j}\right) ,\omega _{j}\left( n_{j}\right) \right) \right) \\
&=&-\frac{1}{\tau }T\left( \left( Z_{i},\theta ^{\left( i\right) }\left(
n_{i}\right) ,\omega _{i}\left( n_{i}\right) \right) ,\left( Z_{j},\theta
^{\left( j\right) }\left( n_{j}\right) ,\omega _{j}\left( n_{j}\right)
\right) \right)  \notag \\
&&+\lambda \left( \hat{T}\left( \left( Z_{i},\theta ^{\left( i\right)
}\left( n_{i}\right) ,\omega _{i}\left( n_{i}\right) \right) ,\left(
Z_{j},\theta ^{\left( j\right) }\left( n_{j}\right) ,\omega _{j}\left(
n_{j}\right) \right) \right) \right) \delta \left( \theta ^{\left( i\right)
}\left( n_{i}\right) -\theta ^{\left( j\right) }\left( n_{j}\right) -\frac{%
\left\vert Z_{i}-Z_{j}\right\vert }{c}\right)  \notag
\end{eqnarray}%
where $\hat{T}$ measures the variation of $T$ due to the signals send from $%
j $ to $i$ and the signals emitted by $i$. It satisfies the following
equation: 
\begin{eqnarray}
&&\nabla _{\theta ^{\left( i\right) }\left( n_{i}\right) }\hat{T}\left(
\left( Z_{i},\theta ^{\left( i\right) }\left( n_{i}\right) ,\omega
_{i}\left( n_{i}\right) \right) ,\left( Z_{j},\theta ^{\left( j\right)
}\left( n_{j}\right) ,\omega _{j}\left( n_{j}\right) \right) \right)
\label{TH} \\
&=&\rho \delta \left( \theta ^{\left( i\right) }\left( n_{i}\right) -\theta
^{\left( j\right) }\left( n_{j}\right) -\frac{\left\vert
Z_{i}-Z_{j}\right\vert }{c}\right)  \notag \\
&&\times \left\{ \left( h\left( Z,Z_{1}\right) -\hat{T}\left( \left(
Z_{i},\theta ^{\left( i\right) }\left( n_{i}\right) ,\omega _{i}\left(
n_{i}\right) \right) ,\left( Z_{j},\theta ^{\left( j\right) }\left(
n_{j}\right) ,\omega _{j}\left( n_{j}\right) \right) \right) \right) C\left(
\theta ^{\left( i\right) }\left( n\right) \right) h_{C}\left( \omega
_{i}\left( n_{i}\right) \right) \right.  \notag \\
&&\left. -D\left( \theta ^{\left( i\right) }\left( n\right) \right) \hat{T}%
\left( \left( Z_{i},\theta ^{\left( i\right) }\left( n_{i}\right) ,\omega
_{i}\left( n_{i}\right) \right) ,\left( Z_{j},\theta ^{\left( j\right)
}\left( n_{j}\right) ,\omega _{j}\left( n_{j}\right) \right) \right)
h_{D}\left( \omega _{j}\left( n_{j}\right) \right) \right\}  \notag
\end{eqnarray}%
where $h_{C}$ and $h_{D}$\ are increasing functions. In the set of equations
(\ref{spn}), (\ref{SPt}), (\ref{spr}), (\ref{spf}): 
\begin{eqnarray*}
h_{C}\left( \omega _{i}\left( n_{i}\right) \right) &=&\omega _{i}\left(
n_{i}\right) \\
h_{D}\left( \omega _{j}\left( n_{j}\right) \right) &=&\omega _{j}\left(
n_{j}\right)
\end{eqnarray*}

We depart slightly from (\cite{IFR}) by the introduction of the function $%
h\left( Z,Z_{1}\right) $ (they choose $h\left( Z,Z_{1}\right) =1$), to
implement some loss due to the distance. We may choose for example:%
\begin{equation*}
h\left( Z,Z_{1}\right) =\exp \left( -\frac{\left\vert Z_{i}-Z_{j}\right\vert 
}{\nu c}\right)
\end{equation*}%
where $\nu $ is a parameter. Equation (\ref{TH}) involves two dynamic
factors $C\left( \theta ^{\left( i\right) }\left( n-1\right) \right) $ and $%
D\left( \theta _{i}\left( n-1\right) \right) $. The factor $C\left( \theta
^{\left( i\right) }\left( n-1\right) \right) $ describes the accumulation of
input spikes. It is solution of the differential equation:%
\begin{eqnarray}
\nabla _{\theta ^{\left( i\right) }\left( n-1\right) }C\left( \theta
^{\left( i\right) }\left( n-1\right) \right) &=&-\frac{C\left( \theta
^{\left( i\right) }\left( n-1\right) \right) }{\tau _{C}} \\
&&+\alpha _{C}\left( 1-C\left( \theta ^{\left( i\right) }\left( n-1\right)
\right) \right) \omega _{j}\left( \theta ^{\left( i\right) }\left(
n-1\right) -\frac{\left\vert Z_{i}-Z_{j}\right\vert }{c}\right)  \notag
\end{eqnarray}%
The term $D\left( \theta _{i}\left( n-1\right) \right) $ is proportional to
the accumulation of output spikes and is solution of:%
\begin{equation}
\nabla _{\theta ^{\left( i\right) }\left( n-1\right) }D\left( \theta
^{\left( i\right) }\left( n-1\right) \right) =-\frac{D\left( \theta ^{\left(
i\right) }\left( n-1\right) \right) }{\tau _{D}}+\alpha _{D}\left( 1-D\left(
\theta ^{\left( i\right) }\left( n-1\right) \right) \right) \omega
_{i}\left( n_{i}\right)
\end{equation}%
For the purpose of field translation, we have to change the variables in the
derivatives by the counting variable $n_{i}$

and replace $\nabla _{\theta ^{\left( i\right) }\left( n_{i}\right) }\simeq
\omega _{i}\left( n_{i}\right) \nabla _{n_{i}}$ in the previous dynamics
equations. We thus rewrite the dynamic equations in the following form:

For the connectivity $T$:

\begin{eqnarray}
&&\nabla _{n_{i}}T\left( \left( Z_{i},\theta ^{\left( i\right) }\left(
n_{i}\right) ,\omega _{i}\left( n_{i}\right) \right) ,\left( Z_{j},\theta
^{\left( j\right) }\left( n_{j}\right) ,\omega _{j}\left( n_{j}\right)
\right) \right)  \label{nqp} \\
&=&-\frac{1}{\tau \omega _{i}\left( n_{i}\right) }T\left( \left(
Z_{i},\theta ^{\left( i\right) }\left( n_{i}\right) ,\omega _{i}\left(
n_{i}\right) \right) ,\left( Z_{j},\theta ^{\left( j\right) }\left(
n_{j}\right) ,\omega _{j}\left( n_{j}\right) \right) \right)  \notag \\
&&+\frac{\lambda }{\omega _{i}\left( n_{i}\right) }\left( \hat{T}\left(
\left( Z_{i},\theta ^{\left( i\right) }\left( n_{i}\right) ,\omega
_{i}\left( n_{i}\right) \right) ,\left( Z_{j},\theta ^{\left( j\right)
}\left( n_{j}\right) ,\omega _{j}\left( n_{j}\right) \right) \right) \right)
\delta \left( \theta ^{\left( i\right) }\left( n_{i}\right) -\theta ^{\left(
j\right) }\left( n_{j}\right) -\frac{\left\vert Z_{i}-Z_{j}\right\vert }{c}%
\right)  \notag
\end{eqnarray}%
For the variation in connectivity $\hat{T}$:%
\begin{eqnarray}
&&\nabla _{n_{i}}\hat{T}\left( \left( Z_{i},\theta ^{\left( i\right) }\left(
n_{i}\right) ,\omega _{i}\left( n_{i}\right) \right) ,\left( Z_{j},\theta
^{\left( j\right) }\left( n_{j}\right) ,\omega _{j}\left( n_{j}\right)
\right) \right)  \label{nqd} \\
&=&\frac{\rho }{\omega _{i}\left( n_{i}\right) }\delta \left( \theta
^{\left( i\right) }\left( n_{i}\right) -\theta ^{\left( j\right) }\left(
n_{j}\right) -\frac{\left\vert Z_{i}-Z_{j}\right\vert }{c}\right)  \notag \\
&&\times \left\{ \left( h\left( Z,Z_{1}\right) -\hat{T}\left( \left(
Z_{i},\theta ^{\left( i\right) }\left( n_{i}\right) ,\omega _{i}\left(
n_{i}\right) \right) ,\left( Z_{j},\theta ^{\left( j\right) }\left(
n_{j}\right) ,\omega _{j}\left( n_{j}\right) \right) \right) \right) C\left(
\theta ^{\left( i\right) }\left( n-1\right) \right) h_{C}\left( \omega
_{i}\left( n_{i}\right) \right) \right.  \notag \\
&&\left. -D\left( \theta ^{\left( i\right) }\left( n-1\right) \right) \hat{T}%
\left( \left( Z_{i},\theta ^{\left( i\right) }\left( n_{i}\right) ,\omega
_{i}\left( n_{i}\right) \right) ,\left( Z_{j},\theta ^{\left( j\right)
}\left( n_{j}\right) ,\omega _{j}\left( n_{j}\right) \right) \right)
h_{D}\left( \omega _{j}\left( n_{j}\right) \right) \right\}  \notag
\end{eqnarray}%
and for the auxiliary variables $C$ and $D$:%
\begin{eqnarray}
\nabla _{n_{i}}C\left( \theta ^{\left( i\right) }\left( n-1\right) \right)
&=&-\frac{C\left( \theta ^{\left( i\right) }\left( n-1\right) \right) }{\tau
_{C}\omega _{i}\left( n_{i}\right) }  \label{nqt} \\
&&+\alpha _{C}\left( 1-C\left( \theta ^{\left( i\right) }\left( n-1\right)
\right) \right) \frac{\omega _{j}\left( Z_{j},\theta ^{\left( i\right)
}\left( n-1\right) -\frac{\left\vert Z_{i}-Z_{j}\right\vert }{c}\right) }{%
\omega _{i}\left( n_{i}\right) }  \notag
\end{eqnarray}%
and:%
\begin{equation}
\nabla _{n_{i}}D\left( \theta ^{\left( i\right) }\left( n-1\right) \right) =-%
\frac{D\left( \theta ^{\left( i\right) }\left( n-1\right) \right) }{\tau
_{D}\omega _{i}\left( n_{i}\right) }+\alpha _{D}\left( 1-D\left( \theta
^{\left( i\right) }\left( n-1\right) \right) \right)  \label{nqQ}
\end{equation}

Then, to describe the connectivity by a field, we have to describe the
connectivity as a set of vectors depending of a set of double indices $kl$
(replacing $ij$) and interacting with the neurons activities, seen as
independent variables indexed by $i,j...$

We thus describe connectivity by a set of matrices:%
\begin{equation*}
\left( T_{kl}\left( n_{kl}\right) ,\hat{T}_{kl}\left( n_{kl}\right) ,\left(
Z_{kl}\left( n_{kl}\right) =\left( Z_{k,},Z_{l}\right) \right) ,\theta
^{\left( kl\right) }\left( n_{kl}\right) ,\omega _{k}\left( n_{kl}\right)
,\omega _{l}^{\prime }\left( n_{kl}\right) ,C_{kl}\left( n_{kl}\right)
,D_{k}\left( n_{kl}\right) \right)
\end{equation*}%
where $n_{kl}$ is an internal parameter given by the average counting
variable for cells or synapses firing simultaneously at point $Z_{k,}$.

Then, we replace the description (\ref{nqp}), (\ref{nqd}), (\ref{nqt}), (\ref%
{nqQ}) by a set of equation in which connectivity $T_{kl}\left(
n_{kl}\right) $ interact with all pairs of neurons at points $Z_{k,}$ and $%
Z_{l}$ whose average activities at time $\theta ^{\left( kl\right) }\left(
n_{kl}\right) $ and $\theta ^{\left( kl\right) }\left( n_{kl}\right) -\frac{%
\left\vert Z_{k}-Z_{l}\right\vert }{c}$ are given by $\omega _{k}\left(
n_{kl}\right) ,\omega _{l}^{\prime }\left( n_{kl}\right) $ respectively. As
a consequence, we replace the notion of connectivity $T_{ij}\left( \left(
n-1,Z_{i}\right) ,\left( m,Z_{j}\right) \right) $ between two specific
neurons $i$ and $j$ by the average connectivity between the two sets of
neurons with identical activities at each extremity of the segment $\left(
Z_{i},Z_{j}\right) $ \ This approximation justifies if we consider that
neurons located at the same place and firing at the same rate can be
considered as closely connected and in average identical.

Stated mathematically, variable is an average $n_{kl}=\bar{n}_{i}$ at a
given time $\theta ^{\left( kl\right) }$ and we assume that in average,
connectivity variable $T_{kl}\left( n_{kl}\right) $ interacts with all
neurons pairs located at $\left( Z_{k,},Z_{l}\right) $ at times $\theta
^{\left( i\right) }\left( n_{i}\right) =\theta ^{\left( kl\right) }\left(
n_{kl}\right) $. Writing $\bar{\omega}\left( Z_{i},n_{i}\right) $ for the
average activity, we impose $\bar{\omega}\left( Z_{i},n_{i}\right) =\omega
_{k}\left( n_{kl}\right) $ and $\bar{\omega}\left( Z_{j},n_{j}\right)
=\omega _{l}^{\prime }\left( n_{kl}\right) $ and $\theta ^{\left( j\right)
}\left( n_{j}\right) =\theta ^{\left( kl\right) }\left( n_{kl}\right) -\frac{%
\left\vert Z_{k}-Z_{l}\right\vert }{c}$ respectively. The densities $%
T_{kl}\left( n_{kl}\right) $ are thus the set of all connections between
points $Z_{k,}$ and $Z_{l}$ between sets of synchronized neurons at $Z_{k}$
and synchronized neurons at $Z_{l}$, i.e. between set of neurons or synapses
at this points. In this point of view, we replace $\nabla _{\theta ^{\left(
i\right) }\left( n_{i}\right) }\simeq \omega _{i}\left( n_{i}\right) \nabla
_{n_{i}}$ by:%
\begin{equation*}
\nabla _{\theta ^{\left( kl\right) }\left( n_{kl}\right) }\simeq \frac{%
\partial n_{kl}}{\partial \theta ^{\left( kl\right) }\left( n_{kl}\right) }%
\nabla _{n_{kl}}=\bar{\omega}\left( Z_{i},n_{i}\right) \nabla _{n_{kl}}
\end{equation*}%
As a consequence, the dynamic equations (\ref{nqp}), (\ref{nqd}), (\ref{nqt}%
), (\ref{nqQ}) are replaced by:%
\begin{eqnarray}
&\nabla _{n_{kl}}T_{kl}\left( n_{kl}\right) =&\left( -\sum_{i,n_{i}}\frac{1}{%
\tau \bar{\omega}\left( Z_{i},n_{i}\right) }T_{kl}\left( n_{kl}\right) +%
\frac{\lambda }{\bar{\omega}\left( Z_{i},n_{i}\right) }\hat{T}_{kl}\left(
n_{kl}\right) \right)  \label{wgd} \\
&&\times \delta \left( \theta ^{\left( i\right) }\left( n_{i}\right) -\theta
^{\left( kl\right) }\left( n_{kl}\right) \right) \delta \left(
Z_{k}-Z_{i}\right) \delta \left( \omega _{k}\left( n_{kl}\right) -\bar{\omega%
}\left( Z_{i},n_{i}\right) \right)  \notag
\end{eqnarray}%
\begin{eqnarray}
&&\nabla _{n_{kl}}\hat{T}\left( n_{kl}\right)  \label{wgt} \\
&=&\left( \sum_{i,n_{i}}\left( h\left( Z_{k},Z_{l}\right) -\hat{T}\left(
n_{kl}\right) \right) C_{kl}\left( n_{kl}\right) h_{C}\left( \omega
_{i}\left( n_{i}\right) \right) -\sum_{j,n_{j}}D_{k}\left( n_{kl}\right) 
\hat{T}\left( n_{kl}\right) h_{D}\left( \omega _{j}\left( n_{j}\right)
\right) \right)  \notag \\
&&\times \frac{\rho }{\bar{\omega}\left( Z_{i},n_{i}\right) }\delta \left(
\theta ^{\left( i\right) }\left( n_{i}\right) -\theta ^{\left( j\right)
}\left( n_{j}\right) -\frac{\left\vert Z_{i}-Z_{j}\right\vert }{c}\right)
\delta \left( \theta ^{\left( i\right) }\left( n_{i}\right) -\theta ^{\left(
kl\right) }\left( n_{kl}\right) \right)  \notag \\
&&\times \delta \left( \left( Z_{k,},Z_{l}\right) -\left(
Z_{i,},Z_{j}\right) \right) \delta \left( \omega _{k}\left( n_{kl}\right) -%
\bar{\omega}\left( Z_{i},n_{i}\right) \right) \delta \left( \omega
_{l}\left( n_{kl}\right) -\bar{\omega}\left( Z_{j},n_{j}\right) \right) 
\notag
\end{eqnarray}%
\begin{eqnarray}
\nabla _{n_{kl}}C\left( n_{kl}\right) &=&\left( -\frac{C\left( n_{kl}\right) 
}{\tau _{C}\bar{\omega}\left( Z_{i},n_{i}\right) }+\sum_{j,n_{j}}\alpha
_{C}\left( 1-C_{kl}\left( n_{kl}\right) \right) \frac{\omega _{j}\left(
n_{j}\right) }{\bar{\omega}\left( Z_{i},n_{i}\right) }\right)  \label{wgq} \\
&&\times \delta \left( \theta ^{\left( i\right) }\left( n_{i}\right) -\theta
^{\left( j\right) }\left( n_{j}\right) -\frac{\left\vert
Z_{i}-Z_{j}\right\vert }{c}\right) \delta \left( \theta ^{\left( i\right)
}\left( n_{i}\right) -\theta ^{\left( kl\right) }\left( n_{kl}\right)
\right) \delta \left( \left( Z_{k,},Z_{l}\right) -\left( Z_{i,},Z_{j}\right)
\right)  \notag \\
&&\times \delta \left( \omega _{k}\left( n_{kl}\right) -\bar{\omega}\left(
Z_{i},n_{i}\right) \right) \delta \left( \omega _{l}\left( n_{kl}\right) -%
\bar{\omega}\left( Z_{j},n_{j}\right) \right)  \notag
\end{eqnarray}%
\begin{eqnarray}
\nabla _{n_{kl}}D_{k}\left( n_{kl}\right) &=&\left( -\frac{D_{k}\left(
n_{kl}\right) }{\tau _{D}\bar{\omega}\left( Z_{i},n_{i}\right) }+\frac{1}{%
\bar{\omega}\left( Z_{i},n_{i}\right) }\sum_{i,n_{i}}\alpha _{D}\left(
1-D_{k}\left( n_{kl}\right) \right) \omega _{i}\left( n_{i}\right) \right)
\label{wgc} \\
&&\times \delta \left( \theta ^{\left( i\right) }\left( n_{i}\right) -\theta
^{\left( kl\right) }\left( n_{kl}\right) \right) \delta \left(
Z_{k}-Z_{i}\right) \delta \left( \omega _{k}\left( n_{kl}\right) -\bar{\omega%
}\left( Z_{i},n_{i}\right) \right) \delta \left( \omega _{l}\left(
n_{kl}\right) -\bar{\omega}\left( Z_{j},n_{j}\right) \right)  \notag
\end{eqnarray}%
Similarly, note that we can also rewrite the currents equation (\ref{crt})
as:%
\begin{equation*}
\hat{J}^{\left( i\right) }\left( \left( n-1\right) \right) =J^{\left(
i\right) }\left( \left( n-1\right) \right) +\frac{\kappa }{N}\sum_{j,m}\frac{%
\omega _{j}\left( m\right) }{\omega _{i}\left( n-1\right) }\delta \left(
\theta ^{\left( i\right) }\left( n-1\right) -\theta ^{\left( j\right)
}\left( m\right) -\frac{\left\vert Z_{i}-Z_{j}\right\vert }{c}\right)
T_{ij}\left( \left( n-1,Z_{i}\right) ,\left( m,Z_{j}\right) \right)
\end{equation*}%
with:%
\begin{equation}
T_{ij}\left( \left( n_{i},Z_{i}\right) ,\left( m_{j},Z_{j}\right) \right)
=\sum_{kl}T_{kl}\left( n_{kl}\right) \delta \left( \theta ^{\left( i\right)
}\left( n_{i}\right) -\theta ^{\left( kl\right) }\left( n_{kl}\right)
\right) \delta \left( \omega _{k}\left( n_{kl}\right) -\bar{\omega}\left(
Z_{i},n_{i}\right) \right) \delta \left( \omega _{l}\left( n_{kl}\right) -%
\bar{\omega}\left( Z_{j},n_{j}\right) \right)  \label{tkl}
\end{equation}

\section{Field theoretic translation of the system}

This section presents the translation of the system neurons+connectivities
dynamics in terms of fields. The detailed derivation was given in (\cite{GLr}%
).

\subsection{translation of Equation (\protect\ref{dnq}) in terms of field
theory}

We have shown in \cite{GL1}\cite{GL2}\cite{GL3} that the probabilistic
description of dynamic system for a large number of degrees of freedom is
equivalent to a statistical field formalism. A concise version of this
method is given in (\cite{GL4}) and this method was applied in (\cite{GLr})
to derive the field theory counterpart of the system presented in section 2.

Within this formalism, the system is collectively described by a field,
which is an element of the Hilbert space of complex functions. The arguments
of these functions correspond to the parameters used to describe an
individual neuron. In this study, we will present the results directly.

The fields action for the neurons activity is a functional for the field $%
\Psi \left( \theta ,Z,\omega \right) $ and encompasses the dynamics (\ref%
{dnq}) along with the activities described by (\ref{MGF}) and (\ref{crt}): 
\begin{eqnarray}
S &=&-\frac{1}{2}\Psi ^{\dagger }\left( \theta ,Z,\omega \right) \nabla
\left( \frac{\sigma _{\theta }^{2}}{2}\nabla -\omega ^{-1}\right) \Psi
\left( \theta ,Z,\omega \right)  \label{lfS} \\
&&+\frac{1}{2\eta ^{2}}\int \left\vert \Psi \left( \theta ,Z,\omega \right)
\right\vert ^{2}\left( \omega ^{-1}-G\left( J\left( \theta ,Z\right) +\int 
\frac{\kappa }{N}\frac{\omega _{1}}{\omega }\left\vert \Psi \left( \theta -%
\frac{\left\vert Z-Z_{1}\right\vert }{c},Z_{1},\omega _{1}\right)
\right\vert ^{2}T\left( Z,\theta ,Z_{1}\right) dZ_{1}d\omega _{1}\right)
\right) ^{2}  \notag
\end{eqnarray}%
$\allowbreak $Using the fact that $\eta ^{2}<<1$, we showed in (\cite{GLr})
that we can restrict the fields to those of the form: 
\begin{equation}
\Psi \left( \theta ,Z\right) \delta \left( \omega ^{-1}-\omega ^{-1}\left(
J,\theta ,Z,\left\vert \Psi \right\vert ^{2}\right) \right)  \label{RS}
\end{equation}%
where $\omega ^{-1}\left( J,\theta ,Z,\left\vert \Psi \right\vert
^{2}\right) $ satisfies:%
\begin{eqnarray}
&&\omega ^{-1}\left( J,\theta ,Z,\left\vert \Psi \right\vert ^{2}\right)
\label{qf} \\
&=&G\left( J\left( \theta ,Z\right) +\int \frac{\kappa }{N}\frac{\omega
\left( J,\theta -\frac{\left\vert Z-Z_{1}\right\vert }{c},Z_{1},\Psi \right)
T\left( Z,\theta ,Z_{1},\theta -\frac{\left\vert Z-Z_{1}\right\vert }{c}%
\right) }{\omega \left( J,\theta ,Z,\left\vert \Psi \right\vert ^{2}\right) }%
\left\vert \Psi \left( \theta -\frac{\left\vert Z-Z_{1}\right\vert }{c}%
,Z_{1}\right) \right\vert ^{2}dZ_{1}\right)  \notag
\end{eqnarray}%
The "classical" effective action becomes (see (\cite{GLr})):%
\begin{equation}
-\frac{1}{2}\Psi ^{\dagger }\left( \theta ,Z\right) \left( \nabla _{\theta
}\left( \frac{\sigma ^{2}}{2}\nabla _{\theta }-\omega ^{-1}\left( J,\theta
,Z,\left\vert \Psi \right\vert ^{2}\right) \right) \right) \Psi \left(
\theta ,Z\right)  \label{nmR}
\end{equation}%
with $\omega ^{-1}\left( J,\theta ,Z,\left\vert \Psi \right\vert ^{2}\right) 
$ given by equation (\ref{qf}). As in (\cite{GL}) we add to this action a
stabilization potential $V\left( \Psi \right) $ ensuring an average activity
of the system. The precise form of this potential is irrelevant here, but we
assume that it has a minimum $\Psi _{0}\left( \theta ,Z\right) $.

\subsection{Translation for connectivity dynamics}

The translation of the four action terms describing the connectivity
dynamics (\ref{wgd}), (\ref{wgt}), (\ref{wgq}) and (\ref{wgc}) is
straightforward. Taking into account the projection (\ref{RS}), we obtain
four terms: $S_{\Gamma }^{\left( 1\right) }$, $S_{\Gamma }^{\left( 2\right)
} $, $S_{\Gamma }^{\left( 3\right) }$, $S_{\Gamma }^{\left( 4\right) }$: 
\begin{equation}
S_{\Gamma }^{\left( 1\right) }=\int \Gamma ^{\dag }\left( T,\hat{T},\theta
,Z,Z^{\prime },C,D\right) \nabla _{T}\left( \frac{\sigma _{T}^{2}}{2}\nabla
_{T}-\left( -\frac{1}{\tau \omega }T+\frac{\lambda }{\omega }\hat{T}\right)
\right) \Gamma \left( T,\hat{T},\theta ,Z,Z^{\prime },C,D\right)  \label{wGD}
\end{equation}%
\begin{eqnarray}
S_{\Gamma }^{\left( 2\right) } &=&\int \Gamma ^{\dag }\left( T,\hat{T}%
,\theta ,Z,Z^{\prime },C,D\right)  \label{wGT} \\
&&\times \nabla _{\hat{T}}\left( \frac{\sigma _{\hat{T}}^{2}}{2}\nabla _{%
\hat{T}}-\frac{\rho }{\omega }\left( \left( h\left( Z,Z^{\prime }\right) -%
\hat{T}\right) C\left\vert \Psi \left( \theta ,Z\right) \right\vert
^{2}h_{C}\left( \omega \right) \right. \right.  \notag \\
&&\left. \left. -D\hat{T}\left\vert \Psi \left( \theta -\frac{\left\vert
Z-Z^{\prime }\right\vert }{c},Z^{\prime }\right) \right\vert ^{2}h_{D}\left(
\omega ^{\prime }\right) \right) \right) \Gamma \left( T,\hat{T},\theta
,Z,Z^{\prime },C,D\right)  \notag
\end{eqnarray}%
\begin{eqnarray}
S_{\Gamma }^{\left( 3\right) } &=&\Gamma ^{\dag }\left( T,\hat{T},\theta
,Z,Z^{\prime },C,D\right)  \label{wGQ} \\
&&\times \nabla _{C}\left( \frac{\sigma _{C}^{2}}{2}\nabla _{C}+\left( \frac{%
C}{\tau _{C}\omega }-\alpha _{C}\left( 1-C\right) \frac{\omega ^{\prime
}\left\vert \Psi \left( \theta -\frac{\left\vert Z-Z^{\prime }\right\vert }{c%
},Z^{\prime }\right) \right\vert ^{2}}{\omega }\right) \right)  \notag \\
&&\times \Gamma \left( T,\hat{T},\theta ,Z,Z^{\prime },C,D\right)  \notag
\end{eqnarray}%
\begin{equation}
S_{\Gamma }^{\left( 4\right) }=\Gamma ^{\dag }\left( T,\hat{T},\theta
,Z,Z^{\prime },C,D\right) \nabla _{D}\left( \frac{\sigma _{D}^{2}}{2}\nabla
_{D}+\left( \frac{D}{\tau _{D}\omega }-\alpha _{D}\left( 1-D\right)
\left\vert \Psi \left( \theta ,Z\right) \right\vert ^{2}\right) \right)
\Gamma \left( T,\hat{T},\theta ,Z,Z^{\prime },C,D\right)  \label{wGC}
\end{equation}%
with:%
\begin{eqnarray*}
\omega &=&\omega \left( J,\theta ,Z,\left\vert \Psi \right\vert ^{2}\right)
\\
\omega ^{\prime } &=&\omega \left( J,\theta -\frac{\left\vert Z-Z^{\prime
}\right\vert }{c},Z^{\prime },\left\vert \Psi \right\vert ^{2}\right)
\end{eqnarray*}%
and:%
\begin{equation*}
h\left( Z,Z^{\prime }\right) =\exp \left( -\frac{\left\vert Z-Z^{\prime
}\right\vert }{\nu c}\right)
\end{equation*}

\subsection{Full action for the system and partition function}

The full action for the system is obtained by gathering the different terms:%
\begin{eqnarray}
S_{full} &=&-\frac{1}{2}\Psi ^{\dagger }\left( \theta ,Z,\omega \right)
\nabla \left( \frac{\sigma _{\theta }^{2}}{2}\nabla -\omega ^{-1}\left(
J,\theta ,Z,\left\vert \Psi \right\vert ^{2}\right) \right) \Psi \left(
\theta ,Z\right) +V\left( \Psi \right)  \label{flt} \\
&&+\frac{1}{2\eta ^{2}}\left( S_{\Gamma }^{\left( 0\right) }+S_{\Gamma
}^{\left( 1\right) }+S_{\Gamma }^{\left( 2\right) }+S_{\Gamma }^{\left(
3\right) }+S_{\Gamma }^{\left( 4\right) }\right) +U\left( \left\{ \left\vert
\Gamma \left( \theta ,Z,Z^{\prime },C,D\right) \right\vert ^{2}\right\}
\right)  \notag
\end{eqnarray}%
with $S_{\Gamma }^{\left( 1\right) }$, $S_{\Gamma }^{\left( 2\right) }$, $%
S_{\Gamma }^{\left( 3\right) }$, $S_{\Gamma }^{\left( 4\right) }$ given by (%
\ref{wGD}), (\ref{wGT}), (\ref{wGQ}), (\ref{wGC}). In (\ref{flt}), we added
a potential:%
\begin{equation*}
U\left( \left\{ \left\vert \Gamma \left( \theta ,Z,Z^{\prime },C,D\right)
\right\vert ^{2}\right\} \right) =U\left( \int T\left\vert \Gamma \left( T,%
\hat{T},\theta ,Z,Z^{\prime },C,D\right) \right\vert ^{2}dTd\hat{T}\right)
\end{equation*}%
that models the constraint about the number of active connections in the
system.

The partition function of the system is given by:%
\begin{equation*}
\exp \left( -S_{full}\left( \left( \Psi ^{\dagger },\Psi \right) ,\left(
\Gamma ,\Gamma ^{\dag }\right) \right) \right) \mathcal{D}\left( \Psi
^{\dagger },\Psi \right) \mathcal{D}\left( \Gamma ,\Gamma ^{\dag }\right)
\end{equation*}

\section{Background state and perturbation}

\subsection{Background state}

We showed in (\cite{GLr}) that $S_{full}$ present several possible minima.
These minima are characterized by the shape of $\left\vert \Psi \left(
\theta ,Z\right) \right\vert ^{2}$ and $\Gamma \left( T,\hat{T},\theta
,Z,Z^{\prime },C,D\right) $ for every point $Z$ and doublet $\left(
Z,Z^{\prime }\right) $. The derivation proceeds in several steps. We first
derive the saddle point $\left\vert \Psi \left( \theta ,Z\right) \right\vert
^{2}$ of:%
\begin{equation*}
-\frac{1}{2}\Psi ^{\dagger }\left( \theta ,Z,\omega \right) \nabla \left( 
\frac{\sigma _{\theta }^{2}}{2}\nabla -\omega ^{-1}\left( J,\theta
,Z,\left\vert \Psi \right\vert ^{2}\right) \right) \Psi \left( \theta
,Z\right) +V\left( \Psi \right)
\end{equation*}%
as a function of the connectivty field $\Gamma $ and then work with an
effective action:%
\begin{equation*}
S_{\Gamma }^{\left( 0\right) }+S_{\Gamma }^{\left( 1\right) }+S_{\Gamma
}^{\left( 2\right) }+S_{\Gamma }^{\left( 3\right) }+S_{\Gamma }^{\left(
4\right) }
\end{equation*}
for this field. In first approximation, the field $\Gamma $ can be
decomposed as a product:%
\begin{equation*}
\Gamma \left( T,\hat{T},\theta ,Z,Z^{\prime },C,D\right) =\Gamma _{C}\left(
\theta ,Z,Z^{\prime },C\right) \Gamma _{D}\left( \theta ,Z,Z^{\prime
},D\right) \Gamma \left( \hat{T},T,\theta ,Z,Z^{\prime }\right)
\end{equation*}%
and this allows to compute the background values for $\Gamma _{C}$ and $%
\Gamma _{D}$, leading ultimately to consider an action $S_{\Gamma }^{\left(
1\right) }+S_{\Gamma }^{\left( 2\right) }$ for:%
\begin{equation*}
\Gamma \left( \hat{T},T,\theta ,Z,Z^{\prime }\right)
\end{equation*}%
with $C$ and $D$ replaced by their background values $\left\langle
C\right\rangle $ and $\left\langle D\right\rangle $. For later purposes we
recall this action here: 
\begin{eqnarray}
&&S\left( \Gamma \left( T,\hat{T},\theta ,Z,Z^{\prime }\right) \right)
\label{RC} \\
&=&\Gamma ^{\dag }\left( T,\hat{T},\theta ,Z,Z^{\prime }\right) \left[
\nabla _{T}\left( \nabla _{T}-\left( \frac{\left( -T+\lambda \hat{T}\right) 
}{\tau \omega _{0}\left( Z\right) }\right) \left\vert \Psi \left( \theta
,Z\right) \right\vert ^{2}\right) \right.  \notag \\
&&+\nabla _{\hat{T}}\left( \nabla _{\hat{T}}-\rho \left( \left( h\left(
Z,Z^{\prime }\right) -\hat{T}\right) \left\langle C\right\rangle \left\vert
\Psi _{0}\left( Z\right) \right\vert ^{2}\frac{h_{C}\left( \omega _{0}\left(
Z\right) +\Delta \omega _{0}\left( Z,\left\vert \Psi \right\vert ^{2}\right)
\right) }{\omega _{0}\left( Z\right) }\right. \right.  \notag \\
&&\left. \left. -\eta H\left( \delta -T\right) -\left\langle D\right\rangle 
\hat{T}\left\vert \Psi _{0}\left( Z^{\prime }\right) \right\vert ^{2}\frac{%
h_{D}\left( \omega _{0}\left( Z^{\prime }\right) \right) }{\omega _{0}\left(
Z\right) }\right) \right) \Gamma \left( T,\hat{T},\theta ,Z,Z^{\prime
}\right)  \notag
\end{eqnarray}%
We derive the saddle-points solutions and compute the associated averages
for a static regime and under some approximations. The background is
determined by two possibilities for $\Gamma $ for all $\left( Z,Z^{\prime
}\right) $. These possibilities describe an activated state $\Gamma
_{a}\left( T,\hat{T},\theta ,Z,Z^{\prime },C,D\right) $ and an unactivatd
one $\Gamma _{u}\left( T,\hat{T},\theta ,Z,Z^{\prime },C,D\right) $.

We also showed how to derive the average connectivities in such background
states. These averages satisfy some set of equations:%
\begin{eqnarray}
\left\langle C_{Z,Z^{\prime }}\right\rangle &=&\frac{\alpha _{C}\omega
^{\prime }\left\vert \Psi \left( \theta -\frac{\left\vert Z-Z^{\prime
}\right\vert }{c},Z^{\prime }\right) \right\vert ^{2}}{\frac{1}{\tau _{C}}%
+\alpha _{C}\omega ^{\prime }\left\vert \Psi \left( \theta -\frac{\left\vert
Z-Z^{\prime }\right\vert }{c},Z^{\prime }\right) \right\vert ^{2}}
\label{CD} \\
\left\langle D_{Z,Z^{\prime }}\right\rangle &=&\frac{\alpha _{D}\omega
\left\vert \Psi \left( \theta ,Z\right) \right\vert ^{2}}{\frac{1}{\tau _{D}}%
+\alpha _{D}\omega \left\vert \Psi \left( \theta ,Z\right) \right\vert ^{2}}
\notag
\end{eqnarray}%
\begin{eqnarray}
\left\langle T\left( Z,Z^{\prime }\right) \right\rangle &=&\lambda \tau
\left\langle \hat{T}\left( Z,Z^{\prime }\right) \right\rangle  \label{Cnv} \\
&=&\frac{\lambda \tau h\left( Z,Z^{\prime }\right) \left\langle
C_{Z,Z^{\prime }}\left( \theta \right) \left\vert \Psi \left( \theta
,Z\right) \right\vert ^{2}\right\rangle }{\left\vert \bar{\Psi}\left( \theta
,Z,Z^{\prime }\right) \right\vert ^{2}}  \notag
\end{eqnarray}%
with:%
\begin{equation}
\left\vert \bar{\Psi}\left( \theta ,Z,Z^{\prime }\right) \right\vert ^{2}=%
\frac{C_{Z,Z^{\prime }}\left( \theta \right) \left\vert \Psi \left( \theta
,Z\right) \right\vert ^{2}h_{C}\left( \omega \left( \theta ,Z\right) \right)
+D_{Z,Z^{\prime }}\left( \theta \right) \left\vert \Psi \left( \theta -\frac{%
\left\vert Z-Z^{\prime }\right\vert }{c},Z^{\prime }\right) \right\vert
^{2}h_{D}\left( \omega \left( \theta -\frac{\left\vert Z-Z^{\prime
}\right\vert }{c},Z^{\prime }\right) \right) }{h_{C}\left( \omega \left(
\theta ,Z\right) \right) }  \label{PB}
\end{equation}%
for $\left( Z,Z^{\prime }\right) $ an $"a"$ (active) doublet, and:%
\begin{equation}
\left\langle T\left( Z,Z^{\prime }\right) \right\rangle =0  \label{CN}
\end{equation}%
\begin{equation}
\left\langle \hat{T}\left( Z,Z^{\prime }\right) \right\rangle =\frac{h\left(
Z,Z^{\prime }\right) \left\langle C_{Z,Z^{\prime }}\left( \theta \right)
h_{C}\left( \omega \left( \theta ,Z\right) \right) \left\vert \Psi \left(
\theta ,Z\right) \right\vert ^{2}\right\rangle -\eta }{\left\langle
h_{C}\left( \omega \left( \theta ,Z\right) \right) \left\vert \bar{\Psi}%
\left( \theta ,Z,Z^{\prime }\right) \right\vert ^{2}\right\rangle }<0
\label{CP}
\end{equation}%
We obtained under some assumptions and in the static case, the possible
averages values for the connectvt functns:%
\begin{eqnarray}
T\left( Z_{-},Z_{+}^{\prime }\right) &=&\frac{\lambda \tau \exp \left( -%
\frac{\left\vert Z-Z^{\prime }\right\vert }{\nu c}\right) \left( \frac{1}{%
\tau _{D}\alpha _{D}}+\frac{1}{b\bar{T}^{2}\left\langle \left\vert \Psi
_{0}\left( Z^{\prime }\right) \right\vert ^{2}\right\rangle _{Z}}\right) }{%
\frac{1}{\tau _{D}\alpha _{D}}+\frac{1}{\alpha _{C}\tau _{C}}+\frac{1}{b\bar{%
T}^{2}\left\langle \left\vert \Psi _{0}\left( Z^{\prime }\right) \right\vert
^{2}\right\rangle _{Z}}+b\bar{T}\left( \bar{T}\left\langle \left\vert \Psi
_{0}\left( Z^{\prime }\right) \right\vert ^{2}\right\rangle _{Z^{\prime
}}^{2}\right) ^{2}}\simeq 0  \label{cnv} \\
T\left( Z_{+},Z_{+}^{\prime }\right) &=&\frac{\lambda \tau \exp \left( -%
\frac{\left\vert Z-Z^{\prime }\right\vert }{\nu c}\right) \left( \frac{1}{%
\tau _{D}\alpha _{D}}+b\bar{T}\left( \bar{T}\left\langle \left\vert \Psi
_{0}\left( Z^{\prime }\right) \right\vert ^{2}\right\rangle _{Z}^{2}\right)
^{2}\right) }{\frac{1}{\tau _{D}\alpha _{D}}+\frac{1}{\alpha _{C}\tau _{C}}+b%
\bar{T}\left( \bar{T}\left\langle \left\vert \Psi _{0}\left( Z^{\prime
}\right) \right\vert ^{2}\right\rangle _{Z}^{2}\right) ^{2}+b\bar{T}\left( 
\bar{T}\left\langle \left\vert \Psi _{0}\left( Z^{\prime }\right)
\right\vert ^{2}\right\rangle _{Z^{\prime }}^{2}\right) ^{2}}\simeq \frac{%
\lambda \tau \exp \left( -\frac{\left\vert Z-Z^{\prime }\right\vert }{\nu c}%
\right) }{2}  \notag \\
T\left( Z_{+},Z_{-}^{\prime }\right) &=&\frac{\lambda \tau \exp \left( -%
\frac{\left\vert Z-Z^{\prime }\right\vert }{\nu c}\right) \left( \frac{1}{%
\tau _{D}\alpha _{D}}+b\bar{T}\left( \bar{T}\left\langle \left\vert \Psi
_{0}\left( Z^{\prime }\right) \right\vert ^{2}\right\rangle _{Z}^{2}\right)
^{2}\right) }{\frac{1}{\tau _{D}\alpha _{D}}+\frac{1}{\alpha _{C}\tau _{C}}+b%
\bar{T}\left( \bar{T}\left\langle \left\vert \Psi _{0}\left( Z^{\prime
}\right) \right\vert ^{2}\right\rangle _{Z}^{2}\right) ^{2}+\frac{1}{b\bar{T}%
^{2}\left\langle \left\vert \Psi _{0}\left( Z^{\prime }\right) \right\vert
^{2}\right\rangle _{Z^{\prime }}}}\simeq \lambda \tau \exp \left( -\frac{%
\left\vert Z-Z^{\prime }\right\vert }{\nu c}\right)  \notag \\
T\left( Z_{-},Z_{-}^{\prime }\right) &\simeq &\frac{\lambda \tau \exp \left(
-\frac{\left\vert Z-Z^{\prime }\right\vert }{\nu c}\right) +\frac{1}{b\bar{T}%
^{2}\left\langle \left\vert \Psi _{0}\left( Z^{\prime }\right) \right\vert
^{2}\right\rangle _{Z}}}{1+\frac{\tau _{D}\alpha _{D}}{\alpha _{C}\tau _{C}}+%
\frac{1}{b\bar{T}^{2}\left\langle \left\vert \Psi _{0}\left( Z^{\prime
}\right) \right\vert ^{2}\right\rangle _{Z}}+\frac{1}{b\bar{T}%
^{2}\left\langle \left\vert \Psi _{0}\left( Z^{\prime }\right) \right\vert
^{2}\right\rangle _{Z^{\prime }}}}\simeq \frac{\lambda \tau \exp \left( -%
\frac{\left\vert Z-Z^{\prime }\right\vert }{\nu c}\right) }{2}  \notag
\end{eqnarray}%
with $\bar{T}=\frac{\lambda \tau \nu cb}{2}$, $b$ a coefficient
characterizing the function $G$ in the linear approximation\footnote{$%
b\simeq G^{\prime }\left( 0\right) $} and $\left\langle \left\vert \Psi
_{0}\left( Z^{\prime }\right) \right\vert ^{2}\right\rangle _{Z}^{2}$, \ $%
\left\langle \left\vert \Psi _{0}\left( Z^{\prime }\right) \right\vert
^{2}\right\rangle _{Z}^{2}$ are some averaged background fields in regions
surrounding $Z$ and $Z^{\prime }$ respctvl. They are determined by a
potential describing some average activity depending on the points. These
results are derived under the assumption of a static field $\Psi _{0}\left(
Z\right) $.

\subsection{source induced perturbation of the static background state}

In (\cite{GLr}), we also showed with qualitative arguments how an external
activation may modify the solutions of equations (\ref{CD}) and (\ref{Cnv})
(or (\ref{CN}) and (\ref{CP})) for connectivity functions averages. Actually
external signals modify the field $\Psi _{0}\left( \theta ,Z\right) $: 
\begin{equation*}
\Psi _{0}\left( \theta ,Z\right) \rightarrow \Psi _{0}\left( \theta
,Z\right) +\delta \Psi \left( \theta ,Z\right)
\end{equation*}%
and induce a modification $\delta \omega \left( J,\theta ,Z,\left\vert \Psi
\right\vert ^{2}\right) $. This modifies in turn the set of equations for $%
T\left( Z,Z^{\prime }\right) $. In (\cite{GL}) we showed that oscillating
signals can induce non linear oscillatory response $\delta \omega \left(
J,\theta ,Z,\left\vert \Psi \right\vert ^{2}\right) $ and produce
interferences phenomena. These phenomena modify the solutions (\ref{cnv}),
linking points where interferences induced perturbations $\delta \omega
\left( J,\theta ,Z,\left\vert \Psi \right\vert ^{2}\right) $ have a large
amplitude, leading to some emerging connected structures. We also discussed
how such structures may interact and activate each other.

However, these results were derived qualitatively in the context of
switching static states. The next sections provide the dynamical study of
these phenomena in terms of field theory.

\part*{II Effective field theoretic approach to transitions of connectivity
states}

In this part, we provide a rationale for adopting the local approach
employed in the initial part of this study (\cite{GLr}) by using the
system's effective action. We elucidate the dynamic aspects of the system
presented in (\cite{GLr}) within the framework of the field model. This
approach is non-local in nature due to the involvement of interactions
between distant interconnected points.

The underlying principle remains consistent with our prior work. Our
objective is to integrate the degrees of freedom of the neuron field with
respect to the connectivity field, denoted as $\Gamma $. This procedure
results in the generation of an effective action for the field $\Gamma $,
enabling the investigation of state transitions in connectivity. However,
within a dynamic context, this integration of degrees of freedom must
account for external perturbations that alter the path integral associated
with the neuron field $\Psi $. We will evaluate this path integral by
developing a time-dependent series expansion for neural activity $\omega
\left( J,\theta ,Z\right) $, which depends both on the connectivity field
and the perturbation component of the field $\Psi $, written $\Delta \Psi
\left( \theta ,Z\right) $.

\section{Principle: Integration of neuron field degrees of freedom in
presence of external perturbation}

As a general method for integrating the $\Psi $ degrees of freedom, we start
with the action functional for cell field alone, along with its partition
function:%
\begin{equation}
\int \exp \left( \frac{1}{2}\Psi ^{\dagger }\left( \theta ,Z,\omega \right)
\nabla \left( \frac{\sigma _{\theta }^{2}}{2}\nabla -\omega ^{-1}\left(
J,\theta ,Z,\left\vert \Psi \right\vert ^{2}\right) \right) \Psi \left(
\theta ,Z\right) +V\left( \Psi \right) \right) \mathcal{D}\Psi \left( \theta
,Z\right)  \label{prc}
\end{equation}%
We then compute this quantity as a function of the connectivity field $%
\Gamma $. Actually, the time scale of neuronal processes is shorter than
that of connectivity dynamics, and perturbations first affect the
equilibrium of the neuronal system. Subsequently, these perturbations
propagate to influence connectivity dynamics.

To model some external perturbation that will propagate in the system, we
modify (\ref{prc}) and consider the insertion, at some point $Z_{i}$ and
time $\theta _{0}$, the factor $a\left( Z_{i},\theta _{0}\right) \left\vert
\Psi \left( Z_{i},\theta _{0}\right) \right\vert ^{2}$ inside (\ref{prc}).
The partition function is thus replaced by: 
\begin{equation}
\int a\left( Z_{i},\theta _{0}\right) \left\vert \Psi \left( Z_{i},\theta
_{0}\right) \right\vert ^{2}\exp \left( \frac{1}{2}\Psi ^{\dagger }\left(
\theta ,Z,\omega \right) \nabla \left( \frac{\sigma _{\theta }^{2}}{2}\nabla
-\omega ^{-1}\left( J,\theta ,Z,\left\vert \Psi \right\vert ^{2}\right)
\right) \Psi \left( \theta ,Z\right) +V\left( \Psi \right) \right) \mathcal{D%
}\Psi \left( \theta ,Z\right)
\end{equation}%
The squared field: 
\begin{equation*}
\left\vert \Psi \left( \theta ,Z\right) \right\vert ^{2}=\Psi \left( \theta
,Z\right) \Psi ^{\dag }\left( \theta ,Z\right)
\end{equation*}%
measures the density of activity at point $Z$ and for a given time $\theta $%
. The presence of the additional contribution:%
\begin{equation*}
\left\vert \Psi \left( Z_{i},\theta _{0}\right) \right\vert ^{2}=\Psi \left(
Z_{i},\theta _{0}\right) \Psi ^{\dag }\left( Z_{i},\theta _{0}\right)
\end{equation*}%
in the integral corresponds, in the field framework, to the introduction of
a ponctual perturbation at time $\theta _{0}$ from the background field
equilibrium, as\ described by the term $\Psi ^{\dagger }\left( Z_{i},\theta
_{0}\right) $, which is immediately switched off, as transcribed by $\Psi
\left( Z_{i},\theta _{0}\right) $.

In other words, the presence of $\left\vert \Psi \left( Z_{i},\theta
_{0}\right) \right\vert ^{2}$ corresponds to a ponctual signal sent from $%
Z_{i}$ and time $\theta _{0}$ that will propagate to the whole thread. The
factor $a\left( Z_{i},\theta _{0}\right) $ is the amplitude of this signal.
To model several sources sending signals at the same moment $\theta _{0}$,
we introduce a product:%
\begin{equation*}
\prod\limits_{i}a\left( Z_{i},\theta _{0}\right) \left\vert \Psi \left(
Z_{i},\theta _{0}\right) \right\vert ^{2}
\end{equation*}

We will consider the possibility that the number of active sources may vary
over time and consider a probabilistic combination of such sources.
Furthermore, we intend to account for the periodic repetition of certain
signals over time. Consequently, we will incorporate these elements into the
path integral by including the factor:

\begin{equation}
\int \exp \left( \sum_{i}a\left( Z_{i},\theta _{0}\right) \left\vert \Psi
\left( Z_{i},\theta _{0}\right) \right\vert ^{2}\right) d\theta _{0}
\label{sgl}
\end{equation}%
The term $\sum_{i}a\left( Z_{i},\theta \right) \left\vert \Psi \left(
Z_{i},\theta \right) \right\vert ^{2}$ creates and cancels some stimulations
at several points $Z_{i}$ which deviate the field $\Psi \left( Z_{i},\theta
\right) $ from the static equilibrium. The exponential factor accounts for
the possibility of multiple similar stimuli occuring at the same location,
as required. The summation over $\theta _{0}$ guarantees the signal's
repetition over a certain time period. Furthermore, to ensure that
perturbations only occur at specific points $Z_{i}$, we assume that the
perturbation is implicitely tensored by:%
\begin{equation*}
\prod\limits_{Z\neq Z_{i}}\delta \left( \left\vert \Psi \left( Z,\theta
_{0}\right) \right\vert ^{2}\right)
\end{equation*}%
This will imply that outside the points $Z_{i}$, there is no initial
perturbation of the system.

The path integral to consider is thus:

\begin{eqnarray}
&&\int \exp \left( -S\left( \Psi \right) \right) \int \exp \left(
\sum_{i}a\left( Z_{i},\theta _{0}\right) \left\vert \Psi \left( Z_{i},\theta
_{0}\right) \right\vert ^{2}\right) d\theta _{0}  \label{prb} \\
&=&\int \exp \left( \frac{1}{2}\Psi ^{\dagger }\left( \theta ,Z\right)
\nabla \left( \frac{\sigma _{\theta }^{2}}{2}\nabla -\omega ^{-1}\right)
\Psi \left( \theta ,Z\right) \right) \int \exp \left( \sum_{i}a\left(
Z_{i},\theta _{0}\right) \left\vert \Psi \left( Z_{i},\theta _{0}\right)
\right\vert ^{2}\right) d\theta _{0}  \notag
\end{eqnarray}

Then, we evaluate the effect of the inserted term (\ref{sgl}) on activities
by expanding $\omega ^{-1}$ as a serie of $\Delta \Psi \left( \theta
,Z\right) $, the fluctuation of the field around the background field $\Psi
_{0}\left( Z\right) $ induced by the perturbation. The knowledge of $\omega
^{-1}$ in the perturbated background state will provide the effective
activity, which will be incorporated into the action for the field $\Gamma $.

We replicate the sequential steps of the derivation as outlined in (\cite{GL}%
) and tailor them to our specific context. This series expansion
subsequently enables us to calculate the transition in the background state
induced by the perturbation (\ref{sgl}) and derive the interferences
phenomena presented in (\cite{GLr}).

\section{Activities $\protect\omega \left( J,\protect\theta ,Z\right) $
series expansion in field in presence of external sources}

This section computes the modification of $\omega ^{-1}$ with respect to its
background value due to an external perturbation. This is achieved by first
computing the expansion of $\omega ^{-1}$ in terms of field, and then by
including the effect of the sources. These sources modify the background
state at specific points, thereby influencing the activities.

\subsection{Formal series expansion}

We showed in (\cite{GL}) that in first approximation, the effective action
for $\Psi $ obtained by replacing:%
\begin{equation}
\left\vert \Psi \left( \theta -\frac{\left\vert Z-Z_{1}\right\vert }{c}%
,Z_{1}\right) \right\vert ^{2}\rightarrow \mathcal{G}_{0}\left(
0,Z_{1}\right) +\left\vert \Psi _{0}\left( Z_{1}\right) +\Psi \left( \theta -%
\frac{\left\vert Z-Z_{1}\right\vert }{c},Z_{1}\right) \right\vert ^{2}
\label{RPLC}
\end{equation}%
in (\ref{qf}), where $\mathcal{G}_{0}\left( 0,Z_{1}\right) $ is the free
Green function computed in (\cite{GL}). In average, $\mathcal{G}_{0}\left(
0,Z_{1}\right) $ is some constant $\frac{1}{\Lambda }$. The field $\Psi
_{0}\left( Z_{1}\right) $ is some static background, while $\Psi \left(
\theta -\frac{\left\vert Z-Z_{1}\right\vert }{c},Z_{1}\right) $ represents
the dynamic part of the background field, i.e. a modification above the
background state, which may be induced by external sources.

Formula (\ref{RPLC}) can be written in a more compact form if we define: 
\begin{equation}
\mathcal{\bar{G}}_{0}\left( 0,Z_{1}\right) =\mathcal{G}_{0}\left(
0,Z_{1}\right) +\left\vert \Psi _{0}\left( Z_{1}\right) \right\vert ^{2}
\label{GR}
\end{equation}
and write:%
\begin{equation*}
\left\vert \Psi \left( \theta -\frac{\left\vert Z-Z_{1}\right\vert }{c}%
,Z_{1}\right) \right\vert ^{2}
\end{equation*}%
as a shorthand for: 
\begin{equation*}
\Psi _{0}^{\dagger }\left( Z_{1}\right) \Psi \left( \theta -\frac{\left\vert
Z-Z_{1}\right\vert }{c},Z_{1}\right) +\Psi _{0}\left( Z_{1}\right) \Psi
^{\dagger }\left( \theta -\frac{\left\vert Z-Z_{1}\right\vert }{c}%
,Z_{1}\right) +\left\vert \Psi \left( \theta -\frac{\left\vert
Z-Z_{1}\right\vert }{c},Z_{1}\right) \right\vert ^{2}
\end{equation*}%
As a consequence, formula (\ref{RPLC}) writes:%
\begin{equation}
\left\vert \Psi \left( \theta -\frac{\left\vert Z-Z_{1}\right\vert }{c}%
,Z_{1}\right) \right\vert ^{2}\rightarrow \mathcal{\bar{G}}_{0}\left(
0,Z_{1}\right) +\left\vert \Psi \left( \theta -\frac{\left\vert
Z-Z_{1}\right\vert }{c},Z_{1}\right) \right\vert ^{2}  \label{RP}
\end{equation}

Taking (\ref{RP}) into account, our starting point the equation for
(inverse) neurons activities:%
\begin{eqnarray}
\omega ^{-1}\left( J,\theta ,Z\right) &=&G\left( J\left( \theta \right) +%
\frac{\kappa }{N}\int \frac{\omega \left( J,\theta -\frac{\left\vert
Z-Z_{1}\right\vert }{c},Z_{1},\Psi \right) T\left( Z,\theta ,Z_{1},\theta -%
\frac{\left\vert Z-Z_{1}\right\vert }{c}\right) }{\omega \left( J,\theta
,Z,\left\vert \Psi \right\vert ^{2}\right) }\right.  \label{qtF} \\
&&\times \left. \left( \mathcal{\bar{G}}_{0}\left( 0,Z_{1}\right)
+\left\vert \Psi \left( \theta -\frac{\left\vert Z-Z_{1}\right\vert }{c}%
,Z_{1}\right) \right\vert ^{2}\right) dZ_{1}\right)  \notag
\end{eqnarray}%
Then, we replace $T\left( Z,\theta ,Z_{1},\theta -\frac{\left\vert
Z-Z_{1}\right\vert }{c}\right) $ in (\ref{qtF}) by its average over the
connectivity, which is, given (\ref{Cnv}):%
\begin{equation}
\left\langle T\left( Z,\theta ,Z_{1},\theta -\frac{\left\vert
Z-Z_{1}\right\vert }{c}\right) \right\rangle =T\left( Z,Z_{1}\right) W\left( 
\frac{\omega \left( \theta ,Z\right) }{\omega \left( \theta -\frac{%
\left\vert Z-Z_{1}\right\vert }{c},Z_{1}\right) }\right)
\end{equation}%
where: 
\begin{equation*}
T\left( Z,Z_{1}\right) =h\left( Z,Z_{1}\right)
\end{equation*}%
and:%
\begin{eqnarray*}
&&W\left( \frac{\omega \left( \theta ,Z\right) }{\omega \left( \theta -\frac{%
\left\vert Z-Z_{1}\right\vert }{c},Z_{1}\right) }\right) \\
&=&\frac{\lambda \tau h\left( Z,Z_{1}\right) \left\langle C_{Z,Z_{1}}\left(
\theta \right) h_{C}\left( \omega \left( \theta ,Z\right) \right) \left\vert
\Psi \left( \theta ,Z\right) \right\vert ^{2}\right\rangle }{h_{C}\left(
\omega \left( \theta ,Z\right) \right) \left\vert \bar{\Psi}\left( \theta
,Z,Z_{1}\right) \right\vert ^{2}}\simeq \frac{\lambda \tau h\left(
Z,Z_{1}\right) \left\langle C_{Z,Z_{1}}\left( \theta \right) \left\vert \Psi
\left( \theta ,Z\right) \right\vert ^{2}\right\rangle }{\left\vert \bar{\Psi}%
\left( \theta ,Z,Z_{1}\right) \right\vert ^{2}}
\end{eqnarray*}%
Thus (\ref{qtF}) writes:%
\begin{eqnarray}
\omega ^{-1}\left( J,\theta ,Z\right) &=&G\left( J\left( \theta \right) +%
\frac{\kappa }{N}\int T\left( Z,Z_{1}\right) \frac{\omega \left( \theta -%
\frac{\left\vert Z-Z_{1}\right\vert }{c},Z_{1}\right) W\left( \frac{\omega
\left( \theta ,Z\right) }{\omega \left( \theta -\frac{\left\vert
Z-Z_{1}\right\vert }{c},Z_{1}\right) }\right) }{\omega \left( \theta
,Z\right) }\right.  \label{QT} \\
&&\times \left. \left( \mathcal{\bar{G}}_{0}\left( 0,Z_{1}\right)
+\left\vert \Psi \left( \theta -\frac{\left\vert Z-Z_{1}\right\vert }{c}%
,Z_{1}\right) \right\vert ^{2}\right) dZ_{1}\right)  \notag
\end{eqnarray}%
and we aim at expansion of the solutions of (\ref{QT}) around the static
background equilibrium. To do so, we write the series expansion in $%
\left\vert \Psi \left( \theta ^{\left( j\right) },Z_{1}\right) \right\vert
^{2}$ of $\omega ^{-1}\left( J,\theta ,Z\right) $ around its background
state value:%
\begin{eqnarray}
\omega ^{-1}\left( J,\theta ,Z\right) &=&\omega ^{-1}\left( \theta ,Z\right)
_{\left\vert \Psi \right\vert ^{2}=0}  \label{psn} \\
&&+\int \sum_{n=1}^{\infty }\left( \frac{\delta ^{n}\omega ^{-1}\left(
J,\theta ,Z\right) }{\dprod\limits_{i=1}^{n}\delta \left\vert \Psi \left(
\theta -l_{i},Z_{i}\right) \right\vert ^{2}}\right) _{\left\vert \Psi
\right\vert ^{2}=0}\dprod\limits_{i=1}^{n}\left\vert \Psi \left( \theta
-l_{i},Z_{i}\right) \right\vert ^{2}  \notag
\end{eqnarray}%
and finding $\omega \left( J,\theta ,Z\right) $ amounts to finding the
derivatives:%
\begin{equation*}
\frac{\delta ^{n}\omega ^{-1}\left( J,\theta ,Z\right) }{\dprod%
\limits_{i=1}^{n}\delta \left\vert \Psi \left( \theta -l_{i},Z_{i}\right)
\right\vert ^{2}}
\end{equation*}%
These derivatives are computed by expanding the right-hand side of (\ref{QT}%
) in $\left\vert \Psi \left( \theta ^{\left( j\right) },Z_{1}\right)
\right\vert ^{2}$. We present the computations in the next paragraphs.

\subsection{First term of the expansion}

The first term in (\ref{psn}), $\omega ^{-1}\left( \theta ^{\left( i\right)
},Z\right) _{\left\vert \Psi \right\vert ^{2}=0}$, is a solution of:%
\begin{eqnarray}
&&\omega ^{-1}\left( \theta ,Z\right) _{\left\vert \Psi \right\vert ^{2}=0}
\label{srf} \\
&=&G\left( J+\frac{\kappa }{N}\int T\left( Z,Z_{1}\right) \frac{\omega
_{\left\vert \Psi \right\vert ^{2}=0}\left( \theta -\frac{\left\vert
Z-Z_{1}\right\vert }{c},Z_{1}\right) }{\omega _{\left\vert \Psi \right\vert
^{2}=0}\left( \theta ,Z\right) }W\left( \frac{\omega _{\left\vert \Psi
\right\vert ^{2}=0}\left( \theta ,Z\right) }{\omega _{\left\vert \Psi
\right\vert ^{2}=0}\left( \theta -\frac{\left\vert Z-Z_{1}\right\vert }{c}%
,Z_{1}\right) }\right) \left( \mathcal{\bar{G}}_{0}\left( 0,Z_{1}\right)
\right) dZ_{1}\right)  \notag
\end{eqnarray}%
To uncover the internal dynamics of the system, we will first consider a
constant external current $J\left( \theta \right) =J$, typically with $J$ $%
=0 $. However, the findings of this section remain applicable even in the
presence of a non-static current $J\left( \theta \right) $. The static
solution of (\ref{srf}) satifies:%
\begin{eqnarray*}
\omega ^{-1}\left( J,Z\right) &=&G\left( J+\frac{\kappa }{N}\int T\left(
Z,Z_{1}\right) \frac{\omega \left( Z_{1}\right) }{\omega \left( Z\right) }%
W\left( \frac{\omega \left( Z\right) }{\omega \left( Z_{1}\right) }\right) 
\mathcal{\bar{G}}_{0}\left( 0,Z_{i}\right) dZ_{1}\right) \\
&\equiv &G\left[ J,\omega ,Z\right]
\end{eqnarray*}%
We assume this solution to be known, and we chose to expand $\omega \left(
J,\theta ,Z\right) $ in (\ref{psn}) around this solution, the dynamics being
determined by the time dependency of $\left\vert \Psi \left( \theta ^{\left(
j\right) },Z_{1}\right) \right\vert ^{2}$. We thus set:%
\begin{equation*}
\omega \left( \theta ,Z\right) _{\left\vert \Psi \right\vert ^{2}=0}=\omega
\left( J,Z\right)
\end{equation*}

\subsection{Computation of the derivatives arising in the series}

Appendices 1 and 2 compute the derivatives $\left( \frac{\delta ^{n}\omega
^{-1}\left( J,\theta ,Z\right) }{\dprod\limits_{i=1}^{n}\delta \left\vert
\Psi \left( \theta -l_{i},Z_{i}\right) \right\vert ^{2}}\right) _{\left\vert
\Psi \right\vert ^{2}=0}$\ in (\ref{psn}).

Defining:%
\begin{eqnarray}
&&\check{T}\left( \theta ,Z,Z_{1},\omega ,\Psi \right)  \label{vdr} \\
&=&-\frac{\frac{\kappa }{N}\omega \left( J,\theta ,Z\right) T\left(
Z,Z_{1},\theta \right) G^{\prime }\left[ J,\omega ,\theta ,Z,\Psi \right] }{%
1-\left( \int \frac{\kappa }{N}\omega \left( J,\theta -\frac{\left\vert
Z-Z^{\prime }\right\vert }{c},Z^{\prime }\right) \left( \mathcal{\bar{G}}%
_{0}\left( 0,Z^{\prime }\right) +\left\vert \Psi \left( \theta -\frac{%
\left\vert Z-Z^{\prime }\right\vert }{c},Z^{\prime }\right) \right\vert
^{2}\right) T\left( Z,Z^{\prime },\theta \right) dZ^{\prime }\right)
G^{\prime }\left[ J,\omega ,\theta ,Z,\Psi \right] }  \notag
\end{eqnarray}%
and the operator with kernel:%
\begin{eqnarray}
\ \check{T}\left( \left( Z^{\left( l-1\right) },\theta ^{\left( l-1\right)
}\right) ,\left( Z^{\left( l\right) },\theta ^{\left( l\right) }\right)
\right) &=&\ \check{T}\left( \theta -\sum_{j=1}^{l-1}\frac{\left\vert
Z^{\left( j-1\right) }-Z^{\left( j\right) }\right\vert }{c},Z^{\left(
l-1\right) },Z^{\left( l\right) },\omega _{0}\right)  \label{rnL} \\
&&\times \delta \left( \left( \theta ^{\left( l\right) }-\theta ^{\left(
l-1\right) }\right) -\frac{\left\vert Z^{\left( l-1\right) }-Z^{\left(
l\right) }\right\vert }{c}\right)  \notag
\end{eqnarray}%
appendix 1 shows that:%
\begin{eqnarray}
&&\frac{\delta \omega ^{-1}\left( J,\theta ,Z\right) }{\delta \left\vert
\Psi \left( \theta -l_{1},Z_{1}\right) \right\vert ^{2}}  \label{rdn} \\
&=&-\sum_{n=1}^{\infty }\int \frac{\omega ^{-1}\left( J,\theta
-\sum_{l=1}^{n}\frac{\left\vert Z^{\left( l-1\right) }-Z^{\left( l\right)
}\right\vert }{c},Z_{1}\right) }{\left( \mathcal{\bar{G}}_{0}\left(
0,Z_{1}\right) +\left\vert \Psi \left( \theta -l_{1},Z_{1}\right)
\right\vert ^{2}\right) }  \notag \\
&&\times \dprod\limits_{l=1}^{n}\ \check{T}\left( \theta -\sum_{j=1}^{l-1}%
\frac{\left\vert Z^{\left( j-1\right) }-Z^{\left( j\right) }\right\vert }{c}%
,Z^{\left( l-1\right) },Z^{\left( l\right) },\omega ,\Psi \right) \delta
\left( l_{1}-\sum_{l=1}^{n}\frac{\left\vert Z^{\left( l-1\right) }-Z^{\left(
l\right) }\right\vert }{c}\right) \dprod\limits_{l=1}^{n-1}dZ^{\left(
l\right) }  \notag
\end{eqnarray}%
and appendix 2 builds on (\ref{rdn}) to compute the derivative arising in
the series expansion (\ref{psn}):%
\begin{equation}
\left( \frac{\delta ^{n}\omega \left( J,\theta ,Z\right) }{%
\dprod\limits_{i=1}^{n}\delta \left\vert \Psi \left( \theta
-l_{i},Z_{i}\right) \right\vert ^{2}}\right) _{\left\vert \Psi \right\vert
^{2}=0}\dprod\limits_{i=1}^{n}\left\vert \Psi \left( \theta
-l_{i},Z_{i}\right) \right\vert ^{2}  \label{rcs}
\end{equation}%
by a graphical representation. We associate the squared field $\left\vert
\Psi \left( \theta -l_{i},Z_{i}\right) \right\vert ^{2}$ to each point $%
Z_{i} $ and draw $m$ lines for $m=1,...,n$. One of them at least is starting
from $Z$. These lines are composed of an arbitrary number of segments and
all the points $Z_{i}$ are crossed by one line. Each line ends at a point $%
Z_{i}$. The starting points of the lines branch either at $Z$ or at some
point of an other line. There are $m$ branching points of valence $k$
including the starting point at $Z$. Apart from $Z$, the branching points
have valence $3,...,n-1$.

To each line $i$ of length $L_{i}$, we associate the factor:%
\begin{eqnarray}
F\left( line_{i}\right) &=&\dprod\limits_{l=1}^{L_{i}}\ \check{T}\left(
\theta -\sum_{j=1}^{l-1}\frac{\left\vert Z^{\left( j-1\right) }-Z^{\left(
j\right) }\right\vert }{c},Z^{\left( l-1\right) },Z^{\left( l\right)
},\omega _{0},\Psi \right)  \label{flN} \\
&&\times \frac{-\omega _{0}\left( J,\theta -\sum_{l=1}^{L_{i}}\frac{%
\left\vert Z^{\left( l-1\right) }-Z^{\left( l\right) }\right\vert }{c}%
,Z_{i}\right) }{\mathcal{\bar{G}}_{0}\left( 0,Z_{i}\right) }  \notag
\end{eqnarray}%
and to each branching point $\left( X,\theta \right) =B$ of valence $k+2$,
we associate the factor:%
\begin{equation}
F\left( \left( X,\theta \right) \right) =\frac{\delta ^{k}\left( \frac{\frac{%
\kappa }{N}T\left( Z,Z^{\left( l\right) }\right) F^{\prime }\left[ J,\theta
,\omega _{0},Z^{\left( l\right) }\right] \mathcal{\bar{G}}_{0}\left(
0,Z^{\left( l\right) }\right) }{\omega _{0}\left( J,\theta ,Z^{\left(
l\right) }\right) }\right) }{\delta ^{k}\omega _{0}\left( J,\theta
,Z^{\left( l\right) }\right) }  \label{fcT}
\end{equation}%
and (\ref{rcs}) writes as a series of lines contributions connected by the
branching points:%
\begin{eqnarray}
&&\left( \frac{\delta ^{n}\omega \left( J,\theta ,Z\right) }{%
\dprod\limits_{i=1}^{n}\delta \left\vert \Psi \left( \theta
-l_{i},Z_{i}\right) \right\vert ^{2}}\right) _{\left\vert \Psi \right\vert
^{2}=0}\dprod\limits_{i=1}^{n}\left\vert \Psi \left( \theta
-l_{i},Z_{i}\right) \right\vert ^{2}  \label{rgr} \\
&=&\left( \sum_{m=1}^{n}\sum_{i=1}^{m}\sum_{\left(
line_{1},...,line_{m}\right) }\dprod\limits_{i}F\left( line_{i}\right)
\dprod\limits_{B}F\left( B\right) \right) \dprod\limits_{i=1}^{n}\left\vert
\Psi \left( \theta -l_{i},Z_{i}\right) \right\vert ^{2}  \notag
\end{eqnarray}%
The graphical representation is generic. The integration over the set of
lines also accounts for the degenerate case of lines that share some
segments.

\subsection{Summing the series expansion for $\protect\omega \left( J,%
\protect\theta ,Z\right) $ in absence of external source: auxiliary path
integral description}

Having obtained the successive derivatives of $\omega \left( J,\theta
,Z\right) $ in (\ref{rgr}), we can now sum the corresponding series
expansion for $\omega \left( J,\theta ,Z\right) $. Appendix 2 uses formula (%
\ref{rgr})\ to derive a non-local formula for the summation of successive
derivatives of $\omega \left( J,\theta ,Z\right) $ and $\omega ^{-1}\left(
J,\theta ,Z\right) $. Actually, equation (\ref{rgr}) can be reformulated to
calculate the expansion (\ref{snp}) as the sum of graphs for an auxiliary
complex field $\Lambda \left( Z_{i},\theta _{i}\right) $. The computation
organizes the graphs in (\ref{rgr}) so that their sum transforms into a
summation over graphs drawn between an arbitrary number of branching points,
viewed as vertices of arbitrary valence $k$ with an associated factor (\ref%
{fcT}). These vertices are connected by the edges of the graph with
associated Green functions $\frac{1}{1-\left( 1+\left\vert \Psi \right\vert
^{2}\right) \ \check{T}}$ where \ $\check{T}$ is the operator whose kernel
is defined in (\ref{rnL}). The factor $\left\vert \Psi \right\vert ^{2}$ is
the operator multiplication by $\left\vert \Psi \left( \theta ,Z\right)
\right\vert ^{2}$ at point $\left( \theta ,Z\right) $.

Appendix 2 shows that the series expansion for activities has the following
auxiliary path integral form:

\begin{equation}
\omega ^{-1}\left( \theta ,Z\right) =\omega _{0}^{-1}\left( J,\theta
,Z\right) +\frac{\int \ \check{T}\digamma ^{\dag }\left( Z,\theta \right)
\exp \left( -S\left( \digamma \right) -\int \digamma \left( X,\theta \right)
\omega _{0}^{-1}\left( J,\theta ,Z\right) \left\vert \Psi \left( J,\theta
,Z\right) \right\vert ^{2}d\left( X,\theta \right) \right) \mathcal{D}%
\digamma }{\int \exp \left( -S\left( \digamma \right) \right) \mathcal{D}%
\digamma }  \label{ft}
\end{equation}%
where the action for the auxiliary fields $\digamma $ and $\digamma ^{\dag }$
is: 
\begin{eqnarray*}
S\left( \digamma \right)  &=&\int \digamma \left( Z,\theta \right) \left(
1-\left\vert \Psi \right\vert ^{2}\ \check{T}\right) \digamma ^{\dag }\left(
Z,\theta \right) d\left( Z,\theta \right)  \\
&&-\int \digamma \left( Z,\theta \right) \ \check{T}\left( \theta -\frac{%
\left\vert Z-Z^{\left( 1\right) }\right\vert }{c},Z,Z^{\left( 1\right)
},\omega _{0}+\ \check{T}\digamma ^{\dag }\right) \digamma ^{\dag }\left(
Z^{\left( 1\right) },\theta -\frac{\left\vert Z-Z^{\left( 1\right)
}\right\vert }{c}\right) dZdZ^{\left( 1\right) }d\theta 
\end{eqnarray*}%
with:%
\begin{eqnarray*}
&&\ \check{T}\left( \theta -\frac{\left\vert Z^{\left( 1\right)
}-Z\right\vert }{c},Z^{\left( 1\right) },Z,\omega _{0}+\ \check{T}\digamma
^{\dag }\right)  \\
&=&\ \check{T}\left( \theta -\frac{\left\vert Z^{\left( 1\right)
}-Z\right\vert }{c},Z^{\left( 1\right) },Z,\omega _{0}\left( Z,\theta
\right) +\int \ \check{T}\left( \theta -\frac{\left\vert Z-Z^{\left(
1\right) }\right\vert }{c},Z^{\left( 1\right) },Z,\omega _{0}\right)
\digamma ^{\dag }\left( Z^{\left( 1\right) },\theta -\frac{\left\vert
Z-Z^{\left( 1\right) }\right\vert }{c}\right) dZ^{\left( 1\right) }\right) 
\end{eqnarray*}

\subsection{Modification of $\protect\omega \left( J,\protect\theta %
,Z\right) $ due to source terms}

Formula (\ref{ft}) was derived without considering the presence of source
terms in the path integral (\ref{prb}). In Appendix 3, we show that these
terms modify the formula (\ref{ft}) which can be replaced by:

\begin{eqnarray}
&&\omega ^{-1}\left( \theta ,Z\right)   \label{wts} \\
&=&\omega _{0}^{-1}\left( J,\theta ,Z\right)   \notag \\
&&+\frac{\int \check{T}\digamma ^{\dag }\left( Z,\theta \right) \exp \left(
-S\left( \digamma \right) -\int \digamma \left( X,\theta \right) \omega
_{0}^{-1}\left( J,\theta ,Z\right) \left\vert \Psi \left( J,\theta ,Z\right)
\right\vert ^{2}d\left( X,\theta \right) +\sum_{i}a\left( Z_{i},\theta
\right) \frac{\omega _{0}^{-1}\left( J,\theta ,Z_{i}\right) }{\Lambda ^{2}}%
\digamma \left( Z_{i},\theta \right) \right) }{\int \exp \left( -S\left(
\digamma \right) \right) \mathcal{D}\digamma }  \notag
\end{eqnarray}%
This integral will be computed by a saddle path approximation.

\subsection{Saddle path approximation}

We then show that the saddle point approximation yields the equations for $%
\digamma ^{\dag }\left( Z,\theta \right) $ and $\digamma \left( Z,\theta
\right) $: 
\begin{equation}
\left( \left( 1-\left\vert \Psi \right\vert ^{2}\check{T}\right) \digamma
^{\dag }\right) \left( Z,\theta \right) -\left( \check{T}_{\omega _{0}^{-1}+%
\check{T}\digamma ^{\dag }}\digamma ^{\dag }\right) \left( Z,\theta \right)
-\sum_{i}a\left( Z_{i},\theta \right) \frac{\omega _{0}^{-1}\left( J,\theta
,Z_{i}\right) }{\Lambda ^{2}}\delta \left( Z-Z_{i}\right) =0  \label{SDp}
\end{equation}%
\begin{equation*}
\digamma \left( Z,\theta \right) =0
\end{equation*}%
In this approximation, equation (\ref{wts}) for $\omega $. becomes:%
\begin{equation}
\omega \left( J,\theta ,Z\right) =\omega _{0}\left( J,\theta ,Z\right) +%
\check{T}\digamma ^{\dag }\left( Z,\theta \right)   \label{SP}
\end{equation}

\subsection{Series expansion for activities in the perturbated state}

We show in appendix 3 that, in first approximation, we can replace $%
\left\vert \Psi \right\vert ^{2}$ with $\frac{1}{\Lambda }$ in (\ref{SDp}).
When considering perturbations around a static backgrund state, it enables
us to rewrite the saddle point equation (\ref{SDp}) as follows:

\begin{equation*}
\left( \left( 1-\frac{1}{\Lambda }\check{T}\right) \digamma ^{\dag }\right)
\left( Z\right) -\left( \check{T}_{\omega _{0}+\check{T}\digamma ^{\dag
}}\digamma ^{\dag }\right) \left( Z\right) -\left( \sum_{i}a\left(
Z_{i},\theta \right) \frac{\omega _{0}^{-1}\left( J,\theta ,Z_{i}\right) }{%
\Lambda ^{2}}\digamma \left( Z_{i},\theta \right) \right) =0
\end{equation*}%
and to find a recursive solution for $\check{T}\digamma ^{\dag }\left(
Z\right) $ and $\omega \left( J,\theta ,Z\right) $ by rewriting:

\begin{eqnarray}
\check{T}\digamma ^{\dag } &=&\check{T}\frac{1}{\left( 1-\frac{1}{\Lambda }%
\check{T}-\check{T}_{\omega _{0}+\check{T}\digamma ^{\dag }}\right) }\left(
\sum_{i}a\left( Z_{i},\theta \right) \frac{\omega _{0}^{-1}\left( J,\theta
,Z_{i}\right) }{\Lambda ^{2}}\digamma \left( Z_{i},\theta \right) \right) 
\label{srx} \\
&=&\check{T}\frac{1}{\left( 1-\left( 1+\frac{1}{\Lambda }\right) \check{T}%
-\left( \check{T}_{\omega _{0}+\check{T}\digamma ^{\dag }}-\check{T}\right)
\right) }\left( \sum_{i}a\left( Z_{i},\theta \right) \frac{\omega
_{0}^{-1}\left( J,\theta ,Z_{i}\right) }{\Lambda ^{2}}\digamma \left(
Z_{i},\theta \right) \right)   \notag
\end{eqnarray}%
Gathering (\ref{qnn}) and (\ref{qnt}), leads to the recursive formula under
some approximations:%
\begin{eqnarray*}
\check{T}\digamma ^{\dag } &\simeq &\sum_{n_{1},...,n_{2}}\frac{\check{T}}{%
1-\left( 1+\frac{1}{\Lambda }\right) \check{T}}\left[ \left( -\frac{\check{T}%
\digamma ^{\dag }\left( Z_{1}\right) }{\omega _{0}\left( Z_{1}\right) }%
\check{T}\right) ^{n_{1}}\right] \frac{1}{1-\left( 1+\frac{1}{\Lambda }%
\right) \check{T}}\left[ \left( -\frac{\check{T}\digamma ^{\dag }\left(
Z_{2}\right) }{\omega _{0}\left( Z_{2}\right) }\check{T}\right) ^{n_{2}}%
\right]  \\
&&...\frac{1}{1-\left( 1+\frac{1}{\Lambda }\right) \ \check{T}}\left(
-\sum_{i}a\left( Z_{i},\theta \right) \frac{\omega _{0}\left( J,\theta
,Z_{i}\right) }{\Lambda ^{2}}\right) 
\end{eqnarray*}%
We explain in appendix 3 how this formula builds recursively $\check{T}%
\digamma ^{\dag }$. Keeping the lowest order solution of the saddle point in
the local approximation $Z^{\prime }\simeq Z$ leads to:%
\begin{eqnarray}
\ \check{T}\digamma ^{\dag } &=&\frac{\check{T}}{\left( 1-\left( 1+\frac{1}{%
\Lambda }\right) \check{T}\right) }\left[ \sum_{i}a\left( Z_{i},\theta
\right) \frac{\omega _{0}\left( J,\theta ,Z_{i}\right) }{\Lambda ^{2}}\right]
\label{srp} \\
&\equiv &\int K\left( Z,\theta ,Z_{i},\theta _{i}\right) \left\{
\sum_{i}a\left( Z_{i},\theta _{i}\right) \frac{\omega _{0}\left( J,\theta
_{i},Z_{i}\right) }{\Lambda ^{2}}\right\} d\theta _{i}  \notag
\end{eqnarray}%
so that the correction to the background state activities (\ref{wts}) due to
the stimuli become:%
\begin{equation}
\omega ^{-1}\left( J,\theta ,Z\right) =\omega _{0}^{-1}\left( J,\theta
,Z\right) +\int K\left( Z,\theta ,Z_{i},\theta _{i}\right) \left\{
\sum_{i}a\left( Z_{i},\theta _{i}\right) \frac{\omega _{0}\left( J,\theta
_{i},Z_{i}\right) }{\Lambda ^{2}}\right\} d\theta _{i}  \label{WT}
\end{equation}%
Appendix 3 computes an estimation of the second term in the RHS at the
lowest order for oscillating signals $a\left( Z_{i},\theta \right) \propto
a\exp \left( i\varpi \theta \right) $. We obtain:

\begin{eqnarray}
\omega ^{-1}\left( J,\theta ,Z\right) &=&\omega _{0}^{-1}\left( J,\theta
,Z\right) +\frac{a\exp \left( -\left\vert Z-Z_{0}\right\vert \right) }{c%
\sqrt{\left( 1+2\alpha \left\vert Z-Z_{0}\right\vert \right) ^{2}+\left( 
\frac{\varpi }{c}\right) ^{2}}}  \label{ntf} \\
&&\times \exp \left( i\left( \frac{\varpi \left( \left\vert
Z-Z_{0}\right\vert \right) }{c}-\arctan \left( \frac{\varpi }{c\left(
1+2\alpha \left\vert Z-Z_{0}\right\vert \right) }\right) \right) \right)
\sum_{i}\exp \left( i\frac{\varpi \left\vert Z_{i}-Z_{0}\right\vert }{%
c\left\vert Z-Z_{0}\right\vert }\right)  \notag
\end{eqnarray}%
and this terms induces interferences. As a consequence, for a large number
of points $Z_{i}$:%
\begin{equation*}
\sum_{i}\exp \left( i\frac{\varpi \left\vert Z_{i}-Z_{0}\right\vert }{%
c\left\vert Z-Z_{0}\right\vert }\right) \simeq 0
\end{equation*}%
except for the maxima of interferences with magnitude:%
\begin{equation*}
\frac{a\exp \left( -\left\vert Z-Z_{0}\right\vert \right) }{c\sqrt{\left(
1+2\alpha \left\vert Z-Z_{0}\right\vert \right) ^{2}+\left( \frac{\varpi }{c}%
\right) ^{2}}}
\end{equation*}%
Note that for these maxima and $N$ large:%
\begin{equation*}
a=\sum_{i}\exp \left( i\frac{\varpi \left\vert Z_{i}-Z_{0}\right\vert }{%
c\left\vert Z-Z_{0}\right\vert }\right)
\end{equation*}%
is proportional to $N$ so that $a>>1$.

\subsection{Graph expansion for the partition function and modified
background activities}

Once the activities $\omega \left( J,\theta ,Z\right) $ are expressed as a
function of the field, we can substitute their form in (\ref{prb}) to
calculate the background state and equilbrium activity. In appendix 3 the
graphs expansion of (\ref{prb}) is derived with $\omega ^{-1}\left( J,\theta
,Z\right) $ given\ by (\ref{WT}). To achieve this, we expand $\Psi \left(
\theta ,Z_{1}\right) $ around a quasi-static background state\footnote{%
As quoted previously, the series expansion for activities is valid for a non
constant background.} $\Psi _{0}\left( \theta ,Z_{1}\right) $ and $%
\left\vert \Psi \left( \theta ,Z_{1}\right) \right\vert ^{2}$ around $%
\left\vert \Psi _{0}\left( Z_{1}\right) \right\vert ^{2}$\ so that we
replace:%
\begin{equation*}
\Psi \left( \theta ,Z_{1}\right) \rightarrow \Psi _{0}\left( \theta
,Z_{1}\right) +\Psi \left( \theta ,Z_{1}\right) 
\end{equation*}%
and:

\begin{equation*}
\left\vert \Psi \left( \theta ,Z_{1}\right) \right\vert ^{2}\rightarrow
\left\vert \Psi _{0}\left( Z_{1}\right) \right\vert ^{2}+\Psi _{0}^{\dagger
}\left( Z_{1}\right) \Psi \left( \theta ,Z_{1}\right) +\Psi _{0}\left(
Z_{1}\right) \Psi ^{\dagger }\left( \theta ,Z_{1}\right) +\left\vert \Psi
\left( \theta ,Z_{1}\right) \right\vert ^{2}
\end{equation*}%
We are thus left with the following form for the path integral with external
perturbations:%
\begin{eqnarray*}
&&\int \exp \left( -S\left( \Psi \right) \right) \int \exp \left(
\sum_{i}a\left( Z_{i},\theta _{0}\right) \left\vert \Psi \left( Z_{i},\theta
_{0}\right) \right\vert ^{2}\right) d\theta _{0} \\
&\equiv &\int \exp \left( \left( \frac{1}{2}\left( \Psi _{0}^{\dagger
}\left( \theta ,Z\right) +\Psi ^{\dagger }\left( \theta ,Z\right) \right)
\nabla \left( \frac{\sigma _{\theta }^{2}}{2}\nabla -\left( \omega _{0}^{-1}-%
\frac{\Omega \left( \theta ,\theta _{0},Z\right) }{\omega _{0}^{2}\left(
Z\right) }\right) \right) \left( \Psi _{0}\left( \theta ,Z\right) +\Psi
\left( \theta ,Z\right) \right) \right) \right) d\theta _{0}
\end{eqnarray*}%
with the correction $\Omega \left( \theta ,\theta _{0},Z\right) $ to the
activities computed by (\ref{srp}):%
\begin{equation*}
\Omega \left( \theta ,\theta _{0},Z\right) =-\sum_{i}\omega _{0}^{2}\left(
Z\right) K\left( Z,\theta ,Z_{i},\theta _{0}\right) \left\{ a\left(
Z_{i},\theta _{0}\right) \frac{\omega _{0}\left( \theta _{0},Z_{0}\right) }{%
\Lambda ^{2}}\right\}
\end{equation*}%
The expansion of:%
\begin{equation}
\exp \int \left( \Psi _{0}^{\dagger }\left( \theta ,Z\right) +\Psi ^{\dagger
}\left( \theta ,Z\right) \right) \nabla _{\theta }\left( \frac{\Omega \left(
\theta ,\theta _{0},Z\right) }{\omega _{0}^{2}\left( Z\right) }\left( \Psi
_{0}\left( \theta ,Z\right) +\Psi \left( \theta ,Z\right) \right) \right)
\label{crp}
\end{equation}%
is performed in appendix 3. We specifically focus on the second-order
expansion to identify the first order corrections to activities only,
although higher orders can be determined using a similar approach. Our
derivation shows that at this level of approximation and under the condition
that $\left\vert \Psi _{0}\left( \theta ,Z\right) \right\vert ^{2}>>\frac{1}{%
\Lambda }$, formula (\ref{crp}) becomes:%
\begin{equation*}
\exp \left( A+B-\frac{1}{2}A^{2}\right)
\end{equation*}%
where:%
\begin{equation*}
A=\frac{1}{\Lambda _{1}\Lambda \omega _{0}^{4}\left( Z\right) }\int \left(
\Psi _{0}^{\dagger }\left( \theta ,Z\right) \nabla \left( 2\left( \left(
\int^{\theta }\int \left( \nabla \Omega \left( \theta ,\theta _{0},Z\right)
\right) ^{2}d\theta _{0}\right) -\Lambda _{1}\left( \int \Omega ^{2}d\theta
_{0}\right) \Psi _{0}\left( \theta ,Z\right) \right) \right) \right) dZ
\end{equation*}%
and:%
\begin{equation*}
B=\left( \int \frac{\Psi _{0}^{\dagger }\left( \theta ,Z\right) }{\omega
_{0}^{4}\left( Z\right) }\sqrt{\int \left( \nabla \Omega \left( \theta
,\theta _{0},Z\right) \right) ^{2}d\theta _{0}}\Psi _{0}\left( \theta
,Z\right) dZ\right) ^{2}
\end{equation*}

The term $B-\frac{1}{2}A^{2}$ is a correction to the potential for the
background field $\Psi _{0}\left( \theta ,Z\right) $. It should thus modify
this background, but this can be neglected in first approximation.

The correction to the activities comes from the term $A$ and leads to switch
the equilibrium inverse activities:%
\begin{equation*}
\omega _{0}^{-1}\left( Z\right) \rightarrow \omega _{0}^{-1}\left( Z\right) -%
\frac{A\omega _{0}^{2}\left( Z\right) }{\omega _{0}^{2}\left( Z\right) }
\end{equation*}%
or, which is equivalent:%
\begin{equation*}
\omega _{0}\left( Z\right) \rightarrow \omega _{0}\left( Z\right) +A\omega
_{0}^{2}\left( Z\right)
\end{equation*}%
In a developped form, the equilibrium activities are modified by the term:%
\begin{equation*}
\omega _{0}\left( Z\right) \rightarrow \omega _{0}\left( Z\right) +\frac{1}{%
\Lambda _{1}\Lambda \omega _{0}^{4}\left( Z\right) }\left( 2\left( \left(
\int^{\theta }\int \left( \nabla \Omega \left( \theta ,\theta _{0},Z\right)
\right) ^{2}d\theta _{0}\right) -\Lambda _{1}\left( \int \Omega ^{2}d\theta
_{0}\right) \right) \right)
\end{equation*}%
for a given background field $\Psi _{0}\left( \theta ,Z\right) $.

\subsection{Interferences}

As explained after formula (\ref{ntf}), the corrections $\Omega \left(
\theta ,\theta _{0},Z\right) $ can be considered equal to zero outside the
points of maximal interferences. At these points $\Omega \left( \theta
,\theta _{0},Z\right) =$\ $\check{T}\digamma ^{\dag }$ is proportional to: 
\begin{equation*}
\bar{\Omega}=\frac{a\exp \left( -\left\vert Z-Z_{0}\right\vert \right) }{c%
\sqrt{\left( 1+2\alpha \left\vert Z-Z_{0}\right\vert \right) ^{2}+\left( 
\frac{\varpi }{c}\right) ^{2}}}
\end{equation*}%
and the correction to activities are:%
\begin{eqnarray*}
\omega _{0}\left( Z\right)  &\rightarrow &\omega _{0}\left( Z\right) +\frac{%
2\left( \left( \int^{\theta }\int \left( \varpi \bar{\Omega}\right)
^{2}d\theta _{0}\right) -\Lambda _{1}\left( \int \bar{\Omega}^{2}d\theta
_{0}\right) \right) }{\Lambda _{1}\Lambda \omega _{0}^{2}\left( Z\right) } \\
&\simeq &\omega _{0}\left( Z\right) +\frac{2\left( \left( T_{\theta }\left(
\varpi \bar{\Omega}\right) ^{2}\right) -\Lambda _{1}\left( \bar{\Omega}%
^{2}\right) \right) T_{\theta }}{\Lambda _{1}\Lambda \omega _{0}^{2}\left(
Z\right) }
\end{eqnarray*}%
where $T_{\theta }$ is the duration of the signals at time $\theta $. Note
that for $T_{\theta }\varpi ^{2}>\Lambda _{1}$, i.e. for long enough
stimulation:%
\begin{equation*}
\frac{2\left( \left( T_{\theta }\left( \varpi \bar{\Omega}\right)
^{2}\right) -\Lambda _{1}\left( \bar{\Omega}^{2}\right) \right) T_{\theta }}{%
\Lambda _{1}\Lambda \omega _{0}^{2}\left( Z\right) }
\end{equation*}%
Moreover, since $a>>1$, we also have $\bar{\Omega}>>1$ so that:%
\begin{equation}
\omega _{0}\left( Z\right) +\frac{2\left( \left( T_{\theta }\left( \varpi 
\bar{\Omega}\right) ^{2}\right) -\Lambda _{1}\left( \bar{\Omega}^{2}\right)
\right) T_{\theta }}{\Lambda _{1}\Lambda \omega _{0}^{2}\left( Z\right) }%
>>\omega _{0}\left( Z\right)   \label{MD}
\end{equation}%
and the stimulated cells have a much higher activity than others.

\section{Implication of modified activities on the emergence of bound states}

The modification in the background activity (\ref{MD}) allows to recover the
existence of bound states as a consequence of interferences. Actually, given
the formula for connectivity functions (\ref{Cnv}):%
\begin{eqnarray*}
\left\langle T\left( Z,Z^{\prime }\right) \right\rangle &=&\lambda \tau
\left\langle \hat{T}\left( Z,Z^{\prime }\right) \right\rangle \\
&=&\frac{\lambda \tau h\left( Z,Z^{\prime }\right) \left\langle
C_{Z,Z^{\prime }}\left( \theta \right) h_{C}\left( \omega \left( \theta
,Z\right) \right) \left\vert \Psi \left( \theta ,Z\right) \right\vert
^{2}\right\rangle }{C_{Z,Z^{\prime }}\left( \theta \right) \left\vert \Psi
\left( \theta ,Z\right) \right\vert ^{2}h_{C}\left( \omega \left( \theta
,Z\right) \right) +D_{Z,Z^{\prime }}\left( \theta \right) \left\vert \Psi
\left( \theta -\frac{\left\vert Z-Z^{\prime }\right\vert }{c},Z^{\prime
}\right) \right\vert ^{2}h_{D}\left( \omega \left( \theta -\frac{\left\vert
Z-Z^{\prime }\right\vert }{c},Z^{\prime }\right) \right) }
\end{eqnarray*}

So that: 
\begin{equation}
\left\langle T\left( Z,Z^{\prime }\right) \right\rangle =0  \label{rcn}
\end{equation}
if $Z^{\prime }$ belongs to the maxima of interferences, but $Z$ does not.
Moreover, if both $Z$ and $Z^{\prime }$ belong to these maxima:%
\begin{equation}
\left\langle T\left( Z,Z^{\prime }\right) \right\rangle \simeq \left\langle
T\left( Z^{\prime },Z\right) \right\rangle  \label{rcm}
\end{equation}

As a consequence these maxima bind together, in a reciprocal manner. They
form a connected set, relatively disconnected from the rest of the thread.
The description of these sets has been given in the first article (\cite{GLr}%
) of this work.

\part*{III Effective formalism for connectivity field}

The preceding section has recovered the findings of the first part of this
work (\cite{GLr}). Certain bound states may emerge due to positive
interferences in activity. To delve deeper into the dynamics of such states,
we establish an effective field formalism for the connectivity field $\Gamma 
$. We also investigate several applications.

\section{Effective field formalism approach to connectivity functions
transitions}

We investigate the modifications in activities, as derived in the preceding
section, and analyze their influence on the background field of
connectivities. Our focus lies in understanding the dynamic processes
involved in transitioning between different background fields, which arises
from perturbations in neural activity. To achieve this, we employ an
effective action for the connectivity field, which is computed through an
expansion of the system's action around a specific background. This
expansion captures a situation in which the background field has been
perturbed by external influences. Consequently, the actual state of the
system at the moment of the transition which is still characterized by the
previous background field, differs from the new equilibrium defined by the
new background field and may undergo a transition to a new state. We
leverage our effective formalism to quantitatively compute these transitions.

It is worth noting that, in this section, we consider the modification in
activities as exogenous, primarily resulting from the sources. However, in
the third part of this study, we will adopt a more comprehensive
perspective, considering that the change in activities is itself contingent
on the connectivity field.

\subsection{Modified action for $\Gamma \left( T,\hat{T},C,D\right) $}

In the perturbated state considered in the previous section, the equilibrium
activity has been shifted by an amount of $\delta \omega _{0}$ and the
action of the system can be approximated by:%
\begin{eqnarray}
&&-\frac{1}{2}\int \Psi _{0}^{\dagger }\left( \theta ,Z\right) \nabla \left( 
\frac{\sigma _{\theta }^{2}}{2}\nabla -\left( \omega _{0}+\delta \omega
_{0}\right) ^{-1}\right) \Psi _{0}\left( \theta ,Z\right) +\int V\left( \Psi
_{0}\left( \theta ,Z\right) \right) +\delta V\left( \Psi _{0}\left( \theta
,Z\right) \right)  \label{hts} \\
&&+\sum_{i=1}^{4}S_{\Gamma }^{\left( i\right) }\left( \Gamma \left( T,\hat{T}%
,\theta ,Z,Z^{\prime },C,D\right) \right)  \notag
\end{eqnarray}%
where $\Psi _{0}\left( \theta ,Z\right) $ is the neuron field equilibrium
background state for $\Psi \left( \theta ,Z\right) $. The deviation in
activity and potential have been obtained in the previous section:%
\begin{equation}
\delta \omega _{0}\propto \frac{2\left( \left( T\left( \varpi \bar{\Omega}%
\right) ^{2}\right) -\Lambda _{1}\left( \bar{\Omega}^{2}\right) \right)
T_{\theta }}{\Lambda _{1}\Lambda \omega _{0}^{2}\left( Z\right) }
\label{frs}
\end{equation}%
and:%
\begin{eqnarray}
&&\delta V\left( \Psi _{0}\left( \theta ,Z\right) \right)  \label{pts} \\
&=&\left( \int \frac{\Psi _{0}^{\dagger }\left( \theta ,Z\right) }{\omega
_{0}^{4}\left( Z\right) }\sqrt{\int \left( \nabla \Omega \left( \theta
,\theta _{0},Z\right) \right) ^{2}d\theta _{0}}\Psi _{0}\left( \theta
,Z\right) dZ\right) ^{2}  \notag \\
&&-\frac{1}{2}\left( \frac{1}{\Lambda _{1}\Lambda \omega _{0}^{4}\left(
Z\right) }\int \left( \Psi _{0}^{\dagger }\left( \theta ,Z\right) \nabla
\left( 2\left( \left( \int^{\theta }\int \left( \nabla \Omega \left( \theta
,\theta _{0},Z\right) \right) ^{2}d\theta _{0}\right) -\Lambda _{1}\left(
\int \Omega ^{2}d\theta _{0}\right) \Psi _{0}\left( \theta ,Z\right) \right)
\right) \right) dZ\right) ^{2}  \notag
\end{eqnarray}

The form of the effective action (\ref{hts}) is justified by the distinct
time scales governing neuronal activities and connectivities. In first
approximation, equilibrium shifts in activities occur before we need to
consider the dynamics of connectivities.

The modification in potential and equilibrium activities, as described in
Equations (\ref{frs}) and (\ref{pts})\ are expected to impact the background
state $\Psi _{0}\left( \theta ,Z\right) $. Nonetheless, given that the shift
is localized at specific points, and considering that this potential
characterizes some collective configuration, we can assume, in first
approximation, that $\Psi _{0}\left( \theta ,Z\right) $ remains unaffected
by the signals. Consequently, the new background field for the connectivity
field is shifted only at the points where $\Omega \left( \theta ,\theta
_{0},Z\right) \neq 0$ and remains unchanged elsewhere.

This change is induced by the activities. Actually, in the perturbed state,
after integrating over $\Psi _{0}\left( \theta ,Z\right) $ the system is
described by an effective action for the connectivity functions $%
\sum_{i=1}^{4}S_{\Gamma }^{\left( i\right) }$ where $\left\vert \Psi \left(
Z\right) \right\vert ^{2}$ is set to $\left\vert \Psi _{0}\right\vert ^{2}$
and $\omega \left( J,\theta ,Z,\left\vert \Psi \right\vert ^{2}\right) $ to $%
\omega _{0}\left( Z\right) +\Delta \omega _{0}\left( Z,\left\vert \Psi
\right\vert ^{2}\right) $ where $\left\vert \Psi _{0}\right\vert ^{2}$ and $%
\omega _{0}\left( Z\right) $\ correspond to the initial background state.

For $\frac{1}{\tau _{C}\alpha _{C}}<<1$ and $\frac{1}{\tau _{D}\alpha _{D}}%
<<1$ we can assume that $C$ and $D$ are close to $\frac{1}{2}$ and we write
the effective action for $\Gamma \left( T,\hat{T},C,D\right) $ by replacing $%
C$ and $D$ with their averages: 
\begin{eqnarray}
C &\rightarrow &\left\langle C\left( \theta \right) \right\rangle =\frac{%
\alpha _{C}\omega ^{\prime }\left\langle \left\vert \Psi \left( \theta -%
\frac{\left\vert Z-Z^{\prime }\right\vert }{c},Z^{\prime }\right)
\right\vert ^{2}\right\rangle }{\frac{1}{\tau _{C}}+\alpha _{C}\omega
^{\prime }\left\langle \left\vert \Psi \left( \theta -\frac{\left\vert
Z-Z^{\prime }\right\vert }{c},Z^{\prime }\right) \right\vert
^{2}\right\rangle }\equiv C\left( \theta \right) \\
D &\rightarrow &\left\langle D\left( \theta \right) \right\rangle =\frac{%
\alpha _{D}\omega \left\langle \left\vert \Psi \left( \theta ,Z\right)
\right\vert ^{2}\right\rangle }{\frac{1}{\tau _{D}}+\alpha _{D}\omega
\left\langle \left\vert \Psi \left( \theta ,Z\right) \right\vert
^{2}\right\rangle }\equiv D\left( \theta \right)
\end{eqnarray}%
so that the effective action becomes:%
\begin{eqnarray}
&&S\left( \Gamma \left( T,\hat{T},\theta ,Z,Z^{\prime },C,D\right) \right)
\label{fct} \\
&=&\Gamma ^{\dag }\left( T,\hat{T},\theta ,Z,Z^{\prime },C,D\right) \left[
\nabla _{T}\left( \nabla _{T}-\left( \frac{\left( -T+\lambda \hat{T}\right) 
}{\tau \omega _{0}\left( Z\right) +\Delta \omega _{0}\left( Z,\left\vert
\Psi \right\vert ^{2}\right) }\right) \left\vert \Psi \left( \theta
,Z\right) \right\vert ^{2}\right) \right.  \notag \\
&&+\nabla _{\hat{T}}\left( \nabla _{\hat{T}}-\rho \left( \left( h\left(
Z,Z^{\prime }\right) -\hat{T}\right) C\left\vert \Psi _{0}\left( Z\right)
\right\vert ^{2}\frac{h_{C}\left( \omega _{0}\left( Z\right) +\Delta \omega
_{0}\left( Z,\left\vert \Psi \right\vert ^{2}\right) \right) }{\omega
_{0}\left( Z\right) +\Delta \omega _{0}\left( Z,\left\vert \Psi \right\vert
^{2}\right) }\right. \right.  \notag \\
&&\left. \left. -\eta H\left( \delta -T\right) -D\hat{T}\left\vert \Psi
_{0}\left( Z^{\prime }\right) \right\vert ^{2}\frac{h_{D}\left( \omega
_{0}\left( Z^{\prime }\right) +\Delta \omega _{0}\left( Z^{\prime
},\left\vert \Psi \right\vert ^{2}\right) \right) }{\omega _{0}\left(
Z\right) +\Delta \omega _{0}\left( Z,\left\vert \Psi \right\vert ^{2}\right) 
}\right) \right) \Gamma \left( T,\hat{T},\theta ,Z,Z^{\prime },C,D\right) 
\notag
\end{eqnarray}%
The minimization of this effective action yields a background field similar
to the one derived using (\ref{RC}), but with modified averages:%
\begin{eqnarray}
\left\langle T\left( Z,Z^{\prime }\right) \right\rangle &=&\lambda \tau
\left\langle \hat{T}\left( Z,Z^{\prime }\right) \right\rangle \\
&=&\frac{\lambda \tau h\left( Z,Z^{\prime }\right) C_{Z,Z^{\prime
}}h_{C}\left( \omega _{0}\left( Z\right) +\Delta \omega _{0}\left( Z\right)
\right) \left\vert \Psi _{0}\left( Z\right) \right\vert ^{2}}{C_{Z,Z^{\prime
}}\left\vert \Psi _{0}\left( Z\right) \right\vert ^{2}h_{C}\left( \omega
\left( Z\right) \right) +D_{Z,Z^{\prime }}\left\vert \Psi _{0}\left(
Z^{\prime }\right) \right\vert ^{2}h_{D}\left( \omega _{0}\left( Z^{\prime
}\right) \right) }  \notag
\end{eqnarray}%
for $\left( Z,Z^{\prime }\right) $ an $"a"$ (active) doublet, and:%
\begin{equation*}
\left\langle T\left( Z,Z^{\prime }\right) \right\rangle =0
\end{equation*}%
\begin{equation}
\left\langle \hat{T}\left( Z,Z^{\prime }\right) \right\rangle =\frac{h\left(
Z,Z^{\prime }\right) C_{Z,Z^{\prime }}h_{C}\left( \omega _{0}\left( Z\right)
+\Delta \omega _{0}\left( Z\right) \right) \left\vert \Psi _{0}\left(
Z\right) \right\vert ^{2}-\eta }{C_{Z,Z^{\prime }}\left\vert \Psi _{0}\left(
Z\right) \right\vert ^{2}h_{C}\left( \omega \left( Z\right) \right)
+D_{Z,Z^{\prime }}\left\vert \Psi _{0}\left( Z^{\prime }\right) \right\vert
^{2}h_{D}\left( \omega _{0}\left( Z^{\prime }\right) \right) }<0
\end{equation}%
for an $"u"$ (unactive) doublet.

Ultimately, remark that the modification:%
\begin{equation*}
\omega _{0}\left( Z\right) +\Delta \omega _{0}\left( Z\right)
\end{equation*}%
may induce some switches in connections in the new background field .
Actually, if:%
\begin{eqnarray*}
h\left( Z,Z^{\prime }\right) C_{Z,Z^{\prime }}h_{C}\left( \omega _{0}\left(
Z\right) \right) \left\vert \Psi _{0}\left( Z\right) \right\vert ^{2}-\eta
&<&0 \\
h\left( Z,Z^{\prime }\right) C_{Z,Z^{\prime }}h_{C}\left( \omega _{0}\left(
Z\right) +\Delta \omega _{0}\left( Z\right) \right) \left\vert \Psi
_{0}\left( Z\right) \right\vert ^{2}-\eta &>&0
\end{eqnarray*}%
the connection becomes active, i.e. a connection is created between $Z$ and $%
Z^{\prime }$. On the other hand , if:%
\begin{eqnarray*}
h\left( Z,Z^{\prime }\right) C_{Z,Z^{\prime }}h_{C}\left( \omega _{0}\left(
Z\right) \right) \left\vert \Psi _{0}\left( Z\right) \right\vert ^{2}-\eta
&>&0 \\
h\left( Z,Z^{\prime }\right) C_{Z,Z^{\prime }}h_{C}\left( \omega _{0}\left(
Z\right) +\Delta \omega _{0}\left( Z\right) \right) \left\vert \Psi
_{0}\left( Z\right) \right\vert ^{2}-\eta &<&0
\end{eqnarray*}%
the connection may be deleted in the new background.

\subsection{Expansion around the background state $\Gamma _{0}\left( T,\hat{T%
},\protect\theta ,Z,Z^{\prime },C,D\right) $ and effective action}

Dynamicaly, the transition between states is achieved by expanding (\ref{fct}%
) around the new background state, after perturbation. The expansion be will
subsequently used to compute the transition functions. The threshold term $%
\eta H\left( \delta -T\right) $ will be neglected in the sequel to consider
only the active connections. The field is expanded as:%
\begin{eqnarray*}
\Gamma \left( T,\hat{T},\theta ,Z,Z^{\prime },C,D\right) &=&\Gamma
_{0}\left( T,\hat{T},\theta ,Z,Z^{\prime },C,D\right) +\Delta \Gamma \left(
T,\hat{T},\theta ,Z,Z^{\prime },C,D\right) \\
\Gamma ^{\dag }\left( T,\hat{T},\theta ,Z,Z^{\prime },C,D\right) &=&\Gamma
_{0}^{\dag }\left( T,\hat{T},\theta ,Z,Z^{\prime },C,D\right) +\Delta \Gamma
^{\dagger }\left( T,\hat{T},\theta ,Z,Z^{\prime },C,D\right)
\end{eqnarray*}%
and the action writes:%
\begin{eqnarray}
&&S\left( \Gamma \left( T,\hat{T},\theta ,Z,Z^{\prime },C,D\right) \right)
\label{fcm} \\
&=&S\left( \Gamma _{0}\left( T,\hat{T},\theta ,Z,Z^{\prime },C,D\right)
\right) +S_{e}\left( \Delta \Gamma \left( T,\hat{T},\theta ,Z,Z^{\prime
},C,D\right) \right)  \notag
\end{eqnarray}%
with:%
\begin{eqnarray}
&&S_{e}\left( \Delta \Gamma \left( T,\hat{T},\theta ,Z,Z^{\prime
},C,D\right) \right)  \label{sft} \\
&=&\Delta \Gamma ^{\dag }\left( T,\hat{T},\theta ,Z,Z^{\prime },C,D\right) %
\left[ \nabla _{T}\left( \nabla _{T}-\frac{\left( \lambda \left( \hat{T}%
-\left\langle \hat{T}\right\rangle \right) -\left( T-\left\langle
T\right\rangle \right) \right) }{\tau \omega _{0}\left( Z\right) +\Delta
\omega _{0}\left( Z,\left\vert \Psi \right\vert ^{2}\right) }\left\vert \Psi
_{0}\left( Z\right) \right\vert ^{2}\right) \right.  \notag \\
&&+\nabla _{\hat{T}}\left( \nabla _{\hat{T}}+\rho \left( C\frac{\left\vert
\Psi _{0}\left( Z\right) \right\vert ^{2}h_{C}\left( \omega _{0}\left(
Z\right) +\Delta \omega _{0}\left( Z,\left\vert \Psi \right\vert ^{2}\right)
\right) }{\omega _{0}\left( Z\right) +\Delta \omega _{0}\left( Z,\left\vert
\Psi \right\vert ^{2}\right) }\right. \right.  \notag \\
&&\left. \left. +D\frac{\left\vert \Psi _{0}\left( Z^{\prime }\right)
\right\vert ^{2}h_{D}\left( \omega _{0}\left( Z^{\prime }\right) +\Delta
\omega _{0}\left( Z^{\prime },\left\vert \Psi \right\vert ^{2}\right)
\right) }{\omega _{0}\left( Z\right) +\Delta \omega _{0}\left( Z,\left\vert
\Psi \right\vert ^{2}\right) }\right) \left( \hat{T}-\left\langle \hat{T}%
\right\rangle \right) \right) \Delta \Gamma \left( T,\hat{T},\theta
,Z,Z^{\prime },C,D\right)  \notag
\end{eqnarray}

\subsection{Individual transition functions}

The effective action (\ref{fcm}) enables the dynamic study of transitions
between different connectivity states. This is based on the calculation of
transition functions for individual states. We write a final state defined
by some given values $\left( T,\hat{T},\theta ,C,D\right) _{f}$ \ between $Z$
and $Z^{\prime }$ as:%
\begin{equation*}
\left. \left( T,\hat{T},\theta ,Z,Z^{\prime },C,D\right) _{f}\right\rangle
\end{equation*}%
and an initial state with given values $\left( T,\hat{T},\theta ,C,D\right)
_{i}$ as:%
\begin{equation*}
\left\langle \left( T,\hat{T},\theta ,Z,Z^{\prime },C,D\right) _{i}\right.
\end{equation*}%
The computation of transition function between two states, initial and
final, is obtained by computing the Green functions:%
\begin{eqnarray*}
&&\left\langle \left( T,\hat{T},\theta ,Z,Z^{\prime },C,D\right) _{i}\right.
\left. \left( T,\hat{T},\theta ,Z,Z^{\prime },C,D\right) _{f}\right\rangle \\
&=&\int \Delta \Gamma \left( \left( T,\hat{T},\theta ,Z,Z^{\prime
},C,D\right) _{f}\right) \Delta \Gamma ^{\dagger }\left( \left( T,\hat{T}%
,\theta ,Z,Z^{\prime },C,D\right) _{i}\right) \\
&&\times \exp \left( -S_{e}\left( \Delta \Gamma \left( T,\hat{T},\theta
,Z,Z^{\prime },C,D\right) \right) +\alpha \left\vert \Delta \Gamma \left( T,%
\hat{T},\theta ,Z,Z^{\prime },C,D\right) \right\vert ^{2}\right) \mathcal{D}%
\Delta \Gamma
\end{eqnarray*}%
where $\alpha $ is the inverse average time of transition of the system as
we will see below.

Given the form of $S_{e}\left( \Delta \Gamma \left( T,\hat{T},\theta
,Z,Z^{\prime },C,D\right) \right) $, the integral is given by:%
\begin{eqnarray*}
&&\left\langle \left( T,\hat{T},\theta ,Z,Z^{\prime },C,D\right) _{i}\right.
\left. \left( T,\hat{T},\theta ,Z,Z^{\prime },C,D\right) _{f}\right\rangle \\
&=&\left\langle \left( T,\hat{T},\theta ,Z,Z^{\prime },C,D\right)
_{i}\right. \frac{1}{\alpha +O}\left. \left( T,\hat{T},\theta ,Z,Z^{\prime
},C,D\right) _{f}\right\rangle
\end{eqnarray*}%
where $O$ is the operator:%
\begin{eqnarray}
O &=&\nabla _{T}\left( \nabla _{T}+\frac{\left( T-\left\langle
T\right\rangle \right) -\left( \lambda \left( \hat{T}-\left\langle \hat{T}%
\right\rangle \right) \right) }{\tau \omega _{0}\left( Z\right) +\Delta
\omega _{0}\left( Z,\left\vert \Psi \right\vert ^{2}\right) }\left\vert \Psi
_{0}\left( Z\right) \right\vert ^{2}\right)  \label{DF} \\
&&+\nabla _{\hat{T}}\left( \nabla _{\hat{T}}+\rho \left( C\frac{\left\vert
\Psi _{0}\left( Z\right) \right\vert ^{2}h_{C}\left( \omega _{0}\left(
Z\right) +\Delta \omega _{0}\left( Z,\left\vert \Psi \right\vert ^{2}\right)
\right) }{\omega _{0}\left( Z\right) +\Delta \omega _{0}\left( Z,\left\vert
\Psi \right\vert ^{2}\right) }\right. \right.  \notag \\
&&\left. \left. +D\frac{\left\vert \Psi _{0}\left( Z^{\prime }\right)
\right\vert ^{2}h_{D}\left( \omega _{0}\left( Z^{\prime }\right) +\Delta
\omega _{0}\left( Z^{\prime },\left\vert \Psi \right\vert ^{2}\right)
\right) }{\omega _{0}\left( Z\right) +\Delta \omega _{0}\left( Z,\left\vert
\Psi \right\vert ^{2}\right) }\right) \right) \left( \hat{T}-\left\langle 
\hat{T}\right\rangle \right)  \notag
\end{eqnarray}%
To interpret the formulas in terms of time transition, we write also:%
\begin{eqnarray*}
&&\left\langle \left( T,\hat{T},\theta ,Z,Z^{\prime },C,D\right)
_{i}\right\vert \left\vert \left( T,\hat{T},\theta ,Z,Z^{\prime },C,D\right)
_{f}\right\rangle \\
&=&\left\langle \left( T,\hat{T},\theta ,Z,Z^{\prime },C,D\right)
_{i}\right\vert \int_{0}^{\infty }\exp \left( -\left( \alpha +O\right)
t\right) dt\left\vert \left( T,\hat{T},\theta ,Z,Z^{\prime },C,D\right)
_{f}\right\rangle \\
&=&\int_{0}^{\infty }\exp \left( -\alpha t\right) \left\langle \left( T,\hat{%
T},\theta ,Z,Z^{\prime },C,D\right) _{i}\right\vert \exp \left( -\left(
\alpha +O\right) t\right) \left\vert \left( T,\hat{T},\theta ,Z,Z^{\prime
},C,D\right) _{f}\right\rangle dt
\end{eqnarray*}%
That is $\left\langle \left( T,\hat{T},\theta ,Z,Z^{\prime },C,D\right)
_{i}\right. \left. \left( T,\hat{T},\theta ,Z,Z^{\prime },C,D\right)
_{f}\right\rangle $ is the Laplace transform of the time transition between
two states defined by the operator $O$. This justifies the interpretation of 
$\alpha $ as the inverse of an average transition time. Moreover, this shows
that the probabilities of transition of the system are defined by $O$.
Before the Laplace transform, the probability of transition of the system
between two states during a time span $t$ is given by:%
\begin{eqnarray}
&&P_{t}\left( \left( T,\hat{T},\theta ,Z,Z^{\prime },C,D\right) _{i},\left(
T,\hat{T},\theta ,Z,Z^{\prime },C,D\right) _{f}\right)  \label{NL} \\
&=&\left\langle \left( T,\hat{T},\theta ,Z,Z^{\prime },C,D\right)
_{i}\right\vert \exp \left( -Ot\right) \left\vert \left( T,\hat{T},\theta
,Z,Z^{\prime },C,D\right) _{f}\right\rangle  \notag
\end{eqnarray}%
This probability satisfies a differential equation given in appendix 4. We
show in this appendix that the transition between $\mathbf{T-}\left\langle 
\mathbf{T}\right\rangle $ and $\mathbf{T}^{\prime }\mathbf{-}\left\langle 
\mathbf{T}\right\rangle $ during a time $t$, written $G_{0}\left( \mathbf{T-}%
\left\langle \mathbf{T}\right\rangle ,\mathbf{T}^{\prime }\mathbf{-}%
\left\langle \mathbf{T}\right\rangle ,t\right) $, is given by:%
\begin{eqnarray}
&&G_{0}\left( \mathbf{T-}\left\langle \mathbf{T}\right\rangle ,\mathbf{T}%
^{\prime }\mathbf{-}\left\langle \mathbf{T}\right\rangle ,t\right)
\label{trss} \\
&=&\left( 2\pi \right) ^{-1}\left( Det\left( \sigma \left( t\right) \right)
\right) ^{-\frac{1}{2}}  \notag \\
&&\times \exp \left( -\left( \left( \mathbf{T-}\left\langle \mathbf{T}%
\right\rangle \right) -M\left( t\right) \left( \mathbf{T}^{\prime }\mathbf{-}%
\left\langle \mathbf{T}\right\rangle \right) \right) ^{t}\frac{\sigma
^{-1}\left( t\right) }{2}\left( \left( \mathbf{T-}\left\langle \mathbf{T}%
\right\rangle \right) -M\left( t\right) \left( \mathbf{T}^{\prime }\mathbf{-}%
\left\langle \mathbf{T}\right\rangle \right) \right) \right)  \notag
\end{eqnarray}%
where the matrices $M\left( t\right) $ and $\sigma \left( t\right) $ are
defined in appendix 4.

For large $t$, the transition simplifies and writes:%
\begin{eqnarray}
G_{0}\left( \mathbf{T-}\left\langle \mathbf{T}\right\rangle ,\mathbf{T}%
^{\prime }\mathbf{-}\left\langle \mathbf{T}\right\rangle \right) &=&\left(
2\pi \right) ^{-1}\left( Det\left( \sigma \left( \infty \right) \right)
\right) ^{-\frac{1}{2}}  \label{tsrr} \\
&&\times \exp \left( -\frac{1}{2}\left( \left( \mathbf{T-}\left\langle 
\mathbf{T}\right\rangle \right) \right) ^{t}\sigma ^{-1}\left( \infty
\right) \left( \left( \mathbf{T-}\left\langle \mathbf{T}\right\rangle
\right) \right) \right)  \notag
\end{eqnarray}%
with:%
\begin{equation*}
\sigma \left( \infty \right) =\left( 
\begin{array}{cc}
\frac{1}{u}+\frac{s^{2}}{uv\left( u+v\right) } & -\frac{s}{v\left(
u+v\right) } \\ 
-\frac{s}{v\left( u+v\right) } & \frac{1-e^{-2tv}}{v}%
\end{array}%
\right)
\end{equation*}

\subsection{$N$ states transition functions}

Formula (\ref{trss}) generalizes directly for the transition of $N$ states: 
\begin{equation*}
\left( \mathbf{T}\left( Z_{i},Z_{j}\right) \mathbf{-}\left\langle \mathbf{T}%
\left( Z_{i},Z_{j}\right) \right\rangle \right) \equiv \left( \mathbf{T-}%
\left\langle \mathbf{T}\right\rangle \right) _{ij}\equiv \mathbf{T}_{ij}%
\mathbf{-}\left\langle \mathbf{T}\right\rangle _{ij}
\end{equation*}%
located at different points $\left( Z_{i},Z_{j}\right) _{i,j}$ fluctuating
around the background state, without interactions. We have:%
\begin{eqnarray*}
&&G_{0}\left( \left( \mathbf{T-}\left\langle \mathbf{T}\right\rangle \right)
_{ij},\left( \left( \mathbf{T}^{\prime }\mathbf{-}\left\langle \mathbf{T}%
\right\rangle \right) \right) _{ij},t\right) \\
&=&\prod\limits_{j}G_{0}\left( \left( \mathbf{T}_{ij}\mathbf{-}\left\langle 
\mathbf{T}_{ij}\right\rangle \right) ,\left( \mathbf{T}_{ij}^{\prime }%
\mathbf{-}\left\langle \mathbf{T}_{ij}\right\rangle \right) ,t\right) \\
&=&\left( 2\pi \right) ^{-N}\left( Det\left( \sigma \left( t\right) \right)
\right) ^{-\frac{N}{2}} \\
&&\prod\limits_{ij}\exp \left( -\left( \left( \mathbf{T}_{ij}\mathbf{-}%
\left\langle \mathbf{T}_{ij}\right\rangle \right) -M\left( t\right) \left( 
\mathbf{T}_{ij}^{\prime }\mathbf{-}\left\langle \mathbf{T}_{ij}\right\rangle
\right) \right) ^{t}\frac{\sigma ^{-1}\left( t\right) }{2}\left( \left( 
\mathbf{T}_{ij}\mathbf{-}\left\langle \mathbf{T}_{ij}\right\rangle \right)
-M\left( t\right) \left( \mathbf{T}_{ij}^{\prime }\mathbf{-}\left\langle 
\mathbf{T}_{ij}\right\rangle \right) \right) \right)
\end{eqnarray*}

\section{Several applications of the effective formalism}

We present several applications of the effective formalism and compute the
transitions between several activated states, including reactivation,
association and sequences of activations. We recover the results presented
in (\cite{GLr}) as consequences of the effective field formalism.

\subsection{Transition function approach to the change in connectivity
background state}

We apply the formalism to the dynamics around a modified background field.
Assume that, due to the change in background activities, the $\Psi
_{0}\left( \theta ,Z\right) $ field action is modified from:%
\begin{equation*}
-\frac{1}{2}\int \Psi _{0}^{\dagger }\left( \theta ,Z\right) \nabla \left( 
\frac{\sigma _{\theta }^{2}}{2}\nabla -\left( \omega _{0}\right)
^{-1}\right) \Psi _{0}\left( \theta ,Z\right) +\int V\left( \Psi _{0}\left(
\theta ,Z\right) \right) +\delta V\left( \Psi _{0}\left( \theta ,Z\right)
\right)
\end{equation*}%
to:%
\begin{equation*}
-\frac{1}{2}\int \Psi _{0}^{\dagger }\left( \theta ,Z\right) \nabla \left( 
\frac{\sigma _{\theta }^{2}}{2}\nabla -\left( \omega _{0}+\delta \omega
_{0}\right) ^{-1}\right) \Psi _{0}\left( \theta ,Z\right) +\int V\left( \Psi
_{0}\left( \theta ,Z\right) \right) +\delta V\left( \Psi _{0}\left( \theta
,Z\right) \right)
\end{equation*}%
where the background activity modification $\delta \omega _{0}$ is equal to
zero except at some given points $\left( Z_{j}\right) _{j\in U}$ where $U$
is a finite set. Consequently, the averages $\left\langle \mathbf{T}%
_{ij}\right\rangle $ are modified only at points $Z_{ij}$, $j\in U$ or $i\in
U$. We define $\bar{U}=\left( ij,j\in U\text{ or }i\in U\right) $ and write $%
\left\langle \mathbf{T}_{ij}\right\rangle ^{old}$ for the old background
state and $\left\langle \mathbf{T}_{ij}\right\rangle ^{new}=\left\langle 
\mathbf{T}_{ij}\right\rangle ^{old}+\Delta \left\langle \mathbf{T}%
_{ij}\right\rangle $ for the new one. We have:%
\begin{eqnarray*}
\Delta \left\langle \mathbf{T}_{ij}\right\rangle &\neq &0\text{ for }ij\in 
\bar{U} \\
\Delta \left\langle \mathbf{T}_{ij}\right\rangle &=&0\text{ otherwise}
\end{eqnarray*}%
Consider the transition from a state corresponding to the previous
background state to a other state We thus set:%
\begin{equation*}
\mathbf{T}_{ij}^{\prime }=\left\langle \mathbf{T}_{ij}\right\rangle ^{old}
\end{equation*}%
As explained in the previous paragraph, the transition for the system of
points $ij\in \bar{U}$ in the new background state is, up the normalization
factor, given by:%
\begin{eqnarray*}
&&\prod\limits_{ij}\exp \left( -\left( \left( \mathbf{T}_{ij}\mathbf{-}%
\left\langle \mathbf{T}_{ij}\right\rangle ^{new}\right) -M\left( t\right)
\left( \mathbf{T}_{ij}^{\prime }\mathbf{-}\left\langle \mathbf{T}%
_{ij}\right\rangle ^{new}\right) \right) ^{t}\frac{\sigma ^{-1}\left(
t\right) }{2}\left( \left( \mathbf{T}_{ij}\mathbf{-}\left\langle \mathbf{T}%
_{ij}\right\rangle ^{new}\right) -M\left( t\right) \left( \mathbf{T}%
_{ij}^{\prime }\mathbf{-}\left\langle \mathbf{T}_{ij}\right\rangle
^{new}\right) \right) \right) \\
&=&\prod\limits_{ij}\exp \left( -\left( \left( \mathbf{T}_{ij}\mathbf{-}%
\left\langle \mathbf{T}_{ij}\right\rangle ^{new}\right) -M\left( t\right)
\Delta \left\langle \mathbf{T}_{ij}\right\rangle \right) ^{t}\frac{\sigma
^{-1}\left( t\right) }{2}\left( \left( \mathbf{T}_{ij}\mathbf{-}\left\langle 
\mathbf{T}_{ij}\right\rangle ^{new}\right) -M\left( t\right) \Delta
\left\langle \mathbf{T}_{ij}\right\rangle \right) \right)
\end{eqnarray*}%
wth:%
\begin{equation*}
M\left( t\right) =\left( 
\begin{array}{cc}
e^{-tu} & s\frac{e^{-tu}-e^{-tv}}{u-v} \\ 
0 & e^{-tv}%
\end{array}%
\right)
\end{equation*}%
As time $t$ increases, the corrections due to the gap between the initial
value and the new background state reduces, so that in average $\mathbf{T}%
_{ij}\mathbf{\rightarrow }\left\langle \mathbf{T}_{ij}\right\rangle ^{new}$.
The higher the values of $u$, $v$ and $s$ the faster the modification in
connectivity functions. Considering that $u$ and $s$ are increasing function
of the modified activity, the higher the average activity in the state, the
lower the modification in the transition functions. Higher activity levels
hinder the system from transitioning to the new equilibrium state.

\subsection{Activation and reactivation of states}

As before, an additional activation at one point for constant connectivities
corresponds to the computation of a transition function. Assume now that,
after stimulation for activities $\omega _{0}\rightarrow \omega _{0}\left(
Z\right) +\delta \omega _{0}\left( Z\right) $, the connectivities have
experienced a transition: $\left\langle \mathbf{T}_{ij}\right\rangle
^{new}=\left\langle \mathbf{T}_{ij}\right\rangle ^{old}+\Delta \left\langle 
\mathbf{T}_{ij}\right\rangle $. This is equivalent to consider the modified
action for the system:

\begin{eqnarray}
&&-\frac{1}{2}\int \Psi _{0}^{\dagger }\left( \theta ,Z\right) \nabla \left( 
\frac{\sigma _{\theta }^{2}}{2}\nabla -\left( \omega _{0}+\delta \omega
_{0}\right) ^{-1}\right) \Psi _{0}\left( \theta ,Z\right) +\int V\left( \Psi
_{0}\left( \theta ,Z\right) \right) +\delta V\left( \Psi _{0}\left( \theta
,Z\right) \right) \\
&&+\sum_{i=1}^{4}S_{\Gamma }^{\left( i\right) }  \notag
\end{eqnarray}

Then, switching off the perturbation $\omega _{0}\left( Z\right) +\delta
\omega _{0}\left( Z\right) \rightarrow \omega _{0}\left( Z\right) $
relatively quickly, the connectivities remain at their new level $%
\left\langle \mathbf{T}_{ij}\right\rangle ^{new}$ for a while. Actually, as
shown in appendix 1, the transmission of the perturbation of activities
includes a factor: 
\begin{equation*}
\int \exp \left( -cl-\alpha \left( \left( cl\right) ^{2}-\left\vert
Z-Z_{i}\right\vert ^{2}\right) \right) \exp \left( i\frac{\varpi \left(
l-\left\vert Z-Z_{i}\right\vert \right) }{c}\right) dl
\end{equation*}%
so that after switching of this perturbation at some time $t_{0}$, the
correction to the activity decays with a factor $\exp \left(
-ct-t_{0}\right) $. Considering again the time scale for the connectivities
to be higher than for activities, the state $\left\langle \mathbf{T}%
_{ij}\right\rangle ^{old}+\Delta \left\langle \mathbf{T}_{ij}\right\rangle $
will decay slowly to $\left\langle \mathbf{T}_{ij}\right\rangle ^{old}$ over
a timespan $T>>1$.

Assume now that at some time $t<<T$, some perturbation raises again:%
\begin{equation*}
\omega _{0}\left( Z_{m}\right) \rightarrow \omega _{0}\left( Z_{m}\right)
+\delta \omega _{0}\left( Z_{m}\right)
\end{equation*}%
at some of the interferences maxima. Given (\ref{rcn}) and (\ref{rcm}) the
perturbation will propagate only along the all set of maxima.

As a consequence, in this particular case, the computation of the series
expansion for the corrections (\ref{srx}) to the activities simplifies.
Actually, replacing the perturbation:%
\begin{equation*}
\sum_{i}a\left( Z_{i},\theta \right) \frac{\omega _{0}^{-1}\left( J,\theta
,Z_{i}\right) }{\Lambda ^{2}}\digamma \left( Z_{i},\theta \right) 
\end{equation*}%
by:%
\begin{equation*}
\delta \omega _{0}\left( Z_{m}\right) \frac{\omega _{0}^{-1}\left(
Z_{m}\right) }{\Lambda ^{2}}\digamma \left( Z_{i},\theta \right) 
\end{equation*}%
representing the activation of one of the maxima, the correction (\ref{srx})
to activities becomes: 
\begin{eqnarray}
\check{T}\digamma ^{\dag } &=&\check{T}\frac{1}{\left( 1-\frac{1}{\Lambda }%
\check{T}-\check{T}_{\omega _{0}+\check{T}\digamma ^{\dag }}\right) }\left(
\delta \omega _{0}\left( Z_{m}\right) \frac{\omega _{0}^{-1}\left(
Z_{m}\right) }{\Lambda ^{2}}\digamma \left( Z_{i},\theta \right) \right)  \\
&=&\check{T}\frac{1}{\left( 1-\left( 1+\frac{1}{\Lambda }\right) \check{T}%
-\left( \check{T}_{\omega _{0}+\check{T}\digamma ^{\dag }}-\check{T}\right)
\right) }\left( \delta \omega _{0}\left( Z_{m}\right) \frac{\omega
_{0}^{-1}\left( Z_{m}\right) }{\Lambda ^{2}}\digamma \left( Z_{i},\theta
\right) \right)   \notag
\end{eqnarray}%
The series expansion in $\check{T}$ has to performed over paths that connect
the maxima since the operator $\check{T}$ is nul outside these paths. Due to
the exponential term in connectivity functions, in first approximation:%
\begin{equation*}
\delta \omega _{0}\left( Z\right) \simeq \check{T}\left( Z,Z_{m}\right) 
\end{equation*}%
if $Z$ is an interference maximum, and:%
\begin{equation*}
\delta \omega _{0}\left( Z\right) \simeq 0
\end{equation*}%
otherwise. As a consequence, the activation of one of these maxima
reactivates the whole set. In turn, doing so and if the stimulation duration
is long enough, the connectivity functions are reactivated towards $%
\left\langle \mathbf{T}_{ij}\right\rangle ^{old}+\Delta \left\langle \mathbf{%
T}_{ij}\right\rangle $.

\subsection{Distant activation}

Consider the sequence of distant signals as in in the first part:%
\begin{eqnarray*}
\left\{ \omega _{0},T_{0}\right\} &\rightarrow &\left\{ T\left(
Z_{M}^{\left( \varepsilon _{1}\right) },Z_{M}^{\left( \varepsilon
_{2}\right) }\right) ,\omega _{M}\right\} \rightarrow \left\{ T\left(
Z_{M}^{\left( \varepsilon _{1}\right) },Z_{M}^{\left( \varepsilon
_{2}\right) }\right) ,\omega _{0}\right\} \\
&\rightarrow &\left\{ T\left( Z_{M}^{\left( \varepsilon _{1}\right)
},Z_{M}^{\left( \varepsilon _{2}\right) }\right) ,\omega _{0}\right\}
+\left\{ T\left( Z_{M}^{\prime \left( \varepsilon _{1}\right)
},Z_{M}^{\prime \left( \varepsilon _{2}\right) }\right) ,\omega _{M}^{\prime
}\right\} \\
&\rightarrow &\left\{ T\left( Z_{M}^{\left( \varepsilon _{1}\right)
},Z_{M}^{\left( \varepsilon _{2}\right) }\right) ,\omega _{0}\right\}
+\left\{ T\left( Z_{M}^{\prime \left( \varepsilon _{1}\right)
},Z_{M}^{\prime \left( \varepsilon _{2}\right) }\right) ,\omega _{0}\right\}
\end{eqnarray*}%
Starting from the equilibrium state, the sequence describes the subsequent
activations of collective states due to some perturbations. This
perturbation initially binds a set $\left( Z_{M}^{\left( \varepsilon
_{1}\right) },Z_{M}^{\left( \varepsilon _{2}\right) }\right) $ with high
activity $\omega _{M}$ and high conncetivity $T\left( Z_{M}^{\left(
\varepsilon _{1}\right) },Z_{M}^{\left( \varepsilon _{2}\right) }\right) $.
Then, the activity dampens and returns to some equilibrium $\omega _{0}$.
This new state may itself induce a transition involving another connected
element $\left\{ T\left( Z_{M}^{\prime \left( \varepsilon _{1}\right)
},Z_{M}^{\prime \left( \varepsilon _{2}\right) }\right) ,\omega _{0}\right\} 
$.

The transition can be described by considering the transitions computed by
the path integrals involving (\ref{prb}) and the action for the connectivity
field:%
\begin{eqnarray}
&&\left\langle \prod\limits_{Z,Z^{\prime }}\left( T,\hat{T},\theta
,Z,Z^{\prime },C,D\right) _{f}\right\vert \left( \exp \left(
-\sum_{i=1}^{4}S_{\Gamma }^{\left( i\right) }\left( \Gamma \left( T,\hat{T}%
,\theta ,Z,Z^{\prime },C,D\right) \right) \right) \right.  \label{Pbv} \\
&&\times \int \mathcal{D}\Psi \left( \theta ,Z\right) \exp \left( \frac{1}{2}%
\Psi ^{\dagger }\left( \theta ,Z\right) \nabla \left( \frac{\sigma _{\theta
}^{2}}{2}\nabla -\omega ^{-1}\right) \Psi \left( \theta ,Z\right) -\int
V\left( \Psi _{0}\left( \theta ,Z\right) \right) \right)  \notag \\
&&\times \int_{\theta _{0}^{\left( 2\right) }}^{\theta _{0}^{\left( 3\right)
}}\exp \left( \sum_{i}a\left( Z_{i},\theta _{0}\right) \left\vert \Psi
\left( Z_{i},\theta _{0}\right) \right\vert ^{2}\right) d\theta _{0}  \notag
\\
&&\left. \int^{\theta _{0}^{\left( 1\right) }}\exp \left( \sum_{i}a\left(
Z_{i},\theta _{0}\right) \left\vert \Psi \left( Z_{i},\theta _{0}\right)
\right\vert ^{2}\right) d\theta _{0}\right) \left\vert \prod_{Z,Z^{\prime
}}\left( T,\hat{T},\theta ,Z,Z^{\prime },C,D\right) _{i}\right\rangle  \notag
\end{eqnarray}%
where the product $\prod\limits_{Z,Z^{\prime }}$ is over points for which
we study the transitions between different states $\left\vert \left( T,\hat{T%
},\theta ,Z,Z^{\prime },C,D\right) _{i}\right\rangle $ and $\left\vert
\left( T,\hat{T},\theta ,Z,Z^{\prime },C,D\right) _{f}\right\rangle $. This
transition is defined as a Green function for the field $\Gamma \left( T,%
\hat{T},\theta ,Z,Z^{\prime },C,D\right) $: 
\begin{eqnarray}
&&\int D\Gamma \left( T,\hat{T},\theta ,Z,Z^{\prime },C,D\right) \mathcal{D}%
\Psi \left( \theta ,Z\right) \left( \prod \Gamma \left( \left( T,\hat{T}%
,\theta ,Z,Z^{\prime },C,D\right) _{f}\right) \right)  \label{Prv} \\
&&\times \left( \prod \Gamma ^{\dagger }\left( \left( T,\hat{T},\theta
,Z,Z^{\prime },C,D\right) _{i}\right) \right)  \notag \\
&&\times \exp \left( -\sum_{i=1}^{4}S_{\Gamma }^{\left( i\right) }\left(
\Gamma \left( T,\hat{T},\theta ,Z,Z^{\prime },C,D\right) \right) +\frac{1}{2}%
\Psi ^{\dagger }\left( \theta ,Z\right) \nabla \left( \frac{\sigma _{\theta
}^{2}}{2}\nabla -\omega ^{-1}\right) \Psi \left( \theta ,Z\right) -\int
V\left( \Psi _{0}\left( \theta ,Z\right) \right) \right)  \notag \\
&&\times \int_{\theta _{0}^{\left( 2\right) }}^{\theta _{0}^{\left( 3\right)
}}\exp \left( \sum_{i}a\left( Z_{i},\theta _{0}\right) \left\vert \Psi
\left( Z_{i},\theta _{0}\right) \right\vert ^{2}\right) d\theta
_{0}\int^{\theta _{0}^{\left( 1\right) }}\exp \left( \sum_{i}a\left(
Z_{i},\theta _{0}\right) \left\vert \Psi \left( Z_{i},\theta _{0}\right)
\right\vert ^{2}\right) d\theta _{0}  \notag
\end{eqnarray}%
with $\theta _{0}^{\left( 2\right) }>>\theta _{0}^{\left( 1\right) }$. The
insertion of the exponential terms corresponds, as in (\ref{prb}), to two
different perturbations distant in time, so that their effect are
disconnected, due to the exponential decay of their persistence.

Expression (\ref{Prv}) thus computes the transition function between two
states where two distant perturbation have been inserted. Given that these
perturbations are independent, the path integral can be computed by
inserting a complete basis of states as border conditions. As a consequence (%
\ref{Pbv}) writes:%
\begin{eqnarray}
&&\left\langle \prod\limits_{Z,Z^{\prime }}\left( T,\hat{T},\theta
,Z,Z^{\prime },C,D\right) _{f}\right\vert \exp \left(
-\sum_{i=1}^{4}S_{\Gamma }^{\left( i\right) }\left( \Gamma \left( T,\hat{T}%
,\theta ,Z,Z^{\prime },C,D\right) \right) +S\left( \Psi \left( \theta
,Z\right) \right) \right)  \label{Pbw} \\
&&\times \int_{\theta _{0}^{\left( 2\right) }}^{\theta _{0}^{\left( 3\right)
}}\exp \left( \sum_{i}a\left( Z_{i},\theta _{0}\right) \left\vert \Psi
\left( Z_{i},\theta _{0}\right) \right\vert ^{2}\right) d\theta
_{0}\left\vert \prod_{Z}\left( Z,\theta \right) \right\rangle \left\vert
\prod_{Z,Z^{\prime }}\left( T,\hat{T},\theta ,Z,Z^{\prime },C,D\right)
\right\rangle  \notag \\
&&\times \left\langle \prod\limits_{Z,Z^{\prime }}\left( T,\hat{T},\theta
,Z,Z^{\prime },C,D\right) \right\vert \left\langle \prod_{Z}\left( Z,\theta
\right) \right\vert \exp \left( -\sum_{i=1}^{4}S_{\Gamma }^{\left( i\right)
}\left( \Gamma \left( T,\hat{T},\theta ,Z,Z^{\prime },C,D\right) \right)
+S\left( \Psi \left( \theta ,Z\right) \right) \right)  \notag \\
&&\times \int^{\theta _{0}^{\left( 1\right) }}\exp \left( \sum_{i}a\left(
Z_{i},\theta _{0}\right) \left\vert \Psi \left( Z_{i},\theta _{0}\right)
\right\vert ^{2}\right) d\theta _{0}\left\vert \prod_{Z,Z^{\prime }}\left(
T,\hat{T},\theta ,Z,Z^{\prime },C,D\right) _{i}\right\rangle  \notag
\end{eqnarray}

The introduction of the complete set:%
\begin{equation*}
\left\vert \prod_{Z}\left( Z,\theta \right) \right\rangle \left\vert
\prod_{Z,Z^{\prime }}\left( T,\hat{T},\theta ,Z,Z^{\prime },C,D\right)
\right\rangle \left\langle \prod\limits_{Z,Z^{\prime }}\left( T,\hat{T}%
,\theta ,Z,Z^{\prime },C,D\right) \right\vert \left\langle \prod_{Z}\left(
Z,\theta \right) \right\vert
\end{equation*}%
in the amplitude(\ref{Pbw}) represents the projection on all possible
states. It includes the possibility of multiple activations for $\left( T,%
\hat{T},\theta ,Z,Z^{\prime },C,D\right) $ and $\left( Z,\theta \right) $,
modeling various potential types of connections at the same point.
Technically this multiple states at the same points correspond to tensor
products of states $\left\vert \left( Z,\theta \right) \right\rangle $ or $%
\left\vert \prod_{Z,Z^{\prime }}\left( T,\hat{T},\theta ,Z,Z^{\prime
},C,D\right) \right\rangle $.

The sum over the inserted states $\left\vert \prod_{Z}\left( Z,\theta
\right) \right\rangle $ can be carried out, yielding two factors of the type
(\ref{prb}). The transition at stake becomes:%
\begin{eqnarray}
&&\left\langle \prod\limits_{Z,Z^{\prime }}\left( T,\hat{T},\theta
,Z,Z^{\prime },C,D\right) _{f}\right\vert \exp \left(
-\sum_{i=1}^{4}S_{\Gamma }^{\left( i\right) }\left( \Gamma \left( T,\hat{T}%
,\theta ,Z,Z^{\prime },C,D\right) \right) +S\left( \Psi \left( \theta
,Z\right) \right) \right) \\
&&\int_{\theta _{0}^{\left( 2\right) }}^{\theta _{0}^{\left( 3\right) }}\exp
\left( \sum_{i}a\left( Z_{i},\theta _{0}\right) \left\vert \Psi \left(
Z_{i},\theta _{0}\right) \right\vert ^{2}\right) d\theta _{0}\left\vert
\prod_{Z,Z^{\prime }}\left( T,\hat{T},\theta ,Z,Z^{\prime },C,D\right)
\right\rangle  \notag \\
&&\times \left\langle \prod\limits_{Z,Z^{\prime }}\left( T,\hat{T},\theta
,Z,Z^{\prime },C,D\right) \right\vert \exp \left( -\sum_{i=1}^{4}S_{\Gamma
}^{\left( i\right) }\left( \Gamma \left( T,\hat{T},\theta ,Z,Z^{\prime
},C,D\right) \right) +S\left( \Psi \left( \theta ,Z\right) \right) \right) 
\notag \\
&&\times \int^{\theta _{0}^{\left( 1\right) }}\exp \left( \sum_{i}a\left(
Z_{i},\theta _{0}\right) \left\vert \Psi \left( Z_{i},\theta _{0}\right)
\right\vert ^{2}\right) d\theta _{0}\left\vert \prod_{Z,Z^{\prime }}\left(
T,\hat{T},\theta ,Z,Z^{\prime },C,D\right) _{i}\right\rangle  \notag
\end{eqnarray}%
and each transition can be computed independently. To describe the
successive activations, we assume that the initial state:%
\begin{equation*}
\left\vert \prod_{Z,Z^{\prime }}\left( T,\hat{T},\theta ,Z,Z^{\prime
},C,D\right) _{i}\right\rangle
\end{equation*}%
is the background state $\left\vert \left\langle \mathbf{T}%
_{ij}\right\rangle ^{old}\right\rangle $ previously described, corresponding
to a product of background states at several points. We have observed that,
given the insertion of a term similar to (\ref{prb}), the state $\left\vert
\left\langle \mathbf{T}_{ij}\right\rangle ^{old}\right\rangle $ undergoes a
transition, with the highest probability, towards $\left\vert \left\langle 
\mathbf{T}_{ij}\right\rangle ^{old}+\Delta \left\langle \mathbf{T}%
_{ij}\right\rangle \right\rangle $. The insertion of the complet set:%
\begin{equation*}
\left\vert \prod_{Z,Z^{\prime }}\left( T,\hat{T},\theta ,Z,Z^{\prime
},C,D\right) \right\rangle \left\langle \prod\limits_{Z,Z^{\prime }}\left(
T,\hat{T},\theta ,Z,Z^{\prime },C,D\right) \right\vert
\end{equation*}%
thus yields a projection:%
\begin{equation*}
\left\vert \left\langle \mathbf{T}_{ij}\right\rangle ^{old}+\Delta
\left\langle \mathbf{T}_{ij}\right\rangle \right\rangle \left\langle
\left\langle \mathbf{T}_{ij}\right\rangle ^{old}+\Delta \left\langle \mathbf{%
T}_{ij}\right\rangle \right\vert
\end{equation*}%
and the transition reduces to:%
\begin{eqnarray}
&&\left\langle \prod\limits_{Z,Z^{\prime }}\left( T,\hat{T},\theta
,Z,Z^{\prime },C,D\right) _{f}\right\vert \exp \left(
-\sum_{i=1}^{4}S_{\Gamma }^{\left( i\right) }\left( \Gamma \left( T,\hat{T}%
,\theta ,Z,Z^{\prime },C,D\right) \right) +S\left( \Psi \left( \theta
,Z\right) \right) \right)  \label{TRS} \\
&&\int_{\theta _{0}^{\left( 2\right) }}^{\theta _{0}^{\left( 3\right) }}\exp
\left( \sum_{i}a\left( Z_{i},\theta _{0}\right) \left\vert \Psi \left(
Z_{i},\theta _{0}\right) \right\vert ^{2}\right) d\theta _{0}\left\vert
\left\langle \mathbf{T}_{ij}\right\rangle ^{old}+\Delta \left\langle \mathbf{%
T}_{ij}\right\rangle \right\rangle \left\langle \left\langle \mathbf{T}%
_{ij}\right\rangle ^{old}+\Delta \left\langle \mathbf{T}_{ij}\right\rangle
\right\vert  \notag \\
&&\left. \exp \left( -\sum_{i=1}^{4}S_{\Gamma }^{\left( i\right) }\left(
\Gamma \left( T,\hat{T},\theta ,Z,Z^{\prime },C,D\right) \right) +S\left(
\Psi \left( \theta ,Z\right) \right) \right) \int^{\theta _{0}^{\left(
1\right) }}\exp \left( \sum_{i}a\left( Z_{i},\theta _{0}\right) \left\vert
\Psi \left( Z_{i},\theta _{0}\right) \right\vert ^{2}\right) d\theta
_{0}\right) \left\vert \left\langle \mathbf{T}_{ij}\right\rangle
^{old}\right\rangle  \notag
\end{eqnarray}%
The second insertion in (\ref{TRS}) shifts the state $\left\vert
\left\langle \mathbf{T}_{ij}\right\rangle ^{old}\right\rangle $ with highest
probability to some:%
\begin{equation*}
\left\vert \left\langle \mathbf{T}_{ij}\right\rangle ^{old}+\Delta
\left\langle \mathbf{T}_{ij}\right\rangle +\Delta ^{\prime }\left\langle 
\mathbf{T}_{ij}\right\rangle \right\rangle
\end{equation*}%
with an amplitude:%
\begin{eqnarray}
&&\left\langle \left\langle \mathbf{T}_{ij}\right\rangle ^{old}+\Delta
\left\langle \mathbf{T}_{ij}\right\rangle +\Delta ^{\prime }\left\langle 
\mathbf{T}_{ij}\right\rangle \right\vert \exp \left(
-\sum_{i=1}^{4}S_{\Gamma }^{\left( i\right) }\left( \Gamma \left( T,\hat{T}%
,\theta ,Z,Z^{\prime },C,D\right) \right) +S\left( \Psi \left( \theta
,Z\right) \right) \right) \\
&&\int_{\theta _{0}^{\left( 2\right) }}^{\theta _{0}^{\left( 3\right) }}\exp
\left( \sum_{i}a\left( Z_{i},\theta _{0}\right) \left\vert \Psi \left(
Z_{i},\theta _{0}\right) \right\vert ^{2}\right) d\theta _{0}\left\vert
\left\langle \mathbf{T}_{ij}\right\rangle ^{old}+\Delta \left\langle \mathbf{%
T}_{ij}\right\rangle \right\rangle \left\langle \left\langle \mathbf{T}%
_{ij}\right\rangle ^{old}+\Delta \left\langle \mathbf{T}_{ij}\right\rangle
\right\vert  \notag \\
&&\left. \exp \left( -\sum_{i=1}^{4}S_{\Gamma }^{\left( i\right) }\left(
\Gamma \left( T,\hat{T},\theta ,Z,Z^{\prime },C,D\right) \right) +S\left(
\Psi \left( \theta ,Z\right) \right) \right) \int^{\theta _{0}^{\left(
1\right) }}\exp \left( \sum_{i}a\left( Z_{i},\theta _{0}\right) \left\vert
\Psi \left( Z_{i},\theta _{0}\right) \right\vert ^{2}\right) d\theta
_{0}\right) \left\vert \left\langle \mathbf{T}_{ij}\right\rangle
^{old}\right\rangle  \notag
\end{eqnarray}%
This computation describes formally the qualitative discussion about
transitions presented in (\cite{GLr}). The distant activation provides two
independent structures of connections that can be reactivated independently
as described by the sequence:

\begin{equation*}
\rightarrow \left\{ T\left( Z_{M}^{\left( \varepsilon _{1}\right)
},Z_{M}^{\left( \varepsilon _{2}\right) }\right) ,\omega _{0}\right\}
+\left\{ T\left( Z_{M}^{\prime \left( \varepsilon _{1}\right)
},Z_{M}^{\prime \left( \varepsilon _{2}\right) }\right) ,\omega _{M}^{\prime
}\right\}
\end{equation*}

\subsection{Subsequent activation}

For a sequence of subsequent activations, as described in (\cite{GLr}), we
consider the scheme:%
\begin{eqnarray}
\left\{ \omega _{0},T_{0}\right\} &\rightarrow &\left\{ T\left(
Z_{M}^{\left( \varepsilon _{1}\right) },Z_{M}^{\left( \varepsilon
_{2}\right) }\right) ,\omega _{M}\right\}  \label{QN} \\
&\rightarrow &\left\{ T\left( Z_{M}^{\left( \varepsilon _{1}\right)
},Z_{M}^{\left( \varepsilon _{2}\right) }\right) ,\omega _{M},T\left(
Z_{M}^{\prime \left( \varepsilon _{1}\right) },Z_{M}^{\prime \left(
\varepsilon _{2}\right) }\right) ,\omega _{M}^{\prime },T\left(
Z_{M}^{\left( \varepsilon _{1}\right) },Z_{M}^{\prime \left( \varepsilon
_{2}\right) }\right) ,T\left( Z_{M}^{\prime \left( \varepsilon _{2}\right)
},Z_{M}^{\left( \varepsilon _{1}\right) }\right) \right\}  \notag \\
&\rightarrow &\left\{ T\left( Z_{M}^{\left( \varepsilon _{1}\right)
},Z_{M}^{\left( \varepsilon _{2}\right) }\right) ,\omega _{0},T\left(
Z_{M}^{\prime \left( \varepsilon _{1}\right) },Z_{M}^{\prime \left(
\varepsilon _{2}\right) }\right) ,\omega _{0},T\left( Z_{M}^{\left(
\varepsilon _{1}\right) },Z_{M}^{\prime \left( \varepsilon _{2}\right)
}\right) ,T\left( Z_{M}^{\prime \left( \varepsilon _{2}\right)
},Z_{M}^{\left( \varepsilon _{1}\right) }\right) \right\}  \notag
\end{eqnarray}%
Here the transition including the second structure is directly caused by the
action of the first structure before its activity dampens.

Technically, the scheme of transitions (\ref{NQ}) implies that the insertion
of perturbations are no longer independent and the transition in (\ref{Pbv})
cannot be shared as a product. It rather writes:%
\begin{eqnarray}
&&\left\langle \prod\limits_{Z,Z^{\prime }}\left( T,\hat{T},\theta
,Z,Z^{\prime },C,D\right) _{f}\right\vert \left( \exp \left(
-\sum_{i=1}^{4}S_{\Gamma }^{\left( i\right) }\left( \Gamma \left( T,\hat{T}%
,\theta ,Z,Z^{\prime },C,D\right) \right) \right) \right. \\
&&\times \int \mathcal{D}\Psi \left( \theta ,Z\right) \exp \left( \frac{1}{2}%
\Psi ^{\dagger }\left( \theta ,Z\right) \nabla \left( \frac{\sigma _{\theta
}^{2}}{2}\nabla -\omega ^{-1}\right) \Psi \left( \theta ,Z\right) -\int
V\left( \Psi _{0}\left( \theta ,Z\right) \right) \right)  \notag \\
&&\left. \times \int_{\theta _{0}^{\left( 2\right) }}^{\theta _{0}^{\left(
3\right) }}\exp \left( \sum_{i}a\left( Z_{i},\theta _{0}\right) \left\vert
\Psi \left( Z_{i},\theta _{0}\right) \right\vert ^{2}\right) d\theta
_{0}\int^{\theta _{0}^{\left( 1\right) }}\exp \left( \sum_{i}a\left(
Z_{i},\theta _{0}\right) \left\vert \Psi \left( Z_{i},\theta _{0}\right)
\right\vert ^{2}\right) d\theta _{0}\right) \left\vert \left\langle \mathbf{T%
}_{ij}\right\rangle ^{old}\right\rangle  \notag
\end{eqnarray}

and projects the final set to some state $\left\vert \left\langle \mathbf{T}%
_{ij}\right\rangle ^{old}+\left\{ \Delta \left\langle \mathbf{T}%
_{ij}\right\rangle +\Delta ^{\prime }\left\langle \mathbf{T}%
_{ij}\right\rangle \right\} \right\rangle $. The additional activations $%
\left\{ \Delta \left\langle \mathbf{T}_{ij}\right\rangle +\Delta ^{\prime
}\left\langle \mathbf{T}_{ij}\right\rangle \right\} $ differ from the set $%
\Delta \left\langle \mathbf{T}_{ij}\right\rangle +\Delta ^{\prime
}\left\langle \mathbf{T}_{ij}\right\rangle $ obtained in the previous
paragraph. Actually, the set $\Delta \left\langle \mathbf{T}%
_{ij}\right\rangle +\Delta ^{\prime }\left\langle \mathbf{T}%
_{ij}\right\rangle $ describes a priori disconected structures that can be
activated independently, while $\left\{ \Delta \left\langle \mathbf{T}%
_{ij}\right\rangle +\Delta ^{\prime }\left\langle \mathbf{T}%
_{ij}\right\rangle \right\} $ encompasses connections between elements of $%
\Delta \left\langle \mathbf{T}_{ij}\right\rangle $ and $\Delta ^{\prime
}\left\langle \mathbf{T}_{ij}\right\rangle $. This implies that reactivation
of one set $\Delta \left\langle \mathbf{T}_{ij}\right\rangle $ or $\Delta
^{\prime }\left\langle \mathbf{T}_{ij}\right\rangle $ will induce the
reactivation of the other one.

\section{Conclusion}

We have shown how the effective field formalism for connectivity enables the
derivation of results pertaining to the activation, association,
reactivation... of states composed of interconnected sets of cells. These
states originate from the deformation of the background fields induced by
external sources. In the subsequent article of this series, we will explore
the internal dynamics between such states as an outcome of the formalism.
Ultimately, in Part IV, we will expand our formalism into a field theory for
groups of interconnected states.

\pagebreak

\section*{Appendix 1 Activities $\protect\omega \left( J,\protect\theta %
,Z\right) $ as functional of the field.}

To obtain the activity $\omega \left( J,\theta ,Z\right) $ as a fld series
expansion, we start with the recursive relation defining $\omega ^{-1}\left(
J,\theta ,Z\right) $. We then proceed in several steps. In this appendix we
compute the first derivative in 1.1. They are estimated through fourier
integrals in 1.2. Then we will obtain the whole series of derivatives by
iterating this result in 2.

\subsection*{1.1 Computation of the first order derivatives in (\protect\ref%
{psn})}

\subsubsection*{1.1.1 General formula}

In the sequel, to simplify the notations:%
\begin{equation*}
\left\vert \Psi \left( \theta -\frac{\left\vert Z-Z_{1}\right\vert }{c}%
,Z_{1}\right) \right\vert ^{2}
\end{equation*}%
stands for:%
\begin{equation*}
\mathcal{\bar{G}}_{0}\left( 0,Z_{1}\right) +\left\vert \Psi \left( \theta -%
\frac{\left\vert Z-Z_{1}\right\vert }{c},Z_{1}\right) \right\vert ^{2}
\end{equation*}%
where\footnote{%
See the discussion after (\ref{RPLC})}:%
\begin{equation*}
\mathcal{\bar{G}}_{0}\left( 0,Z_{1}\right) =\mathcal{G}_{0}\left(
0,Z_{1}\right) +\left\vert \Psi _{0}\left( Z_{1}\right) \right\vert ^{2}
\end{equation*}

Using the recursive definition of $\omega ^{-1}\left( J,\theta ,Z\right) $: 
\begin{equation}
\omega ^{-1}\left( J,\theta ,Z\right) =G\left( J\left( \theta ,Z\right)
+\int \frac{\kappa }{N}\frac{\omega \left( J,\theta -\frac{\left\vert
Z-Z_{1}\right\vert }{c},Z_{1}\right) }{\omega \left( J,\theta ,Z\right) }%
\left\vert \Psi \left( \theta -\frac{\left\vert Z-Z_{1}\right\vert }{c}%
,Z_{1}\right) \right\vert ^{2}T\left( Z,Z_{1},\theta \right) dZ_{1}\right)
\label{btr}
\end{equation}%
we first compute $\frac{\delta \omega ^{-1}\left( J,\theta ,Z\right) }{%
\delta \left\vert \Psi \left( \theta -l_{1},Z_{1}\right) \right\vert ^{2}}$:%
\begin{eqnarray}
&&\frac{\delta \omega ^{-1}\left( J,\theta ,Z\right) }{\delta \left\vert
\Psi \left( \theta -l_{1},Z_{1}\right) \right\vert ^{2}}  \label{vrtft} \\
&=&\frac{\delta G\left( J\left( \theta ,Z\right) +\int \frac{\kappa }{N}%
\frac{\omega \left( J,\theta -\frac{\left\vert Z-Z^{\prime }\right\vert }{c}%
,Z^{\prime }\right) }{\omega \left( J,\theta ,Z\right) }\left\vert \Psi
\left( \theta -\frac{\left\vert Z-Z^{\prime }\right\vert }{c},Z^{\prime
}\right) \right\vert ^{2}T\left( Z,Z^{\prime },\theta \right) dZ^{\prime
}\right) }{\delta \left\vert \Psi \left( \theta -l_{1},Z_{1}\right)
\right\vert ^{2}}  \notag
\end{eqnarray}%
Expanding the right hand side and regrouping $\frac{\delta \omega
^{-1}\left( J,\theta ,Z\right) }{\delta \left\vert \Psi \left( \theta
-l_{1},Z_{1}\right) \right\vert ^{2}}$ on the left yields: 
\begin{eqnarray}
&&\frac{\delta \omega ^{-1}\left( J,\theta ,Z\right) }{\delta \left\vert
\Psi \left( \theta -l_{1},Z_{1}\right) \right\vert ^{2}}  \notag \\
&=&\frac{\frac{1}{\omega \left( J,\theta ,Z\right) }\frac{\kappa }{N}T\left(
Z,Z_{1},\theta \right) G^{\prime }\left[ J,\omega ,\theta ,Z,\Psi \right]
\omega \left( J,\theta -\frac{\left\vert Z-Z^{\prime }\right\vert }{c}%
,Z^{\prime }\right) \delta \left( l_{1}-\frac{\left\vert Z-Z_{1}\right\vert 
}{c}\right) }{1-\left( \int \frac{\kappa }{N}\omega \left( J,\theta -\frac{%
\left\vert Z-Z^{\prime }\right\vert }{c},Z^{\prime }\right) T\left(
Z,Z^{\prime },\theta \right) \left\vert \Psi \left( \theta -\frac{\left\vert
Z-Z^{\prime }\right\vert }{c},Z^{\prime }\right) \right\vert ^{2}dZ^{\prime
}\right) G^{\prime }\left[ J,\omega ,\theta ,Z,\Psi \right] }  \notag \\
&&+\frac{\frac{1}{\omega \left( J,\theta ,Z\right) }\int \frac{\kappa }{N}%
\frac{\delta \omega \left( J,\theta -\frac{\left\vert Z-Z^{\prime
}\right\vert }{c},Z^{\prime }\right) }{\delta \left\vert \Psi \left( \theta
-l_{1},Z_{1}\right) \right\vert ^{2}}T\left( Z,Z^{\prime },\theta \right)
\left\vert \Psi \left( \theta -\frac{\left\vert Z-Z^{\prime }\right\vert }{c}%
,Z^{\prime }\right) \right\vert ^{2}dZ^{\prime }G^{\prime }\left[ J,\omega
,\theta ,Z,\Psi \right] }{1-\left( \int \frac{\kappa }{N}\omega \left(
J,\theta -\frac{\left\vert Z-Z^{\prime }\right\vert }{c},Z^{\prime }\right)
T\left( Z,Z^{\prime },\theta \right) \left\vert \Psi \left( \theta -\frac{%
\left\vert Z-Z^{\prime }\right\vert }{c},Z^{\prime }\right) \right\vert
^{2}dZ^{\prime }\right) G^{\prime }\left[ J,\omega ,\theta ,Z,\Psi \right] }
\notag \\
&&+\frac{\frac{1}{\omega \left( J,\theta ,Z\right) }\int \frac{\kappa }{N}%
\omega \left( J,\theta -\frac{\left\vert Z-Z^{\prime }\right\vert }{c}%
,Z^{\prime }\right) \frac{\partial T\left( Z,Z^{\prime },\theta \right) }{%
\partial \left\vert \Psi \left( \theta -l_{1},Z_{1}\right) \right\vert ^{2}}%
\left\vert \Psi \left( \theta -\frac{\left\vert Z-Z^{\prime }\right\vert }{c}%
,Z^{\prime }\right) \right\vert ^{2}dZ^{\prime }G^{\prime }\left[ J,\omega
,\theta ,Z,\Psi \right] }{1-\left( \int \frac{\kappa }{N}\omega \left(
J,\theta -\frac{\left\vert Z-Z^{\prime }\right\vert }{c},Z^{\prime }\right)
T\left( Z,Z^{\prime },\theta \right) \left\vert \Psi \left( \theta -\frac{%
\left\vert Z-Z^{\prime }\right\vert }{c},Z^{\prime }\right) \right\vert
^{2}dZ^{\prime }\right) G^{\prime }\left[ J,\omega ,\theta ,Z,\Psi \right] }
\end{eqnarray}%
neglecting $\frac{\partial T\left( Z,Z^{\prime },\theta \right) }{\partial
\omega \left( J,\theta -\frac{\left\vert Z-Z^{\prime }\right\vert }{c}%
,Z^{\prime }\right) }$ in first approxmtn, this leads to:%
\begin{eqnarray}
\frac{\delta \omega ^{-1}\left( J,\theta ,Z\right) }{\delta \left\vert \Psi
\left( \theta -l_{1},Z_{1}\right) \right\vert ^{2}} &=&\omega \left(
J,\theta -l_{1},Z_{1}\right) \tilde{T}_{1}\left( \theta ,Z,Z_{1},\omega
,\Psi \right) \delta \left( l_{1}-\frac{\left\vert Z-Z_{1}\right\vert }{c}%
\right)  \label{CR} \\
&&+\int \frac{\delta \omega \left( J,\theta -\frac{\left\vert Z-Z^{\prime
}\right\vert }{c},Z^{\prime }\right) }{\delta \left\vert \Psi \left( \theta
-l_{1},Z_{1}\right) \right\vert ^{2}}\left\vert \Psi \left( \theta -\frac{%
\left\vert Z-Z^{\prime }\right\vert }{c},Z^{\prime }\right) \right\vert ^{2}%
\tilde{T}_{1}\left( \theta ,Z,Z^{\prime },\omega ,\Psi \right) dZ^{\prime } 
\notag
\end{eqnarray}%
where we defined:%
\begin{eqnarray}
&&\tilde{T}_{1}\left( \theta ,Z,Z_{1},\omega ,\Psi \right) =\frac{1}{\omega
\left( J,\theta ,Z\right) }  \label{vrtbb} \\
&&\times \frac{\frac{\kappa }{N}T\left( Z,Z_{1},\theta \right) G^{\prime }%
\left[ J,\omega ,\theta ,Z,\Psi \right] \delta \left( l_{1}-\frac{\left\vert
Z-Z_{1}\right\vert }{c}\right) }{1-\left( \int \frac{\kappa }{N}\omega
\left( J,\theta -\frac{\left\vert Z-Z^{\prime }\right\vert }{c},Z^{\prime
}\right) T\left( Z,Z^{\prime },\theta \right) \left\vert \Psi \left( \theta -%
\frac{\left\vert Z-Z^{\prime }\right\vert }{c},Z^{\prime }\right)
\right\vert ^{2}dZ^{\prime }\right) G^{\prime }\left[ J,\omega ,\theta
,Z,\Psi \right] }  \notag
\end{eqnarray}%
Equation (\ref{CR}) shows that we also need $\frac{\delta \omega \left(
J,\theta ,Z\right) }{\delta \left\vert \Psi \left( \theta
-l_{1},Z_{1}\right) \right\vert ^{2}}$ to compute $\frac{\delta \omega
^{-1}\left( J,\theta ,Z\right) }{\delta \left\vert \Psi \left( \theta
-l_{1},Z_{1}\right) \right\vert ^{2}}$. This is obtained by: 
\begin{eqnarray}
\frac{\delta \omega \left( J,\theta ,Z\right) }{\delta \left\vert \Psi
\left( \theta -l_{1},Z_{1}\right) \right\vert ^{2}} &=&\frac{\delta F\left(
J\left( \theta ,Z\right) +\int \frac{\kappa }{N}\bar{W}\left( \frac{\omega
\left( J,\theta -\frac{\left\vert Z-Z^{\prime }\right\vert }{c},Z^{\prime
}\right) }{\omega \left( J,\theta ,Z\right) }\right) \left\vert \Psi \left(
\theta -\frac{\left\vert Z-Z^{\prime }\right\vert }{c},Z^{\prime }\right)
\right\vert ^{2}T\left( Z,Z^{\prime }\right) dZ^{\prime }\right) }{\delta
\left\vert \Psi \left( \theta -l_{1},Z_{1}\right) \right\vert ^{2}}  \notag
\\
&=&\omega \left( J,\theta -l_{1},Z_{1}\right) \tilde{T}\left( \theta
,Z,Z_{1},\omega ,\Psi \right) \delta \left( l_{1}-\frac{\left\vert
Z-Z_{1}\right\vert }{c}\right)  \notag \\
&&+\int \frac{\delta \omega \left( J,\theta -\frac{\left\vert Z-Z^{\prime
}\right\vert }{c},Z^{\prime }\right) }{\delta \left\vert \Psi \left( \theta
-l_{1},Z_{1}\right) \right\vert ^{2}}\left\vert \Psi \left( \theta -\frac{%
\left\vert Z-Z^{\prime }\right\vert }{c},Z^{\prime }\right) \right\vert ^{2}%
\tilde{T}\left( \theta ,Z,Z^{\prime },\omega ,\Psi \right) dZ^{\prime }
\label{vrttt}
\end{eqnarray}%
with:%
\begin{eqnarray}
&&\tilde{T}\left( \theta ,Z,Z_{1}\omega ,\Psi \right)  \label{vrtftbs} \\
&=&\frac{\frac{\kappa }{N}\omega \left( J,\theta ,Z\right) T\left(
Z,Z_{1}\right) \bar{W}^{\prime }\left( \frac{\omega \left( J,\theta -\frac{%
\left\vert Z-Z_{1}\right\vert }{c},Z_{1}\right) }{\omega \left( J,\theta
,Z\right) }\right) F^{\prime }\left[ J,\omega ,\theta ,Z,\Psi \right] }{%
\omega ^{2}\left( J,\theta ,Z\right) +F^{\prime }\left[ J,\omega ,\theta
,Z,\Psi \right] \int \frac{\kappa \omega \left( J,\theta -\frac{\left\vert
Z-Z^{\prime }\right\vert }{c},Z^{\prime }\right) }{N}\bar{W}^{\prime }\left( 
\frac{\omega \left( J,\theta -\frac{\left\vert Z-Z^{\prime }\right\vert }{c}%
,Z^{\prime }\right) }{\omega \left( J,\theta ,Z\right) }\right) \left\vert
\Psi \left( \theta -\frac{\left\vert Z-Z^{\prime }\right\vert }{c},Z^{\prime
}\right) \right\vert ^{2}T\left( Z,Z^{\prime },\theta \right) dZ^{\prime }} 
\notag
\end{eqnarray}%
Equation (\ref{vrttt}) and (\ref{vrtftbs}) define $\frac{\delta \omega
\left( J,\theta ,Z\right) }{\delta \left\vert \Psi \left( \theta
-l_{1},Z_{1}\right) \right\vert ^{2}}$\ recursively. Actually, writing: 
\begin{eqnarray*}
&&\frac{\delta \omega \left( J,\theta -\frac{\left\vert Z-Z^{\prime
}\right\vert }{c},Z^{\prime }\right) }{\delta \left\vert \Psi \left( \theta
-l_{1},Z_{1}\right) \right\vert ^{2}} \\
&=&\int \omega \left( J,\theta -\frac{\left\vert Z-Z^{\prime }\right\vert }{c%
}-\frac{\left\vert Z^{\prime }-Z^{\prime \prime }\right\vert }{c},Z^{\prime
\prime }\right) \tilde{T}\left( \theta -\frac{\left\vert Z-Z^{\prime
}\right\vert }{c},Z^{\prime },Z^{\prime \prime },\omega ,\Psi \right) \delta
\left( \frac{\left\vert Z-Z^{\prime }\right\vert }{c}+\frac{\left\vert
Z^{\prime }-Z^{\prime \prime }\right\vert }{c}-l_{1}\right) dZ^{\prime
\prime } \\
&&+\int \frac{\delta \omega \left( J,\theta -\frac{\left\vert Z-Z^{\prime
}\right\vert }{c}-\frac{\left\vert Z^{\prime }-Z^{\prime \prime }\right\vert 
}{c},Z^{\prime \prime }\right) }{\delta \left\vert \Psi \left( \theta
-l_{1},Z_{1}\right) \right\vert ^{2}}\left\vert \Psi \left( \theta -\frac{%
\left\vert Z-Z^{\prime }\right\vert }{c}-\frac{\left\vert Z^{\prime
}-Z^{\prime \prime }\right\vert }{c},Z^{\prime \prime }\right) \right\vert
^{2}\tilde{T}\left( \theta -\frac{\left\vert Z-Z^{\prime }\right\vert }{c}%
,Z^{\prime },Z^{\prime \prime },\omega ,\Psi \right) dZ^{\prime \prime }
\end{eqnarray*}%
we have:

\begin{eqnarray*}
&&\frac{\delta \omega \left( J,\theta ,Z\right) }{\delta \left\vert \Psi
\left( \theta -l_{1},Z_{1}\right) \right\vert ^{2}} \\
&=&\int \omega \left( J,\theta -\frac{\left\vert Z-Z^{\prime }\right\vert }{c%
},Z^{\prime }\right) \tilde{T}\left( \theta ,Z,Z_{1},\omega ,\Psi \right)
\delta \left( \frac{\left\vert Z-Z^{\prime }\right\vert }{c}-l_{1}\right)
dZ^{\prime } \\
&&+\int \omega \left( J,\theta -\frac{\left\vert Z-Z^{\prime }\right\vert }{c%
}-\frac{\left\vert Z^{\prime }-Z^{\prime \prime }\right\vert }{c},Z^{\prime
\prime }\right) \tilde{T}\left( \theta -\frac{\left\vert Z-Z^{\prime
}\right\vert }{c},Z^{\prime },Z^{\prime \prime },\omega ,\Psi \right) \\
&&\times \left\vert \Psi \left( \theta -\frac{\left\vert Z-Z^{\prime
}\right\vert }{c},Z^{\prime }\right) \right\vert ^{2}\tilde{T}\left( \theta
,Z,Z^{\prime },\omega ,\Psi \right) \delta \left( \frac{\left\vert
Z-Z^{\prime }\right\vert }{c}+\frac{\left\vert Z^{\prime }-Z^{\prime \prime
}\right\vert }{c}-l_{1}\right) dZ^{\prime }dZ^{\prime \prime } \\
&&+\int \frac{\delta \omega \left( J,\theta -\frac{\left\vert Z-Z^{\prime
}\right\vert }{c}-\frac{\left\vert Z^{\prime }-Z^{\prime \prime }\right\vert 
}{c},Z^{\prime \prime }\right) }{\delta \left\vert \Psi \left( \theta
-l_{1},Z_{1}\right) \right\vert ^{2}}\left\vert \Psi \left( \theta -\frac{%
\left\vert Z-Z^{\prime }\right\vert }{c}-\frac{\left\vert Z^{\prime
}-Z^{\prime \prime }\right\vert }{c},Z^{\prime \prime }\right) \right\vert
^{2} \\
&&\times \tilde{T}\left( \theta -\frac{\left\vert Z-Z^{\prime }\right\vert }{%
c},Z^{\prime },Z^{\prime \prime },\omega ,\Psi \right) \left\vert \Psi
\left( \theta -\frac{\left\vert Z-Z^{\prime }\right\vert }{c},Z^{\prime
}\right) \right\vert ^{2}\hat{T}\left( \theta ,Z,Z^{\prime },\omega ,\Psi
\right) dZ^{\prime }dZ^{\prime \prime }
\end{eqnarray*}%
By a redefinition of $\check{T}$ and $\tilde{T}_{1}$:%
\begin{eqnarray*}
\tilde{T}\left( \theta ,Z,Z^{\prime },\omega ,\Psi \right) \left\vert \Psi
\left( \theta -\frac{\left\vert Z-Z^{\prime }\right\vert }{c},Z^{\prime
}\right) \right\vert ^{2} &\rightarrow &\tilde{T}\left( \theta ,Z,Z^{\prime
},\omega ,\Psi \right) \\
\tilde{T}_{1}\left( \theta ,Z,Z^{\prime },\omega ,\Psi \right) \left\vert
\Psi \left( \theta -\frac{\left\vert Z-Z^{\prime }\right\vert }{c},Z^{\prime
}\right) \right\vert ^{2} &\rightarrow &\tilde{T}_{1}\left( \theta
,Z,Z^{\prime },\omega ,\Psi \right)
\end{eqnarray*}%
we find the series expansion:%
\begin{eqnarray}
\frac{\delta \omega \left( J,\theta ,Z\right) }{\delta \left\vert \Psi
\left( \theta -l_{1},Z_{1}\right) \right\vert ^{2}} &=&\sum_{n=1}^{\infty }%
\frac{1}{\left\vert \Psi \left( \theta -l_{1},Z_{1}\right) \right\vert ^{2}}%
\int \omega \left( J,\theta -\sum_{l=1}^{n}\frac{\left\vert Z^{\left(
l-1\right) }-Z^{\left( l\right) }\right\vert }{c},Z_{1}\right)  \label{xpng}
\\
&&\times \dprod\limits_{l=1}^{n}\tilde{T}\left( \theta -\sum_{j=1}^{l-1}%
\frac{\left\vert Z^{\left( j-1\right) }-Z^{\left( j\right) }\right\vert }{c}%
,Z^{\left( l-1\right) },Z^{\left( l\right) },\omega ,\Psi \right) \delta
\left( l_{1}-\sum_{l=1}^{n}\frac{\left\vert Z^{\left( l-1\right) }-Z^{\left(
l\right) }\right\vert }{c}\right) \dprod\limits_{l=1}^{n-1}dZ^{\left(
l\right) }  \notag
\end{eqnarray}%
and:%
\begin{eqnarray}
\frac{\delta \omega ^{-1}\left( J,\theta ,Z\right) }{\delta \left\vert \Psi
\left( \theta -l_{1},Z_{1}\right) \right\vert ^{2}} &=&\sum_{n=1}^{\infty }%
\frac{1}{\left\vert \Psi \left( \theta -l_{1},Z_{1}\right) \right\vert ^{2}}%
\int \omega \left( J,\theta -\sum_{l=1}^{n}\frac{\left\vert Z^{\left(
l-1\right) }-Z^{\left( l\right) }\right\vert }{c},Z_{1}\right) \tilde{T}%
_{1}\left( \theta ,Z,Z^{\left( 1\right) },\omega ,\Psi \right)  \label{drvt}
\\
&&\times \dprod\limits_{l=2}^{n}\tilde{T}\left( \theta -\sum_{j=1}^{l-1}%
\frac{\left\vert Z^{\left( j-1\right) }-Z^{\left( j\right) }\right\vert }{c}%
,Z^{\left( l-1\right) },Z^{\left( l\right) },\omega ,\Psi \right) \delta
\left( l_{1}-\sum_{l=1}^{n}\frac{\left\vert Z^{\left( l-1\right) }-Z^{\left(
l\right) }\right\vert }{c}\right) \dprod\limits_{l=1}^{n-1}dZ^{\left(
l\right) }  \notag
\end{eqnarray}%
with the convention that $Z^{\left( 0\right) }=Z$ and $Z^{\left( n\right)
}=Z_{1}$.

We can write (\ref{drvt}) in a more symetric way. Defining:%
\begin{equation*}
\check{T}\left( \theta ,Z,Z^{\left( 1\right) },\omega ,\Psi \right) =-\omega
^{2}\left( J,\theta --\frac{\left\vert Z-Z_{1}\right\vert }{c},Z_{1}\right) 
\tilde{T}_{1}\left( \theta ,Z,Z^{\left( 1\right) },\omega ,\Psi \right)
\end{equation*}%
Relation (\ref{CR}) writes:%
\begin{eqnarray*}
\frac{\delta \omega ^{-1}\left( J,\theta ,Z\right) }{\delta \left\vert \Psi
\left( \theta -l_{1},Z_{1}\right) \right\vert ^{2}} &=&-\omega ^{-1}\left(
J,\theta -l_{1},Z_{1}\right) \check{T}\left( \theta ,Z,Z_{1},\omega ,\Psi
\right) \delta \left( l_{1}-\frac{\left\vert Z-Z_{1}\right\vert }{c}\right)
\\
&&+\int \frac{\delta \omega ^{-1}\left( J,\theta -\frac{\left\vert
Z-Z^{\prime }\right\vert }{c},Z^{\prime }\right) }{\delta \left\vert \Psi
\left( \theta -l_{1},Z_{1}\right) \right\vert ^{2}}\left\vert \Psi \left(
\theta -\frac{\left\vert Z-Z^{\prime }\right\vert }{c},Z^{\prime }\right)
\right\vert ^{2}\check{T}\left( \theta ,Z,Z^{\prime },\omega ,\Psi \right)
dZ^{\prime }
\end{eqnarray*}%
and we have:%
\begin{eqnarray}
\frac{\delta \omega ^{-1}\left( J,\theta ,Z\right) }{\delta \left\vert \Psi
\left( \theta -l_{1},Z_{1}\right) \right\vert ^{2}} &=&-\sum_{n=1}^{\infty }%
\frac{1}{\left\vert \Psi \left( \theta -l_{1},Z_{1}\right) \right\vert ^{2}}%
\int \omega ^{-1}\left( J,\theta -\sum_{l=1}^{n}\frac{\left\vert Z^{\left(
l-1\right) }-Z^{\left( l\right) }\right\vert }{c},Z_{1}\right)  \label{dvtr}
\\
&&\times \dprod\limits_{l=1}^{n}\check{T}\left( \theta -\sum_{j=1}^{l-1}%
\frac{\left\vert Z^{\left( j-1\right) }-Z^{\left( j\right) }\right\vert }{c}%
,Z^{\left( l-1\right) },Z^{\left( l\right) },\omega ,\Psi \right) \delta
\left( l_{1}-\sum_{l=1}^{n}\frac{\left\vert Z^{\left( l-1\right) }-Z^{\left(
l\right) }\right\vert }{c}\right) \dprod\limits_{l=1}^{n-1}dZ^{\left(
l\right) }  \notag
\end{eqnarray}

\subsubsection*{1.1.2 Static approximation\protect\bigskip}

We now use a static approximations (\ref{xpng}) and (\ref{drvt}). Actually,
the values of $\tilde{T}_{1}\left( \theta ,Z,Z_{1}\omega ,\Psi \right) $ and 
$\tilde{T}\left( \theta ,Z,Z_{1}\omega ,\Psi \right) $ can be estimated for
the static approximation for activity $\bar{\omega}^{-1}\left( \bar{J}%
,Z\right) $. Moreover, in the limit of small fluctuations, $\bar{\omega}%
^{-1}\left( \bar{J},Z\right) $, $F^{\prime }\left[ J,\bar{\omega},Z,\Psi %
\right] $ and $G^{\prime }\left[ J,\bar{\omega},Z,\Psi \right] $ can be
approximated by their average over $Z$, denoted $\bar{\omega}^{-1}$, $\bar{F}%
^{\prime }$and $\bar{G}^{\prime }$. We also have:%
\begin{equation*}
\frac{\bar{\omega}\left( J,Z^{\prime }\right) }{\bar{\omega}\left(
J,Z\right) }\simeq 1
\end{equation*}%
We also replace $\left\vert \Psi \right\vert ^{2}$ by $\frac{1}{\sqrt{\frac{%
\pi }{2}\left( \frac{1}{\sigma ^{2}\bar{X}_{r}}\right) ^{2}+\frac{2\pi
\alpha }{\sigma ^{2}}}}$. Moreover for $\bar{\omega}$, both $\tilde{T}_{1}$
and $\tilde{T}$ can be considered independent of $\theta $: 
\begin{eqnarray}
\tilde{T}_{1}\left( \theta ,Z,Z_{1}\bar{\omega},\Psi \right) &\simeq &\tilde{%
T}_{1}\left( Z,Z_{1},\bar{\omega}\right)  \label{TNP} \\
&=&\frac{1}{\sqrt{\frac{\pi }{2}\left( \frac{1}{\sigma ^{2}\bar{X}_{r}}%
\right) ^{2}+\frac{2\pi \alpha }{\sigma ^{2}}}}\frac{\frac{\kappa }{N}\bar{%
\omega}^{-1}T\left( Z,Z_{1}\right) \bar{G}^{\prime }}{1-\frac{\bar{G}%
^{\prime }\bar{\omega}\int \frac{\kappa }{N}T\left( Z,Z^{\prime }\right)
dZ^{\prime }}{\sqrt{\frac{\pi }{2}\left( \frac{1}{\sigma ^{2}\bar{X}_{r}}%
\right) ^{2}+\frac{2\pi \alpha }{\sigma ^{2}}}}}  \notag
\end{eqnarray}%
and:%
\begin{eqnarray}
\tilde{T}\left( \theta ,Z,Z_{1}\omega ,\Psi \right) &\simeq &\tilde{T}\left(
Z,Z_{1},\bar{\omega}\right)  \label{TPR} \\
&=&\frac{1}{\sqrt{\frac{\pi }{2}\left( \frac{1}{\sigma ^{2}\bar{X}_{r}}%
\right) ^{2}+\frac{2\pi \alpha }{\sigma ^{2}}}}\frac{\frac{\kappa }{N}%
T\left( Z,Z_{1}\right) \bar{F}^{\prime }}{\bar{\omega}+\frac{\bar{F}^{\prime
}\int \frac{\kappa }{N}T\left( Z,Z^{\prime }\right) dZ^{\prime }}{\sqrt{%
\frac{\pi }{2}\left( \frac{1}{\sigma ^{2}\bar{X}_{r}}\right) ^{2}+\frac{2\pi
\alpha }{\sigma ^{2}}}}}  \notag
\end{eqnarray}%
as a consequence $\tilde{T}_{1}\left( Z,Z_{1},\bar{\omega}\right) $ and $%
\tilde{T}\left( Z,Z_{1},\bar{\omega}\right) $ are functions of $\left\vert
Z-Z_{1}\right\vert $ denoted $\tilde{T}_{1}\left( \left\vert
Z-Z_{1}\right\vert \right) $ and $\tilde{T}\left( \left\vert
Z-Z_{1}\right\vert \right) $. \ As a consequence (\ref{xpng}) becomes:%
\begin{eqnarray}
&&\frac{\delta \omega \left( J,\theta ,Z\right) }{\delta \left\vert \Psi
\left( \theta -l_{1},Z_{1}\right) \right\vert ^{2}}  \notag \\
&=&\sum_{n=1}^{\infty }\frac{1}{\left\vert \Psi \left( \theta
-l_{1},Z_{1}\right) \right\vert ^{2}}\int \omega \left( J,\theta
-\sum_{l=1}^{n}\frac{\left\vert Z^{\left( l-1\right) }-Z^{\left( l\right)
}\right\vert }{c},Z_{1}\right) \dprod\limits_{l=1}^{n}\tilde{T}\left(
\left\vert Z^{\left( l-1\right) }-Z^{\left( l\right) }\right\vert \right) 
\notag \\
&&\times \delta \left( l_{1}-\sum_{l=1}^{n}\frac{\left\vert Z^{\left(
l-1\right) }-Z^{\left( l\right) }\right\vert }{c}\right) \times \delta
\left( Z-Z_{1}-\sum_{l=1}^{n}\left( Z^{\left( l-1\right) }-Z^{\left(
l\right) }\right) \right) \dprod\limits_{l=1}^{n-1}dZ^{\left( l\right) }
\label{xpgk}
\end{eqnarray}%
and (\ref{drvt}) can be estimated by:%
\begin{eqnarray}
&&\frac{\delta \omega ^{-1}\left( J,\theta ,Z\right) }{\delta \left\vert
\Psi \left( \theta -l_{1},Z_{1}\right) \right\vert ^{2}}  \label{vrtftbscc}
\\
&=&\sum_{n=1}^{\infty }\frac{1}{\left\vert \Psi \left( \theta
-l_{1},Z_{1}\right) \right\vert ^{2}}\int \omega \left( J,\theta
-\sum_{l=1}^{n}\frac{\left\vert Z^{\left( l-1\right) }-Z^{\left( l\right)
}\right\vert }{c},Z_{1}\right) \tilde{T}_{1}\left( \left\vert
Z-Z_{1}\right\vert \right)  \notag \\
&&\times \dprod\limits_{l=2}^{n}\tilde{T}\left( \left\vert Z^{\left(
l-1\right) }-Z^{\left( l\right) }\right\vert \right) \delta \left(
l_{1}-\sum_{l=1}^{n}\frac{\left\vert Z^{\left( l-1\right) }-Z^{\left(
l\right) }\right\vert }{c}\right) \delta \left( Z-Z_{1}-\sum_{l=1}^{n}\left(
Z^{\left( l-1\right) }-Z^{\left( l\right) }\right) \right)
\dprod\limits_{l=1}^{n-1}dZ^{\left( l\right) }  \notag
\end{eqnarray}

\subsection*{1.2 Estimation of (\protect\ref{vrtftbscc}) and (\protect\ref%
{xpng}) close to the permanent regime}

The series (\ref{vrtftbscc}) can be computed by using the Fourier transform
of the Dirac functions:%
\begin{eqnarray}
&&\left\vert \Psi \left( \theta -l_{1},Z_{1}\right) \right\vert ^{2}\frac{%
\delta \omega ^{-1}\left( J,\theta ,Z\right) }{\delta \left\vert \Psi \left(
\theta -l_{1},Z_{1}\right) \right\vert ^{2}}  \label{gr} \\
&=&\sum_{n=1}^{\infty }\int \omega \left( J,\theta -l_{1},Z_{1}\right)
\times \tilde{T}_{1}\left( \left\vert Z-Z^{\left( 1\right) }\right\vert
\right) \dprod\limits_{l=2}^{n}\tilde{T}\left( \left\vert Z^{\left(
l-1\right) }-Z^{\left( l\right) }\right\vert \right) \exp \left( i\lambda
\left( cl_{1}-\sum_{l=1}^{n}\left\vert Z^{\left( l-1\right) }-Z^{\left(
l\right) }\right\vert \right) \right)  \notag \\
&&\times \exp \left( i\lambda _{1}.\left( Z-Z_{1}-\sum_{l=1}^{n}\left(
Z^{\left( l-1\right) }-Z^{\left( l\right) }\right) \right) \right) d\lambda
d\lambda _{1}\dprod\limits_{l=1}^{n}\left\vert Z^{\left( l-1\right)
}-Z^{\left( l\right) }\right\vert ^{2}d\left\vert Z^{\left( l-1\right)
}-Z^{\left( l\right) }\right\vert dv_{l}  \notag
\end{eqnarray}%
where the unit vectors $v_{l}$ are defined such that:%
\begin{equation*}
Z^{\left( l-1\right) }-Z^{\left( l\right) }=v_{l}\left\vert Z^{\left(
l-1\right) }-Z^{\left( l\right) }\right\vert
\end{equation*}%
We also define: 
\begin{eqnarray*}
\lambda _{1}.\left( Z-Z_{1}\right) &=&\left\vert \lambda _{1}\right\vert
\left\vert Z-Z_{1}\right\vert \cos \left( \theta _{1}\right) \\
\lambda _{1}.v_{l} &=&\left\vert \lambda _{1}\right\vert \cos \left( \theta
_{l}\right)
\end{eqnarray*}%
The angles $\theta _{l}$ are computed in the plane $\left( \lambda
_{1},Z-Z_{1}\right) $ between the projection of $v_{l}$ and $Z-Z_{1}$.

Before computing the integrals in (\ref{gr}) for arbitrary connectivity
functions, we develop the particular case of an exponential transfer
function.

\subsubsection*{1.2.1 Exponential connectivity function}

We first choose:%
\begin{eqnarray}
\tilde{T}\left( \left\vert Z^{\left( l-1\right) }-Z^{\left( l\right)
}\right\vert \right) &=&C\frac{\exp \left( -c\left\vert Z^{\left( l-1\right)
}-Z^{\left( l\right) }\right\vert \right) }{\left\vert Z^{\left( l-1\right)
}-Z^{\left( l\right) }\right\vert }  \label{srt} \\
\tilde{T}_{1}\left( \left\vert Z^{\left( l-1\right) }-Z^{\left( l\right)
}\right\vert \right) &\simeq &\frac{A_{1}}{A}C\hat{T}\left( \left\vert
Z^{\left( l-1\right) }-Z^{\left( l\right) }\right\vert \right)  \notag
\end{eqnarray}%
where, given (\ref{TNP}) and (\ref{TPR}):%
\begin{eqnarray}
\frac{A_{1}}{A} &=&\frac{\tilde{T}_{1}\left( \theta -\sum_{l=1}^{n}\frac{%
\left\vert Z^{\left( l-1\right) }-Z^{\left( l\right) }\right\vert }{c}%
,Z,Z^{\left( 1\right) },\omega ,\Psi \right) }{\tilde{T}\left( \theta
-\sum_{l=1}^{n}\frac{\left\vert Z^{\left( l-1\right) }-Z^{\left( l\right)
}\right\vert }{c},Z,Z^{\left( 1\right) },\omega ,\Psi \right) }  \label{QR}
\\
&=&\frac{\frac{\kappa }{N}\bar{\omega}^{-1}T\left( Z,Z_{1}\right) \bar{G}%
^{\prime }}{1-\frac{\bar{G}^{\prime }\bar{\omega}\int \frac{\kappa }{N}%
T\left( Z,Z^{\prime }\right) dZ^{\prime }}{\sqrt{\frac{\pi }{2}\left( \frac{1%
}{\sigma ^{2}\bar{X}_{r}}\right) ^{2}+\frac{2\pi \alpha }{\sigma ^{2}}}}}%
\left( \frac{\frac{\kappa }{N}T\left( Z,Z_{1}\right) \bar{F}^{\prime }}{\bar{%
\omega}+\frac{\bar{F}^{\prime }\int \frac{\kappa }{N}T\left( Z,Z^{\prime
}\right) dZ^{\prime }}{\sqrt{\frac{\pi }{2}\left( \frac{1}{\sigma ^{2}\bar{X}%
_{r}}\right) ^{2}+\frac{2\pi \alpha }{\sigma ^{2}}}}}\right) ^{-1}\simeq
-\left( \bar{\omega}^{-1}\right) ^{2}  \notag
\end{eqnarray}%
We will disregard the factor $\frac{A_{1}}{A}$ that will be reintroduced in
the end of the computation.

Using that $\sum_{l=1}^{n}\left( Z^{\left( l-1\right) }-Z^{\left( l\right)
}\right) =cl_{1}$, the right hand side of (\ref{gr}) becomes:

\begin{eqnarray*}
&&\exp \left( -cl_{1}\right) \times \sum_{n=1}^{\infty }\int \exp \left(
i\lambda \left( cl_{1}-\sum_{l=1}^{n}\left\vert Z^{\left( l-1\right)
}-Z^{\left( l\right) }\right\vert \right) \right) \\
&&\times \exp \left( i\lambda _{1}.\left( Z-Z_{1}-\sum_{l=1}^{n}\left(
Z^{\left( l-1\right) }-Z^{\left( l\right) }\right) \right) \right) d\lambda
d\lambda _{1}\dprod\limits_{l=1}^{n}C\left\vert Z^{\left( l-1\right)
}-Z^{\left( l\right) }\right\vert d\left\vert Z^{\left( l-1\right)
}-Z^{\left( l\right) }\right\vert dv_{l}
\end{eqnarray*}%
that can be written in terms of the angles as:%
\begin{eqnarray}
&&\exp \left( -cl_{1}\right) \times \sum_{n=1}^{\infty }\int \exp \left(
i\lambda cl_{1}+i\left\vert \lambda _{1}\right\vert \left\vert
Z-Z_{1}\right\vert \cos \left( \theta _{1}\right) \right)  \label{sh} \\
&&\times \exp \left( -i\sum_{l=1}^{n}\left( \lambda +\left\vert \lambda
_{1}\right\vert \cos \left( \theta _{l}\right) \right) \left\vert Z^{\left(
l-1\right) }-Z^{\left( l\right) }\right\vert \right) d\lambda d\lambda
_{1}\dprod\limits_{l=1}^{n}C\left\vert Z^{\left( l-1\right) }-Z^{\left(
l\right) }\right\vert d\left\vert Z^{\left( l-1\right) }-Z^{\left( l\right)
}\right\vert dv_{l}  \notag
\end{eqnarray}%
The integration over $\theta _{l}$ is:%
\begin{eqnarray*}
&&\pi \int_{0}^{\pi }\exp \left( -i\left( \lambda +\left\vert \lambda
_{1}\right\vert \cos \left( \theta _{l}\right) \right) \left\vert Z^{\left(
l-1\right) }-Z^{\left( l\right) }\right\vert \right) \sin \left( \theta
_{l}\right) d\theta _{l} \\
&=&-\frac{\pi i}{\left\vert \lambda _{1}\right\vert \left\vert Z^{\left(
l-1\right) }-Z^{\left( l\right) }\right\vert }\left( \exp \left( -i\left(
\lambda -\left\vert \lambda _{1}\right\vert \right) \left\vert Z^{\left(
l-1\right) }-Z^{\left( l\right) }\right\vert \right) -\exp \left( -i\left(
\lambda +\left\vert \lambda _{1}\right\vert \right) \left\vert Z^{\left(
l-1\right) }-Z^{\left( l\right) }\right\vert \right) \right) \\
&=&\frac{\pi i}{\left\vert \lambda _{1}\right\vert \left\vert Z^{\left(
l-1\right) }-Z^{\left( l\right) }\right\vert }\left( \exp \left( -i\left(
\lambda +\left\vert \lambda _{1}\right\vert \right) \left\vert Z^{\left(
l-1\right) }-Z^{\left( l\right) }\right\vert \right) -\exp \left( -i\left(
\lambda -\left\vert \lambda _{1}\right\vert \right) \left\vert Z^{\left(
l-1\right) }-Z^{\left( l\right) }\right\vert \right) \right)
\end{eqnarray*}%
and (\ref{sh}) rewrites:%
\begin{eqnarray*}
&&\exp \left( -cl_{1}\right) \times \sum_{n=1}^{\infty }\int \frac{-\pi i}{%
\left\vert \lambda _{1}\right\vert \left\vert Z-Z_{1}\right\vert }\left(
\exp \left( i\lambda cl_{1}+i\left\vert \lambda _{1}\right\vert \left\vert
Z-Z_{1}\right\vert \right) -\exp \left( i\lambda cl_{1}-i\left\vert \lambda
_{1}\right\vert \left\vert Z-Z_{1}\right\vert \right) \right) \\
&&\times \dprod\limits_{l=1}^{n}C\frac{\pi i}{\left\vert \lambda
_{1}\right\vert }\left( \exp \left( -i\left( \lambda +\left\vert \lambda
_{1}\right\vert \right) \left\vert Z^{\left( l-1\right) }-Z^{\left( l\right)
}\right\vert \right) -\exp \left( -i\left( \lambda -\left\vert \lambda
_{1}\right\vert \right) \left\vert Z^{\left( l-1\right) }-Z^{\left( l\right)
}\right\vert \right) \right) \\
&&\times d\left\vert Z^{\left( l-1\right) }-Z^{\left( l\right) }\right\vert
d\lambda \left\vert \lambda _{1}\right\vert ^{2}d\left\vert \lambda
_{1}\right\vert
\end{eqnarray*}%
We can then perform the integrals over the norms $\left\vert Z^{\left(
l-1\right) }-Z^{\left( l\right) }\right\vert $, which yields:

\begin{eqnarray*}
&&\exp \left( -cl_{1}\right) \times \sum_{n=1}^{\infty }\int \frac{-\pi i}{%
\left\vert \lambda _{1}\right\vert \left\vert Z-Z_{1}\right\vert }\left(
\exp \left( i\lambda cl_{1}+i\left\vert \lambda _{1}\right\vert \left\vert
Z-Z_{1}\right\vert \right) -\exp \left( i\lambda cl_{1}-i\left\vert \lambda
_{1}\right\vert \left\vert Z-Z_{1}\right\vert \right) \right) \\
&&\times \dprod\limits_{l=1}^{n}C\frac{\pi }{\left\vert \lambda
_{1}\right\vert }\left( \frac{1}{\lambda +\left\vert \lambda _{1}\right\vert
-i\varepsilon }-\frac{1}{\lambda -\left\vert \lambda _{1}\right\vert
-i\varepsilon }\right) d\lambda \left\vert \lambda _{1}\right\vert
^{2}d\left\vert \lambda _{1}\right\vert
\end{eqnarray*}%
Performing the sum yields then the following expression for (\ref{sh}):

\begin{eqnarray*}
&&\exp \left( -cl_{1}\right) \times \int \frac{-\pi i}{\left\vert \lambda
_{1}\right\vert \left\vert Z-Z_{1}\right\vert }\left( \exp \left( i\lambda
cl_{1}+i\left\vert \lambda _{1}\right\vert \left\vert Z-Z_{1}\right\vert
\right) -\exp \left( i\lambda cl_{1}-i\left\vert \lambda _{1}\right\vert
\left\vert Z-Z_{1}\right\vert \right) \right) \\
&&\times \frac{-C\frac{2\pi }{\left( \lambda +\left\vert \lambda
_{1}\right\vert -i\varepsilon \right) \left( \lambda -\left\vert \lambda
_{1}\right\vert -i\varepsilon \right) }}{1+C\frac{2\pi }{\left( \lambda
+\left\vert \lambda _{1}\right\vert -i\varepsilon \right) \left( \lambda
-\left\vert \lambda _{1}\right\vert -i\varepsilon \right) }}d\lambda
\left\vert \lambda _{1}\right\vert ^{2}d\left\vert \lambda _{1}\right\vert \\
&=&\exp \left( -cl_{1}\right) \times \int \frac{-\pi i}{\left\vert \lambda
_{1}\right\vert \left\vert Z-Z_{1}\right\vert }\left( \exp \left( i\lambda
cl_{1}+i\left\vert \lambda _{1}\right\vert \left\vert Z-Z_{1}\right\vert
\right) -\exp \left( i\lambda cl_{1}-i\left\vert \lambda _{1}\right\vert
\left\vert Z-Z_{1}\right\vert \right) \right) \\
&&\times \frac{-2\pi C}{\left( \lambda +\left\vert \lambda _{1}\right\vert
-i\varepsilon \right) \left( \lambda -\left\vert \lambda _{1}\right\vert
-i\varepsilon \right) +2\pi C}d\lambda \left\vert \lambda _{1}\right\vert
^{2}d\left\vert \lambda _{1}\right\vert
\end{eqnarray*}

Ultimately, the previous formula can be reduced to a single expression, by
performing the change of variable $x=-$ $\left\vert \lambda _{1}\right\vert $
in the term with $\exp \left( i\lambda cl_{1}-i\left\vert \lambda
_{1}\right\vert \left\vert Z-Z_{1}\right\vert \right) $ in factor. We obtain:%
\begin{equation*}
\exp \left( -cl_{1}\right) \times \int \frac{-\pi i}{\left\vert
Z-Z_{1}\right\vert }\exp \left( i\lambda cl_{1}+i\lambda _{1}\left\vert
Z-Z_{1}\right\vert \right) \frac{-2\pi C\lambda _{1}}{\left( \lambda
+\lambda _{1}-i\varepsilon \right) \left( \lambda -\lambda _{1}-i\varepsilon
\right) +2\pi C}d\lambda d\lambda _{1}
\end{equation*}%
where the integral over $\lambda _{1}$ is now performed with $\lambda
_{1}\in 
\mathbb{R}
$. This integral is computed by the residue theorem, where the residues
satisfy:%
\begin{equation*}
\lambda _{1}^{2}=\left( \lambda -i\varepsilon \right) ^{2}+2\pi C
\end{equation*}%
leading to write (\ref{sh}) as: 
\begin{eqnarray}
&&\exp \left( -cl_{1}\right) \times \int \frac{-\pi i}{\left\vert
Z-Z_{1}\right\vert }\exp \left( i\lambda cl_{1}+i\sqrt{\left( \lambda
-i\varepsilon \right) ^{2}+2\pi C}\left\vert Z-Z_{1}\right\vert \right)
d\lambda  \label{hs} \\
&&+\exp \left( -cl_{1}\right) \times \int \frac{-\pi i}{\left\vert
Z-Z_{1}\right\vert }\exp \left( i\lambda cl_{1}-i\sqrt{\left( \lambda
-i\varepsilon \right) ^{2}+2\pi C}\left\vert Z-Z_{1}\right\vert \right)
d\lambda  \notag
\end{eqnarray}%
We then perform the change of variable:%
\begin{eqnarray*}
x &=&\lambda +\sqrt{\lambda ^{2}+2\pi C} \\
dx &=&\left( 1+\frac{\lambda }{\sqrt{\lambda ^{2}+2\pi C}}\right) d\lambda \\
&=&\frac{x}{\sqrt{\lambda ^{2}+2\pi C}}d\lambda =\frac{2x^{2}}{x^{2}+2\pi C}%
d\lambda
\end{eqnarray*}%
and rewrite the exponents in (\ref{hs}) as:%
\begin{eqnarray*}
\lambda cl_{1}+\sqrt{\left( \lambda -i\varepsilon \right) ^{2}+2\pi C}%
\left\vert Z-Z_{1}\right\vert &=&\frac{cl_{1}+\left\vert Z-Z_{1}\right\vert 
}{2}\left( \lambda +\sqrt{\lambda ^{2}+2\pi C}\right) \\
&&+\frac{cl_{1}-\left\vert Z-Z_{1}\right\vert }{2}\left( \lambda -\sqrt{%
\lambda ^{2}+2\pi C}\right) \\
&=&\frac{cl_{1}+\left\vert Z-Z_{1}\right\vert }{2}\left( \lambda +\sqrt{%
\lambda ^{2}+2\pi C}\right) -\frac{cl_{1}-\left\vert Z-Z_{1}\right\vert }{2}%
\frac{2\pi C}{\lambda +\sqrt{\lambda ^{2}+2\pi C}} \\
&=&\frac{cl_{1}+\left\vert Z-Z_{1}\right\vert }{2}x-\frac{cl_{1}-\left\vert
Z-Z_{1}\right\vert }{2}\frac{2\pi C}{x}
\end{eqnarray*}%
and:%
\begin{equation*}
\lambda cl_{1}-\sqrt{\left( \lambda -i\varepsilon \right) ^{2}+2\pi C}%
\left\vert Z-Z_{1}\right\vert =\frac{cl_{1}-\left\vert Z-Z_{1}\right\vert }{2%
}x-\frac{cl_{1}+\left\vert Z-Z_{1}\right\vert }{2}\frac{2\pi C}{x}
\end{equation*}%
As a consequence, expression (\ref{hs}) becomes: 
\begin{eqnarray*}
&&\exp \left( -cl_{1}\right) \times \int \frac{-\pi i}{\left\vert
Z-Z_{1}\right\vert }\exp \left( i\left( \frac{cl_{1}+\left\vert
Z-Z_{1}\right\vert }{2}x-\frac{cl_{1}-\left\vert Z-Z_{1}\right\vert }{2}%
\frac{2\pi C}{x}\right) \right) dx \\
&&+\exp \left( -cl_{1}\right) \times \int \frac{-\pi i}{\left\vert
Z-Z_{1}\right\vert }\exp \left( i\left( \frac{cl_{1}-\left\vert
Z-Z_{1}\right\vert }{2}x-\frac{cl_{1}+\left\vert Z-Z_{1}\right\vert }{2}%
\frac{2\pi C}{x}\right) \right) dx \\
&&+2\pi C\exp \left( -cl_{1}\right) \times \int \frac{-\pi i}{\left\vert
Z-Z_{1}\right\vert }\exp \left( i\left( \frac{cl_{1}+\left\vert
Z-Z_{1}\right\vert }{2}x-\frac{cl_{1}-\left\vert Z-Z_{1}\right\vert }{2}%
\frac{2\pi C}{x}\right) \right) \frac{1}{x^{2}}dx \\
&&+2\pi C\times \int \frac{-\pi i}{\left\vert Z-Z_{1}\right\vert }\exp
\left( i\left( \frac{cl_{1}-\left\vert Z-Z_{1}\right\vert }{2}x-\frac{%
cl_{1}+\left\vert Z-Z_{1}\right\vert }{2}\frac{2\pi C}{x}\right) \right) 
\frac{1}{x^{2}}dx
\end{eqnarray*}%
Performing the change of variable $y=\frac{1}{x}$ in the two last
expressions yields:%
\begin{eqnarray*}
&&\exp \left( -cl_{1}\right) \left( 1+2\pi C\right) \times \left( \int \frac{%
-\pi i}{\left\vert Z-Z_{1}\right\vert }\exp \left( i\left( \frac{%
cl_{1}+\left\vert Z-Z_{1}\right\vert }{2}x-\frac{cl_{1}-\left\vert
Z-Z_{1}\right\vert }{2}\frac{2\pi C}{x}\right) \right) dx\right. \\
&&\left. +\int \frac{-\pi i}{\left\vert Z-Z_{1}\right\vert }\exp \left(
i\left( \frac{cl_{1}-\left\vert Z-Z_{1}\right\vert }{2}x-\frac{%
cl_{1}+\left\vert Z-Z_{1}\right\vert }{2}\frac{2\pi C}{x}\right) \right)
dx\right)
\end{eqnarray*}%
and by analytic continuation $x\rightarrow ix$, this becomes:%
\begin{eqnarray*}
&&\exp \left( -cl_{1}\right) \left( 1+2\pi C\right) \times \left( \int \frac{%
\pi }{\left\vert Z-Z_{1}\right\vert }\exp \left( -\left( \frac{%
cl_{1}+\left\vert Z-Z_{1}\right\vert }{2}x-\frac{cl_{1}-\left\vert
Z-Z_{1}\right\vert }{2}\frac{2\pi C}{x}\right) \right) dx\right. \\
&&\left. +\int \frac{\pi }{\left\vert Z-Z_{1}\right\vert }\exp \left(
-\left( \frac{cl_{1}-\left\vert Z-Z_{1}\right\vert }{2}x-\frac{%
cl_{1}+\left\vert Z-Z_{1}\right\vert }{2}\frac{2\pi C}{x}\right) \right)
dx\right)
\end{eqnarray*}%
Ultimately, reintroducing the constraint $H\left( cl_{1}-\left\vert
Z-Z_{1}\right\vert \right) $ and the factor $\frac{A_{1}}{A}$, (\ref{gr})
writes:%
\begin{eqnarray}
\left\vert \Psi \left( \theta -l_{1},Z_{1}\right) \right\vert ^{2}\frac{%
\delta \omega ^{-1}\left( J,\theta ,Z\right) }{\delta \left\vert \Psi \left(
\theta -l_{1},Z_{1}\right) \right\vert ^{2}} &=&\left( 1+2\pi C\right) \frac{%
A_{1}}{A}\frac{\exp \left( -cl_{1}\right) }{\left\vert Z-Z_{1}\right\vert }%
\left( \sqrt{\frac{cl_{1}-\left\vert Z-Z_{1}\right\vert }{cl_{1}+\left\vert
Z-Z_{1}\right\vert }}+\sqrt{\frac{cl_{1}+\left\vert Z-Z_{1}\right\vert }{%
cl_{1}-\left\vert Z-Z_{1}\right\vert }}\right)  \notag \\
&&\times K_{1}\left( \frac{cl_{1}-\left\vert Z-Z_{1}\right\vert }{2}2\pi C%
\frac{cl_{1}+\left\vert Z-Z_{1}\right\vert }{2}\right) \omega \left(
J,\theta -l_{1},Z_{1}\right)  \notag \\
&=&\left( 1+2\pi C\right) \frac{A_{1}}{A}\frac{\exp \left( -cl_{1}\right) }{%
\left\vert Z-Z_{1}\right\vert }\left( \sqrt{\frac{cl_{1}-\left\vert
Z-Z_{1}\right\vert }{cl_{1}+\left\vert Z-Z_{1}\right\vert }}+\sqrt{\frac{%
cl_{1}+\left\vert Z-Z_{1}\right\vert }{cl_{1}-\left\vert Z-Z_{1}\right\vert }%
}\right)  \notag \\
&&\times K_{1}\left( \pi C\frac{\left( cl_{1}\right) ^{2}-\left\vert
Z-Z_{1}\right\vert ^{2}}{2}\right) \omega \left( J,\theta -l_{1},Z_{1}\right)
\label{cdl}
\end{eqnarray}%
In first approximation, the right hand side of (\ref{cdl}) is:%
\begin{eqnarray}
&&\frac{\exp \left( -cl_{1}\right) \left( cl_{1}+\left\vert
Z-Z_{1}\right\vert \right) }{B\left\vert Z-Z_{1}\right\vert }\exp \left(
-\pi C\frac{\left( cl_{1}\right) ^{2}-\left\vert Z-Z_{1}\right\vert ^{2}}{2}%
\right) \omega \left( J,\theta -l_{1},Z_{1}\right)  \label{rtt} \\
&\sim &\frac{\exp \left( -cl_{1}\right) }{B}\exp \left( -\pi Ccl_{1}\frac{%
cl_{1}-\left\vert Z-Z_{1}\right\vert }{2}\right) H\left( cl_{1}-\left\vert
Z-Z_{1}\right\vert \right) \omega \left( J,\theta -l_{1},Z_{1}\right)  \notag
\end{eqnarray}%
for $cl_{1}>>\left\vert Z-Z_{1}\right\vert $. This can also be replaced by a
simplest form:%
\begin{equation}
\left\vert \Psi \left( \theta -l_{1},Z_{1}\right) \right\vert ^{2}\frac{%
\delta \omega ^{-1}\left( J,\theta ,Z\right) }{\delta \left\vert \Psi \left(
\theta -l_{1},Z_{1}\right) \right\vert ^{2}}\simeq \frac{\exp \left(
-cl_{1}-\alpha \left( \left( cl_{1}\right) ^{2}-\left\vert
Z-Z_{1}\right\vert ^{2}\right) \right) }{B}H\left( cl_{1}-\left\vert
Z-Z_{1}\right\vert \right) \omega \left( J,\theta -l_{1},Z_{1}\right)
\label{rrt}
\end{equation}%
where $B$ and $\alpha $ are constants.

Using (\ref{dvtr}), the same computation can be performed by replacing $\hat{%
T}$ with $\check{T}$ and we obtain: 
\begin{equation}
\left\vert \Psi \left( \theta -l_{1},Z_{1}\right) \right\vert ^{2}\frac{%
\delta \omega ^{-1}\left( J,\theta ,Z\right) }{\delta \left\vert \Psi \left(
\theta -l_{1},Z_{1}\right) \right\vert ^{2}}\simeq \frac{\exp \left(
-cl_{1}-\alpha \left( \left( cl_{1}\right) ^{2}-\left\vert
Z-Z_{1}\right\vert ^{2}\right) \right) }{D}H\left( cl_{1}-\left\vert
Z-Z_{1}\right\vert \right) \omega ^{-1}\left( J,\theta -l_{1},Z_{1}\right)
\label{trr}
\end{equation}%
with $D$ a constant:%
\begin{equation*}
D=\frac{B}{\bar{\omega}^{2}}
\end{equation*}%
with $\bar{\omega}\left( J\right) $ the average activity.

\subsubsection*{1.2.2 General formula}

For an arbitrary connectivity function:%
\begin{equation*}
\tilde{T}\left( \left\vert Z^{\left( l-1\right) }-Z^{\left( l\right)
}\right\vert \right) =C\exp \left( -c\left\vert Z^{\left( l-1\right)
}-Z^{\left( l\right) }\right\vert \right) f\left( \left\vert Z^{\left(
l-1\right) }-Z^{\left( l\right) }\right\vert \right)
\end{equation*}%
we can factor $C\exp \left( -cl\right) $ as in the previous paragraph. It
amounts to replace: 
\begin{equation*}
\tilde{T}\left( \left\vert Z^{\left( l-1\right) }-Z^{\left( l\right)
}\right\vert \right) \rightarrow f\left( \left\vert Z^{\left( l-1\right)
}-Z^{\left( l\right) }\right\vert \right)
\end{equation*}%
We rewrite (\ref{gr}) as:%
\begin{eqnarray}
&&\left\vert \Psi \left( \theta -l_{1},Z_{1}\right) \right\vert ^{2}\frac{%
\delta \omega ^{-1}\left( J,\theta ,Z\right) }{\delta \left\vert \Psi \left(
\theta -l_{1},Z_{1}\right) \right\vert ^{2}}  \label{Srs} \\
&=&\sum_{n=1}^{\infty }\int \omega \left( J,\theta -l_{1},Z_{1}\right)
\times \mathit{T}_{1}^{\prime \prime }\left( \lambda +\lambda
_{1}.v_{1}\right) dv_{1}\dprod\limits_{l=2}^{n}\int \mathit{T}^{\prime
\prime }\left( \lambda +\lambda _{1}.v_{l}\right) dv_{l}\exp \left( i\lambda
cl_{1}+i\lambda _{1}.\left( Z-Z_{1}\right) \right) d\lambda d\lambda _{1} 
\notag \\
&=&\delta \left( \left\vert Z_{1}-Z\right\vert -cl_{1}\right) \tilde{T}%
_{1}\left( \left\vert Z-Z^{\left( 1\right) }\right\vert \right) \omega
\left( J,\theta -l_{1},Z_{1}\right)  \notag \\
&&+\left( -1\right) ^{n}\int \omega \left( J,\theta -l_{1},Z_{1}\right)
\times \frac{\mathit{T}_{1}^{\prime \prime }\left( \lambda +\lambda
_{1}.v_{1}\right) }{2}dv_{1}\dprod\limits_{l=2}^{n}\int \frac{\mathit{T}%
^{\prime \prime }\left( \lambda +\lambda _{1}.v_{l}\right) }{2}dv_{l}\exp
\left( i\lambda cl_{1}+i\lambda _{1}.\left( Z-Z_{1}\right) \right) d\lambda
d\lambda _{1}  \notag
\end{eqnarray}%
With the convention that for $n=1$, the product $\dprod\limits_{l=2}^{n}$ is
set to be equal to $1$. The functions $\mathit{T}_{1}$ and $\mathit{T}$ are
the fourier transform of $\tilde{T}_{1}H$\ and $\tilde{T}H$\ respectively,
and $H$ is the heaviside function. Remark that the first term of (\ref{Srs})
expresses the Dirac function $\delta \left( \left\vert Z_{1}-Z\right\vert
-cl_{1}\right) $ as a Fourier transform:%
\begin{eqnarray*}
&&\exp \left( i\lambda \left( cl_{1}-\sum_{l=1}^{n}\left\vert Z^{\left(
0\right) }-Z^{\left( 1\right) }\right\vert \right) \right) \\
&&\times \exp \left( i\lambda _{1}.\left( Z-Z_{1}-\sum_{l=1}^{n}\left(
Z^{\left( 0\right) }-Z^{\left( 1\right) }\right) \right) \right) d\lambda
d\lambda _{1}\left\vert Z^{\left( 0\right) }-Z^{\left( 1\right) }\right\vert
^{2}d\left\vert Z^{\left( 0\right) }-Z^{\left( 1\right) }\right\vert dv_{l}
\end{eqnarray*}%
Some terms of (\ref{Srs}) can be written in a useful form for the sequel: 
\begin{eqnarray}
\frac{1}{2}\int \mathit{T}^{\prime \prime }\left( \lambda +\lambda
_{1}.v_{l}\right) dv_{l} &=&\pi \int_{0}^{\pi }\mathit{T}^{\prime \prime
}\left( \lambda +\left\vert \lambda _{1}\right\vert \cos \left( \theta
_{l}\right) \right) \sin \left( \theta _{l}\right) d\theta _{l}  \notag \\
&=&\pi \int_{-1}^{1}\mathit{T}^{\prime \prime }\left( \lambda +\left\vert
\lambda _{1}\right\vert u\right) du  \notag \\
&=&\frac{2\pi \left( \mathit{T}^{\prime }\left( \lambda +\left\vert \lambda
_{1}\right\vert \right) -\mathit{T}^{\prime }\left( \lambda -\left\vert
\lambda _{1}\right\vert \right) \right) }{2\left\vert \lambda
_{1}\right\vert }  \notag \\
&\equiv &\mathit{\bar{T}}\left( \lambda ,\left\vert \lambda _{1}\right\vert
\right)  \label{brT}
\end{eqnarray}%
\begin{eqnarray}
\int \mathit{T}_{1}^{\prime \prime }\left( \lambda +\lambda
_{1}.v_{l}\right) dv_{l} &=&\frac{2\pi \left( \mathit{T}_{1}^{\prime }\left(
\lambda +\left\vert \lambda _{1}\right\vert \right) -\mathit{T}_{1}^{\prime
}\left( \lambda -\left\vert \lambda _{1}\right\vert \right) \right) }{%
2\left\vert \lambda _{1}\right\vert }  \notag \\
&\equiv &\mathit{\bar{T}}_{1}\left( \lambda ,\left\vert \lambda
_{1}\right\vert \right)  \label{rbT}
\end{eqnarray}%
\begin{eqnarray}
\exp \left( i\lambda _{1}.\left( Z-Z_{1}\right) \right) d\lambda _{1}
&=&\exp \left( i\cos \left( \theta _{1}\right) \left\vert \lambda
_{1}\right\vert \left\vert Z-Z_{1}\right\vert \right) \sin \left( \theta
_{1}\right) \left\vert \lambda _{1}\right\vert ^{2}d\left\vert \lambda
_{1}\right\vert d\theta _{1}  \label{rF} \\
&=&\exp \left( iu\left\vert \lambda _{1}\right\vert \left\vert
Z-Z_{1}\right\vert \right) \left\vert \lambda _{1}\right\vert
^{2}d\left\vert \lambda _{1}\right\vert du  \notag
\end{eqnarray}%
Remark that the functions of $x$:%
\begin{equation*}
\mathit{\bar{T}}\left( \lambda ,x\right) =\frac{2\pi \left( \mathit{T}%
^{\prime }\left( \lambda +x\right) -\mathit{T}^{\prime }\left( \lambda
-x\right) \right) }{2x}\text{ and }\mathit{\bar{T}}_{1}\left( \lambda
,x\right) =\frac{2\pi \left( \mathit{T}_{1}^{\prime }\left( \lambda
+x\right) -\mathit{T}_{1}^{\prime }\left( \lambda -x\right) \right) }{2x}
\end{equation*}%
are even.

\subsubsection*{1.2.3 Estimation of (\protect\ref{Srs})}

Using (\ref{brT}), (\ref{rbT}) and (\ref{rF}), equation (\ref{Srs}) becomes:%
\begin{eqnarray}
&&\left\vert \Psi \left( \theta -l_{1},Z_{1}\right) \right\vert ^{2}\frac{%
\delta \omega ^{-1}\left( J,\theta ,Z\right) }{\delta \left\vert \Psi \left(
\theta -l_{1},Z_{1}\right) \right\vert ^{2}}  \notag \\
&=&\sum_{n=1}^{\infty }\left( -1\right) ^{n}\int \omega \left( J,\theta
-l_{1},Z_{1}\right) \times \mathit{T}_{1}\left( \lambda +\lambda
_{1}.v_{1}\right) dv_{1}\dprod\limits_{l=2}^{n}\int \mathit{T}\left( \lambda
+\lambda _{1}.v_{l}\right) dv_{l}\exp \left( i\lambda cl_{1}+i\lambda
_{1}.\left( Z-Z_{1}\right) \right) d\lambda d\lambda _{1}  \notag \\
&=&-\int \omega \left( J,\theta -l_{1},Z_{1}\right) \times \frac{\mathit{%
\bar{T}}_{1}\left( \lambda ,\left\vert \lambda _{1}\right\vert \right) }{1+%
\mathit{\bar{T}}\left( \lambda ,\left\vert \lambda _{1}\right\vert \right) }%
\exp \left( i\lambda cl_{1}\right) \int_{-1}^{1}\exp \left( iu\left\vert
\lambda _{1}\right\vert \left\vert Z-Z_{1}\right\vert \right) \left\vert
\lambda _{1}\right\vert ^{2}d\left\vert \lambda _{1}\right\vert dud\lambda 
\notag \\
&=&-\int \omega \left( J,\theta -l_{1},Z_{1}\right) \times \frac{\mathit{%
\bar{T}}_{1}\left( \lambda ,\left\vert \lambda _{1}\right\vert \right) }{1+%
\mathit{\bar{T}}\left( \lambda ,\left\vert \lambda _{1}\right\vert \right) }%
\exp \left( i\lambda cl_{1}\right) \left( 2\frac{\sin \left( \left\vert
\lambda _{1}\right\vert \left\vert Z-Z_{1}\right\vert \right) }{\left\vert
Z-Z_{1}\right\vert }\left\vert \lambda _{1}\right\vert \right) d\left\vert
\lambda _{1}\right\vert d\lambda  \label{tmn}
\end{eqnarray}%
We remark that for even functions $f$, the following identity holds: 
\begin{eqnarray*}
&&\int_{0}^{+\infty }f\left( \left\vert \lambda _{1}\right\vert \right) 2%
\frac{\sin \left( \left\vert \lambda _{1}\right\vert \left\vert
Z-Z_{1}\right\vert \right) }{\left\vert Z-Z_{1}\right\vert }\left\vert
\lambda _{1}\right\vert d\left\vert \lambda _{1}\right\vert \\
&=&\int_{0}^{+\infty }f\left( x\right) \frac{\exp \left( ix\left\vert
Z-Z_{1}\right\vert \right) -\exp \left( -ix\left\vert Z-Z_{1}\right\vert
\right) }{i\left\vert Z-Z_{1}\right\vert }xdx \\
&=&\int_{0}^{+\infty }f\left( x\right) \frac{\exp \left( ix\left\vert
Z-Z_{1}\right\vert \right) }{i\left\vert Z-Z_{1}\right\vert }%
xdx+\int_{-\infty }^{0}f\left( -x\right) \frac{\exp \left( ix\left\vert
Z-Z_{1}\right\vert \right) }{i\left\vert Z-Z_{1}\right\vert }xdx \\
&=&\int_{-\infty }^{+\infty }f\left( x\right) \frac{\exp \left( ix\left\vert
Z-Z_{1}\right\vert \right) }{i\left\vert Z-Z_{1}\right\vert }xdx
\end{eqnarray*}%
so that (\ref{tmn}) becomes:%
\begin{eqnarray}
&&\left\vert \Psi \left( \theta -l_{1},Z_{1}\right) \right\vert ^{2}\frac{%
\delta \omega ^{-1}\left( J,\theta ,Z\right) }{\delta \left\vert \Psi \left(
\theta -l_{1},Z_{1}\right) \right\vert ^{2}}  \label{nmt} \\
&=&-\int \omega \left( J,\theta -l_{1},Z_{1}\right) \times \frac{\mathit{%
\bar{T}}_{1}\left( \lambda ,\lambda _{1}\right) }{1+\mathit{\bar{T}}\left(
\lambda ,\lambda _{1}\right) }\frac{\lambda _{1}}{i\left\vert
Z-Z_{1}\right\vert }\exp \left( i\lambda cl_{1}+i\lambda _{1}\left\vert
Z-Z_{1}\right\vert \right) d\lambda _{1}d\lambda  \notag \\
&=&-\int \omega \left( J,\theta -l_{1},Z_{1}\right) \times \frac{\pi \left( 
\mathit{T}_{1}^{\prime }\left( \lambda +\lambda _{1}\right) -\mathit{T}%
_{1}^{\prime }\left( \lambda -\lambda _{1}\right) \right) }{\lambda _{1}+\pi
\left( \mathit{T}^{\prime }\left( \lambda +\lambda _{1}\right) -\mathit{T}%
^{\prime }\left( \lambda -\lambda _{1}\right) \right) }\frac{\lambda _{1}}{%
i\left\vert Z-Z_{1}\right\vert }  \notag \\
&&\times \exp \left( i\lambda cl_{1}+i\lambda _{1}\left\vert
Z-Z_{1}\right\vert \right) d\lambda _{1}d\lambda  \notag \\
&=&-\int \omega \left( J,\theta -l_{1},Z_{1}\right) \times \frac{\pi \left( 
\mathit{T}_{1}^{\prime }\left( \lambda +\lambda _{1}\right) -\mathit{T}%
_{1}^{\prime }\left( \lambda -\lambda _{1}\right) \right) }{\lambda _{1}+\pi
\left( \mathit{T}^{\prime }\left( \lambda +\lambda _{1}\right) -\mathit{T}%
^{\prime }\left( \lambda -\lambda _{1}\right) \right) }\frac{\lambda _{1}}{%
i\left\vert Z-Z_{1}\right\vert }  \notag \\
&&\times \exp \left( iu\left( cl_{1}+\left\vert Z-Z_{1}\right\vert \right)
\right) \times \exp \left( iv\left( cl_{1}-\left\vert Z-Z_{1}\right\vert
\right) \right) d\lambda _{1}d\lambda  \notag
\end{eqnarray}%
As in the previous paragraph, we also simplify (\ref{nmt}) by writing $%
\mathit{T}_{1}$ as a function of $\mathit{T}$: 
\begin{equation*}
\mathit{T}_{1}^{\prime }\left( \lambda +\lambda _{1}\right) -\mathit{T}%
_{1}^{\prime }\left( \lambda -\lambda _{1}\right) =\frac{A_{1}}{A}\left( 
\mathit{T}^{\prime }\left( \lambda +\lambda _{1}\right) -\mathit{T}^{\prime
}\left( \lambda -\lambda _{1}\right) \right)
\end{equation*}%
and by setting:%
\begin{eqnarray*}
u &=&\frac{\lambda +\lambda _{1}}{2} \\
v &=&\frac{\lambda +\lambda _{1}}{2}
\end{eqnarray*}%
so that we are lead to:%
\begin{eqnarray}
\left\vert \Psi \left( \theta -l_{1},Z_{1}\right) \right\vert ^{2}\frac{%
\delta \omega ^{-1}\left( J,\theta ,Z\right) }{\delta \left\vert \Psi \left(
\theta -l_{1},Z_{1}\right) \right\vert ^{2}} &=&-\int \omega \left( J,\theta
-l_{1},Z_{1}\right) \times \frac{A_{1}}{A}\frac{\pi \left( \mathit{T}%
_{1}^{\prime }\left( \lambda +\lambda _{1}\right) -\mathit{T}_{1}^{\prime
}\left( \lambda -\lambda _{1}\right) \right) }{\lambda _{1}+\pi \left( 
\mathit{T}^{\prime }\left( \lambda +\lambda _{1}\right) -\mathit{T}^{\prime
}\left( \lambda -\lambda _{1}\right) \right) }\frac{\lambda _{1}}{%
i\left\vert Z-Z_{1}\right\vert }  \notag \\
&&\times \exp \left( iu\left( cl_{1}+\left\vert Z-Z_{1}\right\vert \right)
\right) \times \exp \left( iv\left( cl_{1}-\left\vert Z-Z_{1}\right\vert
\right) \right) d\lambda _{1}d\lambda  \label{tnm}
\end{eqnarray}%
Remark that the particular case of the exponential connectivity function is
encompassed in (\ref{nmt}). Actually, if we choose:%
\begin{equation*}
\tilde{T}\left( \left\vert Z^{\left( l-1\right) }-Z^{\left( l\right)
}\right\vert \right) =C\frac{\exp \left( -c\left\vert Z^{\left( l-1\right)
}-Z^{\left( l\right) }\right\vert \right) }{\left\vert Z^{\left( l-1\right)
}-Z^{\left( l\right) }\right\vert }
\end{equation*}%
we have:%
\begin{equation*}
\dprod\limits_{l=1}^{n}\tilde{T}\left( \left\vert Z^{\left( l-1\right)
}-Z^{\left( l\right) }\right\vert \right) =\exp \left( -cl_{1}\right)
\dprod\limits_{l=1}^{n}\frac{C}{\left\vert Z^{\left( l-1\right) }-Z^{\left(
l\right) }\right\vert }
\end{equation*}%
For such a choice, we have formally: $\mathit{T}=-iC\int \left( FH\right) $
where $H$ is the heaviside function. As a consequence:%
\begin{equation*}
\mathit{T}^{\prime }\left( \lambda \right) =CFH=-\frac{C}{\lambda
+i\varepsilon }
\end{equation*}%
and (\ref{tnm}) is equivalent to the expressions of appendix 1.3.2.1.

In the general case, we write $\lambda _{1}^{\left( r\right) }$, $r=1,...$
the solutions to the pole equation of (\ref{tnm}):%
\begin{equation*}
\lambda _{1}+\pi \left( \mathit{T}^{\prime }\left( \lambda +\lambda
_{1}\right) -\mathit{T}^{\prime }\left( \lambda -\lambda _{1}\right) \right)
=0
\end{equation*}%
For regular functions $\mathit{T}^{\prime }\left( \lambda +\lambda
_{1}\right) $ such that for $\lambda \rightarrow \infty $:%
\begin{equation*}
\mathit{T}^{\prime }\left( \lambda +\lambda _{1}\right) \simeq \frac{g\left(
\lambda +\lambda _{1}\right) }{\left( \lambda +\lambda _{1}\right) ^{l}}
\end{equation*}%
\begin{equation*}
\int \frac{1}{\left( \lambda -s\right) ^{l}}\left\vert \Psi \left( s\right)
\right\vert
\end{equation*}%
with $l>0$ given and $g$ bounded, the poles equation implies that for $%
\lambda \rightarrow \infty $:%
\begin{equation*}
\lambda _{1}\simeq \pm \lambda
\end{equation*}%
and as a consequence, we can write: 
\begin{equation}
\lambda _{1}^{\left( r\right) }=\sqrt{\lambda ^{2}+h_{r}\left( \lambda
\right) }  \label{psl}
\end{equation}%
where $h_{r}\left( \lambda \right) $ is bounded.

We can compute the values of the residues at each pole by the first order
expansion of $1+\pi \frac{\mathit{T}^{\prime }\left( \lambda +\lambda
_{1}\right) -\mathit{T}^{\prime }\left( \lambda -\lambda _{1}\right) }{%
\lambda _{1}}$:%
\begin{eqnarray*}
&&1+\pi \frac{\mathit{T}^{\prime }\left( \lambda +\lambda _{1}\right) -%
\mathit{T}^{\prime }\left( \lambda -\lambda _{1}\right) }{\lambda _{1}} \\
&\simeq &\pi \left( \frac{\mathit{T}^{\prime }\left( \lambda +\lambda
_{1}\right) -\mathit{T}^{\prime }\left( \lambda -\lambda _{1}\right) }{%
\lambda _{1}}-\frac{\mathit{T}^{\prime }\left( \lambda +\lambda _{1}^{\left(
r\right) }\right) -\mathit{T}^{\prime }\left( \lambda -\lambda _{1}^{\left(
r\right) }\right) }{\lambda _{1}^{\left( r\right) }}\right) \\
&\simeq &\pi \left( \frac{\frac{1}{\pi }+\mathit{T}^{\prime \prime }\left(
\lambda +\lambda _{1}^{\left( r\right) }\right) +\mathit{T}^{\prime \prime
}\left( \lambda -\lambda _{1}^{\left( r\right) }\right) }{\lambda
_{1}^{\left( r\right) }}\right) \\
&\simeq &\pi \left( \frac{\mathit{T}^{\prime \prime }\left( \lambda +\lambda
_{1}^{\left( r\right) }\right) +\mathit{T}^{\prime \prime }\left( \lambda
-\lambda _{1}^{\left( r\right) }\right) -\frac{\mathit{T}^{\prime }\left(
\lambda +\lambda _{1}^{\left( r\right) }\right) -\mathit{T}^{\prime }\left(
\lambda -\lambda _{1}^{\left( r\right) }\right) }{\lambda _{1}^{\left(
r\right) }}}{\lambda _{1}^{\left( r\right) }}\right)
\end{eqnarray*}%
For regular functions $\mathit{T}^{\prime }\left( \lambda +\lambda
_{1}\right) $, this can be expanded as:%
\begin{equation*}
2\pi \lambda _{1}^{\left( r\right) }\left( \sum_{k\geqslant 1}\frac{\mathit{T%
}^{\left( 2k+2\right) }\left( \lambda \right) }{\left( 2k\right) !}\left(
\lambda _{1}^{\left( r\right) }\right) ^{2k-2}-\sum_{k\geqslant 1}\frac{%
\mathit{T}^{\left( 2k+2\right) }\left( \lambda \right) }{\left( 2k+1\right) !%
}\left( \lambda _{1}^{\left( r\right) }\right) ^{2k-2}\right)
\end{equation*}%
and for relatively slowly varying functions, this reduces to:%
\begin{equation}
1+\pi \frac{\mathit{T}^{\prime }\left( \lambda +\lambda _{1}\right) -\mathit{%
T}^{\prime }\left( \lambda -\lambda _{1}\right) }{\lambda _{1}}\simeq 2\pi
\lambda _{1}^{\left( r\right) }\frac{\mathit{T}^{\left( 4\right) }\left(
\lambda \right) }{3}  \label{spl}
\end{equation}%
and the residue theorem implies to replace:%
\begin{eqnarray}
&&\frac{\pi \left( \mathit{T}_{1}^{\prime }\left( \lambda +\lambda
_{1}\right) -\mathit{T}_{1}^{\prime }\left( \lambda -\lambda _{1}\right)
\right) }{\lambda _{1}+\pi \left( \mathit{T}^{\prime }\left( \lambda
+\lambda _{1}\right) -\mathit{T}^{\prime }\left( \lambda -\lambda
_{1}\right) \right) }\frac{\lambda _{1}}{i\left\vert Z-Z_{1}\right\vert }
\label{dsr} \\
&\rightarrow &-\frac{i}{\pi }\frac{3}{\left\vert Z-Z_{1}\right\vert \mathit{T%
}^{\left( 4\right) }\left( \lambda \right) }  \notag
\end{eqnarray}%
in (\ref{tnm}). Using (\ref{psl}) and (\ref{dsr}) in (\ref{tnm}) leads to:%
\begin{eqnarray*}
&&\left\vert \Psi \left( \theta -l_{1},Z_{1}\right) \right\vert ^{2}\frac{%
\delta \omega ^{-1}\left( J,\theta ,Z\right) }{\delta \left\vert \Psi \left(
\theta -l_{1},Z_{1}\right) \right\vert ^{2}} \\
&\simeq &\sum_{r}\frac{i}{\pi }\frac{1}{\left\vert Z-Z_{1}\right\vert }\int
\omega \left( J,\theta -l_{1},Z_{1}\right) \\
&&\times \frac{3}{\mathit{T}^{\left( 4\right) }\left( \lambda \right) }\exp
\left( iu\left( cl_{1}+\left\vert Z-Z_{1}\right\vert \right) \right) \times
\exp \left( iv\left( cl_{1}-\left\vert Z-Z_{1}\right\vert \right) \right)
d\lambda \\
&=&\sum_{r}\frac{i}{\pi }\frac{1}{\left\vert Z-Z_{1}\right\vert }\int \omega
\left( J,\theta -l_{1},Z_{1}\right) \\
&&\times \frac{3}{\mathit{T}^{\left( 4\right) }\left( \lambda \right) }\exp
\left( iu\left( cl_{1}+\left\vert Z-Z_{1}\right\vert \right) \right) \times
\exp \left( iv\left( cl_{1}-\left\vert Z-Z_{1}\right\vert \right) \right)
d\lambda
\end{eqnarray*}%
where:%
\begin{eqnarray*}
u &=&\frac{\lambda +\lambda _{1}^{\left( r\right) }}{2}=\frac{\lambda
+f^{\left( r\right) }\left( \lambda \right) }{2} \\
v &=&\frac{\lambda +\lambda _{1}^{\left( r\right) }}{2}=\frac{\lambda
-f^{\left( r\right) }\left( \lambda \right) }{2}
\end{eqnarray*}%
As a consequence:%
\begin{eqnarray*}
v &=&\lambda -\sqrt{\lambda ^{2}+h_{r}\left( \lambda \right) } \\
&=&-\frac{h_{r}\left( \lambda \right) }{\lambda +\sqrt{\lambda
^{2}+h_{r}\left( \lambda \right) }} \\
&=&-\frac{h_{r}\left( \lambda \right) }{u}
\end{eqnarray*}%
For $h_{r}\left( \lambda \right) $ varying slowly, we can replace $%
h_{r}\left( \lambda \right) $ by its average $\bar{h}_{r}$, and we have:%
\begin{equation*}
v=-\frac{\bar{h}_{r}}{u}
\end{equation*}%
Replacing $\mathit{T}^{\left( 4\right) }\left( \lambda \right) $ by its
average $\mathit{\bar{T}}^{\left( 4\right) }$, we find:%
\begin{eqnarray*}
\left\vert \Psi \left( \theta -l_{1},Z_{1}\right) \right\vert ^{2}\frac{%
\delta \omega ^{-1}\left( J,\theta ,Z\right) }{\delta \left\vert \Psi \left(
\theta -l_{1},Z_{1}\right) \right\vert ^{2}} &\simeq &\sum_{r}\frac{i}{\pi }%
\frac{1}{\left\vert Z-Z_{1}\right\vert }\int \omega \left( J,\theta
-l_{1},Z_{1}\right) \\
&&\times \frac{3}{\mathit{\bar{T}}^{\left( 4\right) }}\exp \left( iu\left(
cl_{1}+\left\vert Z-Z_{1}\right\vert \right) \right) \times \exp \left( -i%
\frac{\bar{h}_{r}}{u}\left( cl_{1}-\left\vert Z-Z_{1}\right\vert \right)
\right) d\lambda
\end{eqnarray*}%
We can then apply the results of the previous paragraph for each $r$, and
has a consequence, we obtain:%
\begin{eqnarray}
\left\vert \Psi \left( \theta -l_{1},Z_{1}\right) \right\vert ^{2}\frac{%
\delta \omega ^{-1}\left( J,\theta ,Z\right) }{\delta \left\vert \Psi \left(
\theta -l_{1},Z_{1}\right) \right\vert ^{2}} &\simeq &\sum_{r}\left( 1+\bar{h%
}_{r}\right) \frac{3}{\mathit{\bar{T}}^{\left( 4\right) }}\omega \left(
J,\theta -l_{1},Z_{1}\right) \frac{\exp \left( -cl_{1}\right) }{\left\vert
Z-Z_{1}\right\vert }  \label{rgc} \\
&&\times \left( \sqrt{\frac{cl_{1}-\left\vert Z-Z_{1}\right\vert }{%
cl_{1}+\left\vert Z-Z_{1}\right\vert }}+\sqrt{\frac{cl_{1}+\left\vert
Z-Z_{1}\right\vert }{cl_{1}-\left\vert Z-Z_{1}\right\vert }}\right)
K_{1}\left( \bar{h}_{r}\frac{\left( cl_{1}\right) ^{2}-\left\vert
Z-Z_{1}\right\vert ^{2}}{4}\right)  \notag
\end{eqnarray}%
that becomes in first approximation:%
\begin{equation*}
\left\vert \Psi \left( \theta -l_{1},Z_{1}\right) \right\vert ^{2}\frac{%
\delta \omega ^{-1}\left( J,\theta ,Z\right) }{\delta \left\vert \Psi \left(
\theta -l_{1},Z_{1}\right) \right\vert ^{2}}\simeq \sum_{r}\frac{\exp \left(
-cl_{1}-\alpha _{r}\left( \left( cl_{1}\right) ^{2}-\left\vert
Z-Z_{1}\right\vert ^{2}\right) \right) }{B_{r}}H\left( cl_{1}-\left\vert
Z-Z_{1}\right\vert \right) \omega \left( J,\theta -l_{1},Z_{1}\right)
\end{equation*}%
where the $B_{r}$ are constant coefficients and $\alpha _{r}=\frac{\bar{h}%
_{r}}{4}$. As for (\ref{trr}), this also writes:%
\begin{equation}
\left\vert \Psi \left( \theta -l_{1},Z_{1}\right) \right\vert ^{2}\frac{%
\delta \omega ^{-1}\left( J,\theta ,Z\right) }{\delta \left\vert \Psi \left(
\theta -l_{1},Z_{1}\right) \right\vert ^{2}}\simeq \sum_{r}\frac{\exp \left(
-cl_{1}-\alpha _{r}\left( \left( cl_{1}\right) ^{2}-\left\vert
Z-Z_{1}\right\vert ^{2}\right) \right) }{D_{r}}H\left( cl_{1}-\left\vert
Z-Z_{1}\right\vert \right) \omega \left( J,\theta -l_{1},Z_{1}\right)
\label{ttr}
\end{equation}%
for some constants $D_{r}$.

\section*{Appendix 2 Non local expansion for $\protect\omega \left( \protect%
\theta ,Z\right) $ and propagation of signals.}

We can generalize the findings of appendix 1 to calculate and estimate the
successive derivatives of $\omega \left( J,\theta ,Z\right) $.

In 2.1 we will compute the successive derivatives through a graphical
expansion. Section 2.2 that the series expansion in the field of $\omega
\left( J,\theta ,Z\right) $ can be summed and expressed as an auxiliary path
integral depending on the connectivity functions. This integral can be
approximated through a saddle point approximation to obtain a formula for
the activity $\omega \left( J,\theta ,Z\right) $. The results are obtained
without considering the external sources that initiate fluctuations around
the background state. In section 2.3, we include the external sources to
compute the expansion of $\omega \left( J,\theta ,Z\right) $. Once obtained,
we analyze in section 2.4 the effect of the signal propagation on $\omega
\left( J,\theta ,Z\right) $. In section 2.5 we extend these results to
systems with multiple fields, including excitatory and inhibitory
interactions.

\subsection*{2.1 n-th derivatives of $\protect\omega \left( \protect\theta %
,Z\right) $ and $\protect\omega ^{-1}\left( \protect\theta ,Z\right) $ at $%
\left\vert \Psi _{0}\right\vert ^{2}$}

\subsubsection*{2.1.1 General formula}

Based on the results of Appendix 1, we can now compute $\omega \left(
J,\theta ,Z\right) $, $\omega ^{-1}\left( J,\theta ,Z\right) $ and their
derivatives 
\begin{equation*}
\frac{\delta ^{n}\omega \left( J,\theta ,Z\right) }{\dprod\limits_{i=1}^{n}%
\delta \left\vert \Psi \left( \theta -l_{i},Z_{i}\right) \right\vert ^{2}}
\end{equation*}%
and: 
\begin{equation*}
\frac{\delta ^{n}\omega ^{-1}\left( J,\theta ,Z\right) }{\dprod%
\limits_{i=1}^{n}\delta \left\vert \Psi \left( \theta -l_{i},Z_{i}\right)
\right\vert ^{2}}
\end{equation*}%
It allows to compute the expansion of the effective action, and also to
study the solutions of (\ref{btr}) without the locality assumption.

\paragraph*{2.1.1.1 Series expansion for the first order derivative of $%
\protect\omega \left( \protect\theta ,Z\right) $}

Recall that $\omega \left( \theta ,Z\right) $ is solution of\ (\ref{qf}):

\begin{eqnarray}
\omega ^{-1}\left( \theta ,Z\right) &=&G\left( J\left( \theta \right) +\frac{%
\kappa }{N}\int T\left( Z,Z_{1},\theta \right) \frac{\omega \left( \theta -%
\frac{\left\vert Z-Z_{1}\right\vert }{c},Z_{1}\right) }{\omega \left( \theta
,Z\right) }\right.  \label{qtf} \\
&&\times \left. \left( \mathcal{\bar{G}}_{0}\left( 0,Z_{1}\right)
+\left\vert \Psi \left( \theta -\frac{\left\vert Z-Z_{1}\right\vert }{c}%
,Z_{1}\right) \right\vert ^{2}\right) dZ_{1}\right)  \notag
\end{eqnarray}%
where:%
\begin{equation*}
T\left( Z,Z_{1},\theta \right) =\left\langle T\right\rangle \left(
Z,Z_{1}\right)
\end{equation*}%
and where, for the sake of simplicity, the expression:%
\begin{equation*}
\left\vert \Psi \left( \theta -\frac{\left\vert Z-Z_{1}\right\vert }{c}%
,Z_{1}\right) \right\vert ^{2}
\end{equation*}%
will stand for\footnote{%
See discussion after (\ref{GR})}:%
\begin{equation*}
\Psi _{0}^{\dagger }\left( Z_{1}\right) \Psi \left( \theta -\frac{\left\vert
Z-Z_{1}\right\vert }{c},Z_{1}\right) +\Psi _{0}\left( Z_{1}\right) \Psi
^{\dagger }\left( \theta -\frac{\left\vert Z-Z_{1}\right\vert }{c}%
,Z_{1}\right) +\left\vert \Psi \left( \theta -\frac{\left\vert
Z-Z_{1}\right\vert }{c},Z_{1}\right) \right\vert ^{2}
\end{equation*}%
To find the internal dynamics of the system we will consider $J\left( \theta
\right) =J$, a constant external current, usually $J=0$. We use a series
expansion in $\left\vert \Psi \left( \theta ^{\left( j\right) },Z_{1}\right)
\right\vert ^{2}$ of the right hand side of (\ref{qtf}) and write:%
\begin{eqnarray}
\omega \left( \theta ^{\left( i\right) },Z\right) &=&\omega \left( \theta
^{\left( i\right) },Z\right) _{\left\vert \Psi \right\vert ^{2}=0}
\label{snp} \\
&&+\int \sum_{n=1}^{\infty }\left( \frac{\delta ^{n}\omega \left( J,\theta
,Z\right) }{\dprod\limits_{i=1}^{n}\delta \left\vert \Psi \left( \theta
-l_{i},Z_{i}\right) \right\vert ^{2}}\right) _{\left\vert \Psi \right\vert
^{2}=0}\dprod\limits_{i=1}^{n}\left\vert \Psi \left( \theta
-l_{i},Z_{i}\right) \right\vert ^{2}  \notag
\end{eqnarray}%
The first term (\ref{snp}), i.e. $\omega \left( \theta ^{\left( i\right)
},Z\right) _{\left\vert \Psi \right\vert ^{2}=0}$, is a solution of:%
\begin{equation*}
F\left( J+\frac{\kappa }{N}\int T\left( Z,Z_{1},\theta \right) \frac{\omega
\left( \theta -\frac{\left\vert Z-Z_{1}\right\vert }{c},Z_{1}\right) }{%
\omega \left( \theta ,Z\right) }\left\vert \Psi _{0}\left( Z_{1}\right)
\right\vert ^{2}dZ_{1}\right)
\end{equation*}%
One solution is the static frequency (\ref{frs}) solution of:%
\begin{eqnarray*}
\omega \left( J,Z\right) &=&F\left( J+\frac{\kappa }{N}\int T\left(
Z,Z_{1}\right) \frac{\omega \left( Z_{1}\right) }{\omega \left( Z\right) }%
\left\vert \Psi _{0}\left( Z_{1}\right) \right\vert ^{2}dZ_{1}\right) \\
&\equiv &F\left[ J,\omega ,Z\right]
\end{eqnarray*}%
but any time dependent solution for $\left\vert \Psi \right\vert ^{2}=0$ is
also possible. This arises for non constant current $J\left( \theta \right) $%
. Equation (\ref{snp}) is the expansion of $\omega \left( \theta ^{\left(
i\right) },Z\right) $ around this solution, the dynamics depending on $%
\left\vert \Psi \left( \theta ^{\left( j\right) },Z_{1}\right) \right\vert
^{2}$. We set:%
\begin{equation*}
\omega \left( \theta ^{\left( i\right) },Z\right) _{\left\vert \Psi
\right\vert ^{2}=0}=\omega _{0}\left( J,Z\right)
\end{equation*}%
The first derivative $\frac{\delta \omega \left( J,\theta ,Z\right) }{\delta
\left\vert \Psi \left( \theta -l_{1},Z_{1}\right) \right\vert ^{2}}$ \ in (%
\ref{snp})\ has been computed in Appendix 1. It is given by:%
\begin{eqnarray}
\frac{\delta \omega ^{-1}\left( J,\theta ,Z\right) }{\delta \left\vert \Psi
\left( \theta -l_{1},Z_{1}\right) \right\vert ^{2}} &=&-\sum_{n=1}^{\infty }%
\frac{1}{\left( \mathcal{\bar{G}}_{0}\left( 0,Z_{1}\right) +\left\vert \Psi
\left( \theta -l_{1},Z_{1}\right) \right\vert ^{2}\right) }\int \omega
^{-1}\left( J,\theta -\sum_{l=1}^{n}\frac{\left\vert Z^{\left( l-1\right)
}-Z^{\left( l\right) }\right\vert }{c},Z_{1}\right)  \label{rdv} \\
&&\times \dprod\limits_{l=1}^{n}\check{T}\left( \theta -\sum_{j=1}^{l-1}%
\frac{\left\vert Z^{\left( j-1\right) }-Z^{\left( j\right) }\right\vert }{c}%
,Z^{\left( l-1\right) },Z^{\left( l\right) },\omega ,\Psi \right) \delta
\left( l_{1}-\sum_{l=1}^{n}\frac{\left\vert Z^{\left( l-1\right) }-Z^{\left(
l\right) }\right\vert }{c}\right) \dprod\limits_{l=1}^{n-1}dZ^{\left(
l\right) }  \notag
\end{eqnarray}%
where:%
\begin{eqnarray}
&&\check{T}\left( \theta ,Z,Z_{1},\omega ,\Psi \right)  \label{vdR} \\
&=&-\frac{\frac{\kappa }{N}\omega \left( J,\theta ,Z\right) T\left(
Z,Z_{1},\theta \right) G^{\prime }\left[ J,\omega ,\theta ,Z,\Psi \right]
\delta \left( l_{1}-\frac{\left\vert Z-Z_{1}\right\vert }{c}\right) \left(
\left\vert \Psi _{0}\left( Z_{1}\right) \right\vert ^{2}+\left\vert \Psi
\left( \theta -\frac{\left\vert Z-Z_{1}\right\vert }{c},Z_{1}\right)
\right\vert ^{2}\right) }{1-\left( \int \frac{\kappa }{N}\omega \left(
J,\theta -\frac{\left\vert Z-Z^{\prime }\right\vert }{c},Z^{\prime }\right) 
\frac{\partial T\left( Z,Z^{\prime },\theta \right) }{\partial \omega \left(
J,\theta ,Z\right) }\left\vert \Psi \left( \theta -\frac{\left\vert
Z-Z^{\prime }\right\vert }{c},Z^{\prime }\right) \right\vert ^{2}dZ^{\prime
}\right) G^{\prime }\left[ J,\omega ,\theta ,Z,\Psi \right] }  \notag
\end{eqnarray}%
with the convention that $Z^{\left( 0\right) }=Z$ and $Z^{\left( n\right)
}=Z_{1}$. The derivative (\ref{rdv}) was then evaluated in Appendix 5 using
combinations of $K_{1}$ functions, but for the purpose of the computation of
the successive derivatives of $\omega \left( J,\theta ,Z\right) $, we will
work, temporarily, with the general formula (\ref{rdv}). Equation (\ref{rdv}%
) yield recursively $\frac{\delta \omega \left( J,\theta ,Z\right) }{\delta
\left\vert \Psi \left( \theta -l_{1},Z_{1}\right) \right\vert ^{2}}$ in
terms of past activities. Applied to the case $\left\vert \Psi \right\vert
^{2}=0$, the factor (\ref{vdR}) simplifies:%
\begin{eqnarray}
\check{T}\left( \theta ,Z,Z_{1},\omega _{0}\right) &\equiv &\check{T}\left(
\theta ,Z,Z_{1}\omega _{0},0\right)  \label{rvd} \\
&=&-\frac{\frac{\kappa }{N}\omega _{0}\left( J,\theta ,Z\right) T\left(
Z,Z_{1},\theta \right) G^{\prime }\left[ J,\omega ,\theta ,Z,\Psi \right]
\left\vert \Psi _{0}\left( Z_{1}\right) \right\vert ^{2}}{1-\left( \int 
\frac{\kappa }{N}\omega _{0}\left( J,Z^{\prime }\right) \frac{\partial
T\left( Z,Z^{\prime },\theta \right) }{\partial \omega \left( J,\theta
,Z\right) }\left\vert \Psi _{0}\left( Z^{\prime }\right) \right\vert
^{2}dZ^{\prime }\right) G^{\prime }\left[ J,\omega ,\theta ,Z,\Psi \right] }
\notag
\end{eqnarray}%
or in first aproximation:%
\begin{eqnarray}
\check{T}\left( \theta ,Z,Z_{1},\omega _{0},\Psi \right) &\equiv &\check{T}%
\left( Z,Z_{1},\omega _{0}\right)  \label{rvD} \\
&\simeq &-\frac{\frac{\kappa }{N}T\left( Z,Z_{1},\theta \right) G^{\prime }%
\left[ J,\omega ,\theta ,Z,\Psi \right] \left\vert \Psi _{0}\left(
Z_{1}\right) \right\vert ^{2}}{\omega _{0}^{-1}\left( J,Z\right) }  \notag
\end{eqnarray}%
and (\ref{rdv}) becomes:%
\begin{eqnarray}
\left( \frac{\delta \omega ^{-1}\left( J,\theta ,Z\right) }{\delta
\left\vert \Psi \left( \theta -l_{1},Z_{1}\right) \right\vert ^{2}}\right)
_{\left\vert \Psi \right\vert ^{2}=0} &=&-\sum_{n=1}^{\infty }\int \frac{%
\omega _{0}^{-1}\left( J,\theta -\sum_{l=1}^{n}\frac{\left\vert Z^{\left(
l-1\right) }-Z^{\left( l\right) }\right\vert }{c},Z_{1}\right) }{\mathcal{%
\bar{G}}_{0}\left( 0,Z_{1}\right) }  \label{zdv} \\
&&\times \dprod\limits_{l=1}^{n}\check{T}\left( \theta -\sum_{j=1}^{l-1}%
\frac{\left\vert Z^{\left( j-1\right) }-Z^{\left( j\right) }\right\vert }{c}%
,Z^{\left( l-1\right) },Z^{\left( l\right) },\omega _{0},0\right)  \notag \\
&&\times \delta \left( l_{1}-\sum_{l=1}^{n}\frac{\left\vert Z^{\left(
l-1\right) }-Z^{\left( l\right) }\right\vert }{c}\right)
\dprod\limits_{l=1}^{n-1}dZ^{\left( l\right) }  \notag
\end{eqnarray}

\paragraph*{2.1.1.2 Graphical representation of the successive derivatives}

The $n$-th term in (\ref{zdv}) can be understood graphically as a sum over
the set of broken paths with $n$ segments, each path linking $Z^{\left(
l-1\right) }$ and $Z^{\left( l\right) }$ during a timespan of $\frac{%
\left\vert Z^{\left( l-1\right) }-Z^{\left( l\right) }\right\vert }{c}$. To
each point of the segment, we associate the factor: 
\begin{eqnarray}
&&\check{T}\left( \theta -\sum_{j=1}^{l-1}\frac{\left\vert Z^{\left(
j-1\right) }-Z^{\left( j\right) }\right\vert }{c},Z^{\left( l-1\right)
},Z^{\left( l\right) },\omega _{0},\Psi \right)  \label{tf} \\
&\simeq &\frac{\frac{\kappa }{N}\check{T}\left( Z^{\left( l-1\right)
},Z^{\left( l\right) }\right) G^{\prime }\left[ J,\omega _{0},\theta
-\sum_{j=1}^{l-1}\frac{\left\vert Z^{\left( j-1\right) }-Z^{\left( j\right)
}\right\vert }{c},Z^{\left( l-1\right) }\right] \left\vert \Psi _{0}\left(
Z_{l}\right) \right\vert ^{2}}{\omega _{0}^{-1}\left( J,\theta
-\sum_{j=1}^{l-1}\frac{\left\vert Z^{\left( j-1\right) }-Z^{\left( j\right)
}\right\vert }{c},Z^{\left( l-1\right) }\right) }  \notag
\end{eqnarray}%
Ultimately, the product of factor is multiplied by the activity at the last
point:%
\begin{equation}
-\frac{\omega _{0}^{-1}\left( J,\theta -\sum_{l=1}^{n}\frac{\left\vert
Z^{\left( l-1\right) }-Z^{\left( l\right) }\right\vert }{c},Z_{1}\right) }{%
\left\vert \Psi _{0}\left( Z_{1}\right) \right\vert ^{2}}  \label{cf}
\end{equation}%
and by $\left\vert \Psi \left( \theta -l_{1},Z_{1}\right) \right\vert ^{2}$.
The integrals over the points $Z^{\left( l\right) }$ and the sum over $n$,
the length of the broken paths, yield the first order contribution to the
expansion (\ref{snp}).

The next terms in the expansion of (\ref{snp}) are the derivatives $\left( 
\frac{\delta ^{n}\omega ^{-1}\left( J,\theta ,Z\right) }{\dprod%
\limits_{i=1}^{n}\delta \left\vert \Psi \left( \theta -l_{i},Z_{i}\right)
\right\vert ^{2}}\right) _{\left\vert \Psi \right\vert ^{2}=0}$ which are
obtained by successive derivations of (\ref{rdv}) and (\ref{vdR}) by $%
\left\vert \Psi \left( \theta -l_{2},Z_{2}\right) \right\vert ^{2}$ and
evaluated at $\left\vert \Psi \right\vert ^{2}=0$. The $l_{i}$ are ordered
such that $l_{1}<...<l_{n}$. These derivatives are obtained by
differentiating either:%
\begin{equation*}
-\omega _{0}^{-1}\left( J,\theta -\sum_{l=1}^{n}\frac{\left\vert Z^{\left(
l-1\right) }-Z^{\left( l\right) }\right\vert }{c},Z_{n}\right)
\end{equation*}%
or the successive factors:%
\begin{equation*}
\dprod\limits_{l=1}^{n}\check{T}\left( \theta -\sum_{j=1}^{l-1}\frac{%
\left\vert Z^{\left( j-1\right) }-Z^{\left( j\right) }\right\vert }{c}%
,Z^{\left( l-1\right) },Z^{\left( l\right) },\omega ,\Psi \right)
\end{equation*}%
The first possibility amounts to write $\frac{\delta \omega \left( J,\theta
-l_{1},Z_{1}\right) }{\delta \left\vert \Psi \left( \theta
-l_{2},Z_{2}\right) \right\vert ^{2}}$ using (\ref{rdv}). Graphically it
amounts to write broken lines from $Z_{1}$ to $Z_{2}$ and associate to each
broken line the factor (\ref{tf}), (\ref{cf}) and $\left\vert \Psi \left(
\theta -l_{2},Z_{2}\right) \right\vert ^{2}$.

The second possibility is obtained by computing for each $l$:%
\begin{equation}
\frac{\delta \check{T}\left( \theta -\sum_{j=1}^{l-1}\frac{\left\vert
Z^{\left( j-1\right) }-Z^{\left( j\right) }\right\vert }{c},Z^{\left(
l-1\right) },Z^{\left( l\right) },\omega ,\Psi \right) }{\delta \left\vert
\Psi \left( \theta -l_{2},Z_{2}\right) \right\vert ^{2}}  \label{rds}
\end{equation}%
Which can be written as:%
\begin{eqnarray*}
&&\frac{\delta \check{T}\left( \theta -\sum_{j=1}^{l-1}\frac{\left\vert
Z^{\left( j-1\right) }-Z^{\left( j\right) }\right\vert }{c},Z^{\left(
l-1\right) },Z^{\left( l\right) },\omega ,\Psi \right) }{\delta \left\vert
\Psi \left( \theta -l_{2},Z_{2}\right) \right\vert ^{2}} \\
&=&\int d\Delta dZ^{\prime }\frac{\delta \check{T}\left( \theta
-\sum_{j=1}^{l-1}\frac{\left\vert Z^{\left( j-1\right) }-Z^{\left( j\right)
}\right\vert }{c},Z^{\left( l-1\right) },Z^{\left( l\right) },\omega ,\Psi
\right) }{\delta \omega ^{-1}\left( J,\theta -\sum_{j=1}^{l-1}\frac{%
\left\vert Z^{\left( j-1\right) }-Z^{\left( j\right) }\right\vert }{c}%
-\Delta ,Z^{\prime }\right) }\frac{\delta \omega ^{-1}\left( J,\theta
-\sum_{j=1}^{l-1}\frac{\left\vert Z^{\left( j-1\right) }-Z^{\left( j\right)
}\right\vert }{c}-\Delta ,Z^{\prime }\right) }{\delta \left\vert \Psi \left(
\theta -l_{2},Z_{2}\right) \right\vert ^{2}}
\end{eqnarray*}%
This derivative can be described graphically by assigning to some point $%
Z^{\left( l\right) }$ of the initial line the factor:%
\begin{equation*}
\frac{\delta \check{T}\left( \theta -\sum_{j=1}^{l-1}\frac{\left\vert
Z^{\left( j-1\right) }-Z^{\left( j\right) }\right\vert }{c},Z^{\left(
l-1\right) },Z^{\left( l\right) },\omega ,\Psi \right) }{\delta \omega
^{-1}\left( J,\theta -\sum_{j=1}^{l-1}\frac{\left\vert Z^{\left( j-1\right)
}-Z^{\left( j\right) }\right\vert }{c}-\Delta ,Z^{\prime }\right) }
\end{equation*}%
issuing a new succession of segments representing $\frac{\delta \omega
^{-1}\left( J,\theta -\sum_{j=1}^{l-1}\frac{\left\vert Z^{\left( j-1\right)
}-Z^{\left( j\right) }\right\vert }{c}-\Delta ,Z^{\prime }\right) }{\delta
\left\vert \Psi \left( \theta -l_{2},Z_{2}\right) \right\vert ^{2}}$ and
then summing over $\Delta $ and $Z^{\prime }$. In first approximation, we
can set $\Delta =0$ and $Z^{\prime }$, so that the factor is: 
\begin{equation*}
\left( \frac{\delta \check{T}\left( \theta -\sum_{j=1}^{l-1}\frac{\left\vert
Z^{\left( j-1\right) }-Z^{\left( j\right) }\right\vert }{c},Z^{\left(
l-1\right) },Z^{\left( l\right) },\omega ,\Psi \right) }{\delta \omega
^{-1}\left( J,\theta -\sum_{j=1}^{l-1}\frac{\left\vert Z^{\left( j-1\right)
}-Z^{\left( j\right) }\right\vert }{c},Z^{\left( l-1\right) }\right) }%
\right) _{\left\vert \Psi \right\vert ^{2}=0}
\end{equation*}%
and the new succession of segments represents $\frac{\delta \omega
^{-1}\left( J,\theta -\sum_{j=1}^{l-1}\frac{\left\vert Z^{\left( j-1\right)
}-Z^{\left( j\right) }\right\vert }{c},Z^{\left( l-1\right) }\right) }{%
\delta \left\vert \Psi \left( \theta -l_{2},Z_{2}\right) \right\vert ^{2}}$.

More generally, differentiating successively $\hat{T}\left( \theta
,Z,Z_{1}\omega ,\left\vert \Psi \right\vert ^{2}\right) $, corresponds to
insert the vertices: 
\begin{equation*}
\frac{\delta ^{k}\check{T}\left( \theta -\sum_{j=1}^{l-1}\frac{\left\vert
Z^{\left( j-1\right) }-Z^{\left( j\right) }\right\vert }{c},Z^{\left(
l-1\right) },Z^{\left( l\right) },\omega ,\Psi \right) }{\dprod%
\limits_{i=1}^{k}\delta \omega ^{-1}\left( J,\theta -\sum_{j=1}^{l-1}\frac{%
\left\vert Z^{\left( j-1\right) }-Z^{\left( j\right) }\right\vert }{c}%
-\Delta _{l},Z_{l}\right) }\simeq \frac{\delta ^{k}\left( \frac{\check{T}%
\left( Z^{\left( l-1\right) },Z^{\left( l\right) }\right) G^{\prime }\left[
J,\omega _{0},Z^{\left( l\right) }\right] \left\vert \Psi _{0}\left(
Z_{l}\right) \right\vert ^{2}}{\omega _{0}^{-1}\left( J,\theta ,Z^{\left(
l\right) }\right) }\right) }{\delta ^{k}\omega _{0}^{-1}\left( J,\theta
,Z^{\left( l\right) }\right) }
\end{equation*}%
with $k$ new segments representing $\frac{\delta \omega ^{-1}\left( J,\theta
-\sum_{j=1}^{l-1}\frac{\left\vert Z^{\left( j-1\right) }-Z^{\left( j\right)
}\right\vert }{c}-\Delta _{l},Z_{l}\right) }{\delta \left\vert \Psi \left(
\theta -l_{l},Z_{l}\right) \right\vert ^{2}}$.

Gathering the two possibilities forementionned and iterating this procedures
yields a graphical representation for:%
\begin{equation}
\left( \frac{\delta ^{n}\omega ^{-1}\left( J,\theta ,Z\right) }{%
\dprod\limits_{i=1}^{n}\delta \left\vert \Psi \left( \theta
-l_{i},Z_{i}\right) \right\vert ^{2}}\right) _{\left\vert \Psi \right\vert
^{2}=0}\dprod\limits_{i=1}^{n}\left\vert \Psi \left( \theta
-l_{i},Z_{i}\right) \right\vert ^{2}  \label{sl}
\end{equation}%
We associate the squared field $\left\vert \Psi \left( \theta
-l_{i},Z_{i}\right) \right\vert ^{2}$ to each point $Z_{i}$ . For $m=1,...,n$%
, we draw $m$ lines. At least one of them is starting from $Z$. These lines
are composed of an arbitrary number of segments and all the points $Z_{i}$
are crossed by one line. Each line ends at a point $Z_{i}$. The starting
points of the lines have to branch either at $Z$, either at some point of an
other line. There are $m$ branching points of valence $k$ including the
starting point at $Z$ Apart from $Z$ the branching points have valence $%
3,...,n-1$. To each line $i$ of length $L_{i}$, we associate the factor:%
\begin{eqnarray}
F\left( line_{i}\right) &=&\dprod\limits_{l=1}^{L_{i}}\frac{\frac{\kappa }{N}%
T\left( Z^{\left( l-1\right) },Z^{\left( l\right) }\right) G^{\prime }\left[
J,\omega _{0},\theta -\sum_{j=1}^{l-1}\frac{\left\vert Z^{\left( j-1\right)
}-Z^{\left( j\right) }\right\vert }{c},Z^{\left( l-1\right) }\right] 
\mathcal{\bar{G}}_{0}\left( 0,Z^{\left( l\right) }\right) }{\omega
_{0}^{-1}\left( J,\theta -\sum_{j=1}^{l-1}\frac{\left\vert Z^{\left(
j-1\right) }-Z^{\left( j\right) }\right\vert }{c},Z^{\left( l-1\right)
}\right) }  \label{lf} \\
&&\times \frac{-\omega _{0}^{-1}\left( J,\theta -\sum_{l=1}^{L_{i}}\frac{%
\left\vert Z^{\left( l-1\right) }-Z^{\left( l\right) }\right\vert }{c}%
,Z_{i}\right) }{\mathcal{\bar{G}}_{0}\left( 0,Z_{i}\right) }  \notag \\
&=&\dprod\limits_{l=1}^{L_{i}}\check{T}\left( \theta -\sum_{j=1}^{l-1}\frac{%
\left\vert Z^{\left( j-1\right) }-Z^{\left( j\right) }\right\vert }{c}%
,Z^{\left( l-1\right) },Z^{\left( l\right) },\omega _{0},\Psi \right) \frac{%
-\omega _{0}^{-1}\left( J,\theta -\sum_{l=1}^{L_{i}}\frac{\left\vert
Z^{\left( l-1\right) }-Z^{\left( l\right) }\right\vert }{c},Z_{i}\right) }{%
\mathcal{\bar{G}}_{0}\left( 0,Z_{i}\right) }  \notag
\end{eqnarray}%
and to each branching point $\left( X,\theta \right) =B$ of valence $k+2$
arising in the expansion, we associate the factor:%
\begin{equation}
F\left( \left( X,\theta \right) \right) =\frac{\delta ^{k}\left( \frac{%
\check{T}\left( Z^{\left( l-1\right) },Z^{\left( l\right) }\right) G^{\prime
}\left[ J,\omega _{0},Z^{\left( l\right) }\right] \left\vert \Psi _{0}\left(
Z_{l}\right) \right\vert ^{2}}{\omega _{0}^{-1}\left( J,\theta ,Z^{\left(
l\right) }\right) }\right) }{\delta ^{k}\omega _{0}^{-1}\left( J,\theta
,Z^{\left( l\right) }\right) }  \label{rc}
\end{equation}%
and (\ref{sl}) writes:%
\begin{eqnarray}
&&\left( \frac{\delta ^{n}\omega ^{-1}\left( J,\theta ,Z\right) }{%
\dprod\limits_{i=1}^{n}\delta \left\vert \Psi \left( \theta
-l_{i},Z_{i}\right) \right\vert ^{2}}\right) _{\left\vert \Psi \right\vert
^{2}=0}\dprod\limits_{i=1}^{n}\left\vert \Psi \left( \theta
-l_{i},Z_{i}\right) \right\vert ^{2}  \notag \\
&=&\left( \sum_{m=1}^{n}\sum_{i=1}^{m}\sum_{\left(
line_{1},...,line_{m}\right) }\dprod\limits_{i}F\left( line_{i}\right)
\dprod\limits_{B}F\left( B\right) \right) \dprod\limits_{i=1}^{n}\left\vert
\Psi \left( \theta -l_{i},Z_{i}\right) \right\vert ^{2}  \label{rdt}
\end{eqnarray}%
The integral over the branch points is implicit. The factor $F\left(
B\right) $ for a branch point $B$ is defined in (\ref{rc}) The graphical
representation is generic. While integrating over the set of lines, the
degenerate case of lines that share some segments is taken into account.

\subsubsection*{2.1.2 Approximate expression for the $n$-th derivatives of $%
\protect\omega ^{-1}\left( \protect\theta ,Z\right) $}

The results of the section 5 can then be used with (\ref{rdt}) to compute:%
\begin{equation*}
\frac{\delta ^{n}\omega ^{-1}\left( J,\theta ,Z\right) }{\dprod%
\limits_{i=1}^{n}\delta \left\vert \Psi \left( \theta -l_{i},Z_{i}\right)
\right\vert ^{2}}
\end{equation*}%
in the approximation of the dominant contribution. To each line from a
branching point $\theta -l_{j}^{\prime },Z_{j}^{\prime }$ to $\theta
-l_{i},Z_{i}$ (the branching point can be one of the $\theta -l_{i},Z_{i}$)
we associate a factor of the type, as in (\ref{rrt}): 
\begin{equation*}
\frac{\exp \left( -c\left( l_{i}-l_{j}^{\prime }\right) -\gamma \left(
c\left( l_{i}-l_{j}^{\prime }\right) -\left\vert Z_{j}^{\prime
}-Z_{i}\right\vert \right) \right) }{D}H\left( cl_{1}-\left\vert
Z-Z_{1}\right\vert \right)
\end{equation*}%
The dominant contribution is obtained when the set $\left\{ l_{j}^{\prime
},Z_{j}^{\prime }\right\} $ is equal to $\left\{ l_{j},Z_{j}\right\} $ and
the product over the branching points yields a contribution whose form is:%
\begin{eqnarray}
\frac{\delta ^{n}\omega ^{-1}\left( J,\theta ,Z\right) }{\dprod%
\limits_{i=1}^{n}\delta \left\vert \Psi \left( \theta -l_{i},Z_{i}\right)
\right\vert ^{2}} &\simeq &\frac{\exp \left( -cl_{n}-\gamma \left(
\sum_{i=1}^{n-1}\left( \left( c\left( l_{i}-l_{i+1}\right) \right)
^{2}-\left\vert Z_{i}-Z_{i+1}\right\vert ^{2}\right) \right) \right) }{D^{n}}
\label{drnv} \\
&&\times H\left( cl_{n}-\sum_{i=1}^{n-1}\left\vert Z_{i}-Z_{i+1}\right\vert
\right) \dprod\limits_{i=1}^{n}\frac{\omega _{0}^{-1}\left( J,\theta
-l_{i},Z_{i}\right) }{\mathcal{\bar{G}}_{0}\left( 0,Z_{i}\right) }  \notag
\end{eqnarray}%
with $Z_{1}=Z$ and $l_{n}>...>l_{1}$ and $B$ a constant coefficient (see (%
\ref{trr})).

Formula (\ref{dvtr}) shows that the previous computations are also valid for
the derivatives of $\omega \left( J,\theta ,Z\right) $. We thus obtain the
generalization of (\ref{trr}): 
\begin{eqnarray}
\frac{\delta ^{n}\omega \left( J,\theta ,Z\right) }{\dprod\limits_{i=1}^{n}%
\delta \left\vert \Psi \left( \theta -l_{i},Z_{i}\right) \right\vert ^{2}}
&\simeq &\frac{\exp \left( -cl_{n}-\alpha \left( \sum_{i=1}^{n-1}\left(
\left( c\left( l_{i}-l_{i+1}\right) \right) ^{2}-\left\vert
Z_{i}-Z_{i+1}\right\vert ^{2}\right) \right) \right) }{B^{n}}  \label{dnv} \\
&&\times H\left( cl_{n}-\sum_{i=1}^{n-1}\left\vert Z_{i}-Z_{i+1}\right\vert
\right) \dprod\limits_{i=1}^{n}\frac{\omega _{0}\left( J,\theta
-l_{i},Z_{i}\right) }{\mathcal{\bar{G}}_{0}\left( 0,Z_{i}\right) }  \notag
\end{eqnarray}%
The only difference is the appearance of different coefficients $\alpha $
and $B$\ in the expression.

\subsection*{2.2 Series for $\protect\omega ^{-1}\left( \protect\theta %
,Z\right) $}

\subsubsection*{2.2.1 Reordering the graphical sum (\protect\ref{rdt})}

We now sum the series expansion (\ref{snp}):%
\begin{eqnarray}
\omega ^{-1}\left( \theta ^{\left( i\right) },Z\right) &=&\omega ^{-1}\left(
\theta ^{\left( i\right) },Z\right) _{\left\vert \Psi \right\vert ^{2}=0} \\
&&+\int \sum_{n=1}^{\infty }\left( \frac{\delta ^{n}\omega ^{-1}\left(
J,\theta ,Z\right) }{\dprod\limits_{i=1}^{n}\delta \left\vert \Psi \left(
\theta -l_{i},Z_{i}\right) \right\vert ^{2}}\right) _{\left\vert \Psi
\right\vert ^{2}=0}\dprod\limits_{i=1}^{n}\left\vert \Psi \left( \theta
-l_{i},Z_{i}\right) \right\vert ^{2}  \notag
\end{eqnarray}%
by reordering the sums in the RHS of (\ref{rdt}).

To do so, we first compute the sum over the lines between $\left( Z,\theta
\right) $ and $\left( Z_{1},\theta _{1}\right) $ and of given length $%
L_{i}=n $ of the product of factors $\check{T}\left( \theta -\sum_{j=1}^{l-1}%
\frac{\left\vert Z^{\left( j-1\right) }-Z^{\left( j\right) }\right\vert }{c}%
,Z^{\left( l-1\right) },Z^{\left( l\right) },\omega _{0},\Psi \right) $ in $%
F\left( line_{i}\right) $ (see (\ref{lf}) for the definition of $F\left(
line_{i}\right) $). This sum is computed in (\ref{zdv}). We call the result $%
G_{0}^{\left( n\right) }\left( \left( Z,\theta \right) ,\left( Z_{1},\theta
_{1}\right) \right) $, so that: 
\begin{eqnarray*}
&&G_{0}^{\left( n\right) }\left( \left( Z,\theta \right) ,\left(
Z_{1},\theta _{1}\right) \right) =\int \dprod\limits_{l=1}^{n}\check{T}%
\left( \theta -\sum_{j=1}^{l-1}\frac{\left\vert Z^{\left( j-1\right)
}-Z^{\left( j\right) }\right\vert }{c},Z^{\left( l-1\right) },Z^{\left(
l\right) },\omega _{0}\right) \\
&&\times \delta \left( \left( \theta -\theta _{1}\right) -\sum_{l=1}^{n}%
\frac{\left\vert Z^{\left( l-1\right) }-Z^{\left( l\right) }\right\vert }{c}%
\right) \dprod\limits_{l=1}^{n-1}dZ^{\left( l\right) } \\
&=&\int \dprod\limits_{l=1}^{n}\check{T}\left( \theta -\sum_{j=1}^{l-1}\frac{%
\left\vert Z^{\left( j-1\right) }-Z^{\left( j\right) }\right\vert }{c}%
,Z^{\left( l-1\right) },Z^{\left( l\right) },\omega _{0}\right) \delta
\left( \left( \theta ^{\left( l\right) }-\theta ^{\left( l-1\right) }\right)
-\frac{\left\vert Z^{\left( l-1\right) }-Z^{\left( l\right) }\right\vert }{c}%
\right) \dprod\limits_{l=1}^{n-1}dZ^{\left( l\right) }d\theta _{l}
\end{eqnarray*}

with $\left( Z^{\left( 0\right) },\theta ^{\left( 0\right) }\right) =\left(
Z,\theta \right) $ and $\left( Z^{\left( n\right) },\theta ^{\left( n\right)
}\right) =\left( Z_{1},\theta _{1}\right) $.

Then, we sum over the length $n$ of the lines and the factor associated to
the sum of lines, written $G_{0}\left( \left( Z,\theta \right) ,\left(
Z_{1},\theta _{1}\right) \right) $, is:%
\begin{eqnarray*}
G_{0}\left( \left( Z,\theta \right) ,\left( Z_{1},\theta _{1}\right) \right)
&=&\sum_{n=1}^{\infty }\int \dprod\limits_{l=1}^{n}\check{T}\left( \theta
-\sum_{j=1}^{l-1}\frac{\left\vert Z^{\left( j-1\right) }-Z^{\left( j\right)
}\right\vert }{c},Z^{\left( l-1\right) },Z^{\left( l\right) },\omega
_{0}\right) \\
&&\times \delta \left( \left( \theta ^{\left( l\right) }-\theta ^{\left(
l-1\right) }\right) -\frac{\left\vert Z^{\left( l-1\right) }-Z^{\left(
l\right) }\right\vert }{c}\right) \dprod\limits_{l=1}^{n-1}dZ^{\left(
l\right) }d\theta _{l}
\end{eqnarray*}%
The function $G_{0}\left( \left( Z,\theta \right) ,\left( Z_{1},\theta
_{1}\right) \right) $ is a series expansion that can be summed:%
\begin{equation}
G_{0}\left( \left( Z,\theta \right) ,\left( Z_{1},\theta _{1}\right) \right)
=\check{T}\left( 1-\check{T}\right) ^{-1}\left( \left( Z,\theta \right)
,\left( Z_{1},\theta _{1}\right) \right)  \label{prT}
\end{equation}%
with:%
\begin{eqnarray*}
\check{T}\left( \left( Z^{\left( l-1\right) },\theta ^{\left( l-1\right)
}\right) ,\left( Z^{\left( l\right) },\theta ^{\left( l\right) }\right)
\right) &=&\check{T}\left( \theta -\sum_{j=1}^{l-1}\frac{\left\vert
Z^{\left( j-1\right) }-Z^{\left( j\right) }\right\vert }{c},Z^{\left(
l-1\right) },Z^{\left( l\right) },\omega _{0}\right) \\
&&\times \delta \left( \left( \theta ^{\left( l\right) }-\theta ^{\left(
l-1\right) }\right) -\frac{\left\vert Z^{\left( l-1\right) }-Z^{\left(
l\right) }\right\vert }{c}\right)
\end{eqnarray*}%
As a consequence, equation (\ref{rdt}) can be rewritten as a sum over the
branch points.:%
\begin{eqnarray}
&&\omega ^{-1}\left( \theta ^{\left( i\right) },Z\right) -\omega ^{-1}\left(
\theta ^{\left( i\right) },Z\right) _{\left\vert \Psi \right\vert ^{2}=0}
\label{xf} \\
&=&\int \sum_{n}\left( \frac{\delta ^{n}\omega ^{-1}\left( J,\theta
,Z\right) }{\dprod\limits_{i=1}^{n}\delta \left\vert \Psi \left( \theta
-l_{i},Z_{i}\right) \right\vert ^{2}}\right) _{\left\vert \Psi \right\vert
^{2}=0}\dprod\limits_{i=1}^{n}\left\vert \Psi \left( \theta
-l_{i},Z_{i}\right) \right\vert ^{2}dl_{i}dZ_{i}  \notag \\
&=&\left( \sum_{m=1}^{n}\sum_{i=1}^{m}\sum_{B}\sum_{\left( \overline{line_{1}%
},...,\overline{line_{m}}\right) }\dprod\limits_{i}G_{0}\left( \overline{%
line_{i}}\right) \dprod\limits_{B}F\left( B\right) \right)
\dprod\limits_{i=1}^{n}\left\vert \Psi \left( \theta -l_{i},Z_{i}\right)
\right\vert ^{2}  \notag
\end{eqnarray}%
The sum $\sum_{\left( \overline{line_{1}},...,\overline{line_{m}}\right) }$
is over the finite set of $m$\ segments connecting two branch points and
respecting the constraint given above (\ref{lf}). If $\overline{line_{i}}$
connects two branch points $\left( \left( X_{1},\theta _{1}\right) ,\left(
X_{2},\theta _{2}\right) \right) $, then $G_{0}\left( \overline{line_{i}}%
\right) $ is equal to $G_{0}\left( \left( X_{1},\theta _{1}\right) ,\left(
X_{2},\theta _{2}\right) \right) $. At each branch point we insert $\frac{%
\left\vert \Psi \left( \theta -l_{k},Z_{k}\right) \right\vert ^{2}}{\mathcal{%
\bar{G}}_{0}\left( 0,Z_{k}\right) }$ and for a terminal point $-\frac{\omega
_{0}^{-1}\left( J,\theta -l_{k},Z_{k}\right) \left\vert \Psi \left( \theta
-l_{k},Z_{k}\right) \right\vert ^{2}}{\mathcal{\bar{G}}_{0}\left(
0,Z_{k}\right) }$. We will normalize $\left\vert \Psi \right\vert ^{2}$ by $%
\mathcal{\bar{G}}_{0}$, so that $\left\vert \Psi \left( \theta
-l_{k},Z_{k}\right) \right\vert ^{2}$ will stand for $\frac{\left\vert \Psi
\left( \theta -l_{k},Z_{k}\right) \right\vert ^{2}}{\mathcal{\bar{G}}%
_{0}\left( 0,Z_{k}\right) }$.

Now the sums in (\ref{xf}) can be reordered in the following way. We
consider the lines from $\left( \theta ,Z\right) $ to a final point, and sum
over the branch points of valence $2$ crossed by these lines, that is points
crossed or reached only by this line. We then sum the contributions over all
these lines. For instance, if a line crosses only one branch point, the
associated contribution will include two propagators $G_{0}=\check{T}\left(
1-\check{T}\right) ^{-1}$, one between the initial point and the branch
point, one between the branch point and the final point plus the factors
inserted at each point. Summing over all possible branch points crossed by a
line yields the factor associated to the overall sum of single lines
crossing the points $Z_{k}$:%
\begin{eqnarray}
&&\check{T}\left( 1-\check{T}\right) ^{-1}\sum_{n\geqslant 0}\int
\dprod\limits_{l=1}^{n-1}\left\{ \int \left( \left\vert \Psi \left( \theta
-l_{l},Z_{l}\right) \right\vert ^{2}dZ_{l}dl_{l}\right) \check{T}\left( 1-%
\check{T}\right) ^{-1}\right\} \left\vert \Psi \left( \theta
-l_{n},Z_{n}\right) \right\vert ^{2}\frac{-\omega _{0}^{-1}\left( J,\theta
-l_{n},Z_{n}\right) }{\mathcal{\bar{G}}_{0}\left( 0,Z_{n}\right) }  \notag \\
&=&\check{T}\left( 1-\check{T}\right) ^{-1}\frac{1}{1-\left\vert \Psi \left(
\theta ,Z\right) \right\vert ^{2}\check{T}\left( 1-\check{T}\right) ^{-1}}%
\left\vert \Psi \left( \theta -l_{n},Z_{n}\right) \right\vert ^{2}\frac{%
-\omega _{0}^{-1}\left( J,\theta -l_{n},Z_{n}\right) }{\mathcal{\bar{G}}%
_{0}\left( 0,Z_{n}\right) }  \notag \\
&=&\check{T}\frac{1}{1-\left( 1+\left\vert \Psi \right\vert ^{2}\right) 
\check{T}}\left\vert \Psi \left( \theta -l_{n},Z_{n}\right) \right\vert ^{2}%
\frac{-\omega _{0}^{-1}\left( J,\theta -l_{n},Z_{n}\right) }{\mathcal{\bar{G}%
}_{0}\left( 0,Z_{n}\right) }  \label{rl}
\end{eqnarray}%
with $Z_{0}=X_{1}$ and $Z_{k+1}=X_{2}$ and $\dprod\limits_{l=1}^{0}$ is set
to $1$. The $l_{i}$ are ranked such that: $l_{1}<...<l_{k}$ We sum over all
contributions of field insertions between $\left( X_{1},\theta _{1}\right) $
and $\left( X_{2},\theta _{2}\right) $ and integrate over the intermediate
points. The factor $\left\vert \Psi \right\vert ^{2}$ is seen as the
operator multiplication by $\left\vert \Psi \left( \theta ,Z\right)
\right\vert ^{2}$ at the point $\left( \theta ,Z\right) $.

The sum (\ref{rl}) over the single lines is the Green function of the
operator $1-\left( 1+\left\vert \Psi \right\vert ^{2}\right) \check{T}$ with 
$\check{T}$ and $-\left\vert \Psi \left( \theta -l_{n},Z_{n}\right)
\right\vert ^{2}\omega _{0}\left( J,\theta -l_{n},Z_{n}\right) $ inserted at
the starting and ending points. This quantity can be seen as a block $\left[
\left( X_{1},\theta _{1}\right) ,\left( X_{2},\theta _{2}\right) \right] $.

\subsubsection*{2.2.2 Path integral formulation}

The series expansion (\ref{xf}) for $\omega ^{-1}\left( \theta ^{\left(
i\right) },Z\right) $ can ultimately be rewritten as a sum over the number $%
m $ of branch points $\left( X_{i},\theta _{i}\right) $ with valence $%
k_{i}>2 $: we draw all connected graphs whose vertices are the branch points 
$\left( X_{1},\theta _{1}\right) ...\left( X_{m},\theta _{m}\right) $. We
attach $k_{i}$ blocks to the vertex $\left( X_{i},\theta _{i}\right) $, the
endpoint of one of them and the starting point of the others are fixed by
the vertex. To each vertex, the factor $F\left( \left( X_{i},\theta
_{i}\right) \right) $ defined in (\ref{rc}) is associated. The extremities
of the blocks that are not fixed are free and integrated over, except one of
them which is equal to $\left( Z,\theta \right) $. Then the series (\ref{ft}%
) is the sum over $m$ and over all types of graphs with $m$ vertices.

Note that the sum of graph can be computed without ordering in time the
fields. It amounts to replace (\ref{xf}) by: 
\begin{equation*}
\omega ^{-1}\left( \theta ^{\left( i\right) },Z\right) -\omega ^{-1}\left(
\theta ^{\left( i\right) },Z\right) _{\left\vert \Psi \right\vert
^{2}=0}=\int \frac{1}{n!}\sum_{n}\left( \frac{\delta ^{n}\omega ^{-1}\left(
J,\theta ,Z\right) }{\dprod\limits_{i=1}^{n}\delta \left\vert \Psi \left(
\theta -l_{i},Z_{i}\right) \right\vert ^{2}}\right) _{\left\vert \Psi
\right\vert ^{2}=0}\dprod\limits_{i=1}^{n}\left\vert \Psi \left( \theta
_{i},Z_{i}\right) \right\vert ^{2}d\theta _{i}dZ_{i}
\end{equation*}%
As a consequence, the symetry factor of equivalent graphs factored by $%
\dprod\limits_{i=1}^{n}\left\vert \Psi \left( \theta _{i},Z_{i}\right)
\right\vert ^{2}$ and integrated over $\dprod\limits_{i=1}^{n}d\theta
_{i}dZ_{i}$ is:%
\begin{equation*}
\frac{1}{n!}\frac{n!}{\prod\limits_{V}k_{V}!}
\end{equation*}%
where the product is over the vertices of valence $k_{V}$ of the graph. The
factor $n!$ comes from the exchange between the vertices $%
\dprod\limits_{i=1}^{n}\left\vert \Psi \left( \theta _{i},Z_{i}\right)
\right\vert ^{2}$ \ The $k_{V}!$ accounts for the exchange of the $k_{V}$
vertices among the same graph.

The sum of lines connected by vertices can then be computed using the Green
function $\frac{1}{1-\left( 1+\left\vert \Psi \right\vert ^{2}\right) \check{%
T}}$ connecting the vertices of all possible valences.

As a consequence, the generating function for the graphs is equal to the
partition function for an auxiliary complex field $\digamma \left( X,\theta
\right) $ with free Green function equal to $\frac{1}{1-\left( 1+\left\vert
\Psi \right\vert ^{2}\right) \check{T}}$ and interaction terms generating
the various graphs with arbitrary number of vertices. The free part of the
action for $\digamma \left( X,\theta \right) $ is thus:%
\begin{equation*}
\int \digamma \left( X,\theta \right) \left( 1-\left( 1+\left\vert \Psi
\right\vert ^{2}\right) \check{T}\right) \digamma ^{\dag }\left( X,\theta
\right) d\left( X,\theta \right) 
\end{equation*}%
and the interaction terms have to induce the graphs with factor (\ref{rc}).
The $k+2$ valence vertex, with $k\geqslant 1$ is thus described by a term
involving (\ref{rc}) and writes:%
\begin{eqnarray*}
&&\int \digamma \left( Z^{\left( 1\right) },\theta ^{\left( 1\right)
}\right) \frac{\delta ^{k}\left( \check{T}\left( \theta ^{\left( 1\right) }-%
\frac{\left\vert Z^{\left( 1\right) }-Z^{\left( 2\right) }\right\vert }{c}%
,Z^{\left( 1\right) },Z^{\left( 2\right) },\omega _{0}\right) \right) }{%
k!\dprod\limits_{l=3}^{k+2}\delta ^{k}\omega _{0}^{-1}\left( J,\theta
^{\left( l\right) },Z^{\left( l\right) }\right) }\digamma ^{\dag }\left(
Z^{\left( 2\right) },\theta ^{\left( 1\right) }-\frac{\left\vert Z^{\left(
1\right) }-Z^{\left( 2\right) }\right\vert }{c}\right)  \\
&&\times \dprod\limits_{l=3}^{k+2}\check{T}\left( \left( Z^{\left( 1\right)
},\theta ^{\left( 1\right) }\right) ,\left( \theta ^{\left( l\right)
},Z^{\left( l\right) }\right) \right) \left( \digamma ^{\dag }\left( \theta
^{\left( l\right) },Z^{\left( l\right) }\right) \right)
\dprod\limits_{l=1}^{k+2}d\left( \theta ^{\left( l\right) },Z^{\left(
l\right) }\right)  \\
&=&\int \digamma \left( Z^{\left( 1\right) },\theta ^{\left( 1\right)
}\right) \frac{\delta ^{k}\left( \check{T}\left( \theta ^{\left( 1\right) }-%
\frac{\left\vert Z^{\left( 1\right) }-Z^{\left( 2\right) }\right\vert }{c}%
,Z^{\left( 1\right) },Z^{\left( 2\right) },\omega _{0}\right) \right) }{%
k!\dprod\limits_{l=3}^{k+2}\delta ^{k}\omega _{0}^{-1}\left( J,\theta
^{\left( l\right) },Z^{\left( l\right) }\right) }\digamma ^{\dag }\left(
Z^{\left( 2\right) },\theta ^{\left( 1\right) }-\frac{\left\vert Z^{\left(
1\right) }-Z^{\left( 2\right) }\right\vert }{c}\right)  \\
&&\times \dprod\limits_{l=3}^{k+2}\check{T}\left( \theta ^{\left( 1\right) }-%
\frac{\left\vert Z^{\left( 1\right) }-Z^{\left( l\right) }\right\vert }{c}%
,Z^{\left( 1\right) },Z^{\left( l\right) },\omega _{0}\right) \digamma
^{\dag }\left( \theta ^{\left( l\right) },Z^{\left( l\right) }\right)
d\theta ^{\left( 1\right) }\dprod\limits_{l=1}^{k+2}dZ^{\left( l\right) }
\end{eqnarray*}%
Having found the free part of the action and the required vertices, the sum
of all graphs (\ref{xf}) yields, for $\frac{\left\vert \Psi \left( J,\theta
_{i},Z_{i}\right) \right\vert ^{2}}{\mathcal{\bar{G}}_{0}\left(
0,Z_{i}\right) }\rightarrow \left\vert \Psi \left( J,\theta
_{i},Z_{i}\right) \right\vert ^{2}$:%
\begin{eqnarray}
&&\omega _{0}^{-1}\left( J,\theta ,Z\right) +\sum_{n=1}^{\infty }\frac{1}{n!}%
\frac{\int \check{T}\digamma ^{\dag }\left( Z,\theta \right) \int
\dprod\limits_{i=1}^{n}\left( -\omega _{0}^{-1}\left( J,\theta
_{i},Z_{i}\right) \right) \left\vert \Psi \left( J,\theta _{i},Z_{i}\right)
\right\vert ^{2}\digamma \left( Z_{i},\theta _{i}\right) d\left(
Z_{i},\theta _{i}\right) \exp \left( -S\left( \digamma \right) \right) 
\mathcal{D}\digamma }{\exp \left( -S\left( \digamma \right) \right) \mathcal{%
D}\digamma }  \notag \\
&=&\omega _{0}^{-1}\left( J,\theta ,Z\right) +\frac{\int \check{T}\digamma
^{\dag }\left( Z,\theta \right) \exp \left( -S\left( \digamma \right) -\int
\digamma \left( X,\theta \right) \omega _{0}^{-1}\left( J,\theta ,Z\right)
\left\vert \Psi \left( J,\theta ,Z\right) \right\vert ^{2}d\left( X,\theta
\right) \right) \mathcal{D}\digamma }{\int \exp \left( -S\left( \digamma
\right) \right) \mathcal{D}\digamma }  \label{PX}
\end{eqnarray}%
with:%
\begin{eqnarray*}
S\left( \digamma \right)  &=&\int \digamma \left( X,\theta \right) \left(
1-\left( 1+\left\vert \Psi \right\vert ^{2}\right) \check{T}\right) \digamma
^{\dag }\left( X,\theta \right) d\left( X,\theta \right)  \\
&&-\int \digamma \left( Z^{\left( 1\right) },\theta ^{\left( 1\right)
}\right) \sum_{k}\frac{\delta ^{k}\left( \check{T}\left( \theta ^{\left(
1\right) }-\frac{\left\vert Z^{\left( 1\right) }-Z^{\left( 2\right)
}\right\vert }{c},Z^{\left( 1\right) },Z^{\left( 2\right) },\omega
_{0}\right) \right) }{k!\dprod\limits_{l=3}^{k+2}\delta ^{k}\omega
_{0}^{-1}\left( J,\theta ^{\left( l\right) },Z^{\left( l\right) }\right) }%
\digamma ^{\dag }\left( Z^{\left( 2\right) },\theta ^{\left( 1\right) }-%
\frac{\left\vert Z^{\left( 1\right) }-Z^{\left( 2\right) }\right\vert }{c}%
\right)  \\
&&\times \dprod\limits_{l=3}^{k+2}\check{T}\left( \theta ^{\left( 1\right) }-%
\frac{\left\vert Z^{\left( 1\right) }-Z^{\left( l\right) }\right\vert }{c}%
,Z^{\left( 1\right) },Z^{\left( l\right) },\omega _{0}\right) \digamma
^{\dag }\left( \theta ^{\left( l\right) },Z^{\left( l\right) }\right)
d\theta ^{\left( 1\right) }\dprod\limits_{l=1}^{k+2}dZ^{\left( l\right) }
\end{eqnarray*}%
The sum can be computed, and we have the more compact expression:%
\begin{eqnarray}
S\left( \digamma \right)  &=&\int \digamma \left( Z,\theta \right) \left(
1-\left\vert \Psi \right\vert ^{2}\check{T}\right) \digamma ^{\dag }\left(
Z,\theta \right) d\left( Z,\theta \right)   \label{SL} \\
&&-\int \digamma \left( Z,\theta \right) \check{T}\left( \theta -\frac{%
\left\vert Z-Z^{\left( 1\right) }\right\vert }{c},Z,Z^{\left( 1\right)
},\omega _{0}^{-1}+\check{T}\digamma ^{\dag }\right)   \notag \\
&&\times \digamma ^{\dag }\left( Z^{\left( 1\right) },\theta -\frac{%
\left\vert Z-Z^{\left( 1\right) }\right\vert }{c}\right) dZdZ^{\left(
1\right) }d\theta   \notag
\end{eqnarray}%
where:%
\begin{eqnarray*}
&&\check{T}\left( \theta -\frac{\left\vert Z^{\left( 1\right) }-Z\right\vert 
}{c},Z^{\left( 1\right) },Z,\omega _{0}^{-1}+\check{T}\digamma ^{\dag
}\right)  \\
&=&\check{T}\left( \theta -\frac{\left\vert Z^{\left( 1\right)
}-Z\right\vert }{c},Z^{\left( 1\right) },Z,\right.  \\
&&\left. \omega _{0}\left( Z,\theta \right) +\int \check{T}\left( \theta -%
\frac{\left\vert Z-Z^{\left( 1\right) }\right\vert }{c},Z^{\left( 1\right)
},Z,\omega _{0}\right) \digamma ^{\dag }\left( Z^{\left( 1\right) },\theta -%
\frac{\left\vert Z-Z^{\left( 1\right) }\right\vert }{c}\right) dZ^{\left(
1\right) }\right) 
\end{eqnarray*}%
\bigskip 

Integral (\ref{PX}) will be computed in the saddle point approximation. It
is obtained by replacing $\digamma ^{\dag }$ and\ $\digamma $ with their
values minimizing action $S\left( \digamma \right) $ defined in (\ref{SL}).
But before doing so, we will use a perturbation expansion of (\ref{PX}) to
rewrite the source term:%
\begin{equation*}
-\int \digamma \left( X,\theta \right) \omega _{0}^{-1}\left( J,\theta
,Z\right) \left\vert \Psi \left( J,\theta ,Z\right) \right\vert ^{2}d\left(
X,\theta \right) 
\end{equation*}%
as a function of the stimuli:%
\begin{equation*}
\sum_{i}a\left( Z_{i},\theta \right) \left\vert \Psi \left( Z_{i},\theta
\right) \right\vert ^{2}
\end{equation*}

\section*{Appendix 3 Expansion for $\protect\omega ^{-1}\left( J,\protect%
\theta ,Z\right) $ in presence of external sources}

In this appendix, we include the impact of external sources in the
computation of the activity $\omega \left( J,\theta ,Z\right) $. This allows
then to derive the propagation of external signal along the thread.
Ultimately, we generalize the results to severl types of interacting cells.

\subsection*{3. 1 Computation of graphs expansion in stimulated state}

So far, the results for activity $\omega \left( J,\theta ,Z\right) $ are
derived witout external source. We now include these ones in the path
integral to correct our previous expressions.

\subsubsection*{3.1.1 Modified expression for $\protect\omega ^{-1}\left( J,%
\protect\theta ,Z\right) $ in presence of external source}

For given connectivity functions, we want to compute the path integral for $%
\Psi \left( \theta ,Z\right) $ given a series of signals through time at
some particular points. As explained in the text, this amounts to introduce
in the path integral the factor:

\begin{equation*}
\int \exp \left( \sum_{i}a\left( Z_{i},\theta _{0}\right) \left\vert \Psi
\left( Z_{i},\theta _{0}\right) \right\vert ^{2}\right) d\theta _{0}
\end{equation*}%
The term $\sum_{i}a\left( Z_{i},\theta \right) \left\vert \Psi \left(
Z_{i},\theta \right) \right\vert ^{2}$ corresponds to create and cancel some
stimulation that makes the field $\Psi \left( Z_{i},\theta \right) $ to
deviate from the static equilibrium. The exponential factor stands for the
possibility of several similar stimuli at the same point. The sum over $%
\theta $ ensures the repetion of the signal through some period of time.
Recall that the perturbation is implicitely, tensored by:%
\begin{equation*}
\prod\limits_{Z\neq Z_{i}}\delta \left( \left\vert \Psi \left( Z,\theta
_{0}\right) \right\vert ^{2}\right)
\end{equation*}%
to ensure that the perturbation arises only at points $Z_{i}$.

The path integral to consider is thus:

\begin{eqnarray*}
&&\int \exp \left( -S\left( \Psi \right) \right) \int \exp \left(
\sum_{i}a\left( Z_{i},\theta _{0}\right) \left\vert \Psi \left( Z_{i},\theta
_{0}\right) \right\vert ^{2}\right) d\theta _{0} \\
&=&\int \exp \left( \frac{1}{2}\Psi ^{\dagger }\left( \theta ,Z\right)
\nabla \left( \frac{\sigma _{\theta }^{2}}{2}\nabla -\omega ^{-1}\right)
\Psi \left( \theta ,Z\right) \right) \int \exp \left( \sum_{i}a\left(
Z_{i},\theta _{0}\right) \left\vert \Psi \left( Z_{i},\theta _{0}\right)
\right\vert ^{2}\right) d\theta _{0}
\end{eqnarray*}%
with $\omega ^{-1}$ given by the auxiliary path integral (\ref{PX}):%
\begin{equation*}
\omega ^{-1}=\omega _{0}^{-1}\left( J,Z\right) +\frac{\int \check{T}\digamma
^{\dag }\left( Z,\theta \right) \exp \left( -S\left( \digamma \right) -\int
\digamma \left( X,\theta \right) \omega _{0}^{-1}\left( J,\theta ,Z\right)
\left\vert \Psi \left( J,\theta ,Z\right) \right\vert ^{2}d\left( X,\theta
\right) \right) \mathcal{D}\digamma }{\int \exp \left( -S\left( \digamma
\right) \right) \mathcal{D}\digamma }
\end{equation*}%
We start by expanding perturbatively:%
\begin{eqnarray}
&&\int \exp \left( \frac{1}{2}\Psi ^{\dagger }\left( \theta ,Z\right) \nabla
\left( \frac{\sigma _{\theta }^{2}}{2}\nabla -\omega ^{-1}\right) \Psi
\left( \theta ,Z\right) \right) \int \exp \left( \sum_{i}a\left(
Z_{i},\theta _{0}\right) \left\vert \Psi \left( Z_{i},\theta _{0}\right)
\right\vert ^{2}\right) d\theta _{0}  \notag \\
&=&\int \exp \left( \frac{1}{2}\Psi ^{\dagger }\left( \theta ,Z\right)
\nabla \left( \frac{\sigma _{\theta }^{2}}{2}\nabla -\omega _{0}^{-1}\left(
J,Z\right) \right) \Psi \left( \theta ,Z\right) \right)   \notag \\
&&\times \exp \left( -\frac{1}{2}\Psi ^{\dagger }\left( \theta ,Z\right)
\nabla \left( \frac{\int \check{T}\digamma ^{\dag }\left( Z,\theta \right)
\exp \left( -S\left( \digamma \right) -\int \digamma \left( X,\theta \right)
\omega _{0}^{-1}\left( \theta ,Z\right) \left\vert \Psi \left( \theta
,Z\right) \right\vert ^{2}d\left( X,\theta \right) \right) \mathcal{D}%
\digamma }{\int \exp \left( -S\left( \digamma \right) \right) \mathcal{D}%
\digamma }\right) \Psi \left( \theta ,Z\right) \right)   \notag \\
&&\times \int \exp \left( \sum_{i}a\left( Z_{i},\theta _{0}\right)
\left\vert \Psi \left( Z_{i},\theta _{0}\right) \right\vert ^{2}\right)
d\theta _{0}  \notag \\
&=&\int \exp \left( \frac{1}{2}\Psi ^{\dagger }\left( \theta ,Z\right)
\nabla \left( \frac{\sigma _{\theta }^{2}}{2}\nabla -\omega _{0}^{-1}\left(
J,Z\right) \right) \Psi \left( \theta ,Z\right) \right)   \notag \\
&&\times \frac{1}{n!}\left( -\frac{1}{2}\Psi ^{\dagger }\left( \theta
,Z\right) \nabla \left( \frac{\int \check{T}\digamma ^{\dag }\left( Z,\theta
\right) \exp \left( -S\left( \digamma \right) -\int \digamma \left( X,\theta
\right) \omega _{0}^{-1}\left( J,\theta ,Z\right) \left\vert \Psi \left(
\theta ,Z\right) \right\vert ^{2}d\left( X,\theta \right) \right) \mathcal{D}%
\digamma }{\int \exp \left( -S\left( \digamma \right) \right) \mathcal{D}%
\digamma }\right) \Psi \left( \theta ,Z\right) \right) ^{n}  \notag \\
&&\times \int \exp \left( \sum_{i}a\left( Z_{i},\theta _{0}\right)
\left\vert \Psi \left( Z_{i},\theta _{0}\right) \right\vert ^{2}\right)
d\theta _{0}  \label{PR}
\end{eqnarray}

We then compute the graphs associated to the case $n=1$ so that we compute
the graphs associated to:%
\begin{eqnarray}
&&\int \exp \left( \sum_{i}a\left( Z_{i},\theta _{0}\right) \left\vert \Psi
\left( Z_{i},\theta _{0}\right) \right\vert ^{2}\right) d\theta _{0}
\label{CTRCN} \\
&&\times \int \Psi ^{\dag }\left( \theta ,Z\right) \nabla \left[ -\frac{\int 
\check{T}\digamma ^{\dag }\left( Z,\theta \right) \exp \left( -S\left(
\digamma \right) -\int \digamma \left( X,\theta \right) \omega
_{0}^{-1}\left( J,\theta ,Z\right) \left\vert \Psi \left( J,\theta ,Z\right)
\right\vert ^{2}d\left( X,\theta \right) \right) \mathcal{D}\digamma }{\int
\exp \left( -S\left( \digamma \right) \right) \mathcal{D}\digamma }\right]
\Psi \left( \theta ,Z\right)   \notag \\
&=&\int \exp \left( \sum_{i}a\left( Z_{i},\theta _{0}\right) \left\vert \Psi
\left( Z_{i},\theta _{0}\right) \right\vert ^{2}\right) d\theta _{0}\times
\int \Psi ^{\dag }\left( \theta ,Z\right)   \notag \\
&&\times \nabla \left[ -\sum_{n=1}^{\infty }\frac{1}{n!}\frac{\int \check{T}%
\digamma ^{\dag }\left( Z,\theta \right) \int \dprod\limits_{i=1}^{n}\left(
-\omega _{0}^{-1}\left( J,\theta _{i},Z_{i}\right) \right) \left\vert \Psi
\left( J,\theta _{i},Z_{i}\right) \right\vert ^{2}\digamma \left(
Z_{i},\theta _{i}\right) d\left( Z_{i},\theta _{i}\right) \exp \left(
-S\left( \digamma \right) \right) \mathcal{D}\digamma }{\exp \left( -S\left(
\digamma \right) \right) \mathcal{D}\digamma }\right] \Psi \left( \theta
,Z\right)   \notag
\end{eqnarray}

To compute the contractions induced by the Wick theorem, we will use the two
remarks: first, we do not contract $\Psi ^{\dag }\left( \theta ,Z\right) $
and $\Psi \left( \theta ,Z\right) $ outside the brackets in (\ref{CTRCN})
with the source $\int \exp \left( \sum_{i}a\left( Z_{i},\theta _{0}\right)
\left\vert \Psi \left( Z_{i},\theta _{0}\right) \right\vert ^{2}\right)
d\theta _{0}$ since it would imply the appearance of non connected graphs or
vanishing contributions. Actually, the terms inside the brackets are
evaluated at $\theta _{i}<\theta $ and we assume that the perturbation $\Psi
\left( \theta ,Z\right) $ is null before the action of the source.

Second, the loops arising from the series expansion can be neglected.
Actually, we have seen that expanding%
\begin{eqnarray*}
S\left( \digamma \right)  &=&\int \digamma \left( Z,\theta \right) \left(
1-\left\vert \Psi \right\vert ^{2}\check{T}\right) \digamma ^{\dag }\left(
Z,\theta \right) d\left( Z,\theta \right) -\int \digamma \left( Z,\theta
\right) \check{T}\left( \theta -\frac{\left\vert Z-Z^{\left( 1\right)
}\right\vert }{c},Z,Z^{\left( 1\right) },\omega _{0}^{-1}+\check{T}\digamma
^{\dag }\right)  \\
&&\times \digamma ^{\dag }\left( Z^{\left( 1\right) },\theta -\frac{%
\left\vert Z-Z^{\left( 1\right) }\right\vert }{c}\right) dZdZ^{\left(
1\right) }d\theta 
\end{eqnarray*}%
in series of $\left\vert \Psi \right\vert ^{2}$ corresponds to a sum of
lines crossing $\left\vert \Psi \right\vert ^{2}$ at some points that are
integrated over. Contracting two such fields $\left\vert \Psi \left(
Z,\theta \right) \right\vert ^{2}\left\vert \Psi \left( Z^{\prime },\theta
^{\prime }\right) \right\vert ^{2}$ contracts two lines crossing $\left(
Z,\theta \right) $ and $\left( Z^{\prime },\theta ^{\prime }\right) $ This
imposes $Z=Z^{\prime }$ but also, due to the contractions:%
\begin{equation*}
\overbrace{\Psi ^{\dagger }\left( \theta ,Z\right) \Psi \left( \theta
^{\prime },Z\right) }\overbrace{\Psi ^{\dagger }\left( \theta ^{\prime
},Z\right) \Psi \left( \theta ,Z\right) }
\end{equation*}%
that $\theta ^{\prime }=\theta $. Actually, the first propagator imposes $%
\theta ^{\prime }<\theta $ and the second one $\theta <\theta ^{\prime }$.
This means that the loop corresponding to the contraction involves integrals
over a set whose measure is equal to $0$: the integration over the set of
lines forming the loop imposes that the length of these two lines are equal,
but with no $\delta $ function to implement this condition.

Once these remarks made, given the presence of the term $\prod\limits_{Z%
\neq Z_{i}}\delta \left( \left\vert \Psi \left( Z,\theta _{0}\right)
\right\vert ^{2}\right) $ in the path integral, the $n$-th term of the sum:%
\begin{equation}
\frac{1}{n!}\frac{\int \check{T}\digamma ^{\dag }\left( Z,\theta \right)
\int \dprod\limits_{i=1}^{n}\left( -\omega _{0}^{-1}\left( J,\theta
_{i},Z_{i}\right) \right) \left\vert \Psi \left( J,\theta _{i},Z_{i}\right)
\right\vert ^{2}\digamma \left( Z_{i},\theta _{i}\right) d\left(
Z_{i},\theta _{i}\right) \exp \left( -S\left( \digamma \right) \right) 
\mathcal{D}\digamma }{\exp \left( -S\left( \digamma \right) \right) \mathcal{%
D}\digamma }  \label{NT}
\end{equation}%
is contracted, by Wick theorem, with:%
\begin{equation}
\int \frac{1}{n!}\left( \sum_{i}a\left( Z_{i},\theta \right) \left\vert \Psi
\left( Z_{i},\theta \right) \right\vert ^{2}\right) ^{n}d\theta   \label{CT}
\end{equation}%
Actually, given our introductive remarks, contracting the $n$-th term in the
sum with:%
\begin{equation*}
\int \frac{1}{n!}\left( \sum_{i}a\left( Z_{i},\theta \right) \left\vert \Psi
\left( Z_{i},\theta \right) \right\vert ^{2}\right) ^{k}d\theta 
\end{equation*}%
for $k<n$ induces the presence of loops that are negligible, and for $k>n$
induces the presence of disconnected graphs where the sources term are
contracted with themselves. Such graphs are cancelled since the
normalization of the path integral keeps only connected graphs.

To compute the contraction between (\ref{NT}) and (\ref{CT}), recall that:%
\begin{equation*}
\left\vert \Psi \left( \theta ,Z_{1}\right) \right\vert ^{2}
\end{equation*}%
arising in (\ref{NT}) stands for:%
\begin{equation*}
\Psi _{0}^{\dagger }\left( Z_{1}\right) \Psi \left( \theta ,Z_{1}\right)
+\Psi _{0}\left( Z_{1}\right) \Psi ^{\dagger }\left( \theta ,Z_{1}\right)
+\left\vert \Psi \left( \theta ,Z_{1}\right) \right\vert ^{2}
\end{equation*}%
and the contractions are:%
\begin{equation*}
\overbrace{\Psi ^{\dagger }\left( \theta ,Z\right) \Psi \left( \theta
^{\prime },Z\right) }\rightarrow \frac{1}{\Lambda }\overbrace{\Psi ^{\dagger
}\left( \theta ,Z\right) \nabla _{\theta }\Psi \left( \theta ,Z\right) }=%
\frac{1}{\Lambda \Lambda _{1}}
\end{equation*}%
\begin{equation*}
\overbrace{\Psi _{0}^{\dagger }\left( \theta ,Z\right) \Psi \left( \theta
^{\prime },Z\right) }=\overbrace{\Psi ^{\dagger }\left( \theta ,Z\right)
\Psi _{0}\left( \theta ^{\prime },Z\right) }=0
\end{equation*}%
We normalize $\frac{1}{\Lambda \Lambda _{1}}\equiv \frac{1}{\Lambda }$.
Considering $\theta <\theta _{i}$ leads to a contribution and since that due
to the form of the propagator for $\Psi $, there is no loop, it amounts to
replace the contractions $\left\vert \Psi \left( Z_{i},\theta \right)
\right\vert ^{2}\left\vert \Psi \left( J,\theta _{i^{\prime }},Z_{i^{\prime
}}\right) \right\vert ^{2}$ by $\frac{\delta \left( \theta _{i}-\theta
_{i^{\prime }}\right) \delta \left( Z_{i}-Z_{i^{\prime }}\right) }{\Lambda
^{2}}$.and this leads to the expression for the contraction between (\ref{NT}%
) and (\ref{CT}):%
\begin{equation}
\frac{1}{n!}\frac{\sum_{\left( i_{1},...,i_{n}\right) }\int
\dprod\limits_{l=1}^{n}\left( -\frac{\omega _{0}^{-1}\left( J,\theta
,Z_{i_{l}}\right) }{\Lambda ^{2}}\right) a\left( Z_{i_{l}},\theta \right)
\digamma \left( Z_{i_{l}},\theta \right) \exp \left( -S\left( \digamma
\right) \right) \mathcal{D}\digamma }{\exp \left( -S\left( \digamma \right)
\right) \mathcal{D}\digamma }  \label{CRT}
\end{equation}%
Reintroducing $\Psi ^{\dag }\left( \theta ,Z\right) \nabla $ on the left and 
$\Psi \left( \theta ,Z\right) $\ on the right of (\ref{CRT}) and summing
over $n$ these contributions gives the contraction (\ref{CTRCN}):%
\begin{eqnarray*}
&&\overbrace{\Psi ^{\dag }\left( \theta ,Z\right) \nabla \left[ -\frac{\int 
\check{T}\digamma ^{\dag }\left( Z,\theta \right) \exp \left( -S\left(
\digamma \right) -\int \digamma \left( X,\theta \right) \omega
_{0}^{-1}\left( J,\theta ,Z\right) \left\vert \Psi \left( J,\theta ,Z\right)
\right\vert ^{2}d\left( X,\theta \right) \right) \mathcal{D}\digamma }{\int
\exp \left( -S\left( \digamma \right) \right) \mathcal{D}\digamma }\right] }
\\
&&\times \overbrace{\Psi \left( \theta ,Z\right) \int \exp \left(
\sum_{i}a\left( Z_{i},\theta _{0}\right) \left\vert \Psi \left( Z_{i},\theta
_{0}\right) \right\vert ^{2}\right) d\theta _{0}} \\
&\rightarrow &\int \check{T}\digamma ^{\dag }\left( Z,\theta \right) \exp
\left( -S\left( \digamma \right) -\sum_{i}a\left( Z_{i},\theta \right) \frac{%
\omega _{0}^{-1}\left( J,\theta ,Z_{i}\right) }{\Lambda ^{2}}\digamma \left(
Z_{i},\theta \right) \right) 
\end{eqnarray*}%
The contractions $\Psi \left( J,\theta _{i},Z_{i}\right) \Psi ^{\dagger
}\left( J,\theta _{i}^{\prime },Z_{i}^{\prime }\right) $ in the internal
lines can also be replaced by $\frac{\exp \left( -\Lambda _{1}\left( \theta
_{i}-\theta _{i^{\prime }}\right) \right) \delta \left( Z_{i}-Z_{i^{\prime
}}\right) }{\Lambda }$.

The same applies to the powers of:%
\begin{equation*}
\left\{ -\frac{1}{2}\Psi ^{\dagger }\left( \theta ,Z\right) \nabla \left( 
\frac{\int \check{T}\digamma ^{\dag }\left( Z,\theta \right) \exp \left(
-S\left( \digamma \right) -\int \digamma \left( X,\theta \right) \omega
_{0}^{-1}\left( J,\theta ,Z\right) \left\vert \Psi \left( \theta ,Z\right)
\right\vert ^{2}d\left( X,\theta \right) \right) \mathcal{D}\digamma }{\int
\exp \left( -S\left( \digamma \right) \right) \mathcal{D}\digamma }\right)
\Psi \left( \theta ,Z\right) \right\} ^{n}
\end{equation*}%
cntrctd wth (\ref{CT}). The absence of loops allows to replace:%
\begin{equation*}
\left( \exp \left( -S\left( \digamma \right) -\int \digamma \left( X,\theta
\right) \omega _{0}^{-1}\left( J,\theta ,Z\right) \left\vert \Psi \left(
\theta ,Z\right) \right\vert ^{2}d\left( X,\theta \right) \right) \right)
^{n}
\end{equation*}%
by:%
\begin{equation*}
\left( \int \check{T}\digamma ^{\dag }\left( Z,\theta \right) \exp \left(
-S\left( \digamma \right) -\sum_{i}a\left( Z_{i},\theta \right) \frac{\omega
_{0}^{-1}\left( Z_{i}\right) }{\Lambda ^{2}}\digamma \left( Z_{i},\theta
\right) \right) \right) ^{n}
\end{equation*}%
As a consequence, the perturbation expansion (\ref{PR}) rewrites:%
\begin{eqnarray*}
&&\int \exp \left( \frac{1}{2}\Psi ^{\dagger }\left( \theta ,Z\right) \nabla
\left( \frac{\sigma _{\theta }^{2}}{2}\nabla -\omega ^{-1}\right) \Psi
\left( \theta ,Z\right) \right) \int \exp \left( \sum_{i}a\left(
Z_{i},\theta _{0}\right) \left\vert \Psi \left( Z_{i},\theta _{0}\right)
\right\vert ^{2}\right) d\theta _{0} \\
&=&\int d\theta _{0}\int \exp \left( \frac{1}{2}\Psi ^{\dagger }\left(
\theta ,Z\right) \nabla \left( \frac{\sigma _{\theta }^{2}}{2}\nabla -\omega
^{-1}\right) \Psi \left( \theta ,Z\right) \right) 
\end{eqnarray*}%
with:%
\begin{equation}
\omega ^{-1}\left( J,\theta ,Z\right) =\omega _{0}^{-1}\left( J,Z\right) +%
\frac{\int \check{T}\digamma ^{\dag }\left( Z,\theta \right) \exp \left(
-S\left( \digamma \right) -\sum_{i}a\left( Z_{i},\theta _{0}\right) \frac{%
\omega _{0}^{-1}\left( J,\theta _{0},Z_{i}\right) }{\Lambda ^{2}}\digamma
\left( Z_{i},\theta \right) \right) \mathcal{D}\digamma }{\int \exp \left(
-S\left( \digamma \right) \right) \mathcal{D}\digamma }  \label{QV}
\end{equation}%
and $S\left( \digamma \right) $ obtained by replacing $\left\vert \Psi
\left( \theta ,Z\right) \right\vert ^{2}$ with $\frac{1}{\digamma }$:%
\begin{eqnarray}
S\left( \digamma \right)  &=&\int \digamma \left( Z,\theta \right) \left(
1-\left\vert \Psi \right\vert ^{2}\check{T}\right) \digamma ^{\dag }\left(
Z,\theta \right) d\left( Z,\theta \right)   \label{SLp} \\
&&-\int \digamma \left( Z,\theta \right) \check{T}\left( \theta -\frac{%
\left\vert Z-Z^{\left( 1\right) }\right\vert }{c},Z,Z^{\left( 1\right)
},\omega _{0}^{-1}+\check{T}\digamma ^{\dag }\right) \digamma ^{\dag }\left(
Z^{\left( 1\right) },\theta -\frac{\left\vert Z-Z^{\left( 1\right)
}\right\vert }{c}\right) dZdZ^{\left( 1\right) }d\theta   \notag
\end{eqnarray}%
\bigskip 

\subsubsection*{3.1.2 Saddle point approximation for activity auxiliary path
integral}

Integral (\ref{QV}) will be computed in the saddle point approximation. It
is obtained by replacing $\digamma ^{\dag }$ and\ $\digamma $ with their
values minimizing action $S\left( \digamma \right) $ defined in (\ref{SL}).
This yields the equations for $\digamma ^{\dag }\left( Z,\theta \right) $
and $\digamma \left( Z,\theta \right) $:%
\begin{equation}
\left( \left( 1-\left\vert \Psi \right\vert ^{2}\check{T}\right) \digamma
^{\dag }\right) \left( Z,\theta \right) -\left( \check{T}_{\omega _{0}^{-1}+%
\check{T}\digamma ^{\dag }}\digamma ^{\dag }\right) \left( Z,\theta \right)
+\sum_{i}a\left( Z_{i},\theta \right) \frac{\omega _{0}^{-1}\left( J,\theta
,Z_{i}\right) }{\Lambda ^{2}}\delta \left( Z-Z_{i}\right) =0
\end{equation}%
or, using (\ref{SLp}): 
\begin{equation}
\left( \left( 1-\frac{1}{\Lambda }\check{T}\right) \digamma ^{\dag }\right)
\left( Z,\theta \right) -\left( \check{T}_{\omega _{0}^{-1}+\check{T}%
\digamma ^{\dag }}\digamma ^{\dag }\right) \left( Z,\theta \right)
+\sum_{i}a\left( Z_{i},\theta \right) \frac{\omega _{0}^{-1}\left( J,\theta
,Z_{i}\right) }{\Lambda ^{2}}\delta \left( Z-Z_{i}\right) =0  \label{dsn}
\end{equation}%
and:%
\begin{equation*}
\digamma \left( Z,\theta \right) =0
\end{equation*}%
Under the saddle point approximation, equation (\ref{QV}) becomes:%
\begin{equation}
\omega ^{-1}\left( J,\theta ,Z\right) =\omega _{0}^{-1}\left( J,\theta
,Z\right) +\check{T}\digamma ^{\dag }\left( Z,\theta \right)   \label{dst}
\end{equation}%
and (\ref{dsn}) leads to write recursively:%
\begin{equation}
\digamma ^{\dag }=\frac{1}{\left( 1-\left\vert \Psi \right\vert ^{2}\check{T}%
-\check{T}_{\omega _{0}+\check{T}\digamma ^{\dag }}\right) }\left[
-\sum_{i}a\left( Z_{i},\theta \right) \frac{\omega _{0}^{-1}\left( J,\theta
,Z_{i}\right) }{\Lambda ^{2}}\right]   \label{RCS}
\end{equation}%
so that using (\ref{dst}) yields in first approximatn:%
\begin{equation}
\omega ^{-1}\left( J,\theta ,Z\right) =\omega _{0}^{-1}\left( J,\theta
,Z\right) -\check{T}\frac{1}{\left( 1-\left( \left\vert \Psi \right\vert
^{2}+1\right) \check{T}\right) }\left[ \sum_{i}a\left( Z_{i},\theta \right) 
\frac{\omega _{0}^{-1}\left( J,\theta ,Z_{i}\right) }{\Lambda ^{2}}\right] 
\label{CRL}
\end{equation}%
This will be applied to the expansion around the static case.

\subsubsection*{3.1.3 Expansion around static case}

We estimate the correction $\check{T}\digamma ^{\dag }\left( Z,\theta
\right) $ given by (\ref{TRL}), in the static state. We rewrite the
expression of $T\left( Z,Z^{\prime }\right) $: 
\begin{equation*}
T\left( Z,Z^{\prime }\right) =\frac{\lambda \tau \exp \left( -\frac{%
\left\vert Z-Z^{\prime }\right\vert }{\nu c}\right) }{1+\frac{\omega
^{\prime }\left\vert \Psi \left( Z^{\prime },\omega ^{\prime }\right)
\right\vert ^{2}}{\omega \left\vert \Psi \left( Z,\omega \right) \right\vert
^{2}}}
\end{equation*}%
and $\check{T}\left( Z,Z^{\prime }\right) $ is equal to:%
\begin{equation*}
\check{T}\left( Z,Z^{\prime }\right) =-\frac{\frac{\kappa }{N}T\left(
Z,Z_{1}\right) G^{\prime }\left[ J,\omega _{0}Z\right] \left\vert \Psi
_{0}\left( Z_{1}\right) \right\vert ^{2}}{\omega _{0}^{-1}\left( J,Z\right) }
\end{equation*}%
Given our previous choices, $F^{\prime }\left[ J,\omega _{0}Z\right] =-\frac{%
b}{\left( \omega _{0}\left( J,Z\right) \right) ^{2}}$, so that, for $b$
normalized to $1$:%
\begin{equation*}
\check{T}\left( Z,Z^{\prime }\right) =\frac{\frac{\kappa }{N}T\left(
Z,Z^{\prime }\right) \left\vert \Psi _{0}\left( Z^{\prime }\right)
\right\vert ^{2}}{\omega _{0}\left( J,Z\right) }=\frac{\kappa }{N}\frac{%
\lambda \tau \exp \left( -\frac{\left\vert Z-Z^{\prime }\right\vert }{\nu c}%
\right) \left( \left\vert \Psi _{0}\left( Z^{\prime }\right) \right\vert
^{2}\right) ^{2}}{\omega _{0}\left\vert \Psi _{0}\left( Z\right) \right\vert
^{2}+\omega _{0}^{\prime }\left\vert \Psi _{0}\left( Z^{\prime }\right)
\right\vert ^{2}}
\end{equation*}%
We also need to estimate:%
\begin{eqnarray*}
&&\check{T}\left( Z,Z^{\prime },\omega _{0}+\check{T}\digamma ^{\dag
}\right)  \\
&=&\frac{\kappa }{N}\frac{\lambda \tau \exp \left( -\frac{\left\vert
Z-Z^{\prime }\right\vert }{\nu c}\right) \left( \left\vert \Psi _{0}\left(
Z^{\prime }\right) \right\vert ^{2}\right) ^{2}}{\left( \omega _{0}+\check{T}%
\digamma ^{\dag }\left( Z\right) \right) \left\vert \Psi _{0}\left( Z\right)
\right\vert ^{2}+\left( \omega _{0}^{\prime }+\check{T}\digamma ^{\dag
}\left( Z^{\prime }\right) \right) \left\vert \Psi _{0}\left( Z^{\prime
}\right) \right\vert ^{2}} \\
&=&\check{T}\left( Z,Z^{\prime },\omega _{0}\right)  \\
&&-\frac{\lambda \tau \exp \left( -\frac{\left\vert Z-Z^{\prime }\right\vert 
}{\nu c}\right) \left( \left\vert \Psi _{0}\left( Z^{\prime }\right)
\right\vert ^{2}\right) ^{2}\left( \check{T}\digamma ^{\dag }\left( Z\right)
\left\vert \Psi _{0}\left( Z\right) \right\vert ^{2}+\check{T}\digamma
^{\dag }\left( Z^{\prime }\right) \left\vert \Psi _{0}\left( Z^{\prime
}\right) \right\vert ^{2}\right) }{\left( \omega _{0}\left\vert \Psi
_{0}\left( Z\right) \right\vert ^{2}+\omega _{0}^{\prime }\left\vert \Psi
_{0}\left( Z^{\prime }\right) \right\vert ^{2}\right) \left( \left( \omega
_{0}+\check{T}\digamma ^{\dag }\left( Z\right) \right) \left\vert \Psi
_{0}\left( Z\right) \right\vert ^{2}+\left( \omega _{0}^{\prime }+\check{T}%
\digamma ^{\dag }\left( Z^{\prime }\right) \right) \left\vert \Psi
_{0}\left( Z^{\prime }\right) \right\vert ^{2}\right) } \\
&=&\check{T}\left( Z,Z^{\prime },\omega _{0}\right) -\frac{\left( \check{T}%
\digamma ^{\dag }\left( Z\right) \left\vert \Psi _{0}\left( Z\right)
\right\vert ^{2}+\check{T}\digamma ^{\dag }\left( Z^{\prime }\right)
\left\vert \Psi _{0}\left( Z^{\prime }\right) \right\vert ^{2}\right) }{%
\left( \left( \omega _{0}+\check{T}\digamma ^{\dag }\left( Z\right) \right)
\left\vert \Psi _{0}\left( Z\right) \right\vert ^{2}+\left( \omega
_{0}^{\prime }+\check{T}\digamma ^{\dag }\left( Z^{\prime }\right) \right)
\left\vert \Psi _{0}\left( Z^{\prime }\right) \right\vert ^{2}\right) }%
\check{T}\left( Z,Z^{\prime },\omega _{0}\right) 
\end{eqnarray*}%
In the local approximation $Z^{\prime }=Z$, this yields a series expansion:%
\begin{equation}
\check{T}_{\omega _{0}+\check{T}\digamma ^{\dag }}-\check{T}\simeq
\sum_{n\geqslant 1}\left( -\frac{\check{T}\digamma ^{\dag }\left( Z\right) }{%
\omega _{0}\left( Z\right) }\right) ^{n}\check{T}  \label{qnt}
\end{equation}%
The saddle point equation (\ref{dsn}) is rewriten:%
\begin{equation*}
\left( \left( 1-\frac{1}{\Lambda }\check{T}\right) \digamma ^{\dag }\right)
\left( Z\right) -\left( \check{T}_{\omega _{0}+\check{T}\digamma ^{\dag
}}\digamma ^{\dag }\right) \left( Z\right) +\left( \sum_{i}a\left(
Z_{i},\theta \right) \frac{\omega _{0}^{-1}\left( J,\theta ,Z_{i}\right) }{%
\Lambda ^{2}}\digamma \left( Z_{i},\theta \right) \right) =0
\end{equation*}%
and this leads to:%
\begin{eqnarray}
\check{T}\digamma ^{\dag } &=&\check{T}\frac{1}{\left( 1-\frac{1}{\digamma }%
\check{T}-\check{T}_{\omega _{0}+\check{T}\digamma ^{\dag }}\right) }\left(
\sum_{i}-a\left( Z_{i},\theta \right) \frac{\omega _{0}^{-1}\left( J,\theta
,Z_{i}\right) }{\Lambda ^{2}}\digamma \left( Z_{i},\theta \right) \right) 
\label{qnn} \\
&=&\check{T}\frac{1}{\left( 1-\left( 1+\frac{1}{\Lambda }\right) \check{T}%
-\left( \check{T}_{\omega _{0}+\check{T}\digamma ^{\dag }}-\check{T}\right)
\right) }\left( \sum_{i}-a\left( Z_{i},\theta \right) \frac{\omega
_{0}^{-1}\left( J,\theta ,Z_{i}\right) }{\Lambda ^{2}}\digamma \left(
Z_{i},\theta \right) \right)   \notag
\end{eqnarray}%
Gathering (\ref{qnn}) and (\ref{qnt}), leads to the recursive formula:%
\begin{eqnarray*}
\check{T}\digamma ^{\dag } &\simeq &\sum_{n_{1},...,n_{2}}\frac{\check{T}}{%
1-\left( 1+\frac{1}{\Lambda }\right) \check{T}}\left[ \left( -\frac{\check{T}%
\digamma ^{\dag }\left( Z_{1}\right) }{\omega _{0}\left( Z_{1}\right) }%
\check{T}\right) ^{n_{1}}\right] \frac{1}{1-\left( 1+\frac{1}{\Lambda }%
\right) \check{T}}\left[ \left( -\frac{\check{T}\digamma ^{\dag }\left(
Z_{2}\right) }{\omega _{0}\left( Z_{2}\right) }\check{T}\right) ^{n_{2}}%
\right]  \\
&&...\frac{1}{1-\left( 1+\frac{1}{\Lambda }\right) \check{T}}\left(
\sum_{i}-a\left( Z_{i},\theta \right) \frac{\omega _{0}^{-1}\left( J,\theta
,Z_{i}\right) }{\Lambda ^{2}}\right) 
\end{eqnarray*}%
Recursively, (\ref{qtn}) leads to replace $\check{T}_{\omega _{0}+\check{T}%
\digamma ^{\dag }}-\check{T}$ by:%
\begin{equation*}
\left[ \sum_{n_{i}\geqslant 1}\frac{\kappa }{N}\frac{\lambda \tau \left\vert
\Psi _{0}\left( Z_{i}\right) \right\vert ^{2}}{2\Lambda \omega _{0}}\left( -%
\frac{\check{T}\digamma ^{\dag }\left( Z_{i}\right) }{\omega _{0}}\right)
^{n_{i}}\right] 
\end{equation*}

Graphically, we fix $k$ $Z_{i}$ vertices of valence $n_{i}$, $i=1,...,k$
with $k\in 
\mathbb{N}
$, $n_{i}\in 
\mathbb{N}
$. Draw all graphs connected and simply connected crossing these vertices
with the given valence. To each edge, associate $\frac{1}{1-\left( 1+\frac{1%
}{\Lambda }\right) \check{T}}$, to each vertex, associate $%
\prod\limits_{1}^{n}\frac{-\check{T}}{\omega _{0}\left( Z_{i}\right) }$.
The factors associated to the edges are connected to one term of the product
The graphs are ordered in time starting from one edge. The first edge is
composed with $\check{T}$.

\subsubsection*{3.1.4 Lowest order expansion in local approximation}

A first order compact approximation of (\ref{qnn}) can also be obtained by
writing in the local approximation $Z^{\prime }\simeq Z$:%
\begin{eqnarray}
\check{T}\left( Z,Z^{\prime },\omega +\check{T}\digamma ^{\dag }\right) -%
\check{T} &=&-\frac{\kappa }{N}\frac{\left( \check{T}\digamma ^{\dag }\left(
Z\right) \left\vert \Psi _{0}\left( Z\right) \right\vert ^{2}+\check{T}%
\digamma ^{\dag }\left( Z^{\prime }\right) \left\vert \Psi _{0}\left(
Z^{\prime }\right) \right\vert ^{2}\right) }{\left( \left( \omega _{0}+%
\check{T}\digamma ^{\dag }\left( Z\right) \right) \left\vert \Psi _{0}\left(
Z\right) \right\vert ^{2}+\left( \omega _{0}^{\prime }+\check{T}\digamma
^{\dag }\left( Z^{\prime }\right) \right) \left\vert \Psi _{0}\left(
Z^{\prime }\right) \right\vert ^{2}\right) }\check{T}\left( Z,Z^{\prime
},\omega _{0}\right)   \label{RV} \\
&\simeq &-\frac{\left( \check{T}\digamma ^{\dag }\left( Z\right) \right) }{%
\left( \omega _{0}\left( Z\right) +\check{T}\digamma ^{\dag }\left( Z\right)
\right) }\check{T}\left( Z,Z^{\prime },\omega _{0}\right)   \notag
\end{eqnarray}%
and then the solution (\ref{qnn}) of the saddle point equation at zeroth
order leads to:%
\begin{equation}
\check{T}\digamma ^{\dag }=\check{T}_{1}\left[ -\sum_{i}a\left( Z_{i},\theta
\right) \frac{\omega _{0}^{-1}\left( J,\theta ,Z_{i}\right) }{\Lambda ^{2}}%
\right]   \label{LR}
\end{equation}%
with:%
\begin{equation*}
\check{T}_{1}=\frac{\check{T}}{\left( 1-\left( 1+\frac{1}{\Lambda }\right) 
\check{T}\right) }
\end{equation*}%
That is, using (\ref{dst}), we recover a first approximation f (\ref{CRL}):%
\begin{equation}
\omega ^{-1}\left( J,\theta ,Z\right) =\omega _{0}^{-1}\left( J,\theta
,Z\right) +\frac{\check{T}}{\left( 1-\left( 1+\frac{1}{\Lambda }\right) 
\check{T}\right) }\left[ -\sum_{i}a\left( Z_{i},\theta \right) \frac{\omega
_{0}^{-1}\left( J,\theta ,Z_{i}\right) }{\Lambda ^{2}}\right]   \label{CRLPP}
\end{equation}

\subsubsection*{3.1.5 Higher order terms in local approximation}

We can go to higher orders and insert this formula (\ref{LR}) in (\ref{RV}):%
\begin{equation*}
\check{T}\left( Z,Z^{\prime },\omega +\check{T}\digamma ^{\dag }\right) -%
\check{T}\simeq -\frac{\check{T}_{1}\left[ -\sum_{i}a\left( Z_{i},\theta
\right) \frac{\ \omega _{0}^{-1}\left( J,\theta ,Z_{i}\right) }{\Lambda ^{2}}%
\right] }{\omega _{0}\left( Z\right) +\check{T}_{1}\left[ -\sum_{i}a\left(
Z_{i},\theta \right) \frac{\ \omega _{0}^{-1}\left( J,\theta ,Z_{i}\right) }{%
\Lambda ^{2}}\right] }\check{T}
\end{equation*}%
This allows to solve the saddle point equation at the first order by writing
(\ref{qnn}):%
\begin{equation}
\check{T}\digamma ^{\dag }=\check{T}\frac{1}{\left( 1-\frac{1}{\Lambda }%
\check{T}-\check{T}_{\omega _{0}+\check{T}\digamma ^{\dag }}\right) }\left(
\sum_{i}-a\left( Z_{i},\theta \right) \frac{\omega _{0}^{-1}\left( J,\theta
,Z_{i}\right) }{\Lambda ^{2}}\digamma \left( Z_{i},\theta \right) \right) 
\end{equation}%
where we define the effective scale of connectivities:%
\begin{equation*}
\frac{1}{\Lambda _{1}\left( \left( \omega _{0}\left( J,\theta ,Z_{i}\right)
_{i}\right) \right) }=\frac{1}{\Lambda }-\frac{\check{T}_{1}\left[
-\sum_{i}a\left( Z_{i},\theta \right) \frac{\ \omega _{0}^{-1}\left(
J,\theta ,Z_{i}\right) }{\Lambda ^{2}}\right] }{\omega _{0}\left( Z\right) +%
\check{T}_{1}\left[ -\sum_{i}a\left( Z_{i},\theta \right) \frac{\ \omega
_{0}^{-1}\left( J,\theta ,Z_{i}\right) }{\Lambda ^{2}}\right] }
\end{equation*}%
Computing explicitely the integrals, this leads to rewrite (\ref{dst}):%
\begin{eqnarray}
&&\omega ^{-1}\left( J,\theta ,Z\right) =\omega _{0}^{-1}\left( J,\theta
,Z\right)   \label{sdn} \\
&&+\int^{\theta _{i}}\check{T}\left( 1-\left( 1+\frac{1}{\Lambda _{1}\left(
\left( \omega _{0}\left( J,\theta ,Z_{i}\right) _{i}\right) \right) }\right) 
\check{T}\right) ^{-1}\left( Z,\theta ,Z_{i},\theta _{i}\right) \left[
-\sum_{i}a\left( Z_{i},\theta _{i}\right) \frac{\omega _{0}\left( J,\theta
_{i},Z_{i}\right) }{\Lambda ^{2}}d\theta _{i}\right]   \notag \\
&\equiv &\sum_{i}\int K\left( Z,\theta ,Z_{i},\theta _{i}\right) \left\{
a\left( Z_{i},\theta _{i}\right) \frac{\omega _{0}\left( J,\theta
_{i},Z_{i}\right) }{\Lambda ^{2}}\right\} d\theta _{i}  \notag
\end{eqnarray}%
\bigskip Note that we can go recursively to the next orders, by replacing $%
K\left( Z,\theta ,Z_{i},\theta _{i}\right) $ wth:%
\begin{equation}
K\left( Z,\theta ,Z_{i},\theta _{i}\right) \simeq -\check{T}\left( 1-\left(
1+\frac{1}{\Lambda _{2}\left( \left( \omega _{0}\left( J,\theta
,Z_{i}\right) _{i}\right) \right) }\right) \check{T}\right) ^{-1}
\label{SDn}
\end{equation}%
\bigskip wher:%
\begin{equation*}
\frac{1}{\Lambda _{2}\left( \left( \omega _{0}\left( J,\theta ,Z_{i}\right)
_{i}\right) \right) }=\frac{1}{\Lambda }-\frac{\check{T}_{2}\left[
-\sum_{i}a\left( Z_{i},\theta \right) \frac{\ \omega _{0}^{-1}\left(
J,\theta ,Z_{i}\right) }{\Lambda ^{2}}\right] }{\omega _{0}\left( Z\right) +%
\check{T}_{2}\left[ -\sum_{i}a\left( Z_{i},\theta \right) \frac{\ \omega
_{0}^{-1}\left( J,\theta ,Z_{i}\right) }{\Lambda ^{2}}\right] }
\end{equation*}%
along with:%
\begin{equation*}
\check{T}_{2}=\frac{\check{T}}{\left( 1-\left( 1+\frac{1}{\Lambda _{1}\left(
\left( \omega _{0}\left( J,\theta ,Z_{i}\right) _{i}\right) \right) }\right) 
\check{T}\right) }
\end{equation*}%
and so on, the next order being obtained by replacing:%
\begin{equation*}
\left( 1+\frac{1}{\Lambda }\right) \check{T}
\end{equation*}%
by:%
\begin{equation*}
\left( 1+\frac{1}{\Lambda _{2}\left( \left( \omega _{0}\left( J,\theta
,Z_{i}\right) _{i}\right) \right) }\right) \check{T}
\end{equation*}

\subsection*{3.2 Estimation of the propagated signal and correction to
activities.}

Once the kernel $K\left( Z,\theta ,Z_{i},\theta _{i}\right) $ for signals
propagation has been computed, we can study the modification in activity
induced by external signals propagating along the thread.

\subsubsection*{3.2.1 Propagated signal}

We first estimate the transmitted signal from the sources to the points of
the thread given in (\ref{sdn}):%
\begin{equation*}
\int K\left( Z,\theta ,Z_{i},\theta _{i}\right) \left\{ \sum_{i}a\left(
Z_{i},\theta _{i}\right) \frac{\omega _{0}^{-1}\left( J,\theta
_{i},Z_{i}\right) }{\Lambda ^{2}}\right\} d\theta _{i}
\end{equation*}%
at the lowest order. Expanding: 
\begin{equation*}
-\frac{\check{T}}{\left( 1-\left( 1+\frac{1}{\Lambda }\right) \check{T}%
\right) }\sum_{i}a\left( Z_{i},\theta \right) \frac{\ \omega _{0}^{-1}\left(
J,\theta ,Z_{i}\right) }{\Lambda ^{2}}
\end{equation*}%
computes the sum of lines starting at one $Z_{i}$ \ The lines connects
points $Z_{i_{l}}-Z_{i_{l+1}}$ such that $\check{T}\left(
Z_{i_{l}},Z_{i_{l+1}}\right) $ is different from $0$. Considering
oscillating signals $a\left( Z_{i},\theta \right) \propto \exp \left(
i\varpi \theta \right) $, and assuming "quite" straight lines of length $%
\left\vert Z-Z_{i}\right\vert $ from $Z$ to $Z_{i}$, due to the exponential
factor in the transitions, leads to a phase shift proportional to $\exp
\left( i\frac{\varpi \left\vert Z_{i}-Z_{0}\right\vert }{c\left\vert
Z-Z_{0}\right\vert }\right) $ where $Z_{0}\in \left\{ Z_{i}\right\} $ is the
closest point to $Z$. Taking into account corrections due to the length
around $\left\vert Z-Z_{i}\right\vert $ contributes to a phase shift of $%
\exp \left( i\frac{\varpi \left( l-\left\vert Z-Z_{i}\right\vert \right) }{c}%
\right) $ in the integral: 
\begin{equation*}
\int K\left( Z,\theta ,Z_{i},\theta _{i}\right) \left\{ \sum_{i}a\left(
Z_{i},\theta _{i}\right) \frac{\omega _{0}^{-1}\left( J,\theta
_{i},Z_{i}\right) }{\Lambda ^{2}}\right\} d\theta _{i}
\end{equation*}%
Moreover for $K\left( Z,\theta ,Z_{i},\theta _{i}\right) $ proportional to
the product between the average of $T$ along the path and the exponential
factor computed in the previous section, we obtain:%
\begin{eqnarray}
&&\int K\left( Z,\theta ,Z_{i},\theta _{i}\right) \left\{ \sum_{i}a\left(
Z_{i},\theta _{i}\right) \frac{\omega _{0}^{-1}\left( J,\theta
_{i},Z_{i}\right) }{\Lambda ^{2}}\right\} d\theta _{i}  \label{NTG} \\
&\propto &\int \sum_{i}a\left( Z_{i},\theta _{i}\right) \exp \left(
-cl-\alpha \left( \left( cl\right) ^{2}-\left\vert Z-Z_{i}\right\vert
^{2}\right) \right) \exp \left( i\frac{\varpi l}{c}\right) dl\left( \exp
\left( i\frac{\varpi \left( l-\left\vert Z-Z_{i}\right\vert \right) }{c}%
\right) \exp \left( i\frac{\varpi \left\vert Z_{i}-Z_{i+1}\right\vert }{%
c\left\vert Z-Z_{i}\right\vert }\right) \right)  \notag \\
&\propto &\int \sum_{i}a\left( Z_{i},\theta _{i}\right) \exp \left(
-cl-\alpha \left( \left( cl\right) ^{2}-\left\vert Z-Z_{i}\right\vert
^{2}\right) \right) \exp \left( i\frac{\varpi \left( l-\left\vert
Z-Z_{i}\right\vert \right) }{c}\right) \exp \left( i\frac{\varpi \left(
\left\vert Z-Z_{i}\right\vert \right) }{c}\right)  \notag \\
&\simeq &\int \sum_{i}a\left( Z_{i},\theta _{i}\right) \exp \left(
-cl-\alpha \left( \left( cl\right) ^{2}-\left\vert Z-Z_{i}\right\vert
^{2}\right) \right) \exp \left( i\frac{\varpi \left( l-\left\vert
Z-Z_{i}\right\vert \right) }{c}\right) \exp \left( i\frac{\varpi \left(
\left\vert Z-Z_{0}\right\vert \right) }{c}\right) \exp \left( i\frac{\varpi
\left\vert Z_{i}-Z_{0}\right\vert }{c\left\vert Z-Z_{0}\right\vert }\right) 
\notag
\end{eqnarray}%
Now, the integral:%
\begin{equation*}
Z=\int \exp \left( -cl-\alpha \left( \left( cl\right) ^{2}-\left\vert
Z-Z_{i}\right\vert ^{2}\right) \right) \exp \left( i\frac{\varpi \left(
l-\left\vert Z-Z_{i}\right\vert \right) }{c}\right) dl
\end{equation*}%
arisng n (\ref{NTG}) is computed by writing:%
\begin{eqnarray*}
\left( \left( cl\right) ^{2}-\left\vert Z-Z_{i}\right\vert ^{2}\right)
&=&\left( cl-\left\vert Z-Z_{i}\right\vert \right) ^{2}-2\left( \left\vert
Z-Z_{i}\right\vert ^{2}-\left\vert Z-Z_{i}\right\vert cl\right) \\
&=&\left( cl-\left\vert Z-Z_{i}\right\vert \right) ^{2}+2\left\vert
Z-Z_{i}\right\vert \left( cl-\left\vert Z-Z_{i}\right\vert \right)
\end{eqnarray*}%
so that:%
\begin{eqnarray*}
&&Z=\int \exp \left( -\left( cl-\left\vert Z-Z_{i}\right\vert \right)
-\left\vert Z-Z_{i}\right\vert -\alpha \left( \left( cl-\left\vert
Z-Z_{i}\right\vert \right) ^{2}+2\left\vert Z-Z_{i}\right\vert \left(
cl-\left\vert Z-Z_{i}\right\vert \right) \right) \right) \\
&&\exp \left( i\frac{\varpi \left( cl-\left\vert Z-Z_{i}\right\vert \right) 
}{c}\right) dl
\end{eqnarray*}%
and the integral becomes:%
\begin{eqnarray*}
Z &=&\frac{\exp \left( -\left\vert Z-Z_{i}\right\vert \right) }{c}\int \exp
\left( -Y\left( 1+2\alpha \left\vert Z-Z_{i}\right\vert \right) -\alpha
Y^{2}\right) \exp \left( i\frac{\varpi Y}{c}\right) dY \\
&=&\frac{\exp \left( -\left\vert Z-Z_{i}\right\vert +\frac{\left( 1+2\alpha
\left\vert Z-Z_{i}\right\vert \right) ^{2}}{4\alpha }-i\frac{\varpi \left(
1+2\alpha \left\vert Z-Z_{i}\right\vert \right) }{2c\alpha }\right) }{c}%
\int_{\frac{\left( 1+2\alpha \left\vert Z-Z_{i}\right\vert \right) }{2\alpha 
}}\exp \left( -\alpha Y^{2}\right) \exp \left( i\frac{\varpi Y}{c}\right) dY
\end{eqnarray*}%
this result can be approximated by:%
\begin{eqnarray*}
Z &\simeq &\frac{\exp \left( -\left\vert Z-Z_{i}\right\vert \right) }{c}\int
\exp \left( -Y\left( 1+2\alpha \left\vert Z-Z_{i}\right\vert \right) \right)
\exp \left( i\frac{\varpi Y}{c}\right) dY \\
&\simeq &\frac{\exp \left( -\left\vert Z-Z_{i}\right\vert \right) }{c\left(
1+2\alpha \left\vert Z-Z_{i}\right\vert +i\frac{\varpi }{c}\right) }=\frac{%
\exp \left( -\left\vert Z-Z_{i}\right\vert \right) }{c\sqrt{\left( 1+2\alpha
\left\vert Z-Z_{i}\right\vert \right) ^{2}+\left( \frac{\varpi }{c}\right)
^{2}}}\exp \left( -i\arctan \left( \frac{\varpi }{c\left( 1+2\alpha
\left\vert Z-Z_{i}\right\vert \right) }\right) \right)
\end{eqnarray*}

Inserting the reslt in (\ref{NTG}) ylds th prpgtd sgnl:%
\begin{eqnarray*}
&&\int K\left( Z,\theta ,Z_{i},\theta _{i}\right) \left\{ \sum_{i}a\left(
Z_{i},\theta _{i}\right) \frac{\omega _{0}^{-1}\left( J,\theta
_{i},Z_{i}\right) }{\Lambda ^{2}}\right\} d\theta _{i} \\
&\propto &\sum_{i}a\left( Z_{i},\theta _{i}\right) \frac{\exp \left(
-\left\vert Z-Z_{i}\right\vert \right) }{c\sqrt{\left( 1+2\alpha \left\vert
Z-Z_{i}\right\vert \right) ^{2}+\left( \frac{\varpi }{c}\right) ^{2}}} \\
&&\times \exp \left( i\left( \frac{\varpi \left( \left\vert
Z-Z_{0}\right\vert \right) }{c}-\arctan \left( \frac{\varpi }{c\left(
1+2\alpha \left\vert Z-Z_{i}\right\vert \right) }\right) \right) \right)
\exp \left( i\frac{\varpi \left\vert Z_{i}-Z_{0}\right\vert }{c\left\vert
Z-Z_{0}\right\vert }\right)
\end{eqnarray*}%
At the first order in $\frac{\left\vert Z_{i}-Z_{0}\right\vert }{\left\vert
Z-Z_{0}\right\vert }$, this is:%
\begin{eqnarray*}
&&\int K\left( Z,\theta ,Z_{i},\theta _{i}\right) \left\{ \sum_{i}a\left(
Z_{i},\theta _{i}\right) \frac{\omega _{0}^{-1}\left( J,\theta
_{i},Z_{i}\right) }{\Lambda ^{2}}\right\} d\theta _{i} \\
\simeq &&\frac{\exp \left( -\left\vert Z-Z_{0}\right\vert \right) }{c\sqrt{%
\left( 1+2\alpha \left\vert Z-Z_{0}\right\vert \right) ^{2}+\left( \frac{%
\varpi }{c}\right) ^{2}}} \\
&&\times \exp \left( i\left( \frac{\varpi \left( \left\vert
Z-Z_{0}\right\vert \right) }{c}-\arctan \left( \frac{\varpi }{c\left(
1+2\alpha \left\vert Z-Z_{0}\right\vert \right) }\right) \right) \right)
\sum_{i}a\left( Z_{i},\theta _{i}\right) \exp \left( i\frac{\varpi
\left\vert Z_{i}-Z_{0}\right\vert }{c\left\vert Z-Z_{0}\right\vert }\right)
\end{eqnarray*}

For $a\left( Z_{i},\theta _{i}\right) $ constant equal to $a$ ths rdcs t:%
\begin{eqnarray}
&&\int K\left( Z,\theta ,Z_{i},\theta _{i}\right) \left\{ \sum_{i}a\left(
Z_{i},\theta _{i}\right) \frac{\omega _{0}^{-1}\left( J,\theta
_{i},Z_{i}\right) }{\Lambda ^{2}}\right\} d\theta _{i}  \label{KSGL} \\
&\simeq &\frac{a\exp \left( -\left\vert Z-Z_{0}\right\vert \right) }{c\sqrt{%
\left( 1+2\alpha \left\vert Z-Z_{0}\right\vert \right) ^{2}+\left( \frac{%
\varpi }{c}\right) ^{2}}}\exp \left( i\left( \frac{\varpi \left( \left\vert
Z-Z_{0}\right\vert \right) }{c}-\arctan \left( \frac{\varpi }{c\left(
1+2\alpha \left\vert Z-Z_{0}\right\vert \right) }\right) \right) \right)
\sum_{i}\exp \left( i\frac{\varpi \left\vert Z_{i}-Z_{0}\right\vert }{%
c\left\vert Z-Z_{0}\right\vert }\right)  \notag
\end{eqnarray}%
leading to interferences. For large number of points $Z_{i}$:%
\begin{equation*}
\sum_{i}\exp \left( i\frac{\varpi \left\vert Z_{i}-Z_{0}\right\vert }{%
c\left\vert Z-Z_{0}\right\vert }\right) \simeq 0
\end{equation*}%
except for the maxima of interferences with magnitude:%
\begin{equation*}
\frac{a\exp \left( -\left\vert Z-Z_{0}\right\vert \right) }{c\sqrt{\left(
1+2\alpha \left\vert Z-Z_{0}\right\vert \right) ^{2}+\left( \frac{\varpi }{c}%
\right) ^{2}}}
\end{equation*}%
so that (\ref{KSGL}) localizes t ths pnt.

\subsubsection*{3.2.2 Estimation of correction to activities}

Recall that:%
\begin{equation*}
\left\vert \Psi \left( \theta ,Z_{1}\right) \right\vert ^{2}
\end{equation*}%
stands for:%
\begin{equation*}
\Psi _{0}^{\dagger }\left( \theta ,,Z_{1}\right) \Psi \left( \theta
,Z_{1}\right) +\Psi _{0}\left( \theta ,,Z_{1}\right) \Psi ^{\dagger }\left(
\theta ,Z_{1}\right) +\left\vert \Psi \left( \theta ,Z_{1}\right)
\right\vert ^{2}
\end{equation*}%
where $\Psi _{0}\left( \theta ,,Z_{1}\right) $ is quasi static (see the
remark in the text) and ultimately, we are left with the following form for
the path integral over $\Psi $:%
\begin{eqnarray}
&&\int \exp \left( -S\left( \Psi \right) \right) \int \exp \left(
\sum_{i}a\left( Z_{i},\theta _{0}\right) \left\vert \Psi \left( Z_{i},\theta
_{0}\right) \right\vert ^{2}\right) d\theta _{0}  \label{CTN} \\
&\equiv &\int \exp \left( \frac{1}{2}\left( \Psi _{0}^{\dagger }\left(
\theta ,Z\right) +\Psi ^{\dagger }\left( \theta ,Z\right) \right) \nabla
\left( \frac{\sigma _{\theta }^{2}}{2}\nabla -\left( \omega _{0}^{-1}-\frac{%
\Omega \left( \theta ,\theta _{0},Z\right) }{\omega _{0}^{2}\left( Z\right) }%
\right) \right) \left( \Psi _{0}\left( \theta ,Z\right) +\Psi \left( \theta
,Z\right) \right) \right) d\theta _{0}  \notag
\end{eqnarray}%
with:%
\begin{equation*}
\Omega \left( \theta ,\theta _{0},Z\right) =\sum_{i}\omega _{0}^{2}\left(
Z\right) K\left( Z,\theta ,Z_{i},\theta _{0}\right) \left\{ a\left(
Z_{i},\theta _{0}\right) \frac{\omega _{0}^{-1}\left( \theta
_{0},Z_{0}\right) }{\Lambda ^{2}}\right\} 
\end{equation*}%
The expression (\ref{CTN}) includes several contrbtns:

First, the expansion of:%
\begin{equation*}
\frac{1}{2}\left( \Psi _{0}^{\dagger }\left( \theta ,Z\right) +\Psi
^{\dagger }\left( \theta ,Z\right) \right) \nabla \left( \frac{\sigma
_{\theta }^{2}}{2}\nabla -\omega _{0}^{-1}\right) \left( \Psi _{0}\left(
\theta ,Z\right) +\Psi \left( \theta ,Z\right) \right)
\end{equation*}%
around $\Psi _{0}\left( \theta ,Z\right) $ includes the terms:%
\begin{eqnarray*}
&&\frac{1}{2}\Psi ^{\dagger }\left( \theta ,Z\right) \nabla \left( \frac{%
\sigma _{\theta }^{2}}{2}\nabla -\omega _{0}^{-1}\right) \Psi \left( \theta
,Z\right) \\
&&+\frac{1}{2}\Psi _{0}^{\dagger }\left( \theta ,Z\right) \nabla \left( 
\frac{\sigma _{\theta }^{2}}{2}\nabla -\omega _{0}^{-1}\right) \Psi \left(
\theta ,Z\right) +\frac{1}{2}\left( \Psi ^{\dagger }\left( \theta ,Z\right)
\right) \nabla \left( \frac{\sigma _{\theta }^{2}}{2}\nabla -\omega
_{0}^{-1}\right) \left( \Psi _{0}\left( \theta ,Z\right) \right)
\end{eqnarray*}%
The first one is the free action for $\Psi \left( \theta ,Z\right) $, while
the two other terms compute the modifications of the effective action due to
fluctuations $\Psi \left( \theta ,Z\right) $. They contribute to the
effective action above the classical approximation.

The last terms:%
\begin{equation*}
\left( \Psi _{0}^{\dagger }\left( \theta ,Z\right) +\Psi ^{\dagger }\left(
\theta ,Z\right) \right) \nabla _{\theta }\left( \frac{\Omega \left( \theta
,\theta _{0},Z\right) }{\omega _{0}^{2}\left( Z\right) }\left( \Psi
_{0}\left( \theta ,Z\right) +\Psi \left( \theta ,Z\right) \right) \right)
\end{equation*}%
encompass the corrections due to the externl perturbations.

To find these corrections, we will compute the expansion of:%
\begin{equation*}
\exp \left( \int \left( \Psi _{0}^{\dagger }\left( \theta ,Z\right) +\Psi
^{\dagger }\left( \theta ,Z\right) \right) \nabla _{\theta }\left( \frac{%
\Omega \left( \theta ,\theta _{0},Z\right) }{\omega _{0}^{2}\left( Z\right) }%
\left( \Psi _{0}\left( \theta ,Z\right) +\Psi \left( \theta ,Z\right)
\right) \right) \right)
\end{equation*}%
in the pth ntgrl. We consider the second order expansion to find the first
corrections to activities. The field contractions are obtained in the local
approximation:%
\begin{equation*}
\overbrace{\Psi ^{\dagger }\left( \theta ,Z\right) \Psi \left( \theta
^{\prime },Z\right) }\rightarrow \frac{1}{\Lambda }\overbrace{\Psi ^{\dagger
}\left( \theta ,Z\right) \nabla _{\theta }\Psi \left( \theta ,Z\right) }=%
\frac{1}{\Lambda \Lambda _{1}}
\end{equation*}%
\begin{equation*}
\overbrace{\Psi ^{\dagger }\left( \theta ,Z\right) \nabla _{\theta }\Psi
\left( \theta ^{\prime },Z\right) }\rightarrow \frac{1}{\Lambda }\overbrace{%
\Psi ^{\dagger }\left( \theta ,Z\right) \nabla _{\theta }\Psi \left( \theta
,Z\right) }=\frac{\Lambda _{1}}{\Lambda \Lambda _{1}}=\frac{1}{\Lambda }
\end{equation*}%
\begin{equation*}
\overbrace{\Psi _{0}^{\dagger }\left( \theta ,Z\right) \Psi \left( \theta
^{\prime },Z\right) }=\overbrace{\Psi ^{\dagger }\left( \theta ,Z\right)
\Psi _{0}\left( \theta ^{\prime },Z\right) }=0
\end{equation*}%
The first order of the expansion has the contributions:%
\begin{equation*}
\frac{\Psi _{0}^{\dagger }\left( \theta ,Z\right) }{\omega _{0}^{2}\left(
Z\right) }\nabla _{\theta }\left( \frac{\Omega \left( \theta ,\theta
_{0},Z\right) }{\omega _{0}^{2}\left( Z\right) }\right) \Psi _{0}^{\dagger
}\left( \theta ,Z\right) +\frac{\Lambda _{1}}{\Lambda \omega _{0}^{2}\left(
Z\right) }\left( \frac{\Omega \left( \theta ,\theta _{0},Z\right) }{\omega
_{0}^{2}\left( Z\right) }\right) +\frac{1}{\Lambda \omega _{0}^{2}\left(
Z\right) }\nabla _{\theta }\left( \frac{\Omega \left( \theta ,\theta
_{0},Z\right) }{\omega _{0}^{2}\left( Z\right) }\right)
\end{equation*}%
that are equal to $0$ due to the integral over $\theta _{0}$. For
oscillating signals, this integral is equal to $0$.

Second order in the local approximation :%
\begin{eqnarray*}
&&\left( \int \frac{\Psi _{0}^{\dagger }\left( \theta ,Z\right) }{\omega
_{0}^{4}\left( Z\right) }\left( \nabla \Omega \right) \Psi _{0}\left( \theta
,Z\right) dZ\right) ^{2} \\
&&+\frac{2}{\omega _{0}^{4}\left( Z\right) }\left( \int \Psi _{0}^{\dagger
}\left( \theta ,Z\right) \nabla \Omega \overbrace{\Psi \left( \theta
,Z\right) \int \Psi ^{\dagger }\left( \theta ^{\prime },Z\right) }\nabla
\Omega \Psi _{0}\left( \theta ^{\prime },Z\right) dZ\right) \\
&&+\frac{1}{\omega _{0}^{4}\left( Z\right) }\left( \int \Psi ^{\dagger
}\left( \theta ,Z\right) \nabla \Omega \overbrace{\Psi \left( \theta
,Z\right) \int \Psi ^{\dagger }\left( \theta ^{\prime },Z\right) }\nabla
\Omega \Psi \left( \theta ^{\prime },Z\right) dZ\right)
\end{eqnarray*}%
Considering $\left\vert \Psi _{0}\left( \theta ,Z\right) \right\vert ^{2}>>%
\frac{1}{\Lambda }$, this leads to: 
\begin{eqnarray*}
&&\left( \int \frac{\Psi _{0}^{\dagger }\left( \theta ,Z\right) }{\omega
_{0}^{4}\left( Z\right) }\left( \nabla \Omega \right) \Psi _{0}\left( \theta
,Z\right) dZ\right) ^{2} \\
&&+\frac{2}{\Lambda _{1}\Lambda \omega _{0}^{4}\left( Z\right) }\left( \int
\Psi _{0}^{\dagger }\left( \theta ,Z\right) \left( \nabla \Omega \right)
\nabla \Omega \Psi _{0}\left( \theta ,Z\right) dZ\right) -\frac{2}{\Lambda
\omega _{0}^{4}\left( Z\right) }\left( \int \Psi _{0}^{\dagger }\left(
\theta ,Z\right) \Omega \nabla \Omega \Psi _{0}\left( \theta ,Z\right)
dZ\right) \\
&=&\left( \int \frac{\Psi _{0}^{\dagger }\left( \theta ,Z\right) }{\omega
_{0}^{4}\left( Z\right) }\left( \nabla \Omega \right) \Psi _{0}\left( \theta
,Z\right) dZ\right) ^{2} \\
&&+\frac{2}{\Lambda _{1}\Lambda \omega _{0}^{4}\left( Z\right) }\left( \int
\left( \Psi _{0}^{\dagger }\left( \theta ,Z\right) \nabla \left( \left(
\int^{\theta }\left( \nabla \Omega \right) ^{2}\right) \Psi _{0}\left(
\theta ,Z\right) \right) +O\left( \nabla \Psi _{0}\left( \theta ,Z\right)
\right) \right) dZ\right) \\
&&-\frac{1}{\Lambda \omega _{0}^{4}\left( Z\right) }\left( \int \left( \Psi
_{0}^{\dagger }\left( \theta ,Z\right) \nabla \Omega ^{2}\Psi _{0}\left(
\theta ,Z\right) +O\left( \nabla \Psi _{0}\left( \theta ,Z\right) \right)
\right) dZ\right) \\
&\simeq &\left( \int \frac{\Psi _{0}^{\dagger }\left( \theta ,Z\right) }{%
\omega _{0}^{4}\left( Z\right) }\left( \nabla \Omega \right) \Psi _{0}\left(
\theta ,Z\right) dZ\right) ^{2}+\frac{1}{\Lambda _{1}\Lambda \omega
_{0}^{4}\left( Z\right) }\int \left( \Psi _{0}^{\dagger }\left( \theta
,Z\right) \nabla \left( 2\left( \left( \int^{\theta }\left( \nabla \Omega
\right) ^{2}\right) -\Lambda _{1}\Omega ^{2}\Psi _{0}\left( \theta ,Z\right)
\right) \right) \right) dZ
\end{eqnarray*}

first contribtion in frst prxm when integration over $\theta _{0}$:%
\begin{equation*}
B=\left( \int \frac{\Psi _{0}^{\dagger }\left( \theta ,Z\right) }{\omega
_{0}^{4}\left( Z\right) }\sqrt{\int \left( \nabla \Omega \left( \theta
,\theta _{0},Z\right) \right) ^{2}d\theta _{0}}\Psi _{0}\left( \theta
,Z\right) dZ\right) ^{2}
\end{equation*}

second contribution:%
\begin{equation}
A=\frac{1}{\Lambda _{1}\Lambda \omega _{0}^{4}\left( Z\right) }\int \left(
\Psi _{0}^{\dagger }\left( \theta ,Z\right) \nabla \left( 2\left( \left(
\int^{\theta }\int \left( \nabla \Omega \left( \theta ,\theta _{0},Z\right)
\right) ^{2}d\theta _{0}\right) -\Lambda _{1}\left( \int \Omega ^{2}d\theta
_{0}\right) \Psi _{0}\left( \theta ,Z\right) \right) \right) \right) dZ
\label{FR}
\end{equation}

The contributions $A$ and $B$ obtained can be gathered in an exponential and
lead in first approximation to a term:%
\begin{equation*}
\exp \left( A+B-\frac{1}{2}A^{2}\right)
\end{equation*}

The term $B-\frac{1}{2}A^{2}$ is a correction to the potential.

This implies a correction to activities:%
\begin{equation*}
\omega _{0}^{-1}\left( Z\right) -\frac{A\omega _{0}^{2}\left( Z\right) }{%
\omega _{0}^{2}\left( Z\right) }
\end{equation*}%
Using (\ref{FR}), this leads to a modification:%
\begin{equation*}
\omega _{0}\left( Z\right) \rightarrow \omega _{0}\left( Z\right) +A\omega
_{0}^{2}\left( Z\right)
\end{equation*}%
which is: 
\begin{equation*}
\omega _{0}\left( Z\right) \rightarrow \omega _{0}\left( Z\right) +\frac{1}{%
\Lambda _{1}\Lambda \omega _{0}^{2}\left( Z\right) }\left( 2\left( \left(
\int^{\theta }\int \left( \nabla \Omega \left( \theta ,\theta _{0},Z\right)
\right) ^{2}d\theta _{0}\right) -\Lambda _{1}\left( \int \Omega ^{2}d\theta
_{0}\right) \right) \right)
\end{equation*}

As explained above, the corrections $\Omega \left( \theta ,\theta
_{0},Z\right) $ can be considered as nul outside the points of maximal
interferences. At these points $\Omega \left( \theta ,\theta _{0},Z\right) $
is proportional to: 
\begin{equation*}
\bar{\Omega}=\frac{a\exp \left( -\left\vert Z-Z_{0}\right\vert \right) }{c%
\sqrt{\left( 1+2\alpha \left\vert Z-Z_{0}\right\vert \right) ^{2}+\left( 
\frac{\varpi }{c}\right) ^{2}}}
\end{equation*}%
and the correction to drequencies are:%
\begin{eqnarray*}
\omega _{0}\left( Z\right) &\rightarrow &\omega _{0}\left( Z\right) +\frac{%
2\left( \left( \int^{\theta }\int \left( \varpi \bar{\Omega}\right)
^{2}d\theta _{0}\right) -\Lambda _{1}\left( \int \bar{\Omega}^{2}d\theta
_{0}\right) \right) }{\Lambda _{1}\Lambda \omega _{0}^{2}\left( Z\right) } \\
&\simeq &\omega _{0}\left( Z\right) +\frac{2\left( \left( T_{\theta }\left(
\varpi \bar{\Omega}\right) ^{2}\right) -\Lambda _{1}\left( \bar{\Omega}%
^{2}\right) \right) T_{\theta }}{\Lambda _{1}\Lambda \omega _{0}^{2}\left(
Z\right) }
\end{eqnarray*}%
where $T_{\theta }$ is the duration of the signals at time $\theta $.

As a consequence, in presence of the signals, the states is transformed from
the background activities to modification of magnitude:%
\begin{equation*}
\Delta \omega _{0}=\frac{2\left( \left( T_{\theta }\left( \varpi \bar{\Omega}%
\right) ^{2}\right) -\Lambda _{1}\left( \bar{\Omega}^{2}\right) \right)
T_{\theta }}{\Lambda _{1}\Lambda \omega _{0}^{2}\left( Z\right) }
\end{equation*}%
at points of additive interferences, and $0$ elsewhere. The shift may be
positive or negative depending on the parameters of the system.

In the sequel, we consider $T_{\theta }\simeq T$.

\subsection*{3.3 Extension: Excitatory vs inhibitory interaction}

The previous results computing the perturbations in activities may be
extended straightforwarly in the case of two types of interactions. We
consider $n$ populations, each caracterized by their activities $i=1,...,n$.
They interact positively or negatively. Each population is defined by a
field $\Psi _{i}$ and activities $\omega _{i}\left( \theta ,Z\right) $. The
details are given in part 1. Equations for activities are defined by:

\begin{eqnarray}
\omega _{i}^{-1}\left( \theta ,Z\right) &=&G_{i}\left( J\left( \theta
\right) +\frac{\kappa }{N}\int T\left( Z,Z_{1}\right) \frac{\omega
_{j}\left( \theta -\frac{\left\vert Z-Z_{1}\right\vert }{c},Z_{1}\right) }{%
\omega _{i}\left( \theta ,Z\right) }G^{ij}\right.  \label{fml} \\
&&\times \left. W\left( \frac{\omega _{i}\left( \theta ,Z\right) }{\omega
_{j}\left( \theta -\frac{\left\vert Z-Z_{1}\right\vert }{c},Z_{1}\right) }%
\right) \left( \mathcal{\bar{G}}_{0j}\left( 0,Z_{1}\right) +\left\vert \Psi
_{j}\left( \theta -\frac{\left\vert Z-Z_{1}\right\vert }{c},Z_{1}\right)
\right\vert ^{2}\right) dZ_{1}\right)  \notag
\end{eqnarray}

For example, if $i,j=1,2$, a matrix $g$ of the form:%
\begin{equation*}
G=\left( 
\begin{array}{cc}
1 & -g \\ 
-g & 1%
\end{array}%
\right)
\end{equation*}%
represents inhibitory interactions between the two populations. More
generally, the matrix $G$ is $n\times n$ with coefficients in the interval $%
\left[ -1,1\right] $. The sum over indices is understood for $j$. The
resolution of (\ref{fml}) follows the same principle as for (\ref{qtf}),
with a vector of activities. The expansion of the first order derivative is:%
\begin{eqnarray*}
\left( \frac{\delta \omega ^{-1}\left( J,\theta ,Z\right) }{\delta
\left\vert \Psi \left( \theta -l_{1},Z_{1}\right) \right\vert ^{2}}\right)
_{\left\vert \Psi \right\vert ^{2}=0} &=&-\sum_{n=1}^{\infty }\int
\dprod\limits_{l=1}^{n}\check{T}\left( \theta -\sum_{j=1}^{l-1}\frac{%
\left\vert Z^{\left( j-1\right) }-Z^{\left( j\right) }\right\vert }{c}%
,Z^{\left( l-1\right) },Z^{\left( l\right) },\omega _{0},0\right) \\
&&\times \Omega _{0}\left( J,\theta -\sum_{l=1}^{n}\frac{\left\vert
Z^{\left( l-1\right) }-Z^{\left( l\right) }\right\vert }{c},Z_{1}\right)
\times \delta \left( l_{1}-\sum_{l=1}^{n}\frac{\left\vert Z^{\left(
l-1\right) }-Z^{\left( l\right) }\right\vert }{c}\right)
\dprod\limits_{l=1}^{n-1}dZ^{\left( l\right) }
\end{eqnarray*}%
with $\omega _{0}$ a $n$ component vector describing a solution for $%
\left\vert \Psi \right\vert ^{2}=0$. The matrices $\Omega _{0}^{-1}\left(
J,\theta -\sum_{l=1}^{n}\frac{\left\vert Z^{\left( l-1\right) }-Z^{\left(
l\right) }\right\vert }{c},Z_{1}\right) $ and $D\left( \left\vert \Psi
\right\vert ^{2}\right) $ are diagonal with components $\omega
_{0i}^{-1}\left( J,\theta -\sum_{l=1}^{n}\frac{\left\vert Z^{\left(
l-1\right) }-Z^{\left( l\right) }\right\vert }{c},Z_{1}\right) $ and $%
\left\vert \Psi _{i}\right\vert ^{2}$\ respectively on the diagonal. More
generally, for any expression $H\left( \omega _{0i},\left\vert \Psi
_{i}\right\vert ^{2}\right) $, we define $D\left( H\left( \omega
_{0},\left\vert \Psi \right\vert ^{2}\right) \right) $ the diagonal matrix
with components $H\left( \omega _{0i},\left\vert \Psi _{i}\right\vert
^{2}\right) $.

The quantity $\Omega \left( J,\theta ,Z\right) $ $\left\vert \Psi
\right\vert ^{2}$ is a vector with components $\omega _{i}\left( J,\theta
,Z\right) $ $\left\vert \Psi _{i}\right\vert ^{2}$. The expressions $\left( 
\frac{\delta \omega ^{-1}\left( J,\theta ,Z\right) }{\delta \left\vert \Psi
\left( \theta -l_{1},Z_{1}\right) \right\vert ^{2}}\right) _{\left\vert \Psi
\right\vert ^{2}=0}$ and $\check{T}\left( \theta -\sum_{j=1}^{l-1}\frac{%
\left\vert Z^{\left( j-1\right) }-Z^{\left( j\right) }\right\vert }{c}%
,Z^{\left( l-1\right) },Z^{\left( l\right) },\omega _{0},0\right) $ are $%
n\times n$ matrices:%
\begin{equation*}
\left( \left( \frac{\delta \omega ^{-1}\left( J,\theta ,Z\right) }{\delta
\left\vert \Psi \left( \theta -l_{1},Z_{1}\right) \right\vert ^{2}}\right)
_{\left\vert \Psi \right\vert ^{2}=0}\right) _{ij}=\left( \frac{\delta
\omega _{i}^{-1}\left( J,\theta ,Z\right) }{\delta \left\vert \Psi
_{j}\left( \theta -l_{1},Z_{1}\right) \right\vert ^{2}}\right) _{\left\vert
\Psi \right\vert ^{2}=0}
\end{equation*}%
and:%
\begin{eqnarray*}
&&\check{T}_{ij}\left( \theta ,Z,Z_{1}\omega ,\Psi \right)  \\
&=&-\frac{G^{ij}\frac{\kappa }{N}\omega _{i}\left( J,\theta ,Z\right)
T\left( Z,Z_{1}\right) G_{i}^{\prime }\left[ J,\omega ,\theta ,Z,\Psi \right]
}{1+G^{ij}\left( \int \frac{\kappa }{N}\omega _{j}\left( J,\theta -\frac{%
\left\vert Z-Z^{\prime }\right\vert }{c},Z^{\prime }\right) \left( \mathcal{%
\bar{G}}_{0j}\left( 0,Z^{\prime }\right) +\left\vert \Psi _{j}\left( \theta -%
\frac{\left\vert Z-Z^{\prime }\right\vert }{c},Z^{\prime }\right)
\right\vert ^{2}\right) T\left( Z,Z^{\prime }\right) dZ^{\prime }\right)
G_{i}^{\prime }\left[ J,\omega ,\theta ,Z,\Psi \right] }
\end{eqnarray*}%
The factor associated to the sum of single lines (\ref{rl}) crossing the
points $Z_{k}$ generalizes straightforwardly and is given by:%
\begin{eqnarray}
&&-\check{T}\left( 1-\check{T}\right) ^{-1}\dprod\limits_{l=1}^{n-1}\left\{
\left( D\left( \left\vert \Psi \left( \theta -l_{l},Z_{l}\right) \right\vert
^{2}\right) dZ_{l}dl_{l}\right) \check{T}\left( 1-\check{T}\right)
^{-1}\right\} D\left( \left\vert \Psi \left( \theta -l_{n},Z_{n}\right)
\right\vert ^{2}\omega _{0}^{-1}\left( J,\theta -l_{n},Z_{n}\right) \right) 
\notag \\
&=&-\check{T}\left( 1-\check{T}\right) ^{-1}\frac{1}{1-D\left( \left\vert
\Psi \left( \theta ,Z\right) \right\vert ^{2}\right) \check{T}\left( 1-%
\check{T}\right) ^{-1}}D\left( \left\vert \Psi \left( \theta
-l_{n},Z_{n}\right) \right\vert ^{2}\omega _{0}^{-1}\left( J,\theta
-l_{n},Z_{n}\right) \right)   \notag \\
&=&-\check{T}\frac{1}{1-\left( 1+D\left( \left\vert \Psi \right\vert
^{2}\right) \right) \check{T}}D\left( \left\vert \Psi \left( \theta
-l_{n},Z_{n}\right) \right\vert ^{2}\omega _{0}^{-1}\left( J,\theta
-l_{n},Z_{n}\right) \right) 
\end{eqnarray}%
Then, we compute $\omega ^{-1}\left( J,\theta ,Z\right) $ by writing the
action for an auxiliary field which is the same as in appendix 6.2:%
\begin{eqnarray*}
S\left( \digamma \right)  &=&\int \digamma \left( Z,\theta \right) \left(
1-D\left( \left\vert \Psi \right\vert ^{2}\right) \check{T}\right) \digamma
^{\dag }\left( Z,\theta \right) d\left( Z,\theta \right)  \\
&&-\int \digamma \left( Z,\theta \right) \check{T}\left( \theta -\frac{%
\left\vert Z^{\left( 1\right) }-Z\right\vert }{c},Z^{\left( 1\right)
},Z,\omega _{0}+\check{T}\digamma ^{\dag }\right) \digamma ^{\dag }\left(
Z^{\left( 1\right) },\theta -\frac{\left\vert Z-Z^{\left( 1\right)
}\right\vert }{c}\right) dZdZ^{\left( 1\right) }d\theta ^{\left( 1\right) }
\end{eqnarray*}%
where $\digamma \left( Z,\theta \right) $ is a $n$ components vector, and $%
\digamma ^{\dag }\left( Z,\theta \right) $ is the hermitian conjugate. The
activity vector is thus given by the integral:%
\begin{equation*}
\omega ^{-1}\left( J,\theta ,Z\right) =\omega _{0}^{-1}\left( J,\theta
,Z\right) +\frac{\int \check{T}\digamma ^{\dag }\left( Z,\theta \right) \exp
\left( -S\left( \digamma \right) -\int \digamma \left( X,\theta \right)
D\left( \left\vert \Psi \right\vert ^{2}\omega _{0}^{-1}\left( J,\theta
,Z\right) \right) d\left( X,\theta \right) \right) \mathcal{D}\digamma }{%
\exp \left( -S\left( \digamma \right) \right) \mathcal{D}\digamma }
\end{equation*}%
where $\digamma ^{\dag }$ satisfies in the saddle point approximation: 
\begin{equation*}
\left( \left( 1-D\left( \left\vert \Psi \right\vert ^{2}\right) \check{T}%
\right) \digamma ^{\dag }\right) \left( Z,\theta \right) -\left( \check{T}%
_{\omega _{0}+\check{T}\left( \omega _{0}\left\vert \Psi \right\vert
^{2}\right) }\digamma ^{\dag }\right) \left( Z,\theta \right) -D\left(
\left\vert \Psi \right\vert ^{2}\omega _{0}\right) =0
\end{equation*}%
In first approximation:%
\begin{equation*}
\check{T}_{\omega _{0}+\check{T}\left( \omega _{0}\left\vert \Psi
\right\vert ^{2}\right) }\simeq D\left( \frac{\omega \left( J,\theta
,Z\right) }{\omega \left( J,\theta ,Z\right) +\check{T}\left( \left(
\left\vert \Psi \right\vert ^{2}\omega _{0}\right) \right) }\right) \check{T}%
_{\omega _{0}}
\end{equation*}%
and the previous equation becomes:%
\begin{equation}
\left( \left( 1-D\left( \left\vert \Psi \right\vert ^{2}\right) \check{T}%
\right) \digamma ^{\dag }\right) \left( Z,\theta \right) -D\left( \frac{%
\omega _{0}}{\omega _{0}+\check{T}\digamma ^{\dag }}\right) \check{T}%
\digamma ^{\dag }\left( Z,\theta \right) -D\left( \omega _{0}\right)
\left\vert \Psi \right\vert ^{2}\simeq 0  \label{vsr}
\end{equation}%
As for the basic case, under the saddle point approximation:%
\begin{equation*}
\omega \left( J,\theta ,Z\right) =\omega _{0}\left( J,\theta ,Z\right) +%
\check{T}\digamma ^{\dag }\left( Z,\theta \right) 
\end{equation*}%
Equation (\ref{vsr}) can be solved recursively. As in the one component
field case, we find in first approximation:%
\begin{eqnarray}
\check{T}\digamma ^{\dag } &=&A\frac{1}{1-\left( \check{T}_{\omega
_{0}+A\omega _{0}\left\vert \Psi \right\vert ^{2}}-\check{T}\right) \check{T}%
^{-1}A}\omega _{0}\left\vert \Psi \right\vert ^{2}  \label{qvn} \\
&\simeq &A\frac{1}{1-D\left( \frac{\omega _{0}}{\omega _{0}+A\omega
_{0}\left\vert \Psi \right\vert ^{2}}\right) A}\omega _{0}\left\vert \Psi
\right\vert ^{2}  \notag
\end{eqnarray}%
with:%
\begin{eqnarray*}
A &=&\frac{\check{T}}{1-\left( 1+D\left( \left\vert \Psi \right\vert
^{2}\right) \right) \check{T}}=\frac{1}{\left( 1+D\left( \left\vert \Psi
\right\vert ^{2}\right) \right) }\frac{\left( 1+D\left( \left\vert \Psi
\right\vert ^{2}\right) \right) \check{T}}{1-\left( 1+D\left( \left\vert
\Psi \right\vert ^{2}\right) \right) \check{T}} \\
&=&\frac{\check{T}}{1-\check{T}-D\left( \left\vert \Psi \right\vert
^{2}\right) \check{T}} \\
&=&\frac{\check{T}}{1-\check{T}}\sum_{n\geqslant 0}\left( D\left( \left\vert
\Psi \right\vert ^{2}\right) \frac{\check{T}}{1-\check{T}}\right) ^{n}
\end{eqnarray*}%
and the generalization of (\ref{rl}) is obtained by diagonalization of $%
\check{T}$. For two fields, we write: 
\begin{equation*}
\left( 1+D\left( \left\vert \Psi \right\vert ^{2}\right) \right) \check{T}%
=\left( 
\begin{array}{cc}
\check{T}_{1}\left( \left( 1+\left( \left\vert \Psi _{1}\right\vert
^{2}\right) \right) \omega _{01}\right)  & -g\check{T}_{2}\left( \left(
1+\left( \left\vert \Psi _{2}\right\vert ^{2}\right) \right) \omega
_{02}\right)  \\ 
-g\check{T}_{1}\left( \left( 1+\left( \left\vert \Psi _{1}\right\vert
^{2}\right) \right) \omega _{01}\right)  & \check{T}_{2}\left( \left(
1+\left( \left\vert \Psi _{2}\right\vert ^{2}\right) \right) \omega
_{02}\right) 
\end{array}%
\right) 
\end{equation*}%
Assuming $\omega _{01}$ and $\omega _{02}$ changing slowly in time, we have:%
\begin{equation*}
\left( 1+D\left( \left\vert \Psi \right\vert ^{2}\right) \right) \check{T}=U%
\check{T}_{D}U^{-1}
\end{equation*}%
$\allowbreak \allowbreak $%
\begin{eqnarray*}
\check{T}_{D} &=&\left( 
\begin{array}{cc}
\frac{1}{2}\left( \check{T}_{1}+\check{T}_{2}-\sqrt{4g^{2}\check{T}_{1}%
\check{T}_{2}+\left( \check{T}_{1}-\check{T}_{2}\right) ^{2}}\right)  & 0 \\ 
0 & \frac{1}{2}\left( \check{T}_{1}+\check{T}_{2}+\sqrt{4g^{2}\check{T}_{1}%
\check{T}_{2}+\left( \check{T}_{1}-\check{T}_{2}\right) ^{2}}\right) 
\end{array}%
\right)  \\
U &=&\left( 
\begin{array}{cc}
-\frac{1}{2g}\left( \check{T}_{1}-\check{T}_{2}-\sqrt{4g^{2}\check{T}_{1}%
\check{T}_{2}+\left( \check{T}_{1}-\check{T}_{2}\right) ^{2}}\right)  & 
\check{T}_{2} \\ 
\check{T}_{1} & \frac{1}{2g}\left( \check{T}_{1}-\check{T}_{2}-\sqrt{4g^{2}%
\check{T}_{1}\check{T}_{2}+\left( \check{T}_{1}-\check{T}_{2}\right) ^{2}}%
\right) 
\end{array}%
\right) 
\end{eqnarray*}%
As a consequence:%
\begin{equation*}
\check{T}=UD\left( \frac{\exp \left( -cl_{1}-\alpha \left( \check{T}%
_{D}\right) \left( \left( cl_{1}\right) ^{2}-\left\vert Z-Z_{1}\right\vert
^{2}\right) \right) }{B\left( \check{T}_{D}\right) }H\left(
cl_{1}-\left\vert Z-Z_{1}\right\vert \right) \right) U^{-1}
\end{equation*}%
with $\alpha \left( \check{T}\right) $ and $B\left( \check{T}\right) $ are
vectors. That is, given our conventions:%
\begin{equation*}
\check{T}=U\left( 
\begin{array}{cc}
\frac{\exp \left( -cl_{1}-\alpha _{1}\left( \check{T}_{D}\right) \left(
\left( cl_{1}\right) ^{2}-\left\vert Z-Z_{1}\right\vert ^{2}\right) \right) 
}{B_{1}\left( \check{T}\right) } & 0 \\ 
0 & \frac{\exp \left( -cl_{1}-\alpha _{2}\left( \check{T}_{D}\right) \left(
\left( cl_{1}\right) ^{2}-\left\vert Z-Z_{1}\right\vert ^{2}\right) \right) 
}{B_{2}\left( \check{T}\right) }%
\end{array}%
\right) U^{-1}H\left( cl_{1}-\left\vert Z-Z_{1}\right\vert \right) 
\end{equation*}%
For connectivity functions $T_{i}\left( Z,Z_{1}\right) $ that are
proportional $T_{i}\left( Z,Z_{1}\right) =C_{i}T_{0}\left( Z,Z_{1}\right) $,
the change of basis yields the diagonalized connectivity function:%
\begin{equation*}
T_{D}\left( Z,Z_{1}\right) =\left( 
\begin{array}{cc}
\frac{1}{2}\left( C_{1}+C_{2}-\sqrt{4g_{1}^{2}C_{1}C_{2}+\left(
C_{1}-C_{2}\right) ^{2}}\right)  & 0 \\ 
0 & \frac{1}{2}\left( C_{1}+C_{2}+\sqrt{4g_{1}^{2}C_{1}C_{2}+\left(
C_{1}-C_{2}\right) ^{2}}\right) 
\end{array}%
\right) T_{0}\left( Z,Z_{1}\right) 
\end{equation*}

Appendix 5.2 shows that $\alpha _{i}\left( \check{T}\right) $ and $%
B_{i}\left( \check{T}\right) $ are proportional to the averages of $\check{T}%
_{iD}$ and $1+\check{T}_{iD}$, more precisely:%
\begin{eqnarray*}
D\left( \alpha \left( \check{T}\right) \right) &\propto &\left( 
\begin{array}{cc}
\frac{1}{2}\left( C_{1}+C_{2}-\sqrt{4g_{1}^{2}C_{1}C_{2}+\left(
C_{1}-C_{2}\right) ^{2}}\right) & 0 \\ 
0 & \frac{1}{2}\left( C_{1}+C_{2}+\sqrt{4g_{1}^{2}C_{1}C_{2}+\left(
C_{1}-C_{2}\right) ^{2}}\right)%
\end{array}%
\right) \\
D\left( B\left( \check{T}\right) \right) &\propto &\left( 
\begin{array}{cc}
\frac{1}{2}\left( C_{1}+C_{2}-\sqrt{4g_{1}^{2}C_{1}C_{2}+\left(
C_{1}-C_{2}\right) ^{2}}\right) & 0 \\ 
0 & \frac{1}{2}\left( C_{1}+C_{2}+\sqrt{4g_{1}^{2}C_{1}C_{2}+\left(
C_{1}-C_{2}\right) ^{2}}\right)%
\end{array}%
\right)
\end{eqnarray*}

As a consequence, by multiplication with $U$ and $U^{-1}$, we find that:%
\begin{equation*}
\frac{\left( 1+D\left( \left\vert \Psi \right\vert ^{2}\right) \right) 
\check{T}}{1-\left( 1+D\left( \left\vert \Psi \right\vert ^{2}\right)
\right) \check{T}}=\frac{\exp \left( -cl_{1}-\left( 1+D\left( \left\langle
\left\vert \Psi \right\vert ^{2}\right\rangle \right) \right) \Phi \left(
\left( cl_{1}\right) ^{2}-\left\vert Z-Z_{1}\right\vert ^{2}\right) \right) 
}{B}H\left( cl_{1}-\left\vert Z-Z_{1}\right\vert \right)
\end{equation*}%
with: 
\begin{eqnarray*}
\Phi &=&\left( 
\begin{array}{cc}
C_{1} & -gC_{2} \\ 
-gC_{1} & C_{2}%
\end{array}%
\right) \\
B &=&1+2\pi \left( 1+D\left( \left\langle \left\vert \Psi \right\vert
^{2}\right\rangle \right) \right) \Lambda
\end{eqnarray*}%
where the constants $C_{1}$ and $C_{2}$ are as in Appendix 5 to define $%
\check{T}_{1}$ and $\check{T}_{2}$.

Then, the modification to activities to the lowest order writes:%
\begin{eqnarray}
\omega ^{-1}\left( J,\theta ,Z\right) &=&\omega _{0}^{-1}\left( J,\theta
,Z\right) +\hat{T}\Lambda ^{\dag }\left( Z,\theta \right)  \label{TC} \\
&=&\int K\left( Z,\theta ,Z_{i},\theta _{i}\right) \left\{ -\sum_{i}a\left(
Z_{i},\theta _{i}\right) \frac{D\left( \omega _{0}^{-1}\left( J,\theta
_{i},Z_{i}\right) \right) }{\Lambda ^{2}}\right\} d\theta _{i}  \notag \\
&=&\int \frac{\exp \left( -cl_{i}-\left( 1+D\left( \left\langle \left\vert
\Psi \right\vert ^{2}\right\rangle \right) \right) \Phi \left( \left(
cl_{i}\right) ^{2}-\left\vert Z-Z_{i}\right\vert ^{2}\right) \right) }{B}%
\left\{ -\sum_{i}a\left( Z_{i},\theta _{i}\right) \frac{D\left( \omega
_{0}^{-1}\left( J,\theta _{i},Z_{i}\right) \right) }{\Lambda ^{2}}\right\}
d\theta _{i}  \notag
\end{eqnarray}%
The phenomenom of interferences will occur, but will be mitigated by the
intertwining of inhibitory and enhancing interactions that are encompassed
in matrix $\Phi $.

Remark ultimately that the previous results generalizes to a system with $n$
interacting components, and an analogous to (\ref{TC}) holds. If we look to
higher expansion, we consider the expansion of $A\left( \left( Z,\theta
\right) ,\left( Z_{1},\theta -l_{1}\right) \right) $:

\begin{eqnarray}
A\left( \left( Z,\theta \right) ,\left( Z_{1},\theta -l_{1}\right) \right)
&\simeq &D\left( \frac{1}{\left( 1+D\left( \left\langle \left\vert \Psi
\right\vert ^{2}\right\rangle \right) \right) }\right)  \label{rfp} \\
&&\times \left( \frac{\exp \left( -cl_{1}-\left( 1+D\left( \left\langle
\left\vert \Psi \right\vert ^{2}\right\rangle \right) \right) \Lambda \left(
\left( cl_{1}\right) ^{2}-\left\vert Z-Z_{1}\right\vert ^{2}\right) \right) 
}{B}H\left( cl_{1}-\left\vert Z-Z_{1}\right\vert \right) \right)  \notag
\end{eqnarray}

As aconsequence, the expansion of (\ref{qvn}) is:%
\begin{eqnarray}
\omega \left( Z,\theta \right) &=&\omega _{0}\left( J,\theta ,Z\right) +\int
\sum_{k=0}^{\infty }\dprod\limits_{i=0}^{k-1}\frac{\exp \left(
-cl_{i}-\left( 1+D\left( \left\langle \left\vert \Psi \right\vert
^{2}\right\rangle \right) \right) \Lambda \left( \left( cl_{i}\right) ^{2}-%
\frac{\left\vert Z_{i}-Z_{i+1}\right\vert }{c}\right) \right) }{B}  \notag \\
&&\times D\left( \frac{\omega _{0}\left( \theta -l_{i},Z_{i}\right) }{\omega
_{0}\left( \theta -l_{i},Z_{i}\right) +A\omega _{0}\left\vert \Psi
\right\vert ^{2}\left( \theta -l_{i},Z_{i}\right) }\frac{\omega _{0}\left(
J,\theta -l_{k},Z_{k}\right) }{\left( 1+D\left( \left\langle \left\vert \Psi
\right\vert ^{2}\right\rangle \right) \right) }\right)  \notag \\
&&\times \frac{\exp \left( -cl_{k}-\left( 1+D\left( \left\langle \left\vert
\Psi \right\vert ^{2}\right\rangle \right) \right) \Lambda \left( \left(
cl_{i}\right) ^{2}-\frac{\left\vert Z_{k-1}-Z_{k}\right\vert }{c}\right)
\right) }{B}\left\vert \Psi \left( \theta -l_{k},Z_{k}\right) \right\vert
^{2}dZ_{i}dl_{i}
\end{eqnarray}

\section*{Appendix 4 Computation of Green functions}

Given the definition (\ref{DF}) ofoperator $O$, the time-dependent version:%
\begin{equation*}
P_{t}\left( \left( T,\hat{T},\theta ,Z,Z^{\prime },C,D\right) _{i},\left( T,%
\hat{T},\theta ,Z,Z^{\prime },C,D\right) _{f}\right)
\end{equation*}%
of the transition functn defined in (\ref{NL})) satisfies the associated
differential equation:%
\begin{eqnarray}
\frac{\partial }{\partial t}P &=&\nabla _{T}\left( \nabla _{T}+\frac{\left(
T-\left\langle T\right\rangle \right) -\left( \lambda \left( \hat{T}%
-\left\langle \hat{T}\right\rangle \right) \right) }{\tau \omega _{0}\left(
Z\right) +\Delta \omega _{0}\left( Z,\left\vert \Psi \right\vert ^{2}\right) 
}\left\vert \Psi _{0}\left( Z\right) \right\vert ^{2}\right) P  \label{trd}
\\
&&+\nabla _{\hat{T}}\left( \nabla _{\hat{T}}+\rho \left( C\frac{\left\vert
\Psi _{0}\left( Z\right) \right\vert ^{2}h_{C}\left( \omega _{0}\left(
Z\right) +\Delta \omega _{0}\left( Z,\left\vert \Psi \right\vert ^{2}\right)
\right) }{\omega _{0}\left( Z\right) +\Delta \omega _{0}\left( Z,\left\vert
\Psi \right\vert ^{2}\right) }\right. \right.  \notag \\
&&\left. \left. +D\frac{\left\vert \Psi _{0}\left( Z^{\prime }\right)
\right\vert ^{2}h_{D}\left( \omega _{0}\left( Z^{\prime }\right) +\Delta
\omega _{0}\left( Z^{\prime },\left\vert \Psi \right\vert ^{2}\right)
\right) }{\omega _{0}\left( Z\right) +\Delta \omega _{0}\left( Z,\left\vert
\Psi \right\vert ^{2}\right) }\right) \right) \left( \hat{T}-\left\langle 
\hat{T}\right\rangle \right) P  \notag
\end{eqnarray}%
This equation can be writen in the matricial notation:%
\begin{equation}
\frac{\partial }{\partial t}P=\left( \mathbf{\nabla }^{2}+\left( \mathbf{%
\nabla }\right) ^{t}\gamma \mathbf{x}\right) P  \label{trp}
\end{equation}%
with:%
\begin{eqnarray*}
\gamma &=&\left( 
\begin{array}{cc}
\frac{\left\vert \Psi _{0}\left( Z\right) \right\vert ^{2}}{\tau \omega
_{0}\left( Z\right) +\Delta \omega _{0}\left( Z,\left\vert \Psi \right\vert
^{2}\right) } & -\frac{\lambda \left\vert \Psi _{0}\left( Z\right)
\right\vert ^{2}}{\tau \omega _{0}\left( Z\right) +\Delta \omega _{0}\left(
Z,\left\vert \Psi \right\vert ^{2}\right) } \\ 
0 & 
\begin{array}{c}
\rho C\frac{\left\vert \Psi _{0}\left( Z\right) \right\vert ^{2}h_{C}\left(
\omega _{0}\left( Z\right) +\Delta \omega _{0}\left( Z,\left\vert \Psi
\right\vert ^{2}\right) \right) }{\omega _{0}\left( Z\right) +\Delta \omega
_{0}\left( Z,\left\vert \Psi \right\vert ^{2}\right) } \\ 
+\rho D\frac{\left\vert \Psi _{0}\left( Z^{\prime }\right) \right\vert
^{2}h_{D}\left( \omega _{0}\left( Z^{\prime }\right) +\Delta \omega
_{0}\left( Z^{\prime },\left\vert \Psi \right\vert ^{2}\right) \right) }{%
\omega _{0}\left( Z\right) +\Delta \omega _{0}\left( Z,\left\vert \Psi
\right\vert ^{2}\right) }%
\end{array}%
\end{array}%
\right) \\
\mathbf{x} &\mathbf{=}&\left( 
\begin{array}{c}
T-\left\langle T\right\rangle \\ 
\hat{T}-\left\langle \hat{T}\right\rangle%
\end{array}%
\right)
\end{eqnarray*}%
We define the background dependent parameters:%
\begin{eqnarray*}
u &=&\frac{\left\vert \Psi _{0}\left( Z\right) \right\vert ^{2}}{\tau \omega
_{0}\left( Z\right) +\Delta \omega _{0}\left( Z,\left\vert \Psi \right\vert
^{2}\right) } \\
v &=&\rho C\frac{\left\vert \Psi _{0}\left( Z\right) \right\vert
^{2}h_{C}\left( \omega _{0}\left( Z\right) +\Delta \omega _{0}\left(
Z,\left\vert \Psi \right\vert ^{2}\right) \right) }{\omega _{0}\left(
Z\right) +\Delta \omega _{0}\left( Z,\left\vert \Psi \right\vert ^{2}\right) 
}+\rho D\frac{\left\vert \Psi _{0}\left( Z^{\prime }\right) \right\vert
^{2}h_{D}\left( \omega _{0}\left( Z^{\prime }\right) +\Delta \omega
_{0}\left( Z^{\prime },\left\vert \Psi \right\vert ^{2}\right) \right) }{%
\omega _{0}\left( Z\right) +\Delta \omega _{0}\left( Z,\left\vert \Psi
\right\vert ^{2}\right) } \\
s &=&-\frac{\lambda \left\vert \Psi _{0}\left( Z\right) \right\vert ^{2}}{%
\tau \omega _{0}\left( Z\right) +\Delta \omega _{0}\left( Z,\left\vert \Psi
\right\vert ^{2}\right) }
\end{eqnarray*}%
The transition functions are obtained by defining the matricial quantities $%
M\left( t\right) \mathbf{x}$ and $\sigma \left( t\right) $:%
\begin{equation*}
M\left( t\right) \mathbf{x}=\left( 
\begin{array}{cc}
e^{-tu} & s\frac{e^{-tu}-e^{-tv}}{u-v} \\ 
0 & e^{-tv}%
\end{array}%
\right) \left( \mathbf{T-}\left\langle \mathbf{T}\right\rangle \right)
\end{equation*}%
and:%
\begin{eqnarray*}
\sigma \left( t\right) &=&2\int_{0}^{t}\left( 
\begin{array}{cc}
e^{-2tu}+s^{2}\frac{\left( e^{-tu}-e^{-tv}\right) ^{2}}{\left( u-v\right)
^{2}} & se^{-tv}\frac{e^{-tu}-e^{-tv}}{u-v} \\ 
se^{-tv}\frac{e^{-tu}-e^{-tv}}{u-v} & e^{-2tv}%
\end{array}%
\right) dt \\
&=&-\left( 
\begin{array}{cc}
\frac{e^{-2tu}}{u}+s^{2}\frac{\frac{e^{-2tu}}{u}-4\frac{e^{-t\left(
u+v\right) }}{u+v}+\frac{e^{-2tv}}{v}}{\left( u-v\right) ^{2}} & s\frac{2%
\frac{e^{-t\left( u+v\right) }}{u+v}-\frac{e^{-2tv}}{v}}{u-v} \\ 
s\frac{2\frac{e^{-t\left( u+v\right) }}{u+v}-\frac{e^{-2tv}}{v}}{u-v} & 
\frac{e^{-2tv}}{v}%
\end{array}%
\right) \\
&=&\left( 
\begin{array}{cc}
\frac{1-e^{-2tu}}{u}+s^{2}\frac{\frac{\left( u-v\right) ^{2}}{uv\left(
u+v\right) }-\left( \frac{e^{-2tu}}{u}-4\frac{e^{-t\left( u+v\right) }}{u+v}+%
\frac{e^{-2tv}}{v}\right) }{\left( u-v\right) ^{2}} & s\frac{\frac{v-u}{%
v\left( u+v\right) }-\left( 2\frac{e^{-t\left( u+v\right) }}{u+v}-\frac{%
e^{-2tv}}{v}\right) }{u-v} \\ 
s\frac{\frac{v-u}{v\left( u+v\right) }-\left( 2\frac{e^{-t\left( u+v\right) }%
}{u+v}-\frac{e^{-2tv}}{v}\right) }{u-v} & \frac{1-e^{-2tv}}{v}%
\end{array}%
\right)
\end{eqnarray*}

The transition between $\mathbf{T-}\left\langle \mathbf{T}\right\rangle $
and $\mathbf{T}^{\prime }\mathbf{-}\left\langle \mathbf{T}\right\rangle $
during a time $t$ is written $G_{0}\left( \mathbf{T-}\left\langle \mathbf{T}%
\right\rangle ,\mathbf{T}^{\prime }\mathbf{-}\left\langle \mathbf{T}%
\right\rangle ,t\right) $ is obtained as the solution of (\ref{trp}) and is
given directly by:%
\begin{eqnarray}
&&G_{0}\left( \mathbf{T-}\left\langle \mathbf{T}\right\rangle ,\mathbf{T}%
^{\prime }\mathbf{-}\left\langle \mathbf{T}\right\rangle ,t\right) \\
&=&\left( 2\pi \right) ^{-1}\left( Det\left( \sigma \left( t\right) \right)
\right) ^{-\frac{1}{2}}  \notag \\
&&\times \exp \left( -\left( \left( \mathbf{T-}\left\langle \mathbf{T}%
\right\rangle \right) -M\left( t\right) \left( \mathbf{T}^{\prime }\mathbf{-}%
\left\langle \mathbf{T}\right\rangle \right) \right) ^{t}\frac{\sigma
^{-1}\left( t\right) }{2}\left( \left( \mathbf{T-}\left\langle \mathbf{T}%
\right\rangle \right) -M\left( t\right) \left( \mathbf{T}^{\prime }\mathbf{-}%
\left\langle \mathbf{T}\right\rangle \right) \right) \right)  \notag
\end{eqnarray}

Starting from the intial background ste $\mathbf{T}^{\prime }=\left\langle 
\mathbf{T}\right\rangle _{0}$, as $t$ increases, the difference between
initial background state and the new one is progressively reduced, as the
factor:%
\begin{equation*}
M\left( t\right) \left( \mathbf{T}^{\prime }\mathbf{-}\left\langle \mathbf{T}%
\right\rangle \right) =\left( 
\begin{array}{cc}
e^{-tu} & s\frac{e^{-tu}-e^{-tv}}{u-v} \\ 
0 & e^{-tv}%
\end{array}%
\right) \left( \mathbf{T}^{\prime }\mathbf{-}\left\langle \mathbf{T}%
\right\rangle \right)
\end{equation*}%
goes to $0$.

Ultimately, remark that for large $t$, the transition function simplifies
and writes:%
\begin{eqnarray}
G_{0}\left( \mathbf{T-}\left\langle \mathbf{T}\right\rangle ,\mathbf{T}%
^{\prime }\mathbf{-}\left\langle \mathbf{T}\right\rangle \right) &=&\left(
2\pi \right) ^{-1}\left( Det\left( \sigma \left( \infty \right) \right)
\right) ^{-\frac{1}{2}}  \label{tsr} \\
&&\times \exp \left( -\frac{1}{2}\left( \left( \mathbf{T-}\left\langle 
\mathbf{T}\right\rangle \right) \right) ^{t}\sigma ^{-1}\left( \infty
\right) \left( \left( \mathbf{T-}\left\langle \mathbf{T}\right\rangle
\right) \right) \right)  \notag
\end{eqnarray}%
with:%
\begin{equation*}
\sigma \left( \infty \right) =\left( 
\begin{array}{cc}
\frac{1}{u}+\frac{s^{2}}{uv\left( u+v\right) } & -\frac{s}{v\left(
u+v\right) } \\ 
-\frac{s}{v\left( u+v\right) } & \frac{1-e^{-2tv}}{v}%
\end{array}%
\right)
\end{equation*}

\end{document}